%% file: main.tex
\documentclass{packages}

\begin{document}

\renewcommand{\thetable}{\arabic{table}}

\pagenumbering{gobble}
\input{titlepage}

\newpage
\thispagestyle{empty}
\input{second_page}
\newpage
\thispagestyle{empty}
\vspace*{\fill}

\newpage
\pagenumbering{roman}
\chapter*{Acknowledgments}
\addcontentsline{toc}{chapter}{Acknowledgments}

\input{thanks}

\chapter*{Abstract}
\input{abstract}
\addcontentsline{toc}{chapter}{Abstract}
\chapter*{Zusammenfassung}
\input{zusammenfassung}

\addcontentsline{toc}{chapter}{Zusammenfassung}
\newpage
\tableofcontents
\newpage
\thispagestyle{empty}
\listoffigures
\addcontentsline{toc}{chapter}{List of Figures}
\listoftables
\addcontentsline{toc}{chapter}{List of Tables}
\newpage
\thispagestyle{plain}
\printglossary[title={List of Abbreviations}, type=\acronymtype, nonumberlist, nogroupskip, style=list ]
\printglossary
\addcontentsline{toc}{chapter}{Acronyms}


\newpage 
\thispagestyle{empty}
\vspace*{\fill}
\newpage
\pagenumbering{arabic}

\chapter{List of Publications} \label{chapter1}
\input{1_publications}

\chapter{Introduction}
\input{2_introduction}


\chapter{Major Contributions} \label{contributions}
\input{3_contributions}

\chapter{Discussion}
\input{4_discussion}


\appendix
\chapter[Automated Mind Wandering Detection]{Automated Mind Wandering Detection: Investigating Meta-Awareness, Multimodality, and Generalizability}
\renewcommand{\thechapter}{\Alph{chapter}}
\renewcommand{\thesection}{\Alph{chapter}.\arabic{section}}
\renewcommand{\thesubsection}{\Alph{chapter}.\arabic{section}.\arabic{subsection}}
\renewcommand{\thefigure}{\Alph{chapter}.\arabic{figure}}
\renewcommand{\thetable}{\Alph{chapter}.\arabic{table}}
\counterwithin{figure}{chapter}
\counterwithin{table}{chapter}
The following manuscripts are enclosed in this chapter: 

\begin{itemize}
    \item[{[1]}] \printpublication{buhler2024edpsych}.
    \item[{[5]}] \printpublication{buhler2024mm}.
    \item[{[2]}] \printpublication{buhler2024general}.
\end{itemize}

\blfootnote{The publication templates have been slightly adapted to match the formatting of this dissertation. The ultimate versions are accessible via the digital object identifier at the respective publisher. Publication  \cite{buhler2024edpsych} is\copyright American Psychological Association, 2024. This paper is not the copy of record and may not exactly replicate the authoritative document published in the APA journal. The final article is available, upon publication, at: 10.1037/edu0000903. Publication \cite{buhler2024general} is \copyright held by the owner/atuthor(s). This work is licensed under a  Creative Commons CC BY license.}

\input{1_Meta-Awareness_MW}
\input{2_Multimidal_MW}
\input{3_LabToWild}

\chapter{Synchrony as Attention Indicator during Online Learning}
The following publication is enclosed in this chapter:

\begin{itemize}
\item[{[3]}] \printpublication{buhler2024synchr}.
\end{itemize}
\blfootnote{The publication templates have been slightly adapted to match the formatting of this dissertation. The ultimate versions are accessible via the digital object identifier at the respective publisher. Publication \cite{buhler2024synchr} is \copyright 2024 Copyright held by the owner/author(s). This work is licensed under a Creative Commons Attribution-NonCommercial International 4.0 License.}

\input{4_OnTaskInSync}

\chapter{Behavioral Indicator of Engagement in the Classroom}
The following publication is enclosed in this chapter:

\begin{itemize}
\item[{[4]}] \textbf{Babette Bühler}\footnote[1]{Equal Contribution}, Ruikun Hou*, Efe Bozkir, Patricia Goldberg, Peter Gerjets, Ulrich Trautwein, and Enkelejda Kasneci. “Automated hand-raising detection in classroom
videos: A view-invariant and occlusion-robust machine learning approach”. In:
\textit{International Conference on Artificial Intelligence in Education.} Springer. 2023, \newline
pp. 102–113. DOI: \href{https://doi.org/10.1007/978-3-031-36272-9\_9         }{https://doi.org/10.1007/978-3-031-36272-9         \_9}. 
\end{itemize}

\blfootnote{The publication templates have been slightly adapted to match the formatting of this dissertation. The ultimate versions are accessible via the digital object identifier at the respective publisher. Publication \cite{buhler2023hr} is \copyright The Author(s), under exclusive license to Springer Nature Switzerland AG 2023. }

\input{5_Hand-raising}

\newpage
\printbibliography
\end{document}

%% file: titlepage.tex
\begin{titlepage}
   \begin{center}
       \vspace*{1cm}

       {\LARGE\textbf{Multimodal Machine Learning for Automated Assessment of Attention-Related Processes during Learning}}

       \vspace{2cm}

        \textbf{Dissertation}\\
        der Mathematisch-Naturwissenschaftlichen Fakultät\\
        der Eberhard Karls Universität Tübingen\\
        zur Erlangung des Grades eines\\
        Doktors der Naturwissenschaften\\
        (Dr. rer. nat.)

       \vspace{4cm}

       vorgelegt von

       M.Sc. Babette Bühler
       
       aus Fürth

       \vfill

       Tübingen\\
       2024
            
   \end{center}
\end{titlepage}

%% file: second_page.tex
\vspace*{\fill}

{\raggedright
\noindent Gedruckt mit Genehmigung der Mathematisch-Naturwissenschaftlichen Fakultät der
Eberhard Karls Universität Tübingen.\\ \

\noindent Tag der mündlichen Qualifikation: \tab 26.06.2024

\noindent Dekan: \tab Prof. Dr. Thilo Stehle

\noindent 1. Berichterstatterin: \tab Prof. Dr. Enkelejda Kasneci

\noindent 2. Berichterstatter: \tab Prof. Dr. Benjamin Nagengast}

%% file: thanks.tex
I express my sincere appreciation to my supervisors Prof. Dr. Enkelejda Kasneci and Prof. Dr. Ulrich Trautwein, for their invaluable guidance during this interdisciplinary dissertation project. I am very grateful to Prof. Dr. Enkelejda Kasneci for her exceptional academic support, personal encouragement, and methodological guidance. Her mentorship was crucial to my academic and personal development throughout this dissertation. Furthermore, I wish to thank Prof. Dr. Ulrich Trautwein for his extraordinary support, for his pivotal role in deeply integrating my project within the field of education sciences, and for providing an environment and opportunities that have allowed me to learn and grow. Additionally, I would like to thank Prof. Dr. Peter Gerjets for his close collaboration and constant stream of inspiring ideas. My sincere thanks also go to Prof. Dr. Sidney D’Mello for the warm welcome at his lab in Boulder, where I greatly benefited from his expertise and our collaboration, which have shaped my understanding of mind wandering. Sincere thanks to Prof. Dr. Benjamin Nagengast, Jun.-Prof. Dr. Michael Krone and Dr. Shahram Eivazi for evaluating my work. I thank the LEAD Graduate School and all the people involved for their support during this dissertation. 

I sincerely thank Dr. Efe Bozkir for his methodological advice, constant support, and relentless optimism. Moreover, I am truly grateful to Dr. Patricia Goldberg for helping me navigate this doctoral journey, even when it was no longer part of her job description. I thank Dr. Ömer Sümer for sharing valuable lessons learned with me and Ruikun Hou for our fantastic teamwork. I am grateful to all my colleagues at the Hector Research Institute, especially Dr. Wolfgang Wagner, Dr. Tobias Appel, Dr. Alexander Soemer, and Dr. Tim Fütterer, for their assistance and advice. I am particularly grateful to Prof. Dr. Rosa Lavelle-Hill for her encouragement and our friendship. I am also thankful to all my colleagues at the Department of Human-Computer Interaction, with special thanks to Dr. Nora Castner and Dr. Hong Gao.

Moreover, I want to express my heartfelt thanks to Hannah Deininger for being my office mate, co-author, and good friend. I am very happy that our paths crossed again in Tübingen, allowing us to share this journey together. I am incredibly grateful to all my fellow doctoral candidates Aki Schumacher, Fitore Morina, Tosca Daltoé, Wy Ming Lin, Philipp Stark, Alexander Jung, and all the LEADies, who were first and foremost friends before colleagues. The shared moments of joy, lunch breaks on Neckar Island, conference trips, game nights during retreats, and countless pep talks have been invaluable. Without these, navigating this period would have been unimaginably tougher. I thank Annelen Fritz for making Tübingen feel like home. I feel beyond fortunate to have an amazing family and friends who have supported me from near and far. I thank Florian Nachtigall for always encouraging me and giving me the feeling that I can do anything. My deepest gratitude goes to my mother, Christiane Heller-Bühler, for her unconditional and unwavering support. Thank you!

%% file: abstract.tex
Attention is a key factor for successful learning, with research indicating strong associations between (in)attention and learning outcomes. This dissertation advanced the field by focusing on the automated detection of attention-related processes using eye tracking, computer vision, and machine learning, offering a more objective, continuous, and scalable assessment than traditional methods such as self-reports or observations. It introduced novel computational approaches for assessing various dimensions of (in)attention in online and classroom learning settings and addressing the challenges of precise fine-granular assessment, generalizability, and in-the-wild data quality. 

First, this dissertation explored the automated detection of mind-wandering, a shift in attention away from the learning task. Temporal patterns in aware and unaware mind wandering and their associations with learning outcomes were investigated. The two types of mind wandering, previously conflated in detection research, were distinguished using predictive modeling based on gaze data. Based on this, the precision and robustness of aware and unaware mind-wandering detection were enhanced by employing a novel multimodal approach that integrated eye tracking, video, and physiological data and outperformed unimodal approaches. Further, the generalizability of scalable webcam-based mind-wandering detection across diverse tasks, settings, and target groups was examined using a fine-tuned transfer learning approach to address low-quality data in real-world settings. 
Second, this thesis investigated attention indicators during online learning, inferring information from the group level. Eye-tracking analyses revealed significantly greater gaze synchronization among attentive learners. 
Third, it addressed attention-related processes in classroom learning by detecting hand-raising as an indicator of behavioral engagement using a novel view-invariant and occlusion-robust skeleton-based approach. It further explored the correlation between automatically annotated hand-raisings and self-reported learner engagement, interest, and involvement, demonstrating the potential of automated assessments for large-scale video analysis.

This thesis advanced the automated assessment of attention-related processes within educational settings by developing and refining methods for detecting mind wandering, on-task behavior, and behavioral engagement. It bridges educational theory with advanced methods from computer science, enhancing our understanding of attention-related processes that significantly impact learning outcomes and educational practices.


%% file: zusammenfassung.tex
Aufmerksamkeit ist ein zentraler Faktor für erfolgreiches Lernen, denn die Forschung weist auf starke Zusammenhänge zwischen (Un-)Aufmerksamkeit und Lernerfolgen hin. Diese Dissertation entwickelte dieses Forschungsfeld weiter, indem sie sich auf die automatisierte Erkennung von Aufmerksamkeitsprozessen mit Hilfe von Eye Tracking, Computer Vision und maschinellem Lernen konzentrierte und damit, im Gegensatz zu traditionellen Methoden wie Selbstberichten oder Beobachtungen, eine objektive, kontinuierliche und skalierbare Messung  ermöglicht. Es wurden neuartige computergestützte Ansätze zur Erfassung verschiedener Dimensionen der (Un-)Aufmerksamkeit während des Online-Lernens und im Klassenzimmer entwickelt und die Herausforderungen einer präzisen, feingranularen Erfassung, der Generalisierbarkeit und der Datenqualität in naturalistischen Umgebungen adressiert.

Zunächst erforschte diese Dissertation die automatische Erkennung von geistigem Abschweifen, einer Verlagerung der Aufmerksamkeit von der Lernaufgabe weg. Untersucht wurden zeitliche Muster des bewussten und unbewussten Abschweifens und deren Zusammenhang mit Lernergebnissen. Die beiden Arten des geistigen Abschweifens, die bisher in der Erkennungsforschung miteinander vermischt wurden, wurden durch prädiktive Modellierung auf der Grundlage von Blickdaten unterschieden. Darüber hinaus wurden die Präzision und Robustheit der Erkennung von bewusstem und unbewusstem Abschweifen verbessert, indem ein neuartiger multimodaler Ansatz verwendet wurde, der Blickverfolgungs-, Video- und physiologische Daten integriert und unimodale Ansätze übertraf. Darüber hinaus wurde die Generalisierbarkeit der skalierbaren Webcam-basierten Erkennung von geistigem Abschweifen über verschiedene Aufgaben, Umgebungen und Zielgruppen hinweg mit Hilfe eines fein abgestimmten Transfer-Learning-Ansatzes untersucht, um Daten geringer Qualität in realen Umgebungen zu berücksichtigen. 
Anschließend wurden in dieser Arbeit Aufmerksamkeitsindikatoren beim Online-Lernen untersucht, wobei Informationen von der Gruppenebene abgeleitet wurden. Eye Tracking Analysen offenbarten eine signifikant höhere Blicksynchronität bei aufmerksamen Lernenden. 
Zuletzt wurden aufmerksamkeitsbezogene Prozesse beim Lernen im Klassenzimmer untersucht, indem das Handheben als Indikator für das Verhaltensengagement mit Hilfe eines neuartigen ansichtsinvarianten und okklusionsrobusten skelettbasierten Ansatzes erkannt wurde. Darüber hinaus wurde die Korrelation zwischen automatisch annotierten Meldungen und dem selbstberichteten Engagement, dem Interesse und der Beteiligung der Lernenden ermittelt, um das Potenzial automatisierter Bewertungen für eine groß angelegte Videoanalysen zu demonstrieren.

Diese Arbeit verbesserte die automatisierte Bewertung von Aufmerksamkeitsprozessen in Lernumgebungen durch die Entwicklung und Verfeinerung von Methoden zur Erkennung von geistigem Abschweifen, aufgabenorientiertem Verhalten und Verhaltensengagement. 
Sie schlägt eine Brücke zwischen Bildungstheorie und fortschrittlichen Methoden aus der Informatik, um unser Verständnis von aufmerksamkeitsbezogenen Prozessen zu verbessern, die einen erheblichen Einfluss auf Lernergebnisse und Bildungspraxis haben.

%% file: 1_publications.tex
\section{Publications and Accepted Manuscripts Relevant to this Thesis}

\begin{enumerate}[label={[\arabic*]}]
    \item \label{edpsych} \printpublication{buhler2024edpsych}

    \item \printpublication{buhler2024general}
    
    \item \label{ETRA24} \printpublication{buhler2024synchr}    
    \item \label{AIED23} \printpublication{buhler2023hr}

\end{enumerate}

\section{Submitted Manuscripts Relevant to this Thesis}
\begin{enumerate}[label={[\arabic*]}, resume] 
    
    \item \printpublication{buhler2024mm}

\end{enumerate}

\section{Further Publications Not Relevant to this Thesis}
\begin{enumerate}[label={[\arabic*]}, resume]

    \item \printpublication{hou2024automated}
    \item \printpublication{fuhl2023watch}
\end{enumerate}

\section{Scientific Contribution}

This work explores the automatic assessment of different manifestations of (in-)attention in online and classroom learning, leveraging interdisciplinary synergies from research domains such as human-computer interaction, machine learning, research on education, and psychology. It proposes innovative approaches to assess attention while grounding in education and learning theories. This dissertation comprises several important scientific contributions, which are listed below.
\begin{itemize}
    \item Novel exploration of temporal patterns of sub-dimensions of mind-wandering meta-awareness during lecture viewing and their impact on learning.
    \item Initial fine-granular automatic differentiation of aware and unaware mind wandering based on eye gaze.
    \item Multimodal machine learning approach to aware and unaware mind wandering detection, employing a novel combination of modalities: Eye tracking, video, and physiology (electrodermal activity and heart rate).
    \item Examination of generalizability of video-based mind wandering detection to in-the-wild settings, new tasks, and diverse target groups, proposing a new deep learning approach based on facial expression recognition transfer learning.
    \item Investigation of gaze synchrony as a reliable indicator for self-reported attention in online learning and its association to learning outcomes.
    \item Novel approach to assessing behavioral engagement in authentic classroom videos via view-invariant, occlusion-robust skeleton-based hand-raising detection.
\end{itemize}
In doing so, this thesis contributes to a more fine-grained and objective assessment of attention during learning. By bridging educational theory with advanced methods from computer science, it enhances our understanding of attention-related processes, which have important consequences for learning outcomes and educational practices.

%% file: 2_introduction.tex
\section{Attention as a Prerequisite for Learning} \label{learning}

Attention is a core component of the learning process, as it is crucial for learning and knowledge construction \parencite[]{levine1990}. Yet, learners frequently find it challenging to maintain attention over extended periods \cite{Risko.2012}. Therefore, understanding and promoting attention in various learning environments is a fundamental building block for improving learning processes and outcomes. Strategies to support this goal range from enhancing teacher training and learning materials \cite{olney2015attention} to implementing adaptive support in attention-aware learning technologies \cite{dmello2016giving}. 

\textbf{Attention} can be seen as a \textit{multidimensional construct} that varies in intensity \cite{chi2014icap, olney2015attention} and is determined by its task-relatedness. The term attention encompasses a broad spectrum of definitions depending on the field. In psychology, attention is often defined as a selection mechanism, enabling us to focus on certain stimuli while ignoring others and allowing for more efficient information processing given the limited capacity of our cognitive resources  \cite{cohen2014neuropsychology, mcdowd2007}. In neuroscience, attention is considered a complex array of interconnected mechanisms and processes that involve nearly all regions of the human brain \cite{posner2007research, corbetta2002control}. Research on education, specifically in the context of classroom instruction, has focused on overt student behavior, which serves as an indicator for attention-related cognitive processes. Attention is seen as a component of behavioral engagement in school learning \cite{fredricks2004school}.

Research demonstrates the relationship between these attention-related behaviors and learning outcomes. (In)attention can be investigated by using rather generic approaches to cover the construct, revealing a correlation between different manifestations of on- or off-task behavior and learning outcomes \cite{helmke1992}. Other works focus on specific aspects of behavioral engagement closely related to attention. For instance, inattention and disruptive behavior \cite{finn1995disruptive} were negatively associated with academic achievement. In contrast, active participation in classroom discussions was positively related to performance on a reading literacy test \cite{sedova2019}. A recent study showed that the frequency of students’ hand-raising is related to academic achievement and cognitive engagement \cite{boheim2020engagement}. However, studies in classroom contexts often have rather small sample sizes limited to specific grades, age groups, and school subjects \cite{Boheim.2020,boheim2020engagement}, resulting in a lack of generalizability of results.


Whereas such research has concentrated on studying students' visible attention-related behaviors, another ubiquitous phenomenon is that learners struggle to maintain attention without displaying explicit off-task behaviors. Such a state of decoupled attention hinders the learner from processing information from the external environment, inhibiting information integration into internal representations \cite{Smallwood.2007b}. This phenomenon is called \textbf{mind wandering} and is defined as \textit{the shift of attention away from the current task to task-unrelated thought} \cite{smallwood2006}. According to theories on cognitive load and executive functioning, ensuring adequate mental resource allocation is vital for regulating cognitive processes to maintain attention during learning tasks \cite{sweller2010, friedman2017, miyake2000, broadway2010, krumpe2018}. Along these lines, one theory hypothesizes that mind wandering occurs as thoughts compete for the limited resources available in working memory and depend on available executive functions \parencite[]{Smallwood.2006}. The executive control failure hypothesis, as proposed by \textcite{kane2012}, suggests that mind wandering indicates both temporary lapses and persistent deficiencies in executive control functions. This view is reinforced by findings that lower working-memory capacity is linked to more frequent off-task thoughts during challenging tasks \parencite[]{mcvay2012wmc}. 

Learners typically engage around 30\% of the time spent in educational activities, like watching lectures or reading, in mind wandering \cite{wong2022}. The occurrence of task-unrelated thought is consistently linked with lower test performance across various tasks, subjects, and age groups, accounting for approximately 7\% of the variability in learning outcomes \cite{wong2022}. This negative correlation is consistent for surface-level or inference-based learning across subjects, age groups, and various tasks. For instance, mind wandering has been shown to negatively impact reading comprehension \parencite{Smallwood.2011, Feng.2013, dmello2021, bonifacci2022, caruso2023} and lecture retention \parencite{Risko.2012, Szpunar.2013, Hollis.2016, Pan.2020}.

Mind wandering itself is a heterogeneous multidimensional concept rather than a dichotomous state \cite{Seli.2018family}, although often operationalized as such. Definitions of mind wandering employed in research differ in intentionality, task-relatedness, and relationships to external stimuli \parencite[]{Seli.2018family}. Therefore, its definition and conceptualization have been the subject of ongoing controversy in the literature. On the one hand, \citet{Christoff.2018} criticize the use of mind wandering as an umbrella term for disparate mental phenomena. Instead, they propose a dynamic framework in which the defining feature of mind wandering is that it arises and proceeds unconstrained and relatively free. On the other hand, \citet{Seli.2018family, Seli.2018wellclad}, advocate for a family resemblance view under which several manifestations of the phenomenon fall, including, for instance, perseverative and purposeful task-unrelated thoughts. Instead of restricting the concept more strongly, from their point of view, the regarded types of thoughts have to be precisely defined in the respective research. Research on the content of learners' thoughts during learning activities showed the prevalence of arising thoughts that are not focused on the here and now of the lecture but task-related thoughts or thoughts on lecture comprehension \cite{locke1974}. Those are positively related to lecture retention \cite{Kane.2017, Jing.2016}, aligning with research on self-regulated learning showing that the use of cognitive and metacognitive learning strategies, such as elaboration and self-monitoring of one's learning progress, has a positive effect on learning outcomes \parencite[]{dent2016, jansen2019}. Consequently, this work adopts the definition of mind wandering as task-unrelated thoughts \cite{smallwood2006} and applies measurement approaches that disentangle task-unrelated thoughts from other cognitive processes that represent integral parts of learning.

The commonly observed phenomenon of catching oneself mind wandering, utilized in self-caught mind-wandering probes, suggests that people engage in task-unrelated thoughts without being aware of the fact \cite{Smallwood.2007}. This indicates a temporary lack of meta-awareness—a reflection on one’s conscious thoughts and their alignment with one's goals \cite{Schooler.2002}. Awareness of thought digression is essential for detecting and terminating mind-wandering episodes to refocus on the main task. However, research shows that people frequently report the continuance of mind wandering even after gaining awareness of it \cite{seli2017}. \citet{Smallwood.2007} categorized mind wandering into two distinct states: 'tune-outs' with and 'zone-outs' without awareness, noting these can dynamically transition over time. It is critical to distinguish between the dimension of meta-awareness and the dimension of intentionality \parencite[]{seli2017}, the latter defining whether task-unrelated thoughts are deliberate or spontaneous \parencite[e.g.,][]{Seli.2013, Smallwood.2004, smilek2010}. According to \textcite{seli2017}, although a lack of meta-awareness and intentionality may initially coincide at the onset of a mind-wandering episode, they often diverge as the episode progresses. Unaware mind wandering appears to be more strongly linked to failures in response inhibition and deficits in reading comprehension \parencite[]{Smallwood.2007, smallwood2008b} than aware mind wandering, while the two states exhibit distinct neurological activities \parencite[]{Christoff.2009}, indicating varied implications for task performance and underlying cognitive processes. Consequently, the two types of mind wandering require different interventions and solutions. However, meta-awareness of mind wandering and its association with learning have not been thoroughly investigated in educational settings.

Having established the substantial association between different dimensions of student attention and learning success, this thesis employs a cross-disciplinary definition of attention as a multifaceted construct \cite{olney2015attention}. As described above, several research gaps concerning attention and learning persist, including the need for more precise differentiation and continuous measurement. A more fine granular investigation of types of mind wandering and their relation to learning outcomes is warranted. Further, large-scale research on behavioral attention indicators in classroom instruction is still scarce. More research is needed to understand the exact relationships between different dimensions of attention and learning to improve learning processes and outcomes. The following chapters explain to what extent these and other impediments caused by traditional attention measurement methods can be overcome by new automated measurement methods such as those developed in this dissertation. The following Section \ref{tradi} outlines traditional methods for measuring attention and the challenges they present. This is followed by a detailed discussion of automated assessment techniques for attention-related processes in Section \ref{autom} and an overview of the objectives of this thesis in Section \ref{objectives}.

\section{Traditional Attention Measurement Approaches} \label{tradi}

To quantitatively investigate attention and its impact on learning, it is necessary to make the attentional states of learners measurable, typically achieved through external observer ratings by trained experts or the collection of self-reports.

\subsection{Observations} \label{observer}

One way to measure attention is with the help of observation ratings from trained human experts. A significant advantage of this method is that it is neither intrusive nor interruptive and even allows behavioral development to be obtained over time. Elaborate rating manuals to rate student engagement continuously over time have been developed \cite{goldberg2021attentive}. The main drawback of observations is rater bias, which refers to discrepancies among raters that stem from varying interpretations of the rating scales or distinct, individual perceptions of the subject being assessed \cite{hoyt2000rater}. To increase the objectivity of ratings, raters are trained extensively, and information needs to be rated by two or more independent raters. This makes the processes resource-intensive, leading to considerable time and cost expenditures. However, even extensive training might not reduce rater bias in the desired way \cite{praetorius2012observer}. This issue is particularly pronounced when assessing higher-level constructs such as engagement, where even the presence of exemplary behavioral indicators may result in unsatisfactory inter-rater reliability \cite{sumer2021multimodal}. Another limitation is that human observers might have difficulty assessing specific aspects of attention because they do not have access to the student's internal state of mind. Raters must, therefore, rely entirely on students' observable behavioral cues and have enormous difficulty detecting internal cognitive states as mind wandering \cite{bosch2022}. 

\subsection{Self-reports}
To gather information on students' hidden cognitive processes, which are almost entirely of introspective nature, self-reports, i.e., questionnaires in which students indicate their own attention levels, can be employed. Self-reports are comparatively low-cost and easy to administer. However, self-reports depend on students’ compliance and awareness of their attentional state. Further, they are subject to response biases, such as social desirability or acquiescence bias \cite{weinstein2018}. When collected retrospectively, they may be biased by primacy and recency memory effects \cite{whitehill2014faces}. A solution to that problem is experience sampling, which means polling participants during the learning task at multiple points in time \cite{larson1983experience}. The major disadvantage of such an approach is that it is intrusive and disruptive, and the interruption of the activity itself can affect students' attention \cite{Szpunar.2013b}. In research on mind wandering, a frequently employed alternative is the self-caught method, for which learners are instructed to report whenever they have caught themselves drifting off \cite{giambra1989slef}. Both forms of mind-wandering self-reports are deemed reliable in educational settings \cite{VaraoSousa.2019}. However, one disadvantage of the self-caught method is that it requires learners to monitor their attention. Learners need to become aware of their internal state without external probes, which only allows learners to report their mind wandering when they have gained awareness of its occurrence \cite{weinstein2018,bixler2016automatic}. This makes the self-caught method unsuitable for studying different states of meta-awareness in mind wandering. Overall, self-reports suffer from the fact that they offer potentially inaccurate--in case of retrospective assessment--or only discrete rather than continuous information in case of experience sampling.


A potential solution to many of these drawbacks of traditional attention measurement that opens up new possibilities is its automated recognition using machine learning techniques, which will be discussed in the following section.

\section{Automated Assessment of Attention} \label{autom}

Recent developments in the areas of computer vision and machine learning offer a new potential for automated attention measurement. One great opportunity for automated detection of (in)attention is that it significantly reduces the cost of data collection compared to observer ratings. In turn, this enables the analysis of large-scale data, which is a particular advance in quantitative research. Compared to concurrent self-reporting, it is non-obtrusive since there is no need to interrupt the current task, therefore allowing continuous measurement over time. Besides these advantages, similar to the observer-based measurement approach, most automated measurements are also based on the externally visible behavior of students. However, related works show that machine learning algorithms and human observers rely on divergent cues when detecting inattention \cite{bosch2022}. In addition, data collection modules, such as eye tracking or physiological sensors, can be included, capturing information at a much finer granularity than human observers can perceive. Furthermore, automated detection approaches can be implemented in attention-aware learning technologies \cite{dmello2021, hutt2021, Mills.2020}, which can support learners' self-regulation to sustain or redirect their attention. This section gives an overview of different modalities employed in the automated assessment of attention, describes and discusses the use of predictive modeling, and highlights the importance of explainability and fairness of employed methods. 

\subsection{Modalities for Measuring Attention}
Automated attention detection approaches employ a range of modalities, such as video recordings, eye-tracking, or physiological sensors, to gather student data during learning tasks. Typically, multiple indicators are extracted from this data. This section reviews four key modalities: video, eye tracking, physiology, and log files, with an overview of the attention indicators extractable from each modality presented in Table \ref{tab:mods}. 

\begin{table}[h!]
\caption{Modalities Used for Assessing Attention and the Corresponding Attention Indicators Extracted.}
\centering
\label{tab:mods}

\begin{tabular}{|lll|}
\hline
Modality     & Indicators of Attention          &  Exemplary Works\\ \hline
Video        & Facial Action Units (AUs)        &   \cite{stewart2017, Bosch.2021, Hutt.2019} \\
            & Facial Expressions               &   \cite{Lee.2022, sumer2021multimodal, nguyen2022, ngo2024fer}\\
             & Gaze Direction                   & \cite{thomas2017predicting, goldberg2021,hutt2023webcam, zhao2017, Lee.2022}\\
             & Head Pose                        & \cite{thomas2017predicting, Bosch.2021, raca2015head, goldberg2021attentive,sumer2021multimodal, Ahuja.2019}\\
             & Head and Body Motion             & \cite{raca2014sleepers, stewart2017, Bosch.2021, Lee.2022} \\
             & Body Pose                        & \cite{Ahuja.2019, lin2021student, YuTe.2019}\\ \hline
Eye Tracking & Fixations                        & \cite{bixler2014, mills2021, Hutt.2017, Hutt.2019, mills2016}\\
             & Saccades                         & \cite{bixler2014, mills2021, Hutt.2017, Hutt.2019, mills2016}\\
             & Pupillometry                     & \cite{bixler2014,  mills2016, Brishtel.2020}\\
             & Blinks                          & \cite{bixler2014, mills2016, Brishtel.2020}\\ 
             & Area of Interest (AOI)           & \cite{Hutt.2017, mills2016}\\
             & Locality Features               & \cite{bixler2021crossed, Hutt.2019}\\
             & Vergence                      & \cite{huang2019, toates1974vergence}\\ \hline
Physiology   
            & Electrodermal Activity (EDA)         & \cite{Blanchard.2014, Brishtel.2020, bixler2015gazeeda}\\
             & Heart Rate                     &  \cite{Pham.2015, gao2020eda}\\
             & Temperature                    & \cite{Blanchard.2014, bixler2015gazeeda} \\
             & Electroencephalography (EEG)      & \cite{poulsen2017eeg, dikker2017brain, bevilacqua2019brain, conrad2021}                    \\ \hline
Log Files   & Keystrokes                   & \cite{kuvar2023keystroke, kuvar2023partner} \\
            & Mouse Tracking               & \cite{dasilva2020mouse}\\
            & Time Spent       &        \cite{mills2015log}\\
            \hline
\end{tabular}

\end{table}

\subsubsection{Video Recordings for Attention Detection}

Video recordings are a crucial data source for the automated assessment of attention-related processes in educational environments. In the context of teacher training and teaching quality research, education research uses classroom videos \parencite[e.g.,][]{ainley2018teaching}{} that are later, for instance, rated by observers to assess teaching quality \parencite[e.g.,][]{daltoe2024}{}. Similar videos can be employed for attention research, often covering the entire classroom from various angles, capturing multiple students and teachers simultaneously \cite{thomas2017predicting,sumer2021multimodal, goldberg2021, sumer2018teachers}. In computer-mediated settings, webcams offer an accessible opportunity to capture facial videos of learners during educational activities \cite{whitehill2014faces, Bosch.2021}. The process of automatically analyzing and understanding those videos is called computer vision \cite{szeliski2022cv}. The field of computer vision has seen significant advances with the rise of deep learning, for instance, \glspl{cnn} and Autoencoders, and the increased availability of publicly available, large-scale, labeled datasets \cite{voulodimos2018deep}. Core tasks in computer vision are image classification, object recognition, semantic segmentation, action recognition, and human pose estimation \cite{voulodimos2018deep}. 

\begin{figure}
  \centering
  \includegraphics[height=125.5pt]{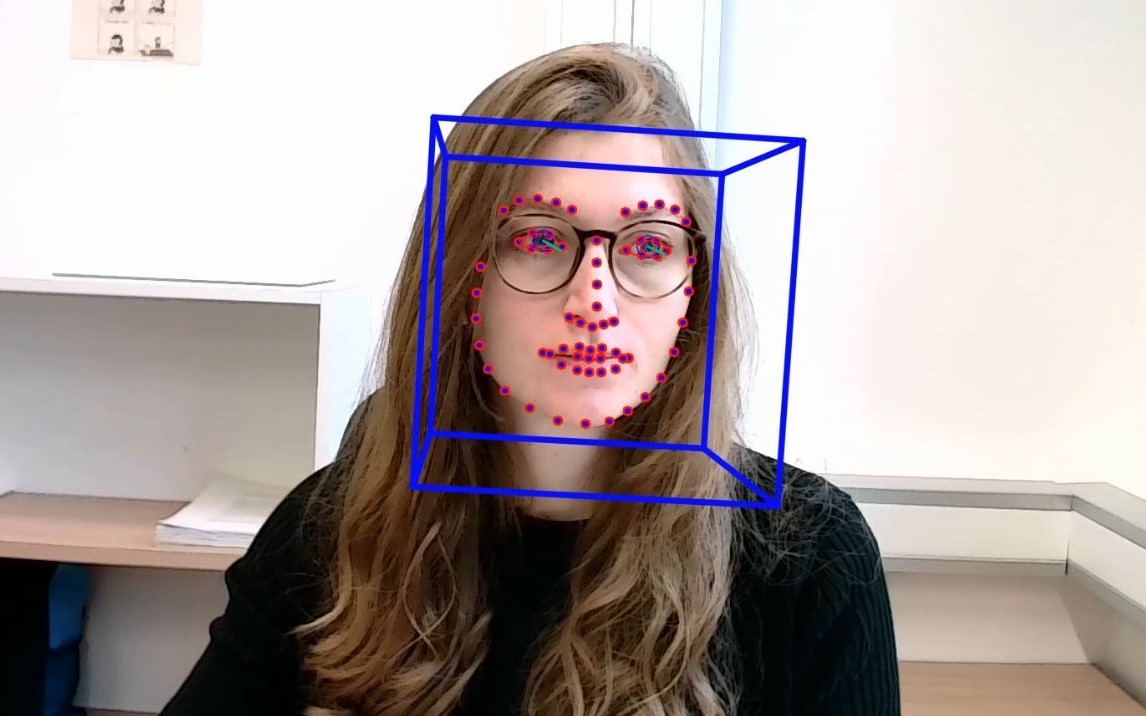}
  \includegraphics[height=125.5pt]{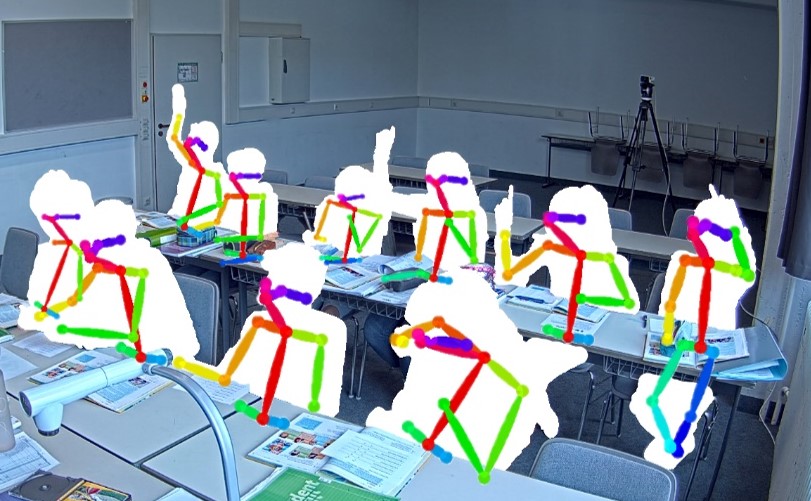}
  \caption{ Webcam Image With Depiction of Openface Head Pose, Gaze and Facial Landmark Features and Classroom Video Frame With Openpose Pose Estimations.}
  \label{fig:modv}
\end{figure}

For the video-based assessment of attention, computer vision techniques are employed to extract behavioral cues, such as facial expressions, that indicate emotional states. Emotions are significantly related to a person's cognitive functions and, consequently, attentional processes \cite{vuilleumier2005emotionalattention}. Studies have demonstrated that \glspl{au}, which encode facial muscle movements \cite{ekman1978facial}, are effective predictors of mind wandering \cite{stewart2017, Bosch.2021, Hutt.2019} and engagement levels \cite{thomas2017predicting, whitehill2014faces}. Additionally, advanced \gls{fer} techniques applied to webcam and classroom videos have proven effective for detecting mind wandering \cite{Lee.2022}, estimating engagement \cite{sumer2021multimodal}, and assessing emotion regulation in remote collaborative settings \cite{nguyen2022, ngo2024fer}. Video analysis also facilitates the extraction of gaze information through appearance-based gaze estimation \cite{thomas2017predicting, goldberg2021}, and webcam-based eye tracking \cite{hutt2023webcam, zhao2017, Lee.2022}. Further, head pose, approximating gaze direction and described by pitch, yaw, roll, and rotation, is a relevant attention marker \cite{thomas2017predicting, Bosch.2021, raca2015head, goldberg2021attentive,sumer2021multimodal, Ahuja.2019}. Researchers have also utilized head and body motion to detect attention-related processes \cite{raca2014sleepers, stewart2017, Bosch.2021, Lee.2022}. In addition to video recordings, Kinect motion sensors can be employed in classrooms to extract such indicators \cite{zaletelj2017kinect}. Figure \ref{fig:modv} features a webcam image on the left showing estimated head pose (blue), gaze vectors (green), and facial landmarks (red) analyzed with the OpenFace toolbox \cite{openface2016}. On the right, body pose estimation algorithms like OpenPose in classroom videos estimate skeletal key points \cite{cao2019openpose}, allowing inference of attention-related behaviors such as hand-raising \cite{Ahuja.2019, lin2021student, YuTe.2019}. In summary, it has been shown from this review that video recordings are a scalable, accessible, and rich data source for the extraction of various attention-related cues. However, processing image or video data involves high computational demands due to the high dimensionality of data and complexity of the operations involved, especially compared to other modalities employed to detect attention, such as eye tracking, which will be reviewed in the following section.

\subsubsection{Eye Tracking for Attention Detection}

\begin{figure}
  \includegraphics[height=125.5pt]{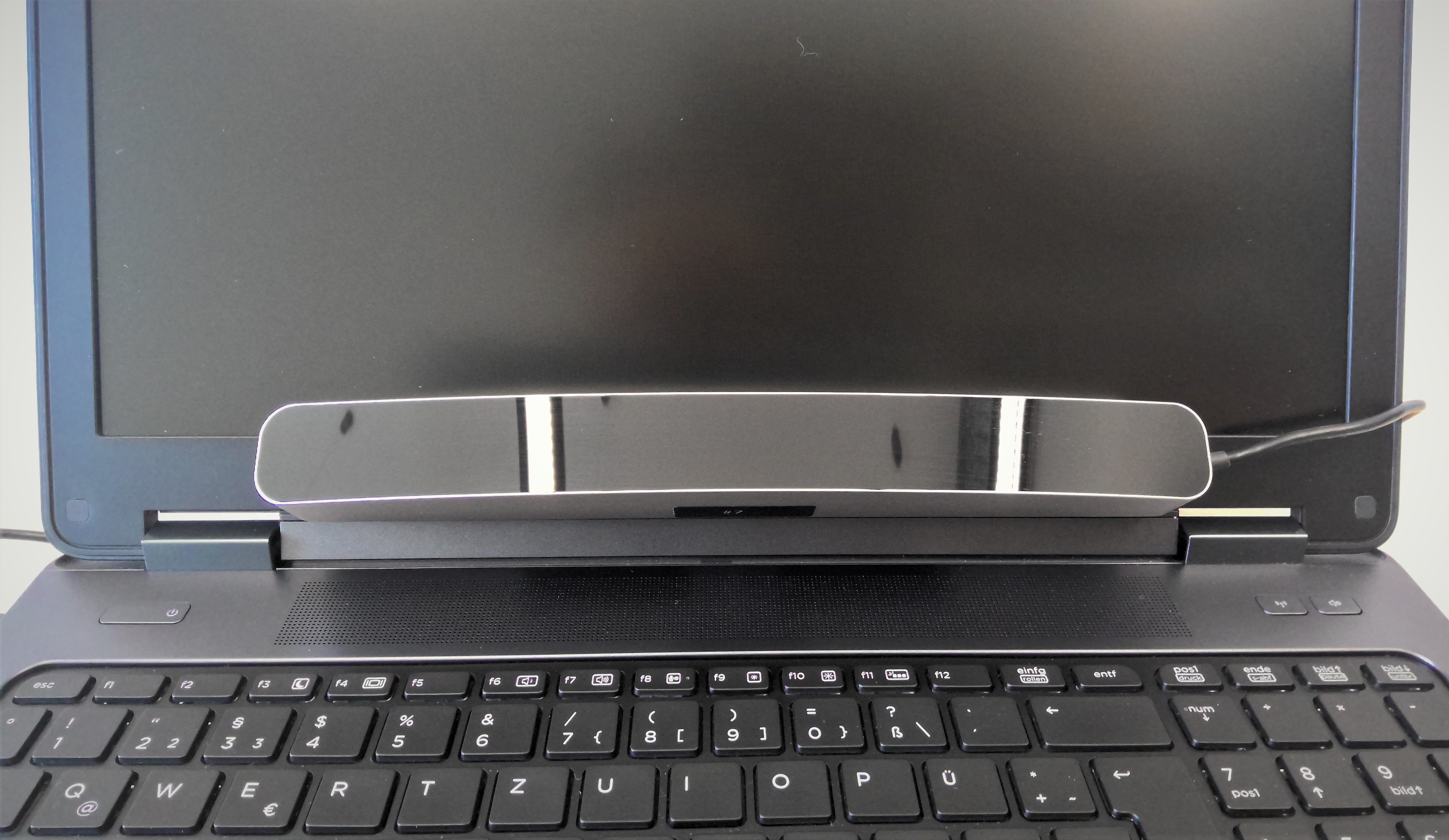}
  \includegraphics[height=125.5pt]{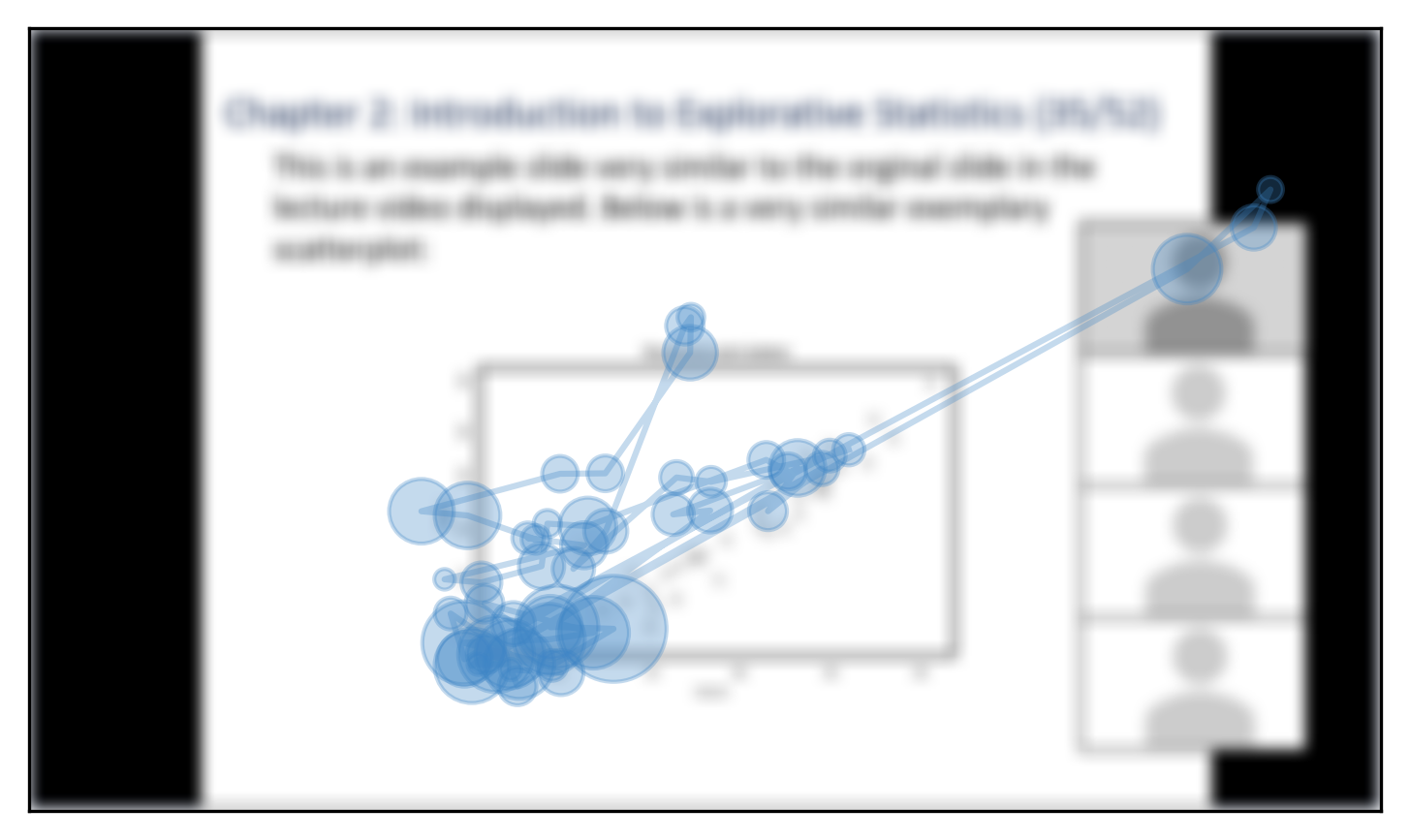}
  \caption{Remote SMI Eye Tracker and Exemplary Scanpath Visualization during Remote Learning.}
  \label{fig:et}
\end{figure}

Eye gaze is a critical feature in understanding visual attention \cite{duchowski2002}, indicating which information is being cognitively processed \cite{rayner1998eye}. According to the mind-eye link, cognitive processes are reflected in eye movements \cite{just1976eye, rayner1998eye, reichle2012using}. It can be most precisely measured with eye trackers, which use various methods to estimate gaze. The most common include video-oculography, which uses video-based tracking with head-mounted or remote visible light cameras, video-based infrared pupil-corneal reflection, and Electrooculography \cite{majaranta2014eye}. These enable the detection of gaze metrics such as fixations (periods of stable gaze) and saccades (rapid eye movements), which in combination result in the visual scanpath \cite{kubler2014subsmatch, geisler2020minhash}. Other measures frequently used include spatial information, for instance, the location of fixations on a particular \gls{aoi}. Figure \ref{fig:et} shows a remote SMI eye tracker on the left and illustrates a scanpath gaze visualization during remote learning on the right.

Eye tracking is often used to assess mental processes \cite{leigh2015neurology} and visual attention in various scenarios, mainly critical use cases, such as driving \parencite[e.g.,][]{braunagel2017ready, bozkir2019assessment, kasneci2015online, baldwin2017detecting} or in medical education \parencite[e.g.,][]{krupinski2006eye, castner2018scanpath, gegenfurtner2013transfer}{}{}. In recent years, eye tracking has been increasingly employed in classroom settings, including virtual reality classrooms, to understand learning processes and interaction during learning and teaching \parencite[e.g.,][]{byrne2023leveraging, gao2021digital, bozkir2021exploiting, hasenbein2023investigating, gao2022evaluating, coskun2021investigation, rosengrant2021, sumer2018teachers, yang2013tracking}. Mobile eye trackers have been employed in real-world classrooms to study student \cite{rosengrant2021, yang2013tracking} and teacher attention \cite{sumer2018teachers, coskun2021investigation}, but their use for automated assessment remains unexplored in this setting. For online learning environments, remote eye trackers attached to computer screens are commonly used. Eye tracking studies on lecture video watching indicate that visualizations of gaze data can help instructors assess learners' attention levels \cite{sauter2023} and understanding of the material \cite{Kok.2023}. In this line, studies suggest that attentive students display similar eye movement patterns when following instructional videos \cite{Madsen.2021}. Beyond the overt attention direction, eye tracking is highly relevant for detecting certain forms of (in-)attention related to hidden cognitive states, such as mind wandering \parencite[e.g.,][]{bixler2015gazeeda, mills2021, hutt2016eyes}{}{}. Research has utilized global gaze features like fixations and saccades, as well as local features including spatial gaze properties, pupil size, and blink dynamics, to predict attentional states \cite{bixler2014, mills2021, Hutt.2017, Hutt.2019}. Studies indicate that the average fixation duration on slides increases during mind wandering, suggesting decreased visual processing efficiency \cite{Zhang.2020} and increased mental workload \cite{degreef2009}. This behavior supports the decoupling hypothesis, which posits that mind wandering impairs the processing of external information \cite{Schooler.2011, smallwood2008c}, and manifests in varied fixation dispersion \cite{Faber.2020}. Mind wandering episodes also feature reduced blink rates \cite{Jang.2020} and increased pupil diameters \cite{franklin2013}, indicative of higher cognitive loads \cite{kahneman1966} and emotional demands \cite{partala2003}, although environmental factors may also influence these metrics \cite{John.2018}. Eye vergence, particularly tonic vergence, which indicates a visually relaxed state where the eyes do not fixate, has been used to enhance predictions of internal thoughts in computational models \cite{huang2019, toates1974vergence}. Despite these insights, no definitive set of gaze behaviors is consistently linked to mind wandering or attention, leading to varied results across studies. Variability in these results may be attributed to task-specific demands that affect spatial allocation and visual processing, suggesting a compensatory adjustment in the visual system during mind-wandering episodes \cite{Faber.2020}. 

\subsubsection{Physiological Data for Attention Detection}
\begin{figure}
   \centering
    \includegraphics[height=125.5pt]{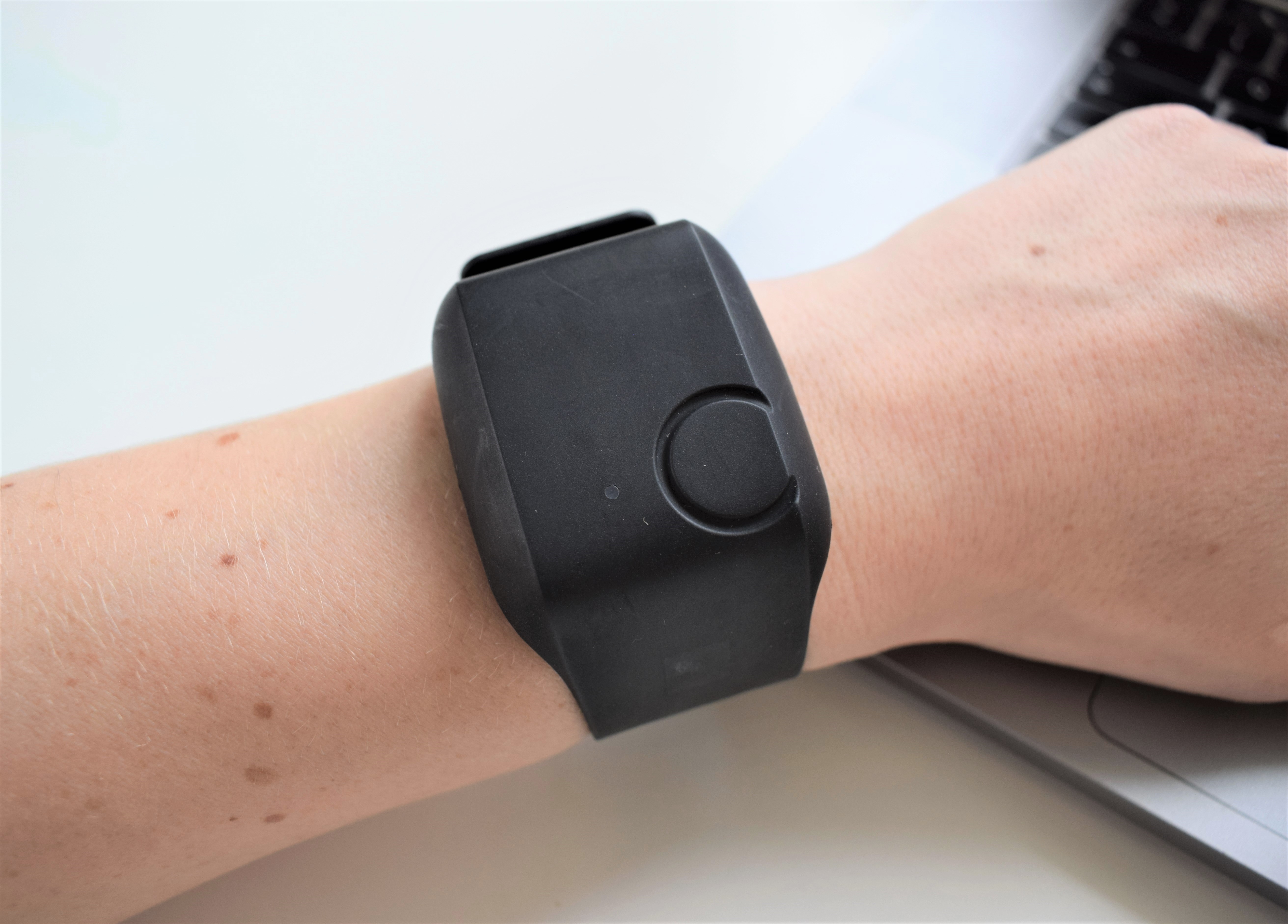}
    \includegraphics[height=125.5pt]{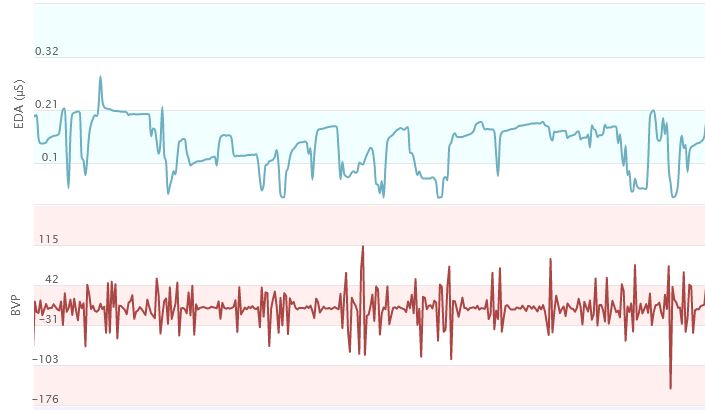}
  \caption{Empatica E4 Wristband and EDA and BVP (Heart Rate) Signals.}
  \label{fig:eda}
\end{figure}

Another approach to automated attention assessment during learning involves physiological sensor data. One of the most accessible physiological sensors is \gls{eda} (also \gls{gsr} or skin conductance), reflecting changes in the skin's electrical properties triggered by sweat production, which alters skin conductivity \cite{greco2015cvxeda}. Additionally, heart rate can be assessed using wearable devices, for instance, by \gls{bvp}, pulse, or photoplethysmography sensing \cite{georgiou2018hr}. Other physiological signals include \gls{eeg} as the primary method to assess brain activity. However, this involves the application of more high-cost, high-stakes sensors requiring controlled settings. On the contrary, \gls{eda}, temperature, and heart rate can be measured using low threshold wearable devices such as the Empatica E4 wristband, shown on the left in Figure \ref{fig:eda}. The right side of the figure displays continuous \gls{bvp} and \gls{eda} signals derived from such wearables.

In the context of attention assessment, signals are meaningful due to the correlation between sympathetic nervous activity, reflected in, for instance, in \gls{gsr} and skin temperature and attentional states \cite{andreassi1980human}.  \gls{gsr}, or \gls{eda}, is linked to increased effort and task engagement \cite{pecchinenda1996affective, appel2019predicting, borisov2021robust}, with lower GSR levels associated with mind wandering \cite{smallwood2004phys}. GSR has been utilized to automatically assess episodes of mind wandering \cite{Blanchard.2014, Brishtel.2020, bixler2015gazeeda} and classroom engagement \cite{gao2020eda, disalvo2022, di2018unobtrusive}. In addition, heart rate is of particular importance as it is related to physiological arousal. This, in turn, is correlated with states of attention such as wandering thoughts, which is presumably due to increased emotional engagement \cite{smallwood2004phys, Smallwood.2007c}. This connection has led to using heart rate in automated systems to detect mind wandering \cite{Pham.2015} to measure engagement \cite{gao2020eda}. Additionally, \gls{eeg} has been used for attention assessment, particularly focusing on the synchronization of brain activity in classroom settings \cite{poulsen2017eeg, dikker2017brain, bevilacqua2019brain}. It has proven effective for detecting mind wandering during lecture video viewing \cite{conrad2021} and in classroom environments \cite{dhindsa2019}. However, as previously mentioned, \gls{eeg} measurements require high-end sensors, and their use is typically restricted to more controlled settings, limiting their large-scale applicability in naturalistic settings.

\subsubsection{Interaction Data for Attention Detection}

Log files record information on user interactions in computer-mediated tasks, for instance, information on keystrokes, mouse tracking, and button presses. Log data has been extensively employed in education research \cite{wang2023log} and is especially relevant when using \glspl{its} and \glspl{mooc}, which require learners to interact with educational technologies \parencite[e.g.,][]{hadwin2007, gobert2013log}{}{}. 

Log data can provide valuable insights for detecting attention, as attention lapses influence reaction times \cite{mcvay2012drifting} and fine motor movements \cite{dias2022motor} captured in this data. For instance, keystrokes have been used to predict attentional states and task-unrelated thought in online conversations \cite{kuvar2023keystroke, kuvar2023partner}. Furthermore, mouse tracking has revealed a correlation between mind wandering and mouse movements during online tasks \cite{dasilva2020mouse}. Also, patterns of interaction with reading interfaces, such as the amount of time spent reading, can indicate mind wandering \cite{mills2015log}. Log files are insightful but only relevant to learning tasks that require active technology use and interaction. They are ineffective for passive activities like lecture viewing, which lack direct system interaction, or learning activities that might not involve technology use, such as classroom instruction.

\subsection{Predictive Modeling of Attention-Related Processes}

The goal of employing behavioral and observable indicators to assess attention is to enable continuous, non-intrusive, and objective measurement of attention-related processes. Numerous studies have used the modalities discussed above \parencite[e.g.,][]{Brishtel.2020, mills2016, hutt2016eyes} in predictive modeling to infer attentional states. These methods mark a shift from traditional statistical analysis primarily because they focus on prediction rather than explanation \cite[]{yarkoni2017}. Unlike conventional models that explain the relationship between variables, machine-learning models are designed to effectively predict future observations, utilizing training and test datasets to optimize and evaluate the model's performance and generalizability \cite[]{yarkoni2017}. Furthermore, the importance of individual predictors can be analyzed using explainable machine-learning techniques after the model has been developed \cite[]{roscher2020}. Another difference lies in the modeling approach: data-centric versus algorithmic \cite[]{breiman2001}. Traditional models rely on pre-established assumptions, which may skew the findings based on the model’s design rather than the actual data. In contrast, machine learning adopts an algorithmic approach that tests multiple models to identify the best fit based on performance, making it ideal for scenarios like sensor-based attention detection where the complex, potentially non-linear relationships between numerous features and the target variable are not predefined. This also constitutes the decisive difference to the traditional use of behavioral measures in education research, such as eye tracking. As reviewed above, even though the indicators extracted from those modalities show associations with attention processes, no clear set of indices correlated to attentional states can be identified. This may be due to the task- and setting dependencies of those indicators \cite{Faber.2020}. However, these findings underscore the relevance of employing machine learning to capture complex, non-linear relationships in attention-related data. 

\begin{figure}
    \centering
  \includegraphics[width=0.9\textwidth]{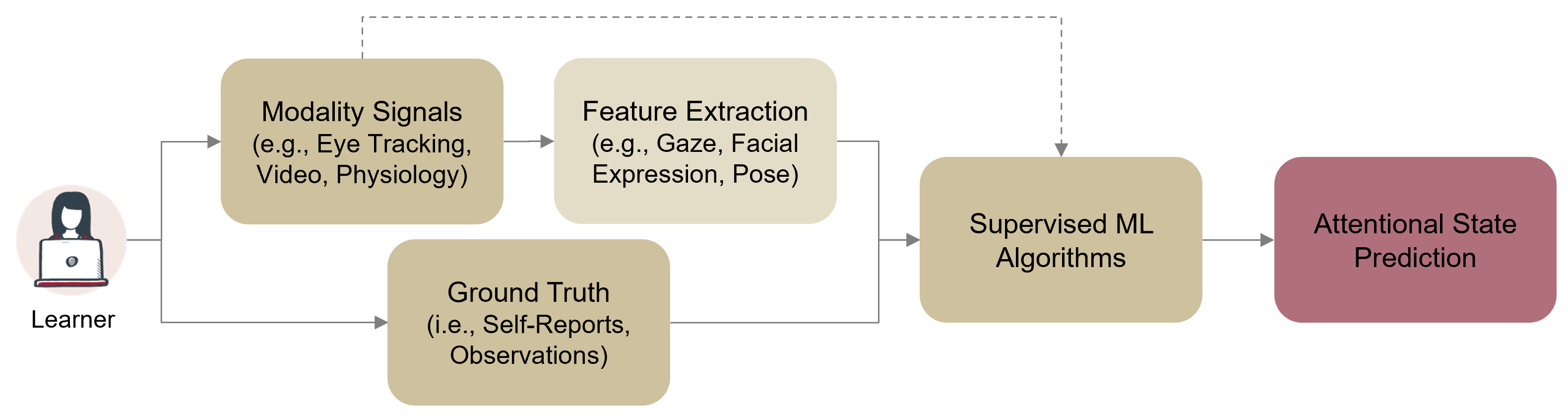}
  \caption{Schematic Pipeline for Predictive Modeling of Attentional States. ML -- Machine Learning.}
  \label{fig:gener}
\end{figure}

The primary goal of predictive modeling is to automatically identify non-directly observable or measurable attentional states, such as mind wandering, from observable indicators like eye gaze \cite{dmello2021} using machine learning algorithms. Figure \ref{fig:gener} outlines the schematic steps of prediction pipelines for attentional states, beginning with a learner engaged in a specific task. Signals are collected via various modalities, such as eye tracking or videos, and features are then extracted from these signals to serve as inputs for a supervised machine-learning model. An alternative approach, made possible by advancements in deep learning, particularly in computer vision, involves feeding raw data, for instance, images, directly into models, as illustrated by the dotted array in Figure \ref{fig:gener}. This allows the models to learn internal representations and has been successfully applied to engagement estimation in classroom settings \cite{sumer2021multimodal, mandia2023eng}. Given the often limited size of educational data sets, these models frequently rely on transfer learning. This approach involves pre-training a deep model on a sufficiently large available dataset that stems from a related task, then fine-tuning the model to the current task at hand, for which only limited data is available.

The ground truth data for training attention detection models typically comes from learner self-reports or observer ratings. These labels are used to train the model to predict specific outcomes. The model’s prediction performance is evaluated using a separate test set containing unseen examples. In the field of learner attention research, datasets often include multiple instances from the same individual. To ensure the model generalizes to new learners, this process is usually conducted in a person-independent manner \cite{dmello2021, kuvar2023}. Additionally, the common issue of highly imbalanced data distribution due to the varied attentiveness levels among learners \cite{bosch2022, Hutt.2019} requires the implementation of balancing techniques during the learning phase.

Predictive modeling using machine learning techniques offers great potential for assessing attention non-intrusively. However, the employment of machine learning in the educational context raises concerns about the lack of transparency in the decision-making processes of many of those models \cite{yang2023xai}. One way to address this concern is by using explainability tools.

\subsection{Explainability and Fairness of Automated Attention Detection}

Employing explainability methods can help account for Fairness, Accountability, Transparency, and Ethics (FATE) concerns,  which is crucial when employing machine learning techniques in educational contexts to develop adaptive and personalized educational interventions \cite{khosravi2024}. Explainability enhances the transparency and fairness of machine learning models by elucidating their internal processes and providing a clear rationale for their outcomes \cite{yang2023xai, arrieta2020explainable}. These explanations are highly relevant for fostering trust and confidence in the models employed. Further, they allow us to gain deeper insights into how specific features impact predictions to better understand the relation between data modalities and attentional states. Highly predictive attention prediction models are typically explained with so-called post-hoc explainablity methods. This adds one more step to the analysis pipeline: after model training, i.e., post-hoc, the model's decision is explained with the help of a separate algorithm.
Two of the most prominent post-hoc explainability methods, which were also employed in this thesis, will be discussed in more detail. First, \gls{shap} values \cite{shap2017} serve to quantify the impact of each feature in a model on the probability of predicting a specific class. These values measure how much each feature contributes positively or negatively to the difference between the current prediction and the model's average prediction across the entire dataset. This method helps to clarify which features are most influential in the model’s decision-making process and how they affect the outcome for each prediction class. 
Another visualization-based explainability tool often employed to make computer vision models more understandable is \gls{lime} \cite{lime}. It identifies and highlights specific areas of an image, called super-pixels, that significantly influence the algorithm's prediction. This process allows users to see which parts of an image lead the model to classify it in a particular way, providing insight into its reasoning. Using explainability methods is crucial for detecting any unintended biases or correlations that the model might have learned, potentially due to artifacts in the data collection process \cite{lime}. Having reviewed relevant modalities and established the potential of predictive modeling and the importance of explainability in automated attention assessment, the following section presents the main objectives of this dissertation.



\section{Objectives} \label{objectives}

As outlined above, there is a strong association between attention and learning success. However, further quantitative research, especially concerning finer distinctions of inattention, is required to understand the underlying processes and interplay between the two. While previous research on automated attention estimation has shown promising results, individual attention estimation and mind-wandering detection methods have not reached sufficient accuracy and have not yet been performed on the granularity level needed in education research or adaptive learning technologies. The automated assessment approaches facilitate the evaluation of phenomena that have not yet been sufficiently researched, ideally progressing in tandem with theoretical development \cite{dmello2021}.

This thesis aimed to advance the state-of-the-art by leveraging machine learning methods, eye-tracking methodologies, and computer vision techniques to assess nuanced aspects and indicators of attention-related processes and facilitate future quantitative research on different aspects of attention. It examined different dimensions of (in)attention in diverse learning settings such as remote, computer-based, and classroom learning. To achieve this overall objective, multiple individual research steps were carried out. Specifically, this dissertation addressed the following challenges and outlined approaches to address them in realistic learning scenarios.

The first part of this thesis aimed to enhance the automated assessment of mind wandering. As described in section \ref{learning}, mind wandering is a multidimensional construct, but its sub-factors have not yet been sufficiently investigated and assessed in the learning context. In advancing the automated detection of mind wandering, this dissertation focused on meta-awareness, multimodality, and generalizability reflected in the following objectives: 

\begin{enumerate}[label=O\arabic*]
    \item  The first objective of this thesis was to differentiate mind-wandering types blended in previous research on learning. To this end, this work investigated the temporal patterns of aware and unaware mind wandering and their relation to learning outcomes.

    \item Previous research focused on the automated detection of mind wandering as a unitary construct. To enable the continuous assessment of meta-awareness in mind wandering over time, the initial differentiation of the two sub-forms--aware and unaware mind wandering--using eye gaze and machine learning were explored. 

    \item To improve the accuracy and robustness of detecting aware and unaware mind wandering, this work used a novel multimodal machine learning approach. Leveraging complementary information encoded in eye tracking, video, and physiology (\gls{eda} and heart rate), the interplay and impact of different modalities in predicting mind wandering by meta-awareness were examined using explainability approaches.

    \item The generalizability of a webcam-based detection approach across different settings (lab to in the wild), tasks (reading to lecture video watching), and culturally diverse target groups (US to Korean) were examined, employing two distinct datasets. To improve generalizability to low quality in the wild data, a novel transfer-deep-learning approach was employed and fine-tuned to the mind-wandering detection problem. 

\end{enumerate}

The second part of this thesis focuses on indicators of attention during computer-based online learning, specifically lecture video watching. Visual attention indicates what information is being cognitively processed; consequently, when paying attention to a lecture, students should focus on parts of the slides that are relevant at that moment. Initial experimental studies suggest that learners' gaze may synchronize during episodes of attentiveness \cite{Madsen.2021} leading to the following objective:

\begin{enumerate}[resume,, label=O\arabic*]

    \item To explore whether gaze synchrony serves as a reliable indicator of self-reported attention during online learning, this work applied and compared three gaze synchrony measures and investigated the relationship between self-reported attention, gaze synchrony, and learning outcomes.
    
\end{enumerate}

The last part of this dissertation addresses attention-related processes in classroom learning. The most studied concept in this domain is behavioral engagement \cite{sumer2021multimodal, goldberg2021attentive, Ahuja.2019}. Due to the challenges of automatically predicting such higher-level constructs, such as low interrater reliability \cite{sumer2021multimodal}, this work focused on detecting one specific indicator of behavioral engagement and active participation in the classroom discourse: hand-raising.

\begin{enumerate}[resume, label=O\arabic*]

\item  To advance previous hand-raising detection research primarily focusing on staged videos and to tackle the challenges of real-world classrooms and authentic student hand-raisings, a novel skeleton-based detection approach in authentic classroom videos was proposed, employing fine-tuned view-invariant, occlusion robust pose embeddings, and temporal models. 

\item Investigating the potential of automated assessment of hand raising as a substitute for manual observations and large-scale analysis of classroom videos, the relationship between automatically annotated hand-raisings and self-reported cognitive engagement, interest, and involvement of learners was investigated. 

\end{enumerate}

All the objectives mentioned were successfully achieved through interdisciplinary efforts and advanced the state-of-the-art in the area of automated assessment of attention during learning. This work led to multiple contributions accepted and published in leading journals and conferences at the intersection of computer science and education research. The methods employed to achieve these objectives are explained in more detail in Chapter \ref{contributions}.

%% file: 3_contributions.tex
\begin{figure}[h!]
    \centering
  \includegraphics[width=\textwidth]{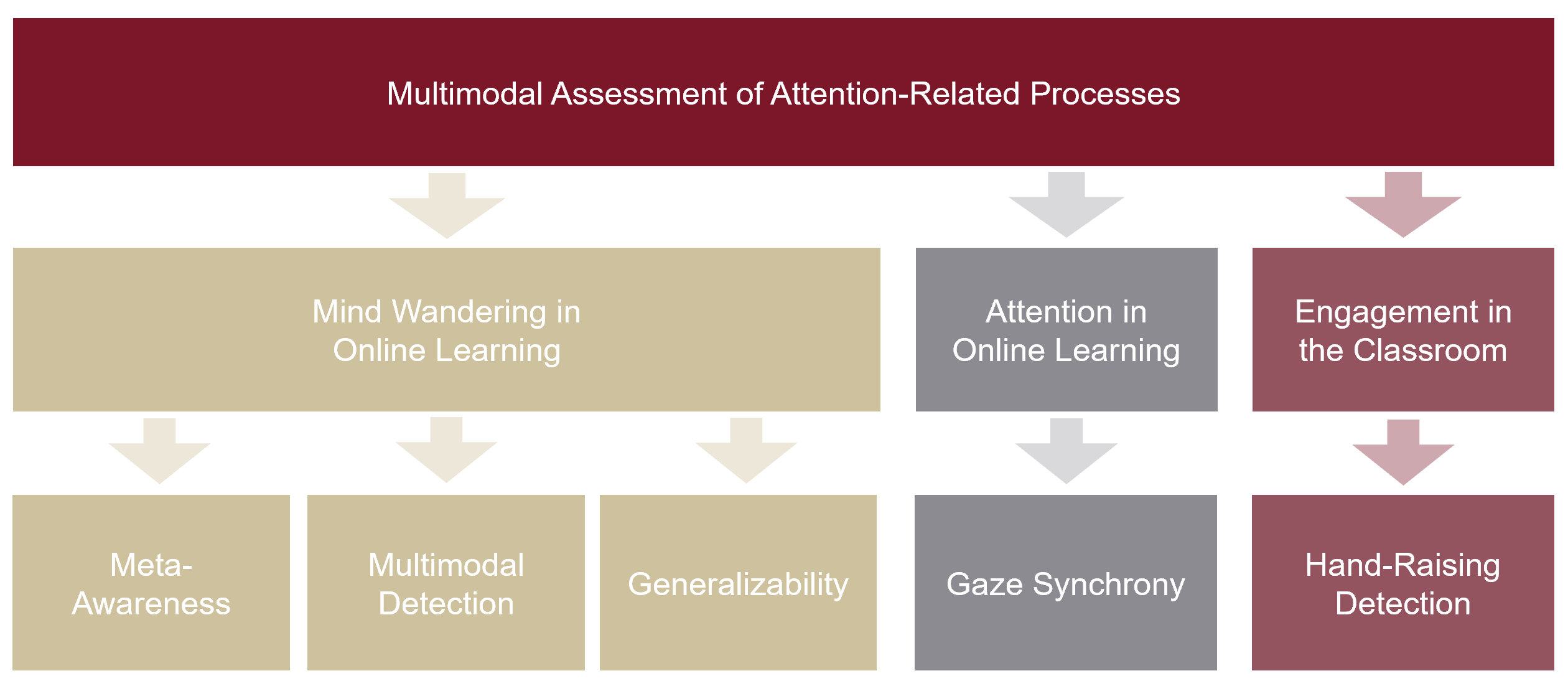}
  \caption{Schematic Overview Over the Contributions in this Dissertation, Structured by Attention-Related Processes and Learning Settings.}
  \label{fig:overall}
\end{figure}

In this chapter, the major contributions toward the objectives outlined previously in Chapter \ref{objectives} are summarized. Initially, the contributions to automated mind wandering are recapped, focusing on meta-awareness, multimodality, and generalizability. Following this, an overview of the examination of gaze synchrony as an attention indicator during video lecture viewing and the automated detection of hand-raising as a behavioral indicator of engagement in classroom settings is provided. Figure \ref{fig:overall} presents a schematic overview of all contributions made in this thesis.


\section{Automated Mind Wandering Detection: Investigating Meta-Awareness, Multimodality and Generalizability}

The first contributions of this thesis revolve around the automated detection of mind wandering, a cognitive process related to inattention. The first section details the contribution towards data-driven theory development, investigating temporal patterns of meta-awareness of mind wandering, its relation to learning outcomes, and discernability employing eye tracking-based automated detection. This is followed by a multimodal machine learning approach, fusing eye-tracking, video, and physiological data for automated aware and unaware mind wandering detection. The third section focuses on examining the generalizability of video-based mind-wandering detection, proposing a transfer learning approach.

\subsection{Meta-awareness of Mind Wandering and its Impact on Learning} \label{meta}

\begin{itemize}
    \item[{[1]}] \printpublication{buhler2024edpsych}.
\end{itemize}

\subsubsection{Motivation}

Several studies have demonstrated that mind-wandering episodes can occur both with and without meta-awareness, indicating variability in whether an individual is aware of their occurrence \cite{Schooler.2002, Smallwood.2007}. However, research on education has not extensively investigated the levels of meta-awareness in mind wandering or how they evolve over time in learning settings. These different manifestations likely stem from different underlying cognitive processes \cite{Christoff.2018}, necessitating tailored interventions and solutions. It is, therefore, essential to further investigate these different forms of mind wandering and their impact on learning within educational contexts. One major challenge is the reliance on experience-sampling self-reports that, although providing reliable assessment, can only capture discrete moments. Using machine learning on sensor data presents a promising method for unobtrusive, continuous, and systematic monitoring of thought processes over time in realistic learning scenarios \cite{Hutt.2019, Mills.2020, Brishtel.2020}. However, current methods primarily focus on binary detection of mind wandering—either occurring or not—and lack the ability to distinguish more nuanced variations. Automated recognition of aware and unaware mind wandering, enabling continuous assessment over time, could allow for in-depth analysis of temporal dynamics of meta-awareness of task-unrelated thoughts over extended periods without the need for disruptive probes. Furthermore, it could open up possibilities for more effective interventions and tailored learning content in attention-aware technologies \parencite[]{Mills.2020, hutt2021}. 

This work determined the meta-awareness and temporal patterns of mind-wandering episodes during a 60-minute prerecorded Zoom lecture using 15 thought probes. The sequences of meta-awareness in mind wandering using clustering to identify typical thought patterns were analyzed. Additionally, the effects of these mind-wandering patterns, both aware and unaware, on both fact-based learning (e.g., memory for facts or details) and inference-level learning (e.g., integration of information with prior knowledge) \cite{mccarthy2018, wong2022} were assessed. To address the limitations of relying solely on discrete self-reports in temporal analysis, a predictive modeling approach based on eye-tracking data and machine learning was applied. This method distinguishes between on-task behavior and aware or unaware task-unrelated thoughts, offering a promising way to assess and differentiate types of mind wandering. Using post-hoc explainability techniques allowed for insights into how specific gaze behaviors relate to different types of mind wandering. This approach not only sheds light on meta-awareness in mind wandering during video lectures but also sets the groundwork for future large-scale research using observable gaze data.

\subsubsection{Principal Methodology}

In this study, data were collected from 96 university students. After excluding those with technical issues and non-native German speakers, 87 students (age 19-33 years, \textit{M} = 23.44, \textit{SD} = 2.6, 19\% male) qualified for analysis. In a lab setting, participants watched a 60-minute lecture video, which was interrupted in 3-5 minute intervals by 15 thought probes assessing mind wandering and its meta-awareness. The probes were composed of two stages to assess thought content at the first question to disentangle being on task, elaborations, and metacognitive monitoring from task-unrelated thoughts. In the second stage, participants were asked whether they were aware of their minds wandering off. Before and after the lecture, they answered questionnaires on person-specific characteristics, as well as pre- and post-knowledge tests. 

To determine the temporal pattern of aware and unaware mind wandering, self-report sequences derived from the 15 thought probes were analyzed by clustering these sequences using the agglomerative clustering algorithm Agnes with the Ward method to identify distinct patterns of mind wandering. The similarity between sequences was quantified using \gls{om}, which calculates the minimal cost of transforming one sequence into another through insertion, deletion, and substitution operations, allowing to cluster based on the general temporal unfolding of thoughts and not just specific timings. Then, a linear regression analysis to investigate the link between meta-awareness of mind-wandering patterns and learning outcomes was conducted, using knowledge post-test scores as independent variables and cluster membership as the dependent variable while adjusting for confounders like age, gender, prior knowledge, and self-concept. 

During the study, participants' eye movements were tracked using remote SMI eye trackers at 250 Hz sampling rates. Eye-movement events like fixations and saccades were analyzed using the BeGaze software \parencite[]{begaze}. Information was extracted from a 10-second window before each thought probe, and additional features like gaze vergence and fixation dispersion were created. Summary statistics from the eye-tracking data (e.g., mean, min, max, std, skew) were compiled to create input features for the machine-learning model to differentiate attention states. It was analyzed how meta-awareness of mind wandering is reflected in eye movements by comparing the gaze features during aware and unaware episodes to those during on-task periods, using \textit{t}-tests for mean comparisons. The investigation was enhanced by implementing various machine-learning classifiers, including Random Forest, \gls{xgb},  \gls{svm}, and  \gls{mlp}, to distinguish between aware and unaware task-unrelated thoughts based on eye-tracking data, addressing data imbalance with techniques like \gls{smote} \cite{smote2002} and random oversampling. Our analysis applied rigorous validation through person-independent three-fold nested cross-validation to ensure the generalizability of our findings, aiming for robust predictions of mind-wandering states from eye movements. Further, \gls{shap} analysis was employed to investigate how individual eye-tracking features contributed to predicting aware and unaware mind wandering.

\subsubsection{Main Findings}
During thought probes, participants reported being on task 35\% of the time, having task-related elaborations 8\% of the time, engaging in metacognitive comprehension monitoring 16\% of the time, and experiencing task-unrelated thoughts 41\% of the time, with 63\% of all task-unrelated thoughts occurring with meta-awareness. Correlation analysis revealed significant correlations between individual characteristics and types of mind wandering, including negative associations between metacognitive self-regulation and unaware task-unrelated thoughts, and positive correlations with self-concept and dispositional interest related to on-task behavior and lecture comprehension; situational interest, self-reported involvement, and cognitive engagement showed negative correlations with both types of task-unrelated thoughts, with stronger associations observed for unaware mind wandering.

\begin{figure}[h!]
    \caption{Attentional State Distribution Plots by Cluster.}
    \includegraphics[width=\textwidth]{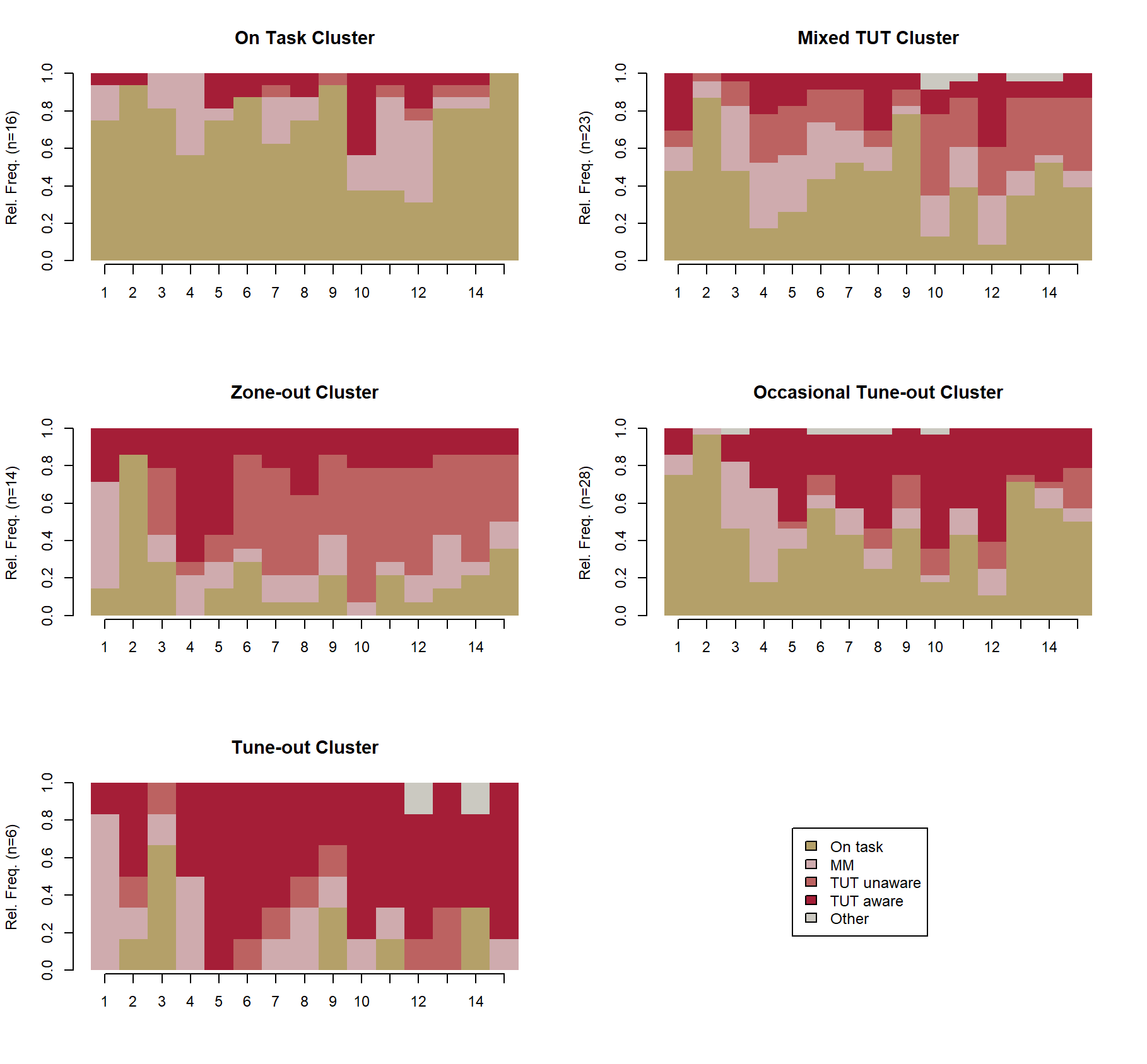}
    \label{fig:clu_stc}
\end{figure}

In the study, hierarchical clustering identified five distinct clusters of mind-wandering patterns during the lecture, depicted in Figure \ref{fig:clu_stc}, labeled as on-task, mixed-TUT, zone-out, occasional tune-out, and tune-out clusters. The on-task cluster demonstrated the highest level of sustained attention with minimal mind-wandering, while the zone-out cluster exhibited consistent unaware mind wandering throughout the lecture. Analysis of variance revealed significant differences between clusters in terms of on-task engagement and types of mind wandering. 

Linear regression analysis to investigate how different thought patterns influenced learning outcomes revealed that participants in the zone-out cluster, characterized by high rates of unaware mind wandering, scored significantly lower on fact-based and deep-level inference learning tests than the on-task cluster. Participants in the tune-out cluster, predominantly engaged in aware-task-unrelated thoughts, also demonstrated lower performance, particularly in inference learning. Additionally, the analysis confirmed that higher levels of self-concept in statistics and prior knowledge positively influenced learning outcomes. A gender effect indicated slightly lower inference test scores for male participants.


To examine how meta-awareness in mind wandering influences gaze behavior, the average levels of gaze features were compared across different self-reported thought groups, revealing significant differences through \textit{t}-tests; notably, unaware mind-wandering episodes had fewer fixations and a higher fixation to saccade duration ratio compared to on-task instances. Further analysis identified no significant differences in gaze features between on-task and aware mind-wandering instances. However, the latter showed modestly higher values for some features, such as saccade velocity and acceleration, with aware mind wandering having a higher fixation count similar to on-task levels.

Multiple state-of-the-art machine-learning classifiers were employed to differentiate instances of aware and unaware mind wandering from a combined on-task and metacognitive monitoring category, where the \gls{rf} model showed the best performance slightly above chance with an F$_{1}$ score of 0.215 for unaware task-unrelated thoughts. The \gls{svc} and \gls{mlp} classifiers demonstrated better prediction capabilities for aware task-unrelated thoughts with F$_{1}$ scores of 0.332 and 0.282, respectively, achieving macro F$_{1}$ scores marginally above chance. Additionally, a binary classification approach distinguishing between on-task and combined task-unrelated thought instances yielded significantly higher detection accuracy with an \gls{mlp} classifier achieving an F$_{1}$ score of 0.529, which is 24\% above chance level.

To understand how gaze features relate to the meta-awareness of task-unrelated thoughts, \gls{shap} values were computed for the best-performing \gls{rf} classifier, revealing that specific saccade, blink, vergence, and fixation duration features were crucial for predicting both aware and unaware task-unrelated thoughts. The variability in these features, especially in vergence and pupil distances, played a significant role, with higher variance often indicating aware task-unrelated thoughts due to phenomena like ''staring into nothingness.'' At the same time, more stable metrics suggested on-task focus. Additionally, factors such as the variability in pupil diameter and saccade velocities contributed differently to each class. The findings underscore the complexity of gaze behavior's relationship with mind-wandering states.

This study investigated the meta-awareness of mind wandering during video lectures, identifying distinct temporal patterns of thought sequences that correlate differently with learning outcomes. By employing probes that inquired about participants' current thoughts and their awareness of them, the study revealed that zone-out patterns negatively impact both fact-based and deep-level inference learning, whereas tune-out patterns primarily affect deep-level understanding. Furthermore, initial efforts in predictive modeling using gaze data showed promise in distinguishing between aware and unaware forms of mind wandering, suggesting potential directions for future research that could enhance understanding and measurement of these phenomena using advanced computational methods and larger data sets. One way to enhance the robustness and precision of predictions is to add indicators derived from additional data sources. The following section presents a novel approach to the multimodal detection of aware and unaware mind wandering.

\subsection{Multimodal Detection of Aware and Unaware Mind Wandering} \label{multi}

\begin{itemize}
    \item[{[5]}]\printpublication{buhler2024mm}.
\end{itemize}

\subsubsection{Motivation}

In research on automated detection of mind wandering, a variety of modalities next to eye tracking \cite{hutt2021, Mills.2020, dmello2017zone, mills2021, hutt2016eyes}, such as physiological signals like skin conductance \cite{Brishtel.2020} and heart rate \cite{Pham.2015}, and facial video recordings \cite{Bosch.2021, Lee.2022} have been employed. 
The integration of different modalities aims to improve performance and ensure robustness by addressing noisy data, resolving ambiguities, and capitalizing on intermodal correlations \cite{baltruvsaitis2018multimodal}. Recent findings from a meta-analysis by \citet{kuvar2023} indicate that multimodal methods surpass unimodal approaches in detecting mind wandering. However, the extent of improvement and its consistency varies across studies and tasks. These results suggest that the benefits of combining features might not be consistently additive, highlighting the need for further research \cite{kuvar2023}. Motivated by these findings, our study introduced a novel approach by merging data from eye tracking, facial video analysis, and physiological sensors. Similar to \citet{buhler2024edpsych} described in Section \ref{meta}, the aim is to detect aware and aware mind wandering, as each type of mind wandering may require different types of interventions to effectively support learning \cite{ Smallwood.2007, Schooler.2011, Christoff.2009, smallwood2008}. 

This study contributes novel insights by employing a combination of eye tracking, video, and physiological sensors (heart rate and electrodermal activity) to detect both aware and unaware mind wandering. Additionally, to facilitate comparisons to previous research, a combined mind-wandering category, including aware and unaware mind wandering, was predicted. This multimodal approach was evaluated for its effectiveness in differentiating mind-wandering categories during lecture viewing. Further, it was analyzed which features from each modality were most influential in recognizing aware, unaware, and combined mind-wandering states. This deeper understanding of the temporal dynamics of mind-wandering meta-awareness and its impacts on learning outcomes provides a critical foundation for further developing adaptive educational technologies and targeted interventions that support learners in remote learning environments. Such advancements are crucial for optimizing engagement and learning efficiency in digital education settings.

\subsubsection{Principal Methods}

For this study, data from 87 university students was employed, parts of which, specifically self-reports and gaze data, have been employed in \citet{buhler2024edpsych}. 
Participants viewed a 60-minute recorded Zoom lecture on the topic of statistics while their eye gaze, facial videos, and physiological responses were tracked using remote SMI eye trackers, standard webcams, and E4 Empatica wristbands. Throughout the session, which included mid-point eye tracker recalibration, 15 thought probes were administered at 3-5 minute intervals. Participants were asked to categorize their thoughts into predefined categories reflecting either task-related or task-unrelated thoughts; task-unrelated thoughts were further classified based on their meta-awareness into aware mind wandering and unaware mind wandering, serving as ground truth for the following machine learning task. Due to occasional eye-tracking failures, including abnormally long blinks, the initial dataset of 1305 instances was reduced by excluding 11 instances with incomplete data. This resulted in a final dataset of 1284 instances used for the analysis.

For our analysis, data from eye-tracking, video, and physiological sensors was extracted for 30-second intervals prior to each thought probe, a duration validated by prior studies for multi-modal mind wandering detection during video lectures \cite{Hutt.2017}. Using proprietary SMI BeGaze Software \cite{begaze}, eye-tracking events such as fixations, saccades, and blinks were extracted. Additional features like fixation dispersion and vergence were computed, focusing on global gaze features to capture broad eye movement patterns that were less dependent on task specifics and to enhance their generalizability. Using the OpenFace toolbox \cite{openface2016}, facial video data was processed to obtain features like facial action units and head pose. Additionally, physiological data from wrist-worn Empatica E4 sensors provided electrodermal activity and heart rate data; these were standardized and cleaned to ensure consistency \cite{Makowski2021neurokit}. The resulting features from each modality were summarized over 30-second intervals, and various statistics for each feature, including minimum, maximum, mean, median, standard deviation, 25\% and 75\% percentiles, skewness, and kurtosis, were computed.

Three binary classification tasks were performed to predict aware and unaware mind wandering, as well as a combined category separately, using both individual and combined feature sets from various modalities. Employing an early fusion approach, aggregated feature arrays were concatenated to integrate these modalities at the feature level, which has been shown to outperform decision-level fusion \cite{bixler2015gazeeda}. \gls{smote} \cite{smote2002} and random oversampling were used during the training phase to mitigate the highly imbalanced dataset. The sklearn package was utilized to train various classifiers—\gls{rf}, \gls{svc}, \gls{mlp}, and \gls{xgb}. To interpret the multimodal models, the \gls{shap} method \cite{shap2017} was used, providing post-hoc explanations by calculating Shapley values, which quantify how individual features influence the prediction relative to the average prediction of the sample. These explanations revealed the impact of specific features on different types of mind wandering. The large feature vectors resulting from multimodal fusion led to potential overfitting, so models were refined to focus only on the top 100 features identified by \gls{shap} as most influential. Validation was conducted using Leave-One-Person-Out Cross-Validation, focusing on the area under the precision-recall curve to robustly assess performance across imbalanced classes \cite{jeni13}.

\subsubsection{Main Findings}

\begin{figure}
    \centering
    \includegraphics[width=0.6\textwidth]{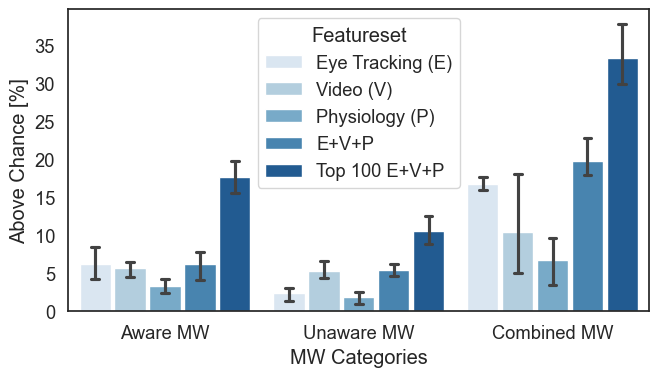}
    \caption{Avergage Classification Results (Above Chance Based on AUC-PR) by Mind Wandering Category and Modality Feature Set.}
    \label{fig:resc}
\end{figure}

In this study, the predictive power of a novel multimodal combination of eye-tracking, facial videos, and physiological signals for detecting aware, unaware, and combined mind wandering during video lectures was investigated. Integrating these three modalities significantly enhanced prediction accuracy compared to using any single modality, with the fused multimodal feature set refined to the 100 most influential features showing the best performance across all mind-wandering categories. Specifically, aware mind wandering was detected at 20\% above chance (\gls{aucpr} = 0.396), unaware mind wandering at 14\% above chance (\gls{aucpr} = 0.267), and the combined category at 40\% above chance (\gls{aucpr} = 0.637). Notably, all top 100 feature subsets contained variables stemming from all three modalities, highlighting each modality's complementary information. Average prediction accuracies by mind-wandering category and feature set are displayed in Figure \ref{fig:resc}. As \citet{kuvar2023} hypothesized, the performances of modality feature sets appear not to be additive. Nevertheless, this study finds significant improvements over unimodal approaches, contrasting with prior research. Our findings revealed that unaware mind wandering was the most challenging to detect, likely due to fewer instances available for training, adding more data could potentially improve accuracies. 

The analysis of unimodal approaches showed that eye-tracking was most effective for predicting aware mind wandering, whereas facial video features excelled in detecting unaware mind wandering, with physiological signals showing lower standalone predictive power but contributing valuable information when combined with other modalities. \gls{shap} analysis of the top-performing multimodal models highlights that for aware mind wandering, the key predictors are saccade velocities, pupil diameter, facial expressions, head pitch, and physiological signals like \gls{bvp} and tonic \gls{eda}. In contrast, the critical features for unaware mind wandering include \glspl{au}, such as the absence of lip sucking and the presence of nose wrinkling, which indicates a strong predictive power of facial expressions. Additionally, gaze-related features such as frequent blinking, albeit video-based, played a significant role in detecting unaware mind wandering.

These results emphasize the importance of multimodal approaches in enhancing the robustness and accuracy of mind-wandering detection systems, suggesting a promising direction for future research and application in educational technologies. Future work should aim to validate these approaches in more ecological, diverse settings to enhance generalizability and examine the effectiveness of sophisticated machine-learning techniques on larger datasets. The potential generalizability of the scalable video-based mind-wandering detection is investigated in the following section.

\subsection{Examining the Generalizability of Video-Based Mind Wandering Detection}\label{generalize}

\begin{itemize}
    \item[{[2]}] \printpublication{buhler2024general}.
\end{itemize}

\subsubsection{Motivation}

In considering the integration of mind-wandering detection into educational technologies to enhance adaptiveness and attention awareness, the scalability and accessibility of the employed modalities emerge as critical factors. Recent advancements in automated mind-wandering detection involve using physiological sensors and eye trackers in controlled environments, which, while insightful, are not scalable or cost-effective for widespread use \cite[]{Hutt.2017, Faber.2018}. Conversely, emerging research indicates that consumer-grade webcams can successfully detect mind wandering in more naturalistic settings, enhancing the potential for broader application in real-world educational environments \cite[]{Bosch.2021, Lee.2022}. However, challenges remain in ensuring the generalizability and fairness of these systems across tasks, different cultures, and demographic groups, necessitating further research into robust, culturally aware models that respect the diversity of learners \cite[]{jack2012facial, kuvar2023}.

Traditional mind-wandering detection has relied on explicit features like Action Units (AUs) and gaze metrics, which are often tailored to specific tasks and may not generalize well across different settings \cite[]{stewart2017, Bosch.2021, Lee.2022}. By leveraging transfer learning from pre-trained \gls{cnn} on large, diverse datasets and employing \gls{fer} technologies, this study examines the potential of latent features derived from facial expressions to improve generalizability and interpretability using explainability tools like \gls{lime} \cite[]{lime}.
Incorporating temporal dynamics, which have been shown to be critical for accurately detecting mind wandering \cite{Bosch.2021, stewart2017}, this research utilizes advanced models capable of processing time-series data to learn temporal relationships directly from video input. 
To test the robustness of these approaches across various environments and cultural contexts, the study extends beyond lab-based settings to naturalistic environments by applying models trained on lab data to 'in-the-wild' scenarios \cite[]{stewart2017generalizability, Bosch.2021, Lee.2022}. This cross-context application not only challenges the universality of facial expressions but also addresses potential cultural biases in facial recognition algorithms, aiming to ensure that these models are effective and fair across diverse global user groups.

\subsubsection{Principal Methods}

Our research utilizes two distinct datasets: one from a laboratory setting involving 135 U.S. university students who self-reported mind wandering while reading scientific texts \cite{Bosch.2021}, and another from an in-the-wild setting with 15 Korean students who were probed about their mind wandering while watching a lecture at home \cite[]{Lee.2022}. The lab data is used for both training and within-dataset evaluations, consisting of 1,031 mind-wandering and 2,406 non-mind-wandering instances, while the in-the-wild data, containing 205 mind-wandering and 1,009 non-mind-wandering instances, is used solely for cross-dataset evaluation to assess generalizability and real-world application.

To extract deep learning-based facial expression features from video frames, a \gls{cnn} with a ResNet50 architecture, pre-trained on the AffectNet dataset \cite{affectnet2017,}, where faces are detected and aligned using RetinaFace \cite{retinaface2020, He_2016_CVPR_resnet50} was employed. This process generated a 2048-digit long latent feature vector per frame, which can be utilized by downstream classifiers to predict mind wandering. In comparison, explicit features such as Action Units (AUs), facial landmarks, head pose, gaze direction, and eye region landmarks are extracted using the OpenFace toolkit \cite[]{openface2016} from each frame for potential use in alternative models. Recurrent neural networks, specifically \gls{lstm} and \gls{bilstm}, were employed to capture temporal dynamics in the data, using these models alongside traditional machine learning methods like  \gls{svm} and \gls{xgb}, with performance comparisons made through person-independent 4-fold validation splits \cite{Cortes.1995, Chen.2016}. To handle the diverse data inputs, features were aggregated across frames and refined through feature selection, with class imbalance addressed by techniques such as \gls{smote} \cite{Chawla.2002}. Additionally, a CNN-LSTM model was finetuned to better adapt the feature extraction to our specific task, utilizing a two-step training process that initially freezes and then partially trains the \gls{cnn} to refine its understanding of mind-wandering-related features.

\begin{figure}[h]
  \centering
  \includegraphics[width=0.9\textwidth]{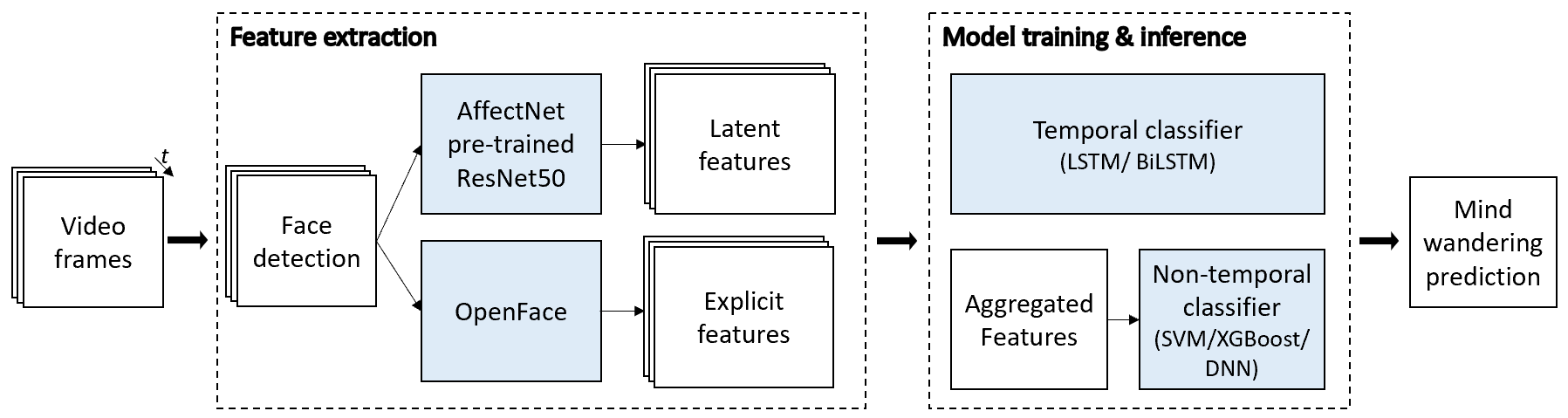}
  \caption{Mind wandering detection pipeline.}
  \label{fig_pipelinec}
\end{figure}

To evaluate the generalizability of our mind-wandering detection model across different settings and tasks, it was trained using the lab-based reading task data and then applied to predict mind wandering during a lecture in a naturalistic setting on the in-the-wild data. To gain a better understanding of the model, \gls{lime} \cite[]{lime}, an explainability algorithm to highlight the most informative image areas, was applied. Comparative analysis for prediction accuracies across genders was conducted to examine the fairness of predictions.

\subsubsection{Main Findings}


Our study demonstrated that latent features from a pre-trained \gls{cnn} on the AffectNet facial expression dataset effectively predict mind wandering, showing an approximately 14\% improvement over chance when integrated with deep learning models like \gls{dnn}, \gls{lstm}, and \gls{bilstm}. These transfer-learned features performed comparably to explicit feature sets (AUs, facial landmarks, gaze vectors) extracted via OpenFace, indicating the richness of information contained in the latent representations of basic emotional expressions for mind-wandering detection. Visualization of the model’s focus areas revealed significant encoding of eye and mouth regions, validating the relevance of these facial areas in recognizing mind wandering \cite{Bosch.2021}. 
Transfer learning can mitigate the issue of insufficient data for training deep learning models on this task. While fine-tuning these features in an end-to-end model slightly enhanced performance, the degree of improvement heavily depends on the quantity of available data.

When applying these models to a new dataset that varied in task, setting, and cultural background, the latent features generalized effectively, achieving results approximately 11.4\% above chance, surpassing performance on explicit features and indicating better generalizability without manual feature engineering. This suggests that facial expressions, being less task-specific than, for instance, gaze metrics, are more robust across diverse settings, including naturalistic environments \cite{Lee.2022, stewart2017generalizability}. Moreover, despite cultural differences in facial expression interpretation between Western and East Asian contexts, our transfer-learned features performed well across these diverse groups, underscoring their potentially universal applicability despite ongoing debates about the cultural specificity of facial expressions \cite{russell1994, ekman1994, jack2012facial, dailey2010evidence, Benitez2017, ali2020artificial}. 

Additionally, our findings indicated gender-based differences in prediction accuracy, with models performing better for females, possibly due to the higher proportion of women in the datasets and their higher mind-wandering rates, highlighting the importance of unbiased data sets. The selection of an optimal predictive model also proved critical; adjusting decision-making thresholds could enhance prediction efficacy, particularly in cross-dataset scenarios. This customization allows for a better balance between recall and precision, essential for tailoring applications to specific needs such as material testing or intervention deployment without disrupting learners who are already attentive \cite{Bosch.2021}.

The preceding sections have shown the potential of multimodal automated assessment approaches to detect mind wandering, i.e., lapses of attention during computer-based learning. However, the modalities discussed particularly eye tracking, can also be leveraged to examine behavioral indicators during attentive learning, which is explored in the following section.

\section{Synchrony as Attention Indicator during Online Learning} 

Turning towards the detection of attention during online learning the following section presents a summary of the investigation of gaze synchrony in attentive learners. 

\subsection{Synchronization of Attentive Learners' Gaze during Video Lecture Watching} \label{sync}

\begin{itemize}
    \item[{[3]}] \printpublication{buhler2024synchr}.
\end{itemize}

\subsubsection{Motivation}

Detecting attention to a given task based on behavioral indicators is challenging. Attention direction is primarily manifested through the mind-eye connection reflected in our gaze patterns \cite{just1976eye, rayner1998eye, reichle2012using}. Where we direct our gaze indicates where our attention lies, allowing our cognitive system to absorb and process information. In educational contexts such as lecture videos, one might infer attentiveness by whether individuals focus on currently relevant slide components being discussed. However, this inference greatly depends on the stimulus. To ascertain what is deemed relevant can be achieved by inferring information from the group level; attentive learners viewing the same content concurrently should exhibit similar eye movements, thus synchronizing their gaze \cite{Madsen.2021}. The forthcoming study investigated whether this synchronization truly occurs among learners in naturalistic attention fluctuation.

Recent studies have utilized visualizations of eye gaze data to enable instructors to gauge the attention levels of learners and adapt instructional strategies accordingly \cite{sauter2023, Kok.2023}. Furthermore, studies have examined the potential of synchrony in eye movements among students during video lectures as an indicator of attention levels and comprehension \cite{Madsen.2021}. These studies suggest that attentive students exhibit similar gaze patterns, which could be predictive of their learning outcomes as measured by test scores. However, attempts to replicate these findings using webcam-based eye-tracking have highlighted reliability issues, underscoring the challenges of implementing this technology in educational settings \cite{Sauter.2022, Liu.2023}. Moreover, previous research methods often involve experimental manipulations, such as introducing a secondary task to simulate inattention, which may not accurately represent the natural dynamics of learner distraction and engagement. Real-world learning experiences involve more subtle and varied attention fluctuations, influenced by both external distractions \cite{wammes2019} and internal cognitive processes like mind wandering \cite{lindquist2011}.

This study aims to deepen the understanding of how gaze synchrony relates to naturalistic fluctuations of attention and learning outcomes in educational settings. By examining the relationship between gaze synchrony measured by two distinct methodologies and self-reported attention during a pre-recorded Zoom lecture, it is investigated whether gaze synchrony can reliably indicate attention levels and predict learning outcomes. This approach allows us to assess the practical implications of using gaze data to enhance learning experiences in digital environments, moving beyond the constraints of controlled experimental setups to determine the real-time dynamics of learner engagement.

\subsubsection{Principal Methods}

For this study, data stemming from the same data collection as the data in publications \cite{buhler2024edpsych} and \cite{buhler2024mm} were employed. After exclusions, data from 84 university students aged 19 to 33 (19\% male) were used for analysis. Gaze data was collected using an SMI Red remote eye tracker in a laboratory setting where participants, after calibration and initial questionnaires, watched a 60-minute pre-recorded Zoom lecture on statistics, followed by a post-test. The video stimulus used in the study featured a typical Zoom layout with lecture slides and a webcam display of the lecturer, where the slides were primarily static, with a cursor used for pointing at specific locations. 15 experience sampling thought probes, administered every three to five minutes, were used to assess attention during the lecture; participants responded to a screen prompt asking about their current focus, with responses categorized from ''I was on task, following the lecture'' to ''Everyday personal concerns,'' allowing for analysis of attentive versus inattentive states. This method revealed that while 36\% of responses indicated attentiveness, most thoughts related to personal concerns or lecture content elaborations, with attention levels showing significant fluctuation throughout the lecture. Participants completed a pre-test assessing prior knowledge on the topic and a post-video knowledge test featuring 14 questions—seven on factual memory and seven on deep understanding—tailored to the lecture content on linear regression analysis, scoring an average of 5.63 out of 14.

To extract eye movement events such as fixations, saccades, and blinks from gaze data, utilizing the BeGaze software. A 10-second window cut before each of the 15 probes was used for synchrony analysis, yielding 1335 instances. The average calibration error of our nine-point calibration procedure was 0.31°, ensuring precise gaze data for analysis. blinks over 500 ms \cite{castner2020, schiffman2001} were excluded and gaze sequences with less than a 75\% tracking ratio, which, due to quality thresholds, reduced the dataset to 785 instances from 84 participants. 

To assess gaze synchrony, two primary measures were employed: the \gls{kld} and the MultiMatch scanpath comparison algorithm. The \gls{kld} measures discrepancies between gaze density maps by comparing gaze distributions, quantifying the degree of visual attention pattern differences between two sets of gaze data \cite{rajashekar2004, tatler2005, le2013methods}. Additionally, the MultiMatch algorithm evaluates scanpath similarities in terms of shape, direction, length, position, and duration, offering a comprehensive assessment of gaze alignment \cite{jarodzka2010, dewhurst2012, foulsham2012}. Additionally, the \gls{isc} was calculated as a baseline, aggregating data on gaze position and pupil diameter \cite{Madsen.2021, sauter2023, Liu.2023}. These measures were applied to 10-second video segments preceding attention probes and analyzed to discern the synchrony between groups classified as attentive or inattentive based on self-reported attention, with results standardized across video sequences.

The relationship between gaze synchrony and attention self-reports was analyzed. To address multiple measurement points per participant, mixed linear regression was employed, treating participant ID as a random effect and self-reported attention as a fixed effect, with the probe number included as a categorical variable to account for variability. In a separate analysis, the potential of gaze synchrony measures as indicators for attention was investigated by correlating these measures with learning outcomes. Average gaze synchrony scores and the proportion of on-task self-reports for each participant were calculated. Linear regression analyses were employed to examine the relationships between these aggregated values and post-test scores, adjusting for pre-test scores to control for prior knowledge. 
This approach allowed us to examine how well attentional states are reflected in synchronized gaze behavior and to assess the effectiveness of gaze synchrony as an indicator of attention by analyzing its relation to learning outcomes.

\subsubsection{Main Findings}

\begin{figure*}[h]
  \centering
  \includegraphics[width=0.9\linewidth]{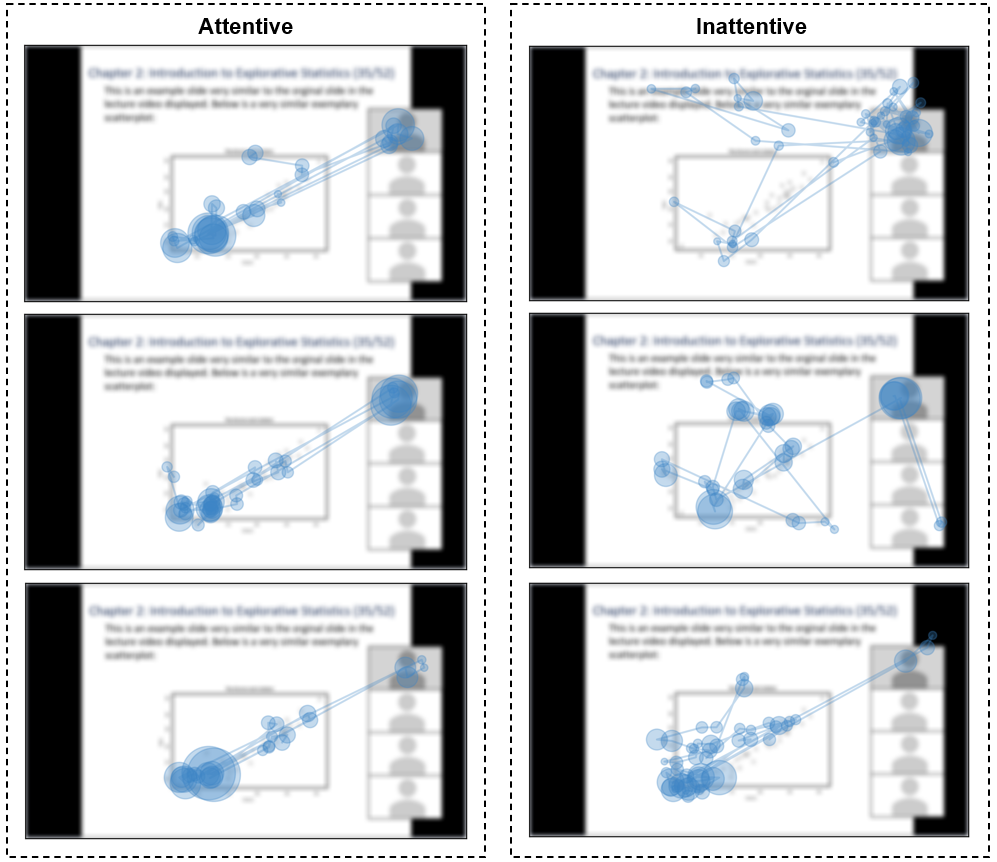}
  \caption{Example scannpath visualizations of one 10-second video sequence by self-reported attention.}
  \label{fig:spsync}
\end{figure*}

This study investigated the relationship between gaze synchrony and self-reported attention during video lectures, revealing that participants who reported being attentive exhibited higher gaze synchrony. Although the differences in gaze synchrony by self-reported attention were statistically significant, they were relatively small in magnitude, suggesting a nuanced relationship between natural attentiveness and gaze alignment during educational tasks. Figure \ref{fig:spsync} shows example scanpaths for self-reported attentive and inattentive episodes. Interestingly, of all MultiMatch dimensions, the position similarity—how similarly participants focused their gaze—showed the strongest link to reported attentiveness, emphasizing the importance of specific visual engagement with content as a component of cognitive engagement \cite{dewhurst2012, foulsham2012}. In contrast, saccade length similarity, which measures the movement distance between fixations, did not correlate significantly with self-reported attention, suggesting that the extent of eye movement does not necessarily reflect attentiveness \cite{foulsham2012}.

Despite these insights, the study did not find that average levels of gaze synchrony were predictive of post-test scores, which contrasts with previous research suggesting gaze synchrony could reflect learning effectiveness \cite{Madsen.2021, Liu.2023}. However, self-reported attentiveness did correlate significantly with learning outcomes, affirming its importance as a predictor of educational success \cite{Sauter.2022}.

These findings highlight that while gaze synchrony provides valuable insights into attentional engagement, its direct linkage to learning outcomes remains unclear. The variation in gaze synchrony's predictive power might be influenced by the type of video content used in the study, which featured more traditional, slide-based lectures rather than dynamic or interactive content \cite{Sauter.2022, zhang2020}. This suggests that the educational context and the nature of the video material might significantly affect gaze behavior and its synchrony. Further research is needed to investigate these relationships more deeply, particularly considering different types of educational videos and extending the analysis to longer time windows to potentially enhance the robustness of gaze synchrony measures in predicting learning outcomes.

This section has described the examination of gaze synchrony as an indicator for attention during lecture video watching. The following section will focus on the detection of behavioral indicators in a different learning setting, the classroom.

\section{Hand-Raising as an Indicator of Behavioral Engagement in the Classroom} 

Transitioning to the complex environment of the classroom, assessing attention and engagement presents unique challenges. One difficulty arises from the reliance on observer ratings of higher-level constructs, which can introduce subjectivity, and a stronger need for inferences, becoming evident from low inter-rater reliability \cite{sumer2021multimodal} as detailed in Section \ref{observer}. Such imprecise ground truth poses challenges for the training of machine learning models. One way to overcome this is to focus automated detection on specific indicators of the concept under consideration that are easier for human raters to annotate. Therefore this work concentrates on a specific and more distinctively observable indicator of behavioral engagement in the classroom: hand-raising \cite{boheim2020engagement}. The following section presents a novel approach to hand-raising detection and automated annotation in authentic classroom videos. 

\subsection{Occlusion-Roboust and View-Invariant Hand-raising Detection in Authentic Classroom Videos}\label{hr}

\begin{itemize}
    \item[{[4]}] \printpublication{buhler2023hr}.
\end{itemize}


\subsubsection{Motivation}
This study is motivated by the recognition that students' active participation, particularly through hand-raising, plays a critical role in attention and engagement and, consequently, academic success within classroom environments. Hand-raising is a visible measure of participation and engagement, linked to students' achievement, cognitive engagement, emotional support from teachers, and motivation \cite{boheim2020engagement, Boheim.2020}. Despite its importance, research on hand-raising has been limited by the labor-intensive and costly process of manual observation, which restricts sample sizes and reduces the generalizability of findings \cite{Boheim.2020,boheim2020engagement}. Advancements in computer vision and machine learning now offer the potential to automate the detection of hand-raising events, thereby facilitating more extensive and less intrusive studies of classroom interaction. The goal of this research is to develop a robust automated approach that can reliably detect hand-raising activities in classroom videos. This approach aims to substitute manual annotation methods, enabling large-scale research applications and providing a more efficient tool for assessing student engagement. 

To address the complexities of real-world classroom environments, such as diverse filming angles and potential occlusions by classmates or objects, this study introduces a novel approach that employs view-invariant and occlusion-robust techniques. By focusing on body pose estimations \cite{ YuTe.2019} rather than directly employing video frames \cite{Si.2019, Ahuja.2019}, this method also upholds student privacy by eliminating the need to store sensitive video data. Ultimately, this research not only aims to validate the effectiveness of automated hand-raising detection but also investigates its correlation with cognitive engagement and other learning-related activities. This approach could significantly enhance our understanding of the dynamics of classroom participation and its impact on learning processes, setting the stage for future research that can utilize these automated techniques.

\subsubsection{Principal Methods}
In this study, data gathered from 36 real-world classroom sessions at a German secondary school involving 127 students across various grades (5-12) and subjects was employed. Each session was video-recorded and followed by a student questionnaire to assess learning activities in the lesson, including self-reported involvement \cite{frank2014presence}, cognitive engagement \cite{rimm2015extent}, and situational interest \cite{knogler2015situational}, resulting in 323 student-lesson instances. Hand-raisings were manually annotated by two raters with high inter-rater reliability (ICC = 0.96), resulting in 2442 events recorded. Constructing the automated hand-raising detection model required detailed annotations. For this purpose, hand-raising in half of the data, 18 videos, was annotated using the VIA software \cite{dutta2019via}, noting the start and end times and bounding boxes with a joint agreement probability of 83.36\% for two annotators. The annotations from both raters were combined to create ground truth by intersecting their temporal data, resulting in 1584 hand-raising instances with an average duration of 15.6 seconds.

\begin{figure}[h]
\centering
\includegraphics[width=\textwidth]{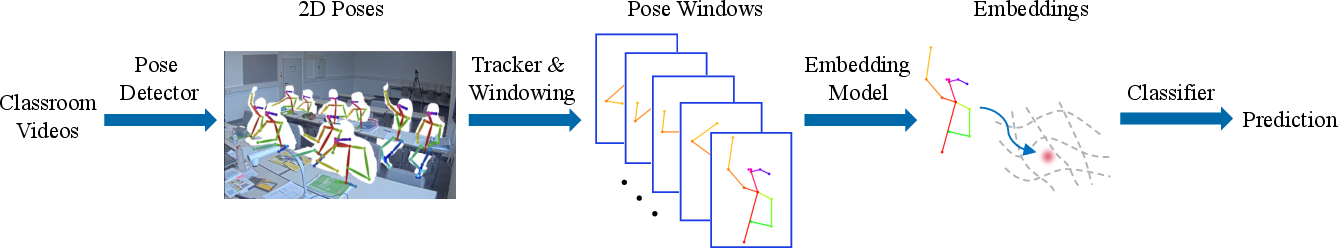}
\caption{Hand-raising detection pipeline.} \label{fig:pipelinehr}
\end{figure}

To develop a skeleton-based hand-raising detector, the OpenPose \cite{cao2019openpose} library was used to estimate 25 body key points per student in each frame, focusing on 13 upper body key points that represent the head, torso, and arms due to frequent occlusions of the lower body in classroom settings. To track students' movements over time, an intersection-over-union (IOU) tracker was applied based on the torso key points. This process created skeleton tracks for each student across frames. The 18 videos that were annotated for both temporal duration and spatial location of hand-raising events constituted the dataset for classifier training and testing. Annotations were segmented into subsequences labeled as ''hand-raising'' or ''non-hand-raising'' and were further broken down into 2-second windows without overlap. This process generated a large-scale dataset comprising 243,069 instances, with approximately 5\% (12,839 instances) representing hand-raising events, highlighting the challenge of imbalance in the dataset.

To address challenges related to viewpoint changes and partial occlusions, an approach by \citet{liu2022prvipe} to develop view-invariant and occlusion-robust pose embeddings was adapted. These embeddings are probabilistic, forming a Gaussian distribution that accurately captures ambiguities in 2D pose projection from 3D, enhanced by synthetic keypoint occlusion for training. The training, conducted on the Human3.6M dataset\cite{h36m_pami}, focused on upper body key points to optimize the embeddings for classroom settings, resulting in a combined 64D feature vector of mean and variance for downstream classification. To predict hand-raisings \gls{lstm} models performing binary classification of sequential inputs, using sequences of 2-second frame-wise feature vectors. Alongside \gls{lstm}s, also non-temporal \gls{rf} models for baseline comparisons were trained, adjusting for imbalanced data during training. Model performance was assessed independently across different videos, using a set for training and another for testing to ensure the robustness and generalizability of our detection approach. The detection pipeline is displayed in Figure \ref{fig:pipelinehr}.

In the last step, the proposed hand-raising detection method was applied to estimate the frequency of hand-raisings for students across the 22 unseen classroom videos. Using a sliding window technique on tracks generated for each student, hand-raising occurrences were annotated by applying the trained classifier on pose embeddings and merging adjacent detections. This approach allowed us to robustly detect and count hand-raisings and correlate them to self-reported learning-related activities.

\subsubsection{Main Findings}

Our study introduced a novel automated detection method for hand-raising using view-invariant and occlusion-robust pose embeddings, which proved more effective than simpler geometric features previously used~\cite{lin2021student}. Further temporal \gls{lstm} models outperformed shallow Random Forrest classifiers. Temporal models based on body pose embeddings reached an F$_{1}$ score of 0.76. Analysis of misclassified instances showed that similar skeletal positions, such as scratching the head or resting it on a hand, are often mistakenly identified as hand-raising. In contrast, more subtle gestures like a raised index finger or low hand raises were more frequently not recognized as hand-raising. 

Furthermore, this work confirmed that student hand-raising significantly correlates with self-reported learning activities across a broad range of subjects and grades, supporting its use as an indicator of classroom engagement. The automated instance-wise annotation of hand-raisings resulted in a \gls{mae} of 3.76, with students raising their hands on average 6.10 times per lesson. It was possible to improve the performance to a \gls{mae} of 3.34 when the analysis was limited to data, allowing for consistent student tracking and addressing issues like complete occlusions and pose estimation failures. Despite overestimations in individual hand-raising counts, our automated hand-raising instance annotations aligned closely with manual methods in correlating hand-raising to learning-related activities, cognitive engagement, situational interest, and involvement, demonstrating its utility for large-scale educational research. 

This method not only enhances the feasibility of large-scale studies by eliminating the labor-intensive process of manual coding but also offers a robust, privacy-preserving tool, adaptable to various classroom environments without the need to store video data. Future improvements could focus on enhancing pose estimation accuracy and incorporating hand key point detection to refine hand-raising identification, especially in subtle instances.

%% file: 4_discussion.tex
This chapter discusses the findings and contributions presented in the previous Chapter \ref{contributions} and aligns them with the objectives of the thesis outlined in Section \ref{objectives}. The central aspects are data-driven theory development \cite{buhler2024edpsych, buhler2024synchr} and the use of various modalities for objectively and non-intrusively assessing attention, including eye tracking \cite{buhler2024edpsych, buhler2024mm, buhler2024synchr}, video \cite{buhler2024mm, buhler2024general, buhler2023hr}, and physiological data \cite{buhler2024mm}. Moreover, advanced machine learning strategies based on transfer and deep learning \cite{buhler2024general, buhler2023hr} were applied to address challenges in authentic, in-the-wild educational environments and limitations in model selection due to restricted dataset sizes. This is followed by a discussion of challenges encountered in the automated assessment of attention-related assessment, a review of the potential applications of such methods and implications for educational practice, as well as a summary of resulting ethical implications. Finally, an outlook on potential future research is given. 

\section{The Potential of Automated Assessment of Attention-Related Processes During Learning}

This dissertation significantly advanced the current state-of-the-art in automated assessment of attentional processes. The concepts employed are strongly rooted in educational theories and contributed to an improved understanding of these processes using objective and fine-granular data derived from multiple modalities. By developing and refining automated methods for detecting aware and unaware mind wandering, gaze synchrony, and hand-raising, this thesis enhanced the capacity to assess and understand how students engage in learning and how this is linked to learning outcomes.

The first part of this thesis advanced the assessment of \textit{mind wandering} during learning along three dimensions: \textit{Fine-granular assessment}, differentiating by meta-awareness,  \textit{robustness and precision}, leveraging multimodality and \textit{generalizability}, improving cross-dataset predictions. 
The first study revealed complex associations of distinct thought patterns during lecture viewing with learning outcomes, highlighting a negative correlation between patterns dominated by unaware mind wandering and both fact-based and deep-level understanding, while persistent aware mind-wandering patterns specifically impaired deep-level understanding. These findings reinforce the need to investigate mind-wandering processes at this level of granularity and suggest that interventions should be customized based on a detailed understanding of meta-awareness and its manifestations. Further, it was demonstrated for the first time that gaze-based machine learning could differentiate aware and unaware minds wandering above chance. 
Advancing the accuracy and robustness of fine granular mind wandering assessment, a multimodal approach was employed that combined video, gaze, and physiological data. In line with previous findings \cite{kuvar2023}, the multimodal approach outperformed all unimodal approaches, demonstrating that different modalities can provide complementary information. \gls{shap} analysis further highlighted the importance of all three modalities for aware and, specifically, video-based facial expression features for unaware mind wandering detection. A novel transfer-deep-learning approach, fine-tuned to the mind-wandering detection problem, showed promising results for cross-dataset predictions. It overcomes restrictions set by often limited dataset sizes \parencite[e.g.,][]{Lee.2022}{} in education research and highlights the potential of advanced deep learning methods to generalize across different settings, specifically in-the-wild learning tasks and culturally diverse, potentially global user groups.

The second part of this thesis advances the scientific understanding and assessment of attentive gaze behavior in online learning environments by rigorously investigating  \textit{gaze synchrony}. Overcoming previous limitations of experimental attention manipulation \cite{Madsen.2021} using self-reports and analyzing three eye-tracking synchrony measures, the findings support the hypothesis that attentive learners exhibit similar eye movements \cite{Kok.2023, Madsen.2021, sauter2023}. However, synchrony scores were not significantly associated with learning outcomes, contrary to the self-reports of attention, underscoring the complexity of using synchrony as a direct indicator of attention.

The third part of this thesis targeted attention-related processes in real-world classroom environments by focussing on the automated assessment of an observable indicator of behavioral engagement \textit{hand-raising}. Using fine-tuned view-invariant, occlusion-robust pose embeddings, overcoming challenges such as occlusions by peers and various camera angles \cite{lin2021student}, and temporal models, capturing temporal dynamics, this method could substantively improve hand-raising detection in authentic classroom videos. 
It could further show promising results when employing those modes for automated annotations of hand-raising instances in a new set of videos. They showed high correlations with self-reported cognitive engagement and involvement, similar to manual annotations. This supports the hypothesis that hand-raising is a significant indicator in this context \cite{boheim2020engagement} and emphasizes the potential of automated annotation tools for large-scale educational research. 

Overall, this thesis leveraged substantive and methodological synergies (an expression coined by \citet{marsh2023synergies}) and contributed to data-driven theory development, showing the discernability of aware and unaware mind wandering based on objective measures \cite{buhler2024edpsych} and supporting the gaze synchronization hypothesis during attentive lecture watching \cite{buhler2024synchr}. Further, it examined the employment of a range of modalities to objectively and non-intrusively assess attention, including eye tracking \cite{buhler2024edpsych, buhler2024mm, buhler2024synchr}, video \cite{buhler2024mm, buhler2024general, buhler2023hr}, and physiological data \cite{buhler2024mm}. Additionally, advanced machine learning approaches based on transfer and deep learning were proposed \cite{buhler2024general, buhler2023hr} to encounter challenges posed by authentic, in-the-wild educational settings and restraints in model selection due to limited dataset sizes, advancing state-of-the-art research in this domain.

\section{Challenges for Automated Assessment}

This research underscores the significant potential of automated assessments in measuring attention but also delineates the prevailing limitations and challenges. The detection accuracies for the cognitive and behavioral constructs studied are rather moderate, which may be due to several potential reasons.
One of them being that for nuanced \textit{covert cognitive processes} such as aware and unaware mind wandering, there might exist limits to the precision with which these can be inferred from observable indicators. Computational models currently outperform human observers in recognizing these processes \cite{bosch2022}, which indicates the potential of automated methods yet highlights the complexity of detecting these cognitive phenomena.

Another major challenge is the \textit{reliability of ground-truth labels} used for training machine learning models. These labels are primarily derived from self-reports, necessitating interruptions of learners or requiring them to self-monitor their attention. Such interventions could potentially alter the learning process, although the extent of this impact remains unclear. Additionally, although temporal data is collected, pinpointing the exact onset of phenomena such as mind wandering is challenging. Moreover, from a theoretical standpoint, the measured constructs might not be distinctly enough defined in research on automated detection. The presented research introduces a more refined differentiation, demonstrating that aware and unaware mind wandering can be distinguished. However, as discussed above, wandering is a multidimensional construct and has potentially overlapping dimensions, such as intentionality. The reliance on observer ratings further complicates the accuracy of labels, evidenced by low inter-rater reliability \cite{sumer2018teachers}. Consequently, hand-raising was chosen as a more observable indicator. However, agreement between raters is still imperfect, leading to ambivalence regarding the exact start of a signal or the adequacy of certain gestures, such as only extending the index finger. These ambiguities introduce additional complexities in labeling for machine learning models.

Another major challenge faced is that datasets for detecting attention-related processes are typically smaller compared to those used to train machine learning models in other domains. This arises from the labor-intensive and costly data collection process via self-reports or observations, which is discussed in Section \ref{tradi}. This \textit{limited dataset size} tends to lead to overfitting more quickly and does not provide the model with sufficient examples to detect rare patterns and weak signals in the data. Concurrently, it restricts the application of more complex models, such as deep learning and temporal models, which are too complex for the small amounts of data available. In an attempt to address this issue, for example, transfer learning \cite{buhler2024general, buhler2023hr} was employed. However, large-scale datasets suitable for transfer learning are unavailable for some modalities and tasks. In fine-grained mind-wandering studies, sensors with high sampling rates are utilized that capture extensive information within brief measurement intervals. However, the total number of measurement points remains too small to effectively use recurrent neural networks, which are designed to process this type of sequential data. Consequently, these signals are aggregated over time windows, which risks losing potentially important information.

Arguably, one of the biggest challenges, especially for the generalizability of results, is the limited \textit{data quality} of data collected in naturalistic settings. As shown in Section \ref{generalize} \cite{buhler2024general}, low data quality, for example, due to low luminance and reduced resolutions, may impair feature extraction and reduce prediction accuracies. Consequently, detection accuracies on lab-based data might not be directly transferable to naturalistic settings, highlighting the need for further research to enable real-world applications, as described in the next section.

\section{Applications and Implications for Educational Practice}

The proposed methods in this dissertation allow for continuous and unobtrusive assessment of attention, setting the basis for future quantitative, large-scale research on attentional processes during learning. They further enable research on temporal dynamics, moderators of attention, and the effectiveness of interventions. Specifically, the automated annotation of hand-raising in classroom videos could be meaningful in research on teacher-learner interactions, which is still limited and heavily relies on manual annotations.

Furthermore, the proposed automated detection techniques, based on low-threshold sensors, could be employed in attention-aware learning technologies to adapt to the learner's attentional state in real-time. This would enable the customization of learning content and targeted interventions, enhancing intelligent learning systems' ability to support self-regulated learning effectively. Such interventions could include feedback mechanisms, content adaptation, and interactive elements like quizzes, which have traditionally been generic \cite{Kane.2017, Szpunar.2013} but can now be tailored to the learner's attentional state. Initial studies on real-time interventions triggered during instances of automatically detected mind wandering, such as during reading or within an \gls{its}, have demonstrated promising outcomes for enhancing long-term retention and comprehension \cite{hutt2021, Mills.2020}, despite relying on imperfect prediction accuracies. The nuanced recognition of, for instance, specific mind-wandering states allows for interventions aimed explicitly at combating frequent zone-outs by enhancing metacognitive strategies or alleviating persistent tune-outs by adjusting the difficulty and engagement level of the content. 

Given the current limitations in predicting attentional states and the risk of false positive detections, interventions should focus on non-disruptive methods. Nonintrusive interventions, such as follow-up prompts to review potentially missed learning content, help enhance self-regulated learning while minimizing distractions. Setting thresholds based on prediction confidence or duration could also improve the quality of interventions. 

However, the effectiveness of such measures requires thorough empirical research, and a comprehensive assessment of the impact of such technologies on education is needed. Recent studies suggest a trend where students overly rely more on provided \gls{ai} support rather than learning from it \parencite[]{darvishi2023}. Hence, it is paramount to give learners agency and responsibility over their own learning \cite{berger2014leaders, winne1995} in attention-aware learning environments. Further, the potential implementation of real-time attention detection necessitates consideration of ethics, learner privacy, and data use, as discussed in the following section.

\section{Ethical Implications}

Detecting attention-related processes offers promising potential for research on education and enhancements for educational technologies but raises significant privacy concerns. For instance, videos capturing students' faces require careful data management to protect privacy. Methods developed for automated attention detection should facilitate the real-time processing of features, thus eliminating the need to store sensitive data, which is crucial when dealing with vulnerable groups, such as minors, who stand to benefit from attention detection technologies such as \glspl{its}. Privacy-preserving techniques previously developed for face recognition \cite{Erkin.2009}, gaze estimation \cite{Bozkir.2020}, and classroom contexts \cite{Suemer.2020} provide a framework for the safe deployment of attention detection systems. Another practical strategy involves employing federated learning systems \cite{Li.2020, wang2024turbosvm}, where machine learning models are trained locally on user devices. This method keeps sensitive data private by only sharing the trained models with a central server for aggregation rather than the data itself. Ensuring privacy by design is paramount, and any data processing must be fully transparent to users, with clear communication about how and why data is used \cite{nguyen2022ethical}. The tools developed should not be employed to monitor students. Informed consent is essential, ensuring that students are fully aware of how their data are used and have the autonomy to opt out \cite{nguyen2022ethical}.

Further, ensuring Fairness, Accountability, Transparency, and Ethics (FATE) principles when employing \gls{ai} in education is crucial \cite{khosravi2024}. In this regard, the inclusiveness of data and algorithms needs to be ensured \cite{nguyen2022ethical}. To promote fairness and equality, it is essential for these algorithms to be developed and validated using diverse and unbiased datasets to prevent biases that could adversely affect underrepresented learner groups. Furthermore, the models used should be explainable \cite{nguyen2022ethical}, as this is essential in offering clear explanations and justifications for the decisions made by \gls{ai} systems \cite{khosravi2024}.


\section{Outlook}

Looking ahead, this research can expand in several promising directions. First, collecting larger datasets will enable the use of more complex models, potentially increasing the accuracy of detecting attention-related processes. Also, the initial use of transfer learning techniques leveraging related large-scale datasets of \gls{fer} and action recognition can be further expanded and explored. Moreover, current developments in the field of generative \gls{ai} open up new possibilities for data analysis. The rapid development of foundation models beyond large language models, expanding into various other domains such as vision \cite{wang2023internimage} and multimodal \cite{radford2021learning, team2023gemini}, combining text, image, and audio, along with their potential to analyze sequential data, could significantly enhance attention recognition efforts. Additionally, their potential future capability to generate synthetic supplementary training data, as already utilized, for instance, in medical research \parencite[e.g.,][]{huang2023chatgpt}, could help overcome challenges associated with limited dataset sizes, offering promising improvements in model training.
In parallel, further investigation into the ground truth and the concepts that are being measured will enhance the validity and applicability of the findings. Employing the proposed detection techniques to study the temporal dynamics of attention can inform data-driven theory development, contributing to a more nuanced understanding of educational engagement over time. Further, detection methods should be evaluated based on datasets that are more diverse in target groups and settings to ensure fairness and generalizability. Moreover, research should investigate whether current levels of detection accuracy are sufficient to be integrated into adaptive learning technologies to support learners' self-regulated learning through targeted, non-intrusive interventions aimed at improving learning experiences and outcomes while ensuring fairness, accountability, transparency, and privacy. Also, in this context, generative \gls{ai} opens up new opportunities for advancing attention-aware educational technologies, such as developing optimally tailored interventions, automatically adapting and creating learning content, and providing personalized learning experiences \cite{kasneci2023chatgpt}. With the increasing integration of adaptive generative \gls{ai} in educational technology, the investigation of attention, self-regulation, and mind wandering in the collaboration between learners and \gls{ai} is moving to the center of attention.

%% file: 1_Meta-Awareness_MW.tex
\section[Temporal Dynamics of Meta-Awareness of Mind Wandering]{Temporal Dynamics of Meta-Awareness of Mind Wandering During Lecture Viewing: Implications for Learning and Automated Assessment Using Machine Learning}

\subsection{Abstract}
Remote learning settings require students to self-regulate their behavioral, affective, and cognitive processes, including preventing mind wandering. Such engagement in task-unrelated thoughts (TUTs) has a negative impact on learning outcomes and can occur with or without students’ awareness of it. However, research on the meta-awareness of mind wandering in education remains limited, predominantly relying on self-report measures that capture discrete information at specific time points. Therefore, there is a need to investigate and measure temporal dynamics in the meta-awareness of mind wandering continuously over time. 
This study examined the temporal patterns of 15 mind-wandering and meta-awareness probes in a sample of university students (\textit{N = 87}) while they watched a video lecture. We found that the majority (60\%) of mind wandering occurred with meta-awareness. Cluster analysis identified five distinct thought sequence clusters. Thought patterns dominated by unaware mind wandering were negatively associated with fact- and inference-based learning, whereas persistent aware mind-wandering patterns were linked to reduced deep-level understanding. Initial exploration into predictive modeling, based on eye gaze features, revealed that the models could distinguish between aware and unaware mind-wandering instances above the chance level (Macro F1 = 0.387). Model explainability methods were employed to investigate the intricate relationship between gaze and mind wandering. It revealed the importance of eye vergence and saccade velocity in distinguishing mind wandering types. The findings contribute to understanding mind-wandering meta-awareness dynamics and highlight the capacity of continuous assessment methods to capture and address mind wandering in remote learning environments. 

\subsection{Introduction}

Learner attention is crucial for learning and knowledge construction \parencite[]{levine1990}. However, when following a class or lecture, commonly 45 to 90 min, it can be challenging to sustain attention over a longer period. This task becomes even more demanding when shifted to remote learning settings, such as watching school lessons or university lectures online (either live-streamed or recorded) \parencite[]{Wammes.2017}. In such settings, learners receive less adaptive support from instructors to maintain attention. Students' experiences may vary depending on the lecture style and format, particularly in terms of opportunities for teacher-student interactions, discussions, or assessments of understanding, which are, for instance, impossible in asynchronous online formats. Thus, learners' ability to self-regulate their behavioral, affective, and cognitive processes becomes even more crucial. Besides learners being likely to be exposed to greater external distractions \parencite[e.g., media use][]{ Hollis.2016}, they are at higher risk of digression \parencite[]{Wammes.2017}.  This shift of attention away from the current task to task-unrelated thoughts (TUTs) is known in the literature as mind wandering \parencite[]{Smallwood.2006}. Learners tend to engage in mind wandering for about 30\% of the time spent in educational activities, such as lecture viewing or reading, \parencite{wong2022}. Overall, TUTs are significantly related to poorer test performance across tasks, topics, and age groups \parencite[]{wong2022} and are therefore a serious concern.

Whereas the exact nature and definition of mind wandering are still being debated \parencite{Seli.2018family, Christoff.2018}, several studies have shown that mind-wandering episodes can occur with and without meta-awareness, which means that they differ in whether a person is aware of their occurrence or not \parencite[]{Schooler.2002, Smallwood.2007}. However, the levels of meta-awareness and its temporal unfolding over time have not yet been studied extensively in research on education. These manifestations are likely related to different underlying mechanisms, which in turn require different interventions and solutions. It is therefore crucial to investigate these manifestations of mind wandering and their relationships with learning in educational settings.
One challenge in research is the reliance on self-report measures such as probe-caught or self-caught methods, which, although they provide robust and reliable assessment, can capture only discrete moments in time \parencite{Weinstein.2018}. However, a promising way to achieve unobtrusive, continuous, and systematic observations of thought processes over time during realistic learning scenarios is to use machine learning on sensor data \parencite[e.g.,][]{Hutt.2019,Mills.2020, Brishtel.2020}. 
Especially gaze behavior, due to its property of reflecting cognitive processes \parencite[]{just1976eye, rayner1998eye, reichle2012using}, has been shown to be indicative of episodes of mind wandering \parencite[e.g.,][]{Hutt.2017, bixler2014}.
Recent advancements have predominantly considered mind wandering to be a unitary state. 
However, based on the premise of distinct cognitive mechanisms underlying aware and unaware TUTs, discernible eye movements should manifest accordingly. Consequently, it is imperative to investigate the potential of eye-tracking-based computational modeling of aware and unaware TUTs in educational contexts.
A continuous, objective, and noninterruptive assessment would further allow researchers to investigate the temporal dynamics of meta-awareness during TUTs for a better understanding of underlying processes, their influence on learning, and effective interventions. 

In this study, we aimed to shed light on the meta-awareness and temporal patterns of mind-wandering episodes during a 60-min prerecorded Zoom lecture by employing 15 thought probes. Related to the dynamic nature of mind wandering, we examined the sequences of mind-wandering meta-awareness, as reported in thought probes, using clustering to identify typical patterns of thought. Furthermore, we investigated the impact of the unaware and aware mind-wandering patterns we identified on fact-based (e.g., memory for facts or details) and inference-level (e.g., integration of information with prior knowledge) learning \parencite[]{mccarthy2018}. 
To investigate the distinctiveness of forms of mind wandering based on eye tracking and potentially overcome a limitation of temporal analyses (i.e., their reliance on discrete self-reports), we chose to apply a predictive modeling approach that was based on eye-tracking features and machine learning. This approach can be applied to distinguish between on-task, aware, and unaware TUTs, thus offering promising opportunities to potentially assess and differentiate aware and unaware mind wandering and take a critical first step for future large-scale research. Employing posthoc explainability techniques, we aimed to gain deeper insights into the complex relationships of specific gaze indicators and mind-wandering types. Therefore, in this study, we uniquely explored meta-awareness in mind wandering during video lectures, unveiling temporal patterns and their association with learning outcomes while also pioneering predictive modeling using observable gaze behavior.

\subsubsection{Self-Regulation, Cognitive Control, and Mind Wandering}

Remote, asynchronous video learning requires learners to engage more intensively in self-regulated learning (SRL) because they are primarily responsible for their own learning. SRL can be seen as an expansive set of skills that enable learners to systematically initiate, sustain, monitor, and regulate their cognitive, motivational, behavioral, and affective states and processes in pursuit of their learning objectives \parencite[]{schunk2017}. Prominent models of SRL \parencite[]{winnie1998, boekaerts1996, efklides2011, greene2007, pintrich2004} describe the monitoring and control of learning activities, alongside the motivational and emotional processes that either foster or impede active involvement in these activities. The use of cognitive and metacognitive learning strategies, which are part of the skills that constitute SRL, such as the elaboration and self-monitoring of one's learning progress, has been shown to have a positive effect on learning outcomes \parencite[]{dent2016, jansen2019}. 

Furthermore, more low-level self-regulation abilities extending beyond the mere utilization of strategies, such as executive functions and attention control in working memory, have also been shown to be highly relevant for successful learning in academic settings \parencite[]{rutherford2018, titz2014}. Accordingly, recent studies have tried to bridge cognitive load theories and self-regulated learning frameworks \parencite[]{debruin2017}. Moreover, more fine-grained aspects of SRL, for instance, fundamental executive functioning \parencite[EF;][]{friedman2017, miyake2000}, enable individuals to regulate their thoughts and behaviors in alignment with goals by maintaining, manipulating, or inhibiting the contents that require the focus of attention and occupy limited working-memory resources \parencite[]{broadway2010, krumpe2018}. Consequently, ensuring proper mental resource investment is a crucial part of regulating cognitive processes to sustain attention during a learning task. In this vein, one theory of mind wandering suggests that thoughts compete for limited resources in working memory capacity and depend on available executive resources \parencite[]{Smallwood.2006}. According to the executive control failure hypothesis proposed by \textcite{kane2012}, the occurrence of mind wandering is considered to be indicative of both momentary failures of and enduring deficiencies in executive-control functions. This account is supported by the finding that lower working-memory capacity (WMC) is associated with more frequent off-task thoughts during demanding tasks \parencite[]{mcvay2012wmc}. 

\subsubsection{Meta-Awareness of Mind Wandering}

Definitions of mind wandering employed in research vary in content, intentionality, task-relatedness, and relationships to external stimuli \parencite[]{Seli.2018family}. The most widely adopted definition of mind wandering is a shift of attention away from the current task to TUTs \parencite[]{Smallwood.2006}. However, mind wandering is a heterogeneous, multidimensional construct rather than a dichotomous state, the definition and conceptualization of which has been the subject of ongoing controversy in the literature \parencite{Seli.2018family, Christoff.2018}. On the one hand, \Textcite{Christoff.2018} criticized the use of mind wandering as an umbrella term for disparate mental phenomena. Instead, they proposed a dynamic framework in which the defining feature of mind wandering is that it arises and proceeds in an unconstrained and relatively free fashion. On the other hand, \Textcite{Seli.2018family,Seli.2018b} advocated for a family resemblance view under which several manifestations of the phenomenon fall, including, for instance, perseverative and purposeful TUTs. Instead of restricting the concept more strongly, from their point of view, the types of thoughts under consideration have to be precisely defined in the respective research. In this work, we adopted the most widely used definition of mind wandering as TUTs, which might vary in terms of meta-awareness \parencite[]{Smallwood.2006, Smallwood.2007, Schooler.2011}.

The extensively reported experience of catching oneself in the act of mind wandering, also exploited when using self-caught mind-wandering probes, indicates that engaging in TUTs often happens without awareness \parencite{Smallwood.2007}. This lack of awareness suggests that mind wandering is associated with a temporary lack of meta-awareness, which refers to the ability to reflect on the content of one's basic consciousness and its alignment with one's goals \parencite[]{Schooler.2002}. \textcite{Smallwood.2007} defined mind wandering with and without awareness as two different states and referred to them as tune-outs and zone-outs, respectively, which unfold dynamically over time and can also transition into each other. It is important to note that the dimension of meta-awareness is distinct from the dimension of intentionality \parencite[]{seli2017}, which describes whether engagement in TUTs happens deliberately/intentionally or spontaneously/unintentionally \parencite[e.g.,][]{Seli.2013, Smallwood.2004, smilek2010}. \textcite{seli2017} reported that whereas the two dimensions might overlap at the ignition of a mind-wandering episode, as awareness and unawareness are prerequisites for intentional and unintentional engagement in TUTs, respectively, they might fluctuate in opposite directions as the episode continues. 

Being aware of the digression of one's thoughts is the prerequisite for being able to catch one's mind wandering and terminating an episode to redirect one's attention back to the primary task. However, studies have shown that people also allow their mind-wandering episodes to continue after becoming aware that an episode has occurred \parencite[]{seli2017}. This finding leads to the assumption that the two manifestations of mind wandering (i.e., unaware and aware mind wandering) are based on different cognitive processes. 
Consistent with this account, studies have shown that unaware mind wandering is more strongly associated with response inhibition failures \parencite[]{Smallwood.2007, smallwood2008b} and reading comprehension deficits \parencite[]{smallwood2008}. Furthermore, the two states are neurologically dissociable, showing different brain activity patterns \parencite[]{Christoff.2009}. These results suggest that the two manifestations of mind wandering have different implications for task performance and different underlying cognitive processes. 

When individuals are aware that their mind is wandering, they should consequently be able to self-catch and terminate the episode. However, as individuals have reported aware mind wandering in thought probes, meaning that they gained meta-awareness before being interrupted by the probe, people seem to allow their minds to wander \parencite[]{seli2017}. 
This phenomenon could indicate that participants, who are learners in this context, might not be fully attempting to focus on the primary task. Therefore, they might not be trying to ignore distractors \parencite[]{seli2017}, especially when those distracting thoughts have a great deal of personal importance for them \parencite[]{scheiter2014}. Learners might consequently then end up pursuing a competing goal. Such pursuit of a competing goal is especially relevant in an experimental setting but might still hold in real-world learning contexts. 

An alternative account posits that when aware mind wandering potentially overlaps with intentionality, it may be considered a deliberate allocation of cognitive resources away from the primary task rather than a failure of control \parencite[]{jefferies2008}. For example, aware mind wandering may serve as a regulatory mechanism for managing affective states, such as lack of motivation or heightened frustration during task performance \parencite[]{Risko.2012}. The effect of motivation on lecture retention has been shown to be mediated by TUTs \parencite[]{Seli.2016}. Additionally, TUTs have been shown to reduce the negative effect of a boring task on participants' moods \parencite[]{baird2011}. 

In contrast to episodes of aware TUTs, individuals engaging in unaware TUTs do so without conscious awareness. In terms of typical SRL frameworks, one could argue that this state signifies, at least temporarily, a suspension of the (successful) monitoring of one's own cognitive processes (which would mean catching oneself in the act of mind wandering), thereby rendering control and adjustment (i.e., redirecting attention to the primary task) unfeasible. Consequently, we argue that the two manifestations of TUTs can best be understood as representing distinct facets within the domain of self-regulation during learning. Aware TUTs potentially indicate deficiencies in regulating motivational or affective states or signify the pursuit of conflicting personal goals. Conversely, unaware TUTs may signal deficiencies in the monitoring and regulation of cognitive processes.

\subsubsection{Mind Wandering While Viewing a Lecture} 

A recent meta-analysis by \textcite{wong2022} demonstrated that the frequent occurrence of TUTs during learning is significantly associated with lower test performance and explains about 7\% of the variability in learning outcomes. This negative relationship holds equally for fact-based and inference-level learning and is consistent across learning tasks and topics (i.e., STEM vs. non-STEM). For instance, mind wandering has been shown to have a negative effect on reading comprehension~\parencite[][]{Smallwood.2011, Feng.2013, dmello2021, bonifacci2022} and lecture retention in classrooms \parencite[]{Lindquist.2011} as well as video-based, asynchronous online-learning settings~\parencite[][]{Risko.2012, Szpunar.2013, Hollis.2016, Pan.2020}. A comparison of mind-wandering rates during a live versus a video lecture by \textcite{Wammes.2017} revealed that the technology-mediated form of a video lecture presentation led to more off-task thoughts in students. Remote live lectures, potentially offering remote lecturer support, represent a critical intermediate synchronous format between in-person and video lectures. However, limited research has been conducted in this area. Whereas the effect sizes that have been reported are nontrivial, \textcite{wong2022} argued that not all TUTs are uniformly bad, and the effect might partially stem from the heterogeneity of off-task thoughts. Most of the studies covered by \textcite{wong2022} captured mind wandering as a unitary state and did not look at more fine-grained dimensions, such as thought content. Another limitation of previous research on the relationship between TUTs and learning outcomes was the focus on fact-based memory representations rather than inference-level learning outcomes \parencite[]{wong2022}.

The content of mind wandering varies with task type and is therefore very context-sensitive \parencite[]{Faber.2018b}. In educational contexts, open-ended thought probes have revealed a large proportion of thoughts that were stimulus-independent but task-related \parencite[i.e., not to the here and now of the lecture, but to lecture-related content or thoughts about metacognitive lecture comprehension][]{locke1974, schoen1970}. Such thoughts might be confounded with TUTs when mind wandering is assessed in a binary fashion. Studies by \textcite{Kane.2017, Jing.2016} distinguished different types of off-task thoughts by their content and showed that thoughts about lecture-related topics were positively related to lecture retention, as they might indicate cognitive strategies, such as elaboration, an integral part of SRL. By contrast, TUTs (i.e., lecture-unrelated thoughts, e.g., personal concerns or daydreams) were negatively associated with learning outcomes. Whereas \textcite{Kane.2017} included task-relevant cognition (e.g., lecture-related and comprehension-related thoughts) in their mind-wandering definition, in this study, we employed the more common, more restrictive definition of TUTs.

With regard to the intentionality of mind wandering, which has been studied more extensively in the educational context, diverging results have been presented in the literature. In a study asking about intentional and unintentional mind wandering during a video lecture, both forms were negatively associated with lecture retention and mediated the influence of motivation on retention \parencite[]{Seli.2016}. Other investigations have revealed that intentional mind wandering was negatively associated with short-term academic performance in the form of in-class quizzes, whereas unintentional mind wandering affected long-term learning outcomes manifested in course grades \parencite[]{Wammes.2016}.
However, the meta-awareness of mind wandering has not received much attention in educational contexts, and—to our knowledge—has not yet been considered in lecture viewing. Differentiating between different types of mind wandering in learning situations is crucial for understanding their associations with learning success. Whereas studies on simple attention tasks indicate a stronger influence of unaware mind wandering on performance deficits \parencite[]{Smallwood.2007}, and research on reading comprehension indicates a poorer mental model \parencite[]{smallwood2008}, their occurrence and implications in the context of video-lecture watching have yet to be investigated. To develop appropriate interventions and to improve learning materials accordingly, it is necessary to distinguish between different manifestations of mind wandering with diverging underlying processes, one of them being the dimension of meta-awareness: Learners who engage in mind wandering without noticing it due to failed executive control require a different kind of support than learners who allow themselves to dwell on TUTs because they already know the learning content. To properly support learners in their SRL, we need to understand the mechanisms that underlie why they lose focus in specific learning situations. 

Several studies have shown that mind-wandering rates increase with time on task \parencite[]{zanesco2024}, for example, during lecture viewing \parencite[]{Risko.2012}. Regarding lecture viewing as a sustained attention task, two accounts can be considered for an explanation. As a prolonged sustaining of attention occupies resources for executive control, the likelihood of executive control failure should also increase with time on task \parencite[]{warm2008,warm2018}. Other underlying reasons for an increase in mind wandering could be decreased motivation and increased frustration as a function of time \parencite[]{Risko.2012}. In this context, it is particularly interesting to look at the temporal dynamics of aware and unaware mind wandering during the continuation of a lecture and to investigate whether both forms increase to the same extent over time. In addition, it could be intriguing to delve into the distinct temporal patterns that each form exhibits over time. Analyzing these typical patterns and temporal unfoldings may offer insights into how the awareness of mind wandering evolves throughout a lecture. Such an analysis will be especially interesting, given the theoretical account that these patterns might transition into each other \parencite[]{Smallwood.2007}. Furthermore, it may uncover whether the two forms, aware and unaware mind wandering, follow similar trajectories or exhibit different progression rates throughout the task. This investigation into the temporal evolution of types of mind wandering could provide valuable nuances in the understanding of their cognitive underpinnings and differences in occurrence patterns during sustained attention tasks, such as lecture viewing.

\subsubsection{Gaze Indicators of Mind Wandering}

So far, meta-awareness in mind wandering has largely been measured only with the help of explicit thought probes. This approach entails the repeated interruption of students at irregular time intervals throughout a lecture, followed by inquiries about where their attention is presently focused \parencite[]{Weinstein.2018}. However, another approach is to employ observable behaviors to make inferences about a person's cognitive state. One data source that is frequently used to assess mind wandering is eye movement data. The approach builds upon the \textit{mind-eye-link} concept, which suggests that eye movements can provide insights into underlying cognitive processes \parencite[][]{just1976eye, rayner1998eye, reichle2012using}.
Mind wandering is described as a state in which attention is decoupled from external stimuli, and internal thoughts are prioritized over the processing of external information \parencite[]{Schooler.2011}. Thus, mind wandering could potentially hinder visual processing by depleting executive resources \parencite[]{mcvay2012wmc} or by diminishing the visual \parencite[]{smallwood2008c, baird2014decoupled, kam2011} and cognitive \parencite[]{barron2011} processing of information.
Global gaze features, such as fixations (periods of stable gaze) and saccades (rapid eye movements between two consecutive fixations), local features that characterize the spatial properties of gaze, as well as pupil size and blink properties, have been employed extensively in research aiming to predict attentional states \parencite[e.g.,][]{bixler2014, mills2021, Hutt.2017, Hutt.2019}.  
These studies have supported the idea that mind wandering appears to be reflected in gaze. However, no specific set of gaze behaviors that are indicative of mind wandering can be defined. Thus, inconsistent results have sometimes been reported in empirical studies. \textcite{Faber.2020}, for example, argued that the observed diversity can be attributed to task-inherent properties that require different levels of spatial allocation and visual processing, representing a compensatory shift in the functioning of the visual system during instances of mind wandering. Hence, we emphasize studies that focused on a lecture task in our subsequent review of the established correlations between gaze characteristics and mind wandering. 

Previous studies have revealed that the average duration of participants' fixation on slides increased during mind wandering \parencite[]{Zhang.2020, Jang.2020}. This extended duration implies a potential decrease in visual processing efficiency, likely due to participants requiring more time to process information at each location before transitioning to the next one \parencite[]{Zhang.2020}. Increased fixation durations are also related to increased mental workload \parencite{degreef2009}, in line with findings on the relationship between longer fixation durations and task difficulty~\parencite[]{Pomplun.2013}. 
According to the decoupling hypothesis of mind wandering impairing the processing of external information \parencite[]{Schooler.2011, smallwood2008c}, this gaze behavior could express that the learner is gazing at slides without processing the corresponding perceptual information \parencite[]{Zhang.2020}.
Accompanying the longer fixation durations, the dispersion of fixations across the slides decreased during mind-wandering episodes, signifying that the fixations were restricted to a smaller part of the screen \parencite[]{Zhang.2020, Jang.2020}. Contradictory results were found by \textcite{Faber.2020}, where mind wandering was associated with a larger fixation dispersion and no varying fixation durations during lecture viewing. 
The investigation of saccades (i.e., rapid eye movements between two fixations) by \textcite{Faber.2020} revealed that mind wandering was associated with larger saccade amplitudes during a lecture task. Contrasting results were found by \textcite{Jang.2020}, indicating smaller saccade amplitudes and a lower total number of saccades, with lower saccade peak velocity during mind wandering. 

Further research is needed to get to the bottom of these sometimes contradictory results on global gaze features. However, it is noteworthy that the presentation of the lectures in the previous studies was fundamentally different. Whereas \textcite{Faber.2020} used a recording of an in-person lecture in a lecture hall, depicting slides projected onto a wall and a lecturer in one frame, the stimulus employed by \textcite{Zhang.2020} consisted of two online lectures, which comprised a screen recording of the slides and a facial video of the lecturer. This discrepancy highlights the importance of the differences in visual processing even within similar tasks with different stimuli \parencite[]{Faber.2020}.

When analyzing the frequency of fixations in specific areas of interest (AOIs) during video lecture watching, \textcite{Zhang.2020} found that learners increasingly looked at the instructor's image during mind wandering. When employing a predictive modeling approach, \textcite{Hutt.2017} found that employing grid-based locality gaze features did not improve the prediction accuracy of mind wandering above global gaze features, such as fixations, saccade durations, and saccade amplitude.

Additionally, previous research has examined a range of oculometric characteristics, such as blinks. 
A decreased blink count was identified for self-reported mind-wandering episodes during lecture viewing \cite{Jang.2020}. 

Another oculometric indicator, the pupil diameter, was found to be significantly larger during mind-wandering episodes when participants watched a lecture video \parencite[]{Jang.2020}. Moreover, pupil dilation is associated with a high cognitive processing load \parencite[]{kahneman1966} and emotional demands \parencite[]{partala2003}. Additionally, increased pupil diameter and its associated standard error were observed when participants provided incorrect responses in working memory tasks, indicating a momentary diversion or inattention to the ongoing task \parencite[]{smallwood2011pupillometric}. Yet, it is known that pupillary responses change according to the brightness of the environment and stimulus as well, making it challenging to identify whether the measured changes were due to the cognitive states of the participants or environmental conditions~\parencite[]{John.2018}. 
Another gaze measure that was found to improve the prediction of internal thought and was employed as an indicator in a computational modeling paradigm is eye vergence \parencite[]{huang2019}. Vergence measures the rotation of both eyes either inwards or outwards, thereby capturing the phenomenon of the visually relaxed state of not fixating on anything, referred to as tonic vergence \parencite[]{toates1974vergence}.

Part of the heterogeneity found in gaze indices of mind wandering could be attributed not only to the varying idiosyncratic task-processing demands but also to the heterogeneity of mind wandering itself. The gaze patterns associated with mind wandering might also differ across subdimensions, such as whether somebody is aware of their mind wandering \parencite[]{Faber.2020}. Given this inherent heterogeneity in the relationship between gaze features and mind wandering, we investigated gaze indices through the lens of the meta-awareness of mind wandering during lecture viewing. In line with previous findings, we included previously employed fixation-, saccade-, blink-, and pupil-derived feature groups to investigate their relationships with the meta-awareness of mind wandering, thereby contributing to the state of the art.

\subsubsection{Prediction of Mind Wandering Using Machine Learning}

Several studies have employed physiological sensors, such as eye-tracking \parencite[e.g.,][]{mills2016} or electrodermal activity (EDA) data \parencite[e.g.,][]{Brishtel.2020}, in predictive modeling approaches to make inferences about a unitary mind-wandering state and potentially allowing researchers to measure it non obtrusively and continuously over time. These approaches diverge from traditional statistical analyses in two key dimensions. First, they differ in their primary objective, focusing on prediction rather than explanation (see \textcite{yarkoni2017} for a more extensive discussion). Whereas traditional models aim to explain the relationship between variables and outcomes, machine-learning models prioritize finding the most suitable model for predicting future observations \parencite[]{yarkoni2017}. This priority is evident in the typical practice of dividing data into training and test sets to train the algorithm on one set and evaluate it on unseen data. 
The second dimension of differentiation lies in the modeling paradigms: data-centric versus algorithmic \parencite[]{breiman2001}. In the former paradigm, models are chosen on the basis of prior assumptions, potentially leading to conclusions that are based on the model's mechanism rather than on actual data patterns. By contrast, machine learning employs an algorithmic modeling approach by exploring multiple models and selecting the one that best fits the data based on performance metrics. This property makes machine-learning models particularly appealing when there is a need to handle many features, where the intricate interplay and direct association with the target variable cannot be hypothesized in advance, aligning with the scenario of sensor-based mind-wandering prediction \parencite[]{dmello2022comp}. Please refer to \textcite{dmello2021} for a brief tutorial on computational modeling for mind wandering.

However, machine learning models are often referred to as ”black box models” due to the lack of transparency in how these models make decisions \parencite{yang2023xai}. In educational settings, specifically learning analytics research, when applying machine learning to create adaptive and personalized interventions, concerns related to Fairness, Accountability, Transparency, and Ethics (FATE) are increasingly being debated, leading to a focus on Explainable Artificial Intelligence \parencite[XAI;][]{khosravi2022explainable}. Explainability methods can help ensure the transparency and fairness of machine learning models by clarifying their internal processes and providing clear reasons for their outcomes \parencite{yang2023xai, arrieta2020explainable}. These explanations are crucial for building trust and confidence in the models used. Black-box machine learning models can be explained using post-hoc explainability methods. This involves an additional step in the analysis pipeline: after model training, a separate algorithm explains the model’s decisions. This offers deeper insights into how specific features influence predictions \parencite{roscher2020}, helping us understand the relationship between input variables and outcome, i.e., attentional states. The interpretability of the provided explanations, however, relies on human-interpretable input variables. 

The automated assessment of mind wandering has been investigated based on a range of physiological and behavioral modalities, including electrodermal activity \parencite[]{Blanchard.2014} or facial videos \parencite[]{Bosch.2021}. However, gaze is the most prominent sensor and has been employed in the predictive modeling of mind wandering across a diverse range of learning tasks. These tasks encompass activities such as reading \parencite[][]{bixler2014, dmello2016, Faber.2018b, mills2021}, and Intelligent Tutoring System (ITS) \parencite[][]{Hutt.2019}.
Only a few works have investigated the predictive modeling of mind wandering with gaze behavior while learning with video lectures \parencite[][]{Hutt.2017, Zhao.2017}. 
\textcite{Hutt.2017} employed two sets of eye-tracking features to predict probe-caught mind wandering while participants watched a lecture video. The first set included global gaze features, such as fixation and saccade duration. The second set of local gaze features consisted of saliency-based AOI fixation information. 
Model performance is evaluated by employing the F1 score, which is a measure in machine learning that combines the precision (accuracy of positive predictions) and recall (how many actual positives were predicted correctly) into a single number, which is helpful for evaluating how well a model performs in finding the target outcomes, such as mind wandering. This approach is particularly valuable when dealing with unbalanced data (please refer to the "Mind Wandering Assessment by Meta-Awareness Employing Gaze Features" section for detailed information). 
The best results were achieved (F1 = 0.47) by employing global gaze features in a Bayesian network classifier, constituting a 24\% improvement above chance (F1 = 0.3). 
In an attempt to develop a more scalable, online detection of mind wandering, \textcite{Zhao.2017} compared predictions based on gaze features from an eye tracker with webcam-based gaze-tracking features. They used 6 - 8-min-long lecture videos as learning material. The best results, 11\% above chance (F1 = 0.29), were achieved when predicting self-reported mind wandering employing webcam-based global gaze features with a Naive Bayes classifier (F1 = 0.4).
A recent study by \textcite{bixler2021crossed}, who investigated the potential for cross-domain predictions of mind-wandering models trained on one task to predict mind wandering in another, also included a lecture-viewing task. For the within-data-set prediction based on global gaze features, they achieved a 21\% (F1 = 0.57) improvement above chance (F1 = 0.45).

However, to our knowledge, existing approaches are limited to the binary detection of mind wandering (i.e., mind wandering vs. not mind wandering) and cannot make more fine-grained distinctions. In general, the automated recognition of mind wandering offers the potential to assess mind-wandering states continuously over time and consequently to study and understand the temporal unfolding of meta-awareness and TUTs over longer periods of time without employing intrusive thought probes. This new approach would allow us to study mind wandering in ecologically valid educational settings, such as schools or homes, and to collect large-scale data. 

Additionally, it would provide new opportunities for effective interventions and adaptations of learning content in attention-aware learning technology \parencite[]{Mills.2020, hutt2021}. Nevertheless, these kinds of interventions can be designed more effectively if the different mind-wandering states and their possible relationships with learning are considered adequately.

\subsubsection{The Present Study}

In this study, we investigated metacognitive awareness of mind-wandering episodes during lecture viewing and their patterns over time. We aimed to disentangle TUTs from other stimulus-independent thoughts by employing a thought-content probe, followed by a meta-awareness probe. We investigated the person-specific sequences of thought self-reports to investigate the temporal unfolding of meta-awareness in TUTs throughout a video lecture. Additionally, we employed a clustering approach in these sequences to identify distinctive temporal patterns that each form exhibits. We furthermore studied how the identified patterns were related to fact-based and inference-level learning outcomes. This approach constitutes a novel addition to the existing literature, as prior studies have predominantly concentrated on a binary differentiation between being on task and TUTs and have motivated the main contribution of the paper: employing observable behavioral measures for the continuous assessment of meta-awareness in mind wandering.

To this end, we investigated the relationships between these subdimensions of mind wandering with behavioral eye gaze measures by employing eye tracking. Research has highlighted the significance of eye gaze as a valuable objective indicator of cognitive processes such as mind wandering. Its intricate and multilayered nature, coupled with the absence of clear linear associations, advocates for the adoption of more complex algorithms within computational models to enhance the prediction of mind wandering \parencite[]{dmello2022comp}, specifically when distinguishing it via meta-awareness. Identifying these differences, derived from objective gaze behavior measures, can enhance and progress the theory by distinguishing between these two states of mind wandering by collecting data-driven evidence. The adoption of explainability techniques for machine learning models enabled a detailed examination of the intricate connections between eye gaze and meta-awareness of mind wandering.
The present study addressed the following research questions:

\begin{enumerate}
    \item What different kinds of mind wandering along the dimensions of thought content and meta-awareness can be observed during video-lecture viewing?
    \item Can typical patterns of aware and unaware mind-wandering sequences be identified over the course of time?
    \item How do different patterns of aware and unaware mind wandering affect fact-based learning and inference learning during lecture viewing? 
    \item Can eye-tracking-based machine learning models differentiate between aware and unaware mind wandering to achieve a continuous measurement of these states?
\end{enumerate}

\subsection{Methods}

The ethics committee of the Faculty of Economics and Social Sciences, University of T\"ubingen (Date of approval January 13th, 2022, approval \#A2.5.4-210\textunderscore ns) 
approved our study procedures, and all participants gave written consent to the data collection. 

\subsubsection{Participants}
In this study, we collected data from \textit{N} = 96 university students. Six participants had to be excluded due to technical errors during the study and three because they did not speak German at the native level, leading to a final sample of \textit{N} = 87 participants for analyses. They were between 19 and 33 years old (\textit{M} = 23.44, \textit{SD} = 2.6), and 19\% of the participants were male. Additionally, we assessed information about participants' subjects of study, with the three predominant groups being psychology, teacher training, and languages, along with details about the study year (\textit{M} = 2.88, \textit{SD} = 1.7), which was coded numerically from 1 to 8, corresponding to the year of university study.

\subsubsection{Procedure}
\begin{figure}[h]
    \centering
    \caption{Zoom Lecture Video Layout.}
    \includegraphics[width=0.6\textwidth]{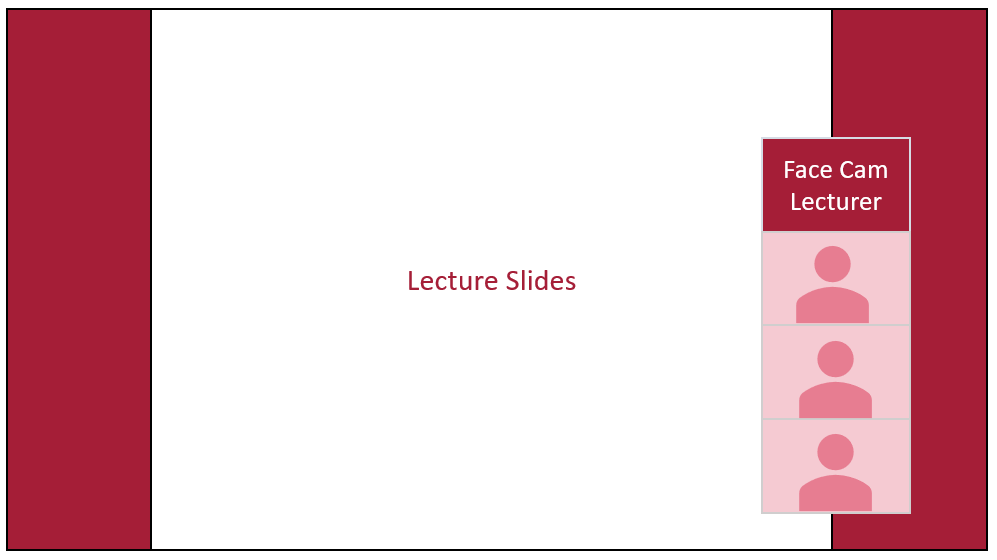}
    \label{fig:stimuluslayout}
\end{figure}

First, participants were asked to complete questionnaires on individual prerequisites and a pretest on the topic of linear regression analysis. We then began recording gaze data with a brief 9-point pulsating calibration of the remote eye tracker. Participants then watched a prerecorded video of a real-world lecture held over Zoom, namely, "Statistics 1." The students could see the lecture slides and a webcam image of the lecturer's face on the right, as shown in Figure~\ref{fig:stimuluslayout}. The slides incorporated a mix of text, formulas, and simple visuals (i.e., scatterplots) typically seen in standard lecture formats. The total duration of the video lecture was approximately 60 min. We included 15 quasirandomized thought probes that appeared on the screen in 3- to 5-min intervals. The video was paused after approximately 30 min to recalibrate the eye trackers. After the session, participants were asked to complete a knowledge test on the topic of the session and a questionnaire on situation-specific variables. Along with the time for the general instructions, the total time came to about 120 min per participant, for which they were compensated with 20€.

\subsubsection{Instruments}

\begin{figure}[h]
    \caption{Two-Stage Mind Wandering Probe.}
    \includegraphics[width=\textwidth]{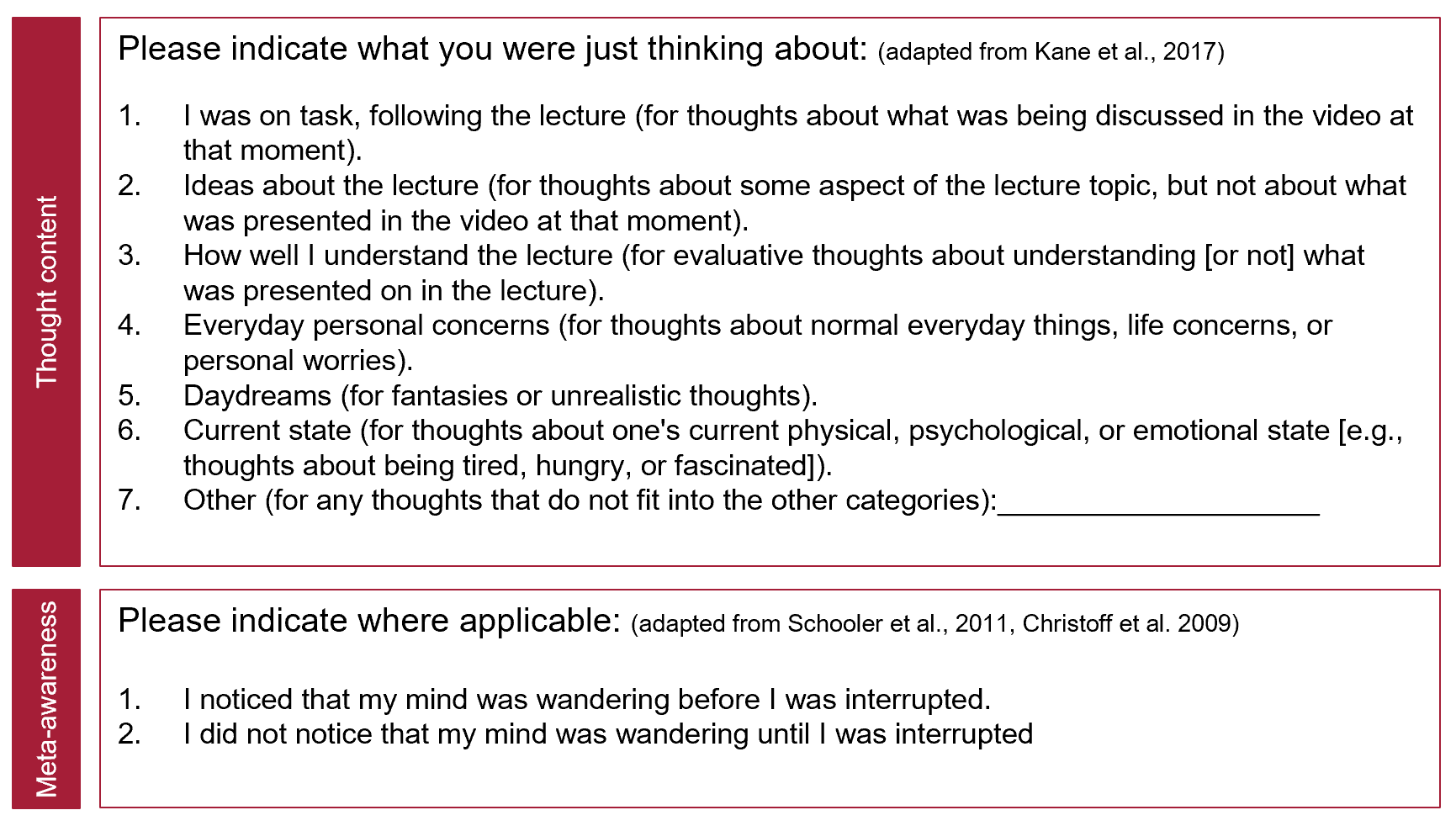}
    \label{fig:probe}
\end{figure}

\paragraph{Mind-Wandering Probes}
Participants' mind wandering was measured with the probe-caught method, interrupting participants at quasirandomized points in time to ask them for mind-wandering self-reports. The 60-min lecture was interrupted by 15 of these thought probes at intervals ranging from 3 to 5 min. This repeated assessment over a long period allowed us to draw a quasicontinuous picture of the time course of the mind wandering over a lecture. Because we were interested in the thought content of mind wandering and whether participants were aware of their TUTs, we administered a two-stage probe at each interruption. We first asked participants to indicate what they were thinking about and offered six answer categories adapted from \textcite{Kane.2017} plus a seventh open-ended answer option. The exact probes can be found in Figure \ref{fig:probe}. 

\begin{figure}[hbt!]
    \caption{Content Categories for Coding Open-Ended Answers.}
    \includegraphics[width=\textwidth]{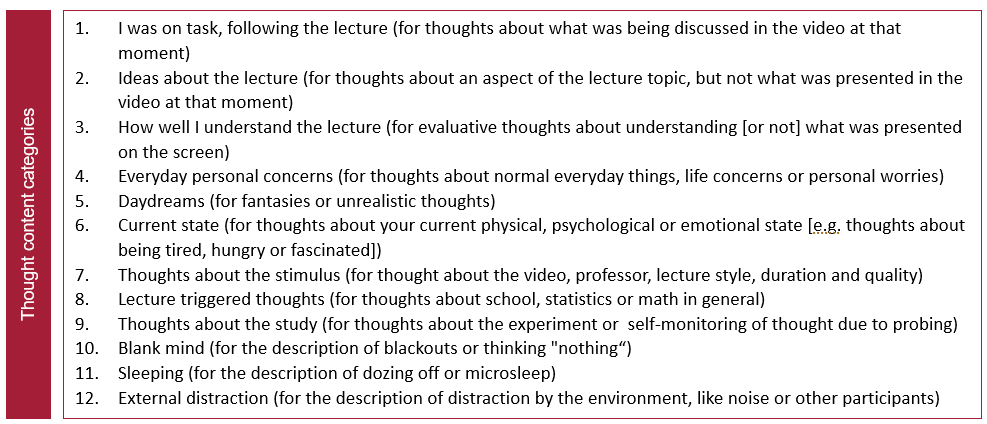}
    \label{fig:fin}
\end{figure}

In accordance with the definition of mind wandering as TUTs, we defined Categories 3–6 as mind wandering. For further analysis, we additionally combined Category 2, "task-related thought" (TRT), with Category 1, "on task," because Category 2 was comparatively small (\textit{n} = 107) and because it represented elaborations to the lecture content, which is very closely related to Catergory 1. Answers to the open-ended category (\textit{n }= 117) were independently manually coded by two raters in an iterative approach in which new categories were defined and then the answers were assigned to the categories. The inter-rater reliability (Cohen's $\kappa = .64$) was limited. This may be due to the fine-grained approach to coding involving a large number of 12 categories (see Figure \ref{fig:fin}). When disagreement occurred (\textit{n }= 39), the raters discussed their coding results to reach a consensus. In this process, four additional TUT categories were identified: thoughts about the video stimulus (e.g., quality of the lecture), thoughts about the study (e.g., when the next interruption will happen), lecture-triggered but TUTs (e.g., lectures in one's own curricula), and blank mind. Few participants reported being distracted by their environment (\textit{n }= 10) or falling asleep (\textit{n }= 2). Those answers were allocated to an "other" category. In a second step, participants were asked to indicate, if applicable, whether they were aware that their mind was wandering before they were interrupted (adapted from \textcite{Schooler.2011, Christoff.2009}; see Figure \ref{fig:probe}). The answer to the second question was then utilized to divide the previously defined TUT category into aware and unaware TUTs. Following these guidelines resulted in five categories: on task, lecture comprehension (MM), aware TUTs, unaware TUTs, and other.

\begin{figure}[h]
    \caption{Fact-Based and Inference Example Questions From the Posttest.}
    \includegraphics[width=\textwidth]{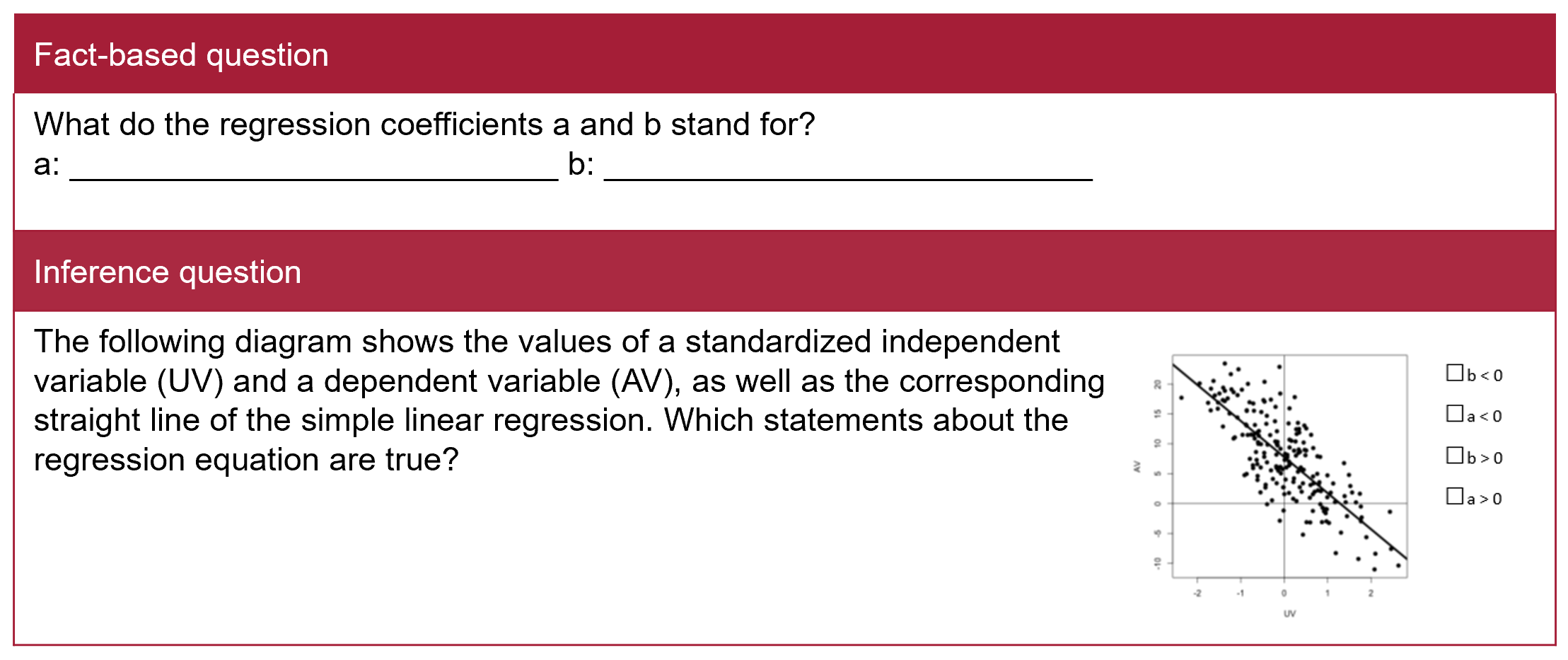}
    \label{fig:test}
\end{figure}

\paragraph{Learning Outcomes}
The postvideo knowledge test we administered consisted of 14 questions that tested participants' understanding of the video lecture content in the form of seven fact-based memory and seven inference questions. The test was designed specifically for the prerecorded lecture on the topic of linear regression analysis and included questions that referred to topics such as empirical covariance, method of local averaging, and least squares estimation. Examples of fact-based and inference questions are presented in Figure \ref{fig:test}. The sum scores we calculated for each participant ranged from 0 to 14, with each accurately answered question contributing 1 point. The test's internal consistency was acceptable (Kuder and Richardson Formula 20 $(KR-20) = .72$). We computed a sum score due to the independent nature of the individual questions on the test, because the test itself was not designed to evaluate a unified competency but rather to assess knowledge of specific facts and concepts presented at distinct points in the 60-min video. Consequently, the sum score of correct answers could potentially serve as a quantitative measure that captured participants' levels of attentiveness over time during the lecture video. 

\paragraph{Scales and Participant Characteristics}

As a person-specific characteristic, \textit{self-concept in statistics}, referring to one's self-perception of competencies and skills in this field of study, was assessed with five items (e.g., "Some topics in statistics are just so difficult that I already know beforehand that I won't understand them"; $\alpha = .83$; adapted from \cite{marsh2006}). We assessed \textit{metacognitive self-regulation} with a 12-item subtest from the MLSQ (e.g., "When I'm reading for a course, I think of questions to help me concentrate while reading"; $\alpha = .67$; \cite[]{pintrich1991manual, pintrich1993, duncan2005}). \textit{Dispositional mind wandering} was assessed with the five items from the mind wandering questionnaire (e.g., "I have difficulty concentrating on simple or repetitive tasks"; $\alpha = .75$; \cite[]{mrazek2013}). We measured \textit{dispositional interest} in statistics with four items (e.g., "I like statistics"; $\alpha = .94$; \cite[]{gaspard2017}). Additionally, \textit{situational interest} was measured with the situational interest scale (seven items, e.g., "Today's lecture captured my attention"; $\alpha = .86$; \cite{knogler2015}). As a further situation-specific variable, we measured involvement with four items (e.g., "During the lecture, I focused heavily on the situation";  $\alpha = .68$; \cite[]{frank2014}). \textit{Cognitive engagement} was measured with six items (e.g., "I tried as hard as I could during the session"; $\alpha = .69$; \cite[]{rimm2015}). We assessed \textit{emotions} with the seven items from the emotion scale by \textcite{pekrun2017} (e.g., "How intensely did you experience the feeling of "enthusiasm" during the lecture?"). Because the internal consistency for the negative emotion subscale was too low, we included only the positive emotion subscale in our analyses (positive emotions: $\alpha = .69$, negative emotions: $\alpha = .46$).

We assessed previous knowledge of the lecture topic before the lecture started with a pretest consisting of 8 general questions on the topic of regression analysis. This test covered a wide range of topics in statistics and questions of varying difficulty. We created the sum score for the pretest, ranging from 0 to 8, with every correctly answered question scoring 1 point. Further, we asked for demographic information on age and gender as well as information on study subject and study year (refer to the Participants section for detailed information).

\paragraph{Eye Tracking} \label{eyetracking}

While participants viewed the lecture, we additionally collected participants' eye movement data by employing remote eye trackers (SMI, 250 HZ sampling rate) that were attached to the bottom of the laptop screens on which the lecture video was watched. We employed analysis software BeGaze provided by SMI to extract eye-movement events, such as fixations (dispersion-based threshold) and saccades (velocity-based threshold) from the raw gaze data \parencite[]{begaze}. Fixations are defined as stable eye movements across a certain area in the stimulus, whereas saccades are high-speed eye movements that reflect movement of the eyes from one fixation location to another. After detecting fixations and saccades, we then extracted the information right before a thought probe, choosing a 10-s time window derived from the related literature on mind-wandering detection. 

As we used this data to train a machine learning model, we had to ensure high data quality, as poor-quality eye tracking data, characterized by excessive noise or invalid measurements, has the potential to compromise or entirely invert the outcomes \parencite[]{dunn2023minimal}. We employ quality criteria of data loss, expressed in tracking ratio and gaze offset, expressed in angular distance between fixation location and intended fixation target \parencite{holmqvist2017} in a validation calibration at the end of the study. Data quality may be influenced by participant-specific factors such as wearing glasses or mascara \parencite{holmqvist2017, carter2020best}. Concerning spatial accuracy \textcite{holmqvist2017} report that many studies employ an accuracy of 0.5°. A high spatial precision is of less importance when stimulus-independent measures, like in this study, are used. Therefore, we excluded participants with less than 1.0° accuracy during a validation test at the end of the videos exhibiting large gaze offsets and, consequently, invalid gaze measurements. In our study a total of 17 of 87 participants was affected, leading to the exclusion of 255 instances from predictive analysis. As we noted spectacle wearers in the experiment protocol, we were able to determine that half of these excluded participants wore spectacles during the experiment, whereas the overall proportion of glass wearers was 33\%. This further supports our assumption that invalid measurements and the associated exclusion are related to technical requirements of the eye tracker and participant criteria that are not systematically related to mind wandering itself. \textcite{holmqvist2017} assume an expected data loss of 3-10\% under perfect laboratory conditions. To balance high data quality and minimize exclusion we excluded instances with less than 70\% tracking ratio for the respective ten-second window, constituting a data loss of more than 30\%. This threshold led to the exclusion of 23 instances.

In addition, it is a known deficiency of the eye trackers we used that tracking failures are sometimes recorded as unusually long blinks \parencite{castner2020}. Therefore, blinks longer than 500 ms (i.e., those exceeding an expected blink duration range between 100 to 400 ms; \cite[]{schiffman2001}) were excluded. 
Following this criterion led to a total exclusion of 278 examples, resulting in a data set containing 1,027 examples, 622 of which were labeled as on task, TRT, or MM; 249 as aware TUTs; and 147 as unaware TUTs, based on the subsequent probe-based self-reports. Following the exclusion process, the general distribution across self-reported attentional states remained unchanged (on task: 61\%, aware TUTs: 24\%, and unaware TUTs: 15\%). This outcome suggests the absence of the systematic exclusion of crucial instances, particularly those related to mind wandering.

To create gaze features that would serve as input for our machine-learning model, we created summary statistics (i.e., minimum, maximum, mean, median, standard deviation, skewness, kurtosis, and range) for all eye-tracking events. Based on the event statistics, we created additional features, such as the number of fixations and blinks, fixation dispersion, pupil diameter, and vergence. Table \ref{tab:etfeatures} reports a complete list and description of the features we employed.

\begin{table}[h]
\caption{Eye Tracking Features.}
  {\setlength{\tabcolsep}{1.2pt} 
    \begin{tabular}{ll} 
\toprule
 Feature & Description\\
\midrule 
        \textbf{Fixation Count}         & Number of fixations \\
        \textbf{Fixation Duration}      & Duration of fixations in ms\\
        \textbf{Fixation Dispersion} & Square root of distances between individual fixations  \\ & to average fixation position in pixels \\       
        \textbf{Fixation Saccade Ratio} & Ratio between fixation duration and saccade duration \\
        \textbf{Saccade Duration}       & Duration of saccades in ms\\
        \textbf{Saccade Amplitude}       & Visual angle in degrees \\
        \textbf{Saccade Length}          & Saccade distance in pixels \\   
        \textbf{Saccade Velocity Average} & Average saccade length/saccade duration in °/s\\
        \textbf{Saccade Velocity Peak}   & Peak saccade length/saccade duration in °/s\\        
        \textbf{Saccade Acceleration Average} & Average derivative of saccade length/saccade\\ & duration in °/s²\\
        \textbf{Saccade Acceleration Peak} & Peak derivative of saccade length/saccade duration\\ & in °/s²\\ 
        \textbf{Blink Count}            & Number of blinks \\       
        \textbf{Blink Duration}        & Duration of Blinks in ms\\
        \textbf{Vergence}            & Angle between gaze vectors in degrees and                      
 distances \\ & between the pupil's position of the left and right eye\\ &in pixels\\ 
        \textbf{Pupil Diameter}        & Pupil diameter during fixations in mm, with \\ &  person-specific subtractive baseline correction \\ & (baseline length:  50 ms) \\ 

\bottomrule
\end{tabular}}
\label{tab:etfeatures}
\end{table}

\subsubsection{Analysis}

\paragraph{Mind Wandering Sequence Clustering}
To investigate the pattern of mind wandering and its association with meta-awareness over the course of a lecture, we created thought sequences based on the 15 thought probes administered during the lecture. Consequently, one person's sequence consisted of 15 time points, with information on whether their attention was directed to the task, they were thinking about understanding the lecture, or they engaged in aware or unaware TUTs. The resulting sequences yielded information on the temporal unfolding of meta-awareness in TUTs. These sequences were then clustered to identify distinct patterns the two forms of mind wandering might exhibit. We applied the agglomerative hierarchical clustering algorithm Agnes with the Ward method, frequently used for sequence clustering. We employed Optimal Matching (OM) to assess the similarity between two sequences on which the clustering algorithm based its clustering decisions. OM quantifies the similarity between two sequences by determining the minimal cost required to transform one sequence into the other \parencite[]{abbott1986}. This process involves the consideration of costs for insertion (the addition of an element at a specific position) and deletion (the removal of an element from a given position), collectively referred to as indel operations. Additionally, the analysis encompasses substitution operations, which entail replacing one element with another. We applied a constant cost of one for all aforementioned operations. The OM distance measure considers small time shifts in the sequences and is less sensitive to exact timing than other distance measures, thus rendering our measure particularly suitable for comparing thought probe sequences with an interest in general temporal unfolding.

\begin{figure}
    \centering
    \caption{Agnes Clustering Dendogram.}
    \includegraphics[width=0.8\textwidth]{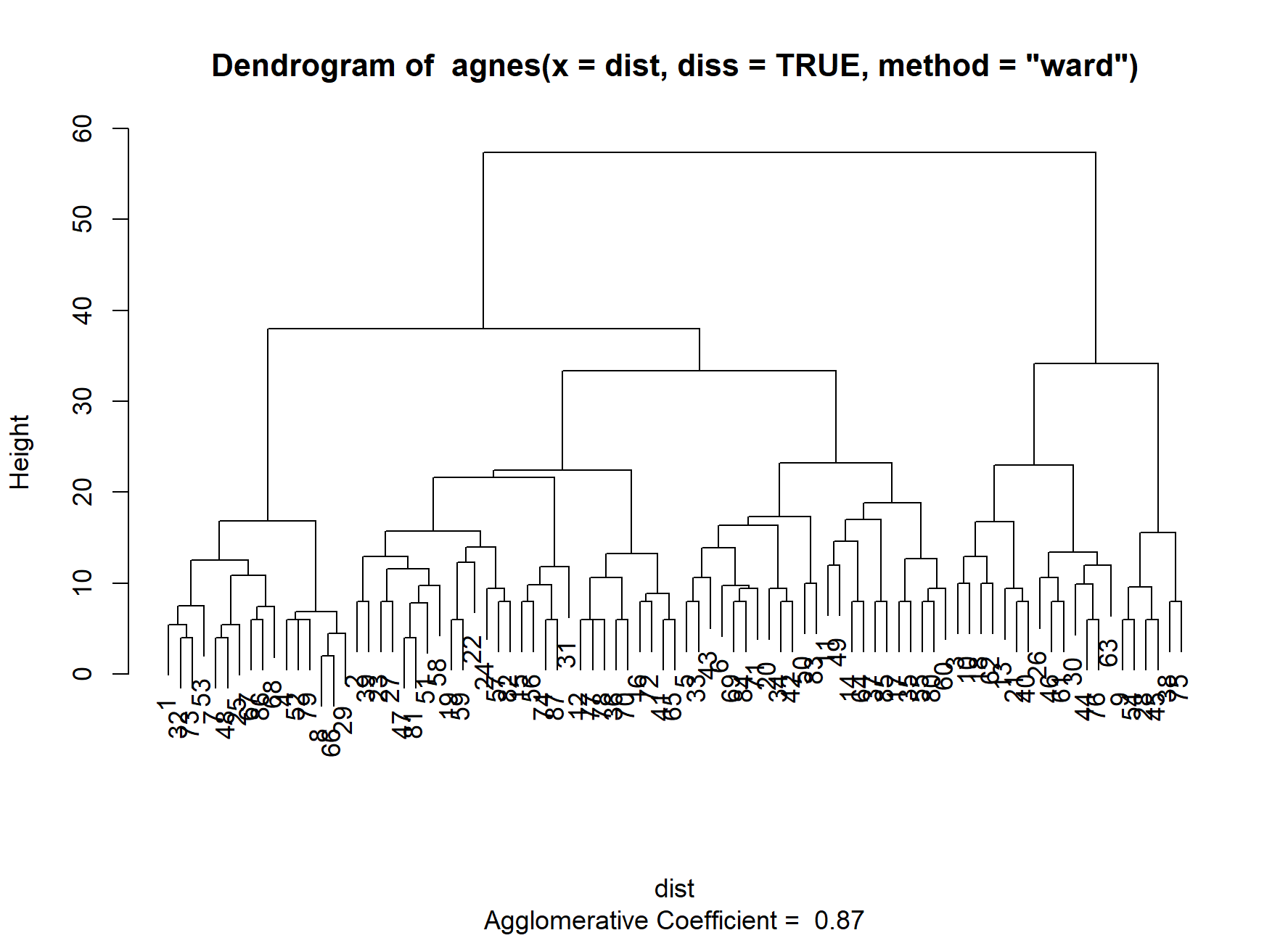}
    \label{fig:dendo}
\end{figure}

\begin{figure}
    \centering
    \caption{Average Silhouette Width (ASW), Hubert’s C Coefficient (HC), and Point Biserial Correlation (PBC) for 2-8 Clusters.}
    \includegraphics[ width=0.7\textwidth]{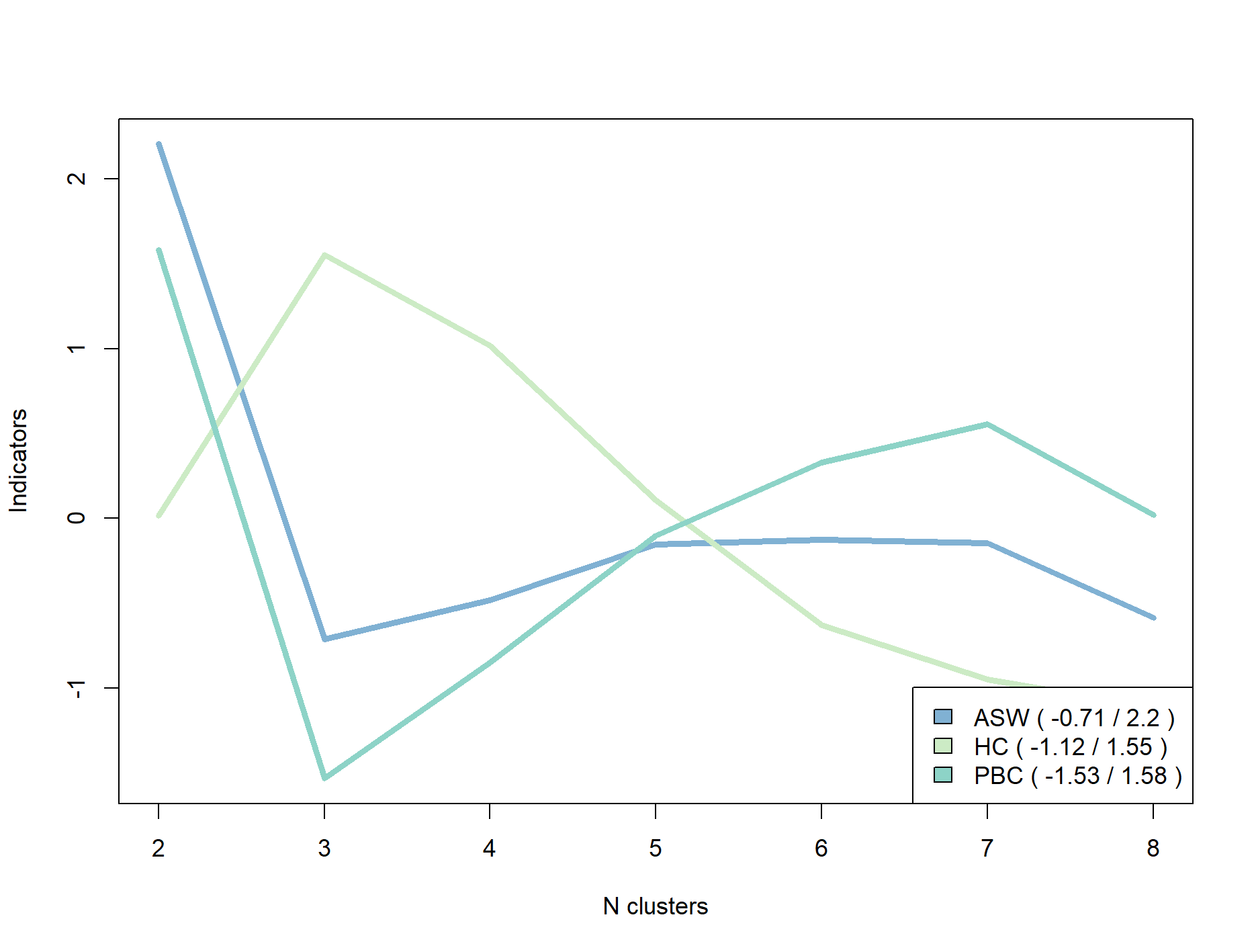}
    \label{fig:shil}
    
\end{figure}

We identified the optimal number of clusters by inspecting the cluster dendrogram (see Figure \ref{fig:dendo}), average silhouette width, Hubert’s C coefficient, and the point biserial correlation (see Figure \ref{fig:shil}). 
To investigate the differences between clusters, we compared the person-level proportion of each thought category by employing one-way multivariate analysis of variance (ANOVA) with Pillai’s Trace statistic and follow-up univariate Welch ANOVAs, using a Bonferroni-adjusted alpha level of .0125. We then conducted pairwise comparisons between the clusters using the Games-Howell post hoc test with Tukey’s studentized range distribution to compute the \textit{p}-values.

\paragraph{Predicting Learning Outcomes From Cluster-Belonging}

With the goal of examining the association between the meta-awareness of mind-wandering patterns and learning outcomes, we conducted a linear regression analysis. The post knowledge test total sum scores and the fact-based and inference sum scores served as independent variables for the analyses, whereas cluster-belonging was a categorical dependent variable. We additionally included person-specific characteristics, such as age, gender, previous knowledge, and self-concept in statistics in the model, as they might have a confounding impact on test scores. 
Dispositional interest in statistics, another potential confounder, could not be included due to its high collinearity with self-concept (\textit{r} = .57, \textit{p} < .01). 
 To compare the indications of our clustering approach, we additionally computed regression analyses to investigate the influence of thought category rates, summarized over the whole sequence and not entailing information about temporal dynamics, on posttest scores.

\paragraph{Mind Wandering Assessment by Meta-Awareness Employing Gaze Features}
We explored how meta-awareness in mind wandering manifests in eye movements. For descriptive comparisons of average gaze features extracted during aware and unaware mind wandering instances with those that occurred during on-task instances, see Appendix Section \ref{etsum}.
To emphasize the potential of predictive modeling methods to capture complex, nonlinear relationships, we trained several machine-learning classifiers based on the eye-tracking features extracted from 10-s windows before each thought probe and labeled those based on the self-reports. As the goal was to detect mind-wandering episodes and specifically distinguish between aware and unaware TUTs, we conducted a three-class classification, summarizing the previous thought categories on task, TRT, and MM into one category. Missing values were imputed by using constant value imputation, and values were \textit{z}-standardized. Algorithms tend to exhibit a bias toward the majority class (i.e., in our study, the on-task state), particularly in data sets characterized by a high level of imbalance, as observed in our data. To address this issue, balancing techniques are commonly employed \parencite[e.g.,][]{Faber.2018, Hutt.2017}. These methods are aimed at mitigating the imbalance by equalizing the class distribution within the training set, allowing machine-learning models to effectively learn representations from the minority classes. Our study utilized various balancing methods, including random oversampling. This technique involves randomly duplicating instances from the minority class within the training set. Additionally, we employed SMOTE \parencite[]{smote2002}, a method that generates synthetic samples to augment the representation of the minority classes during training. These techniques collectively ensured a more balanced and representative training environment for our study, contributing significantly to the robustness and reliability of our findings.

In line with standard practices in machine-learning research, we utilized various classification algorithms, each carefully chosen for their unique strengths and compatibility with our specific problem. These algorithms adopt distinct strategies for learning patterns within the eye-tracking data, thus allowing us to compare the achieved classification outcomes across different approaches.
We employed Random Forest, XGBoost, Support Vector Machine, and Multilayer Perceptron models.
Random Forest is renowned for its robustness and ability to handle nonlinear data, making it particularly suitable for complex data sets such as ours. XGBoost, known for its efficiency and performance, is ideal for large data sets and has demonstrated effectiveness in reducing overfitting. We chose the Support Vector Machine (SVM) for its effectiveness in high-dimensional spaces, crucial for handling the intricate patterns in eye-tracking data. Lastly, the Multilayer Perceptron (MLP), a type of neural network, offers the advantage that it can learn nonlinear relationships, an advantage that was essential for capturing the subtleties in our data.
This selection of standard machine-learning classifiers has also been employed in previous research on binary mind-wandering classification based on eye-tracking data \parencite[e.g.,][]{hutt2016eyes, Faber.2018b}. Their demonstrated track record in similar studies provided a solid foundation for their application in our research, ensuring that we were building upon established methods while also exploring the unique aspects of our data set. This combination of demonstrated and innovative approaches was expected to yield comprehensive insights into the patterns present in the eye-tracking data, contributing significantly to the field of machine learning and cognitive research.

We then applied person-independent three-fold nested cross-validation for hyperparameter optimization, model training, and evaluation. Cross-validation is a process in which distinct subsets of the data are used repeatedly to train and test a model to prevent overfitting on a single test set \parencite[]{kohavi1995}. By creating these folds in a person-independent manner, we ensured that the prediction accuracies we reported would generalize to new, unseen participants. Each classifier allowed us to set specific hyperparameters, such as the number of decision trees learned in a Random Forest classifier. Hyperparameter tuning, which describes the process of finding the optimal setting of those parameters for a specific classification problem, was conducted by a grid search. A table of the tested hyperparameter grids per classifier can be found in Table \ref{tab:hyper} in the Appendix. We employed nested cross-validation to test the best hyperparameter settings of every single classifier within each fold on a validation set to avoid overfitting to the test set. 

Due to our sample's highly imbalanced data distribution, we evaluated the prediction performance by reporting F1 scores, representing the harmonic mean of precision and the recall of predictions. This measure is commonly used as a performance metric in cases in which one class (e.g., being on task in these data) greatly outnumbers one or more minority classes (i.e., aware/unaware TUTs) that are of specific interest in terms of prediction performance. This measure has repeatedly been employed in previous research on mind-wandering detection \parencite[e.g.,][]{Hutt.2019, Hutt.2017}. We computed the F1 scores for each of the single predicted classes and the macro score, which is the average of all F1 scores. Please find more details on the evaluation metrics in the Appendix Section \ref{evalmet}.

However, comparing the performance of models across data sets with different underlying class distributions is difficult, as a higher F1 score becomes more likely as the proportion of the minority class increases, which occurs because the chance level of predicting the minority class becomes higher. Consequently, a useful measure, particularly when comparing the performance of data sets with different class distributions, is the model's improvement over the chance level. This improvement is calculated as follows:
\[
Above Chance Level = \frac{Actual Performance - Chance}{Perfect Performance - Chance}
\]


To investigate the relationship between gaze and mind-wandering predictions as modeled by the machine learning approach, we applied explainability methods. Specifically, we computed SHAP (SHapley Additive exPlanations) values \parencite{shap2017}. These values quantify the contribution of each feature to the likelihood of predicting each class, relative to the model’s average prediction across the dataset. 

\label{sec:LA}
 
\subsubsection{Transparency and Openness}
We report how we determined our sample size, all data exclusions, all manipulations, and measures in the study, and we follow JARS \parencite[]{kazak2018apa}. We plan to make the survey and sensor data collected in this project, as well as all analysis scripts, openly available at a later point in time. Supplemental materials are available at \url{https://osf.io/4qkvx/?view_only=096cc5bc171c496283f85064f82067d6}. We conducted statistical analyses using R, version 4.2.2 \parencite[]{r2021}, and clustering by employing the package TraMineR, version 2.2.7 \parencite[]{gabadinho2011analyzing}. Machine-learning approaches were conducted using Python, version 3.8.5 \parencite[]{van1995python} and the package scikit-learn, version 1.3.2 \parencite[]{scikit-learn}. This study’s design and analysis were not preregistered. Parts of this data, specifically eye-tracking data, were analyzed regarding gaze synchrony during self-reported on-task behavior \parencite{buhler2024synchr}.

\subsection{Results}

Descriptive statistics and correlations between self-reports, person characteristics, and the knowledge tests administered before and after the learning sequence can be found in Table \ref{tab:corr_tab}.

\subsubsection{Mind-Wandering Meta-Awareness During Lecture Viewing}

\begin{figure}[h]
    \centering
    \caption{Content Probes.}
    \includegraphics[width=0.8\textwidth]{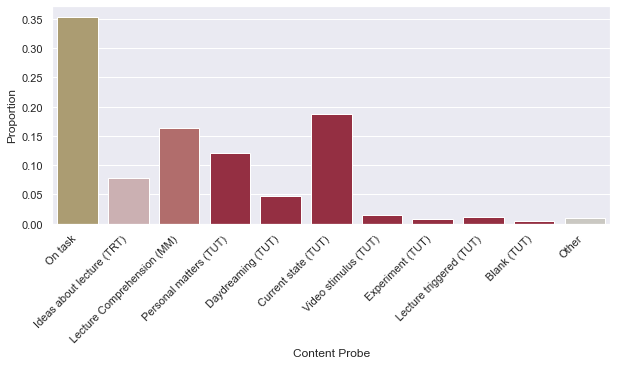}
    \label{fig:con}
\end{figure}

\begin{figure}[h]
    \centering
    \caption{Content Probe TUTs by Meta-Awareness.}
    \includegraphics[width=0.8\textwidth]{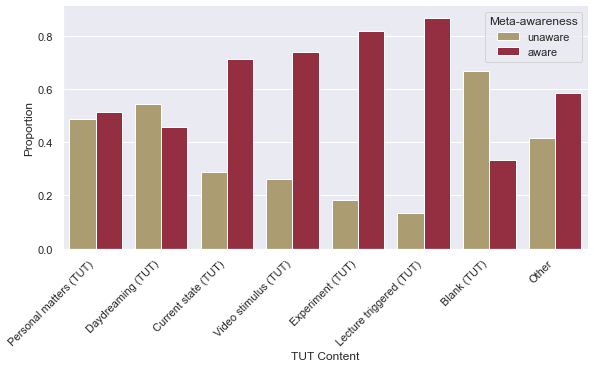}
    \label{fig:conaw}
\end{figure}

Overall, when probed, participants reported the following proportions of answers: 35\% being on task, 8\% task-related thoughts, 16\% metacognitive comprehension monitoring, and 41\% TUTs. The last category is a combination of seven thought content categories. The overall distribution of the thought content categories is presented in Figure \ref{fig:con} (please see Table \ref{tab:con} for the absolute and relative numbers). The largest proportion of TUTs consisted of thoughts about participants' current physical, psychological, or emotional states (19\%; e.g., whether somebody was hungry or cold). This category was followed by personal matters (12\%; e.g., thoughts about normal everyday things, life concerns, or personal worries). Furthermore, 63\% of all combined TUTs occurred with meta-awareness, indicating that participants had already realized they had drifted off before the interruption. Figure \ref{fig:conaw} displays all content TUT categories by meta-awareness (refer to Table \ref{tab:conaw} for the absolute and relative numbers). Whereas for most categories, the majority were reported to occur with awareness, the ratio was almost balanced for the personal matters category. Only daydreaming and a blank mind predominantly occurred without meta-awareness. 

Upon analyzing the correlations between the different types of mind wandering and individual characteristics (see Table \ref{tab:corr_tab}), several significant associations were observed. We found a significant negative relationship between metacognitive self-regulation and unaware TUTs, whereas no correlation was detected with aware mind wandering. Conversely, positive correlations were noted between self-concept, dispositional interest in the lecture subject, statistics, on-task rates, and reflecting on lecture comprehension. Notably, no significant correlations were found between TUT categorized by meta-awareness and the dispositional mind-wandering scale.
Regarding situational variables surveyed immediately after the video lecture, negative correlations were identified between situational interest in the lecture and both aware and unaware TUTs. Similarly, self-reported involvement in the lecture exhibited negative correlations, with a slightly stronger association seen in relation to aware mind wandering frequency. By contrast, cognitive engagement was significantly negatively correlated with both types of TUTs, demonstrating a more pronounced relationship with unaware mind wandering. Furthermore, positive emotions experienced during the lecture exhibited a significant negative correlation only with the incidence of aware TUTs, while showing a notable positive association with the frequency of on-task reports.

\subsubsection{Temporal Patterns of Mind-Wandering Meta-Awareness}

\begin{figure}[h]
    \caption{Thought Category Sequence Index and Distribution Plots.}
    \includegraphics[width=0.5 \textwidth]{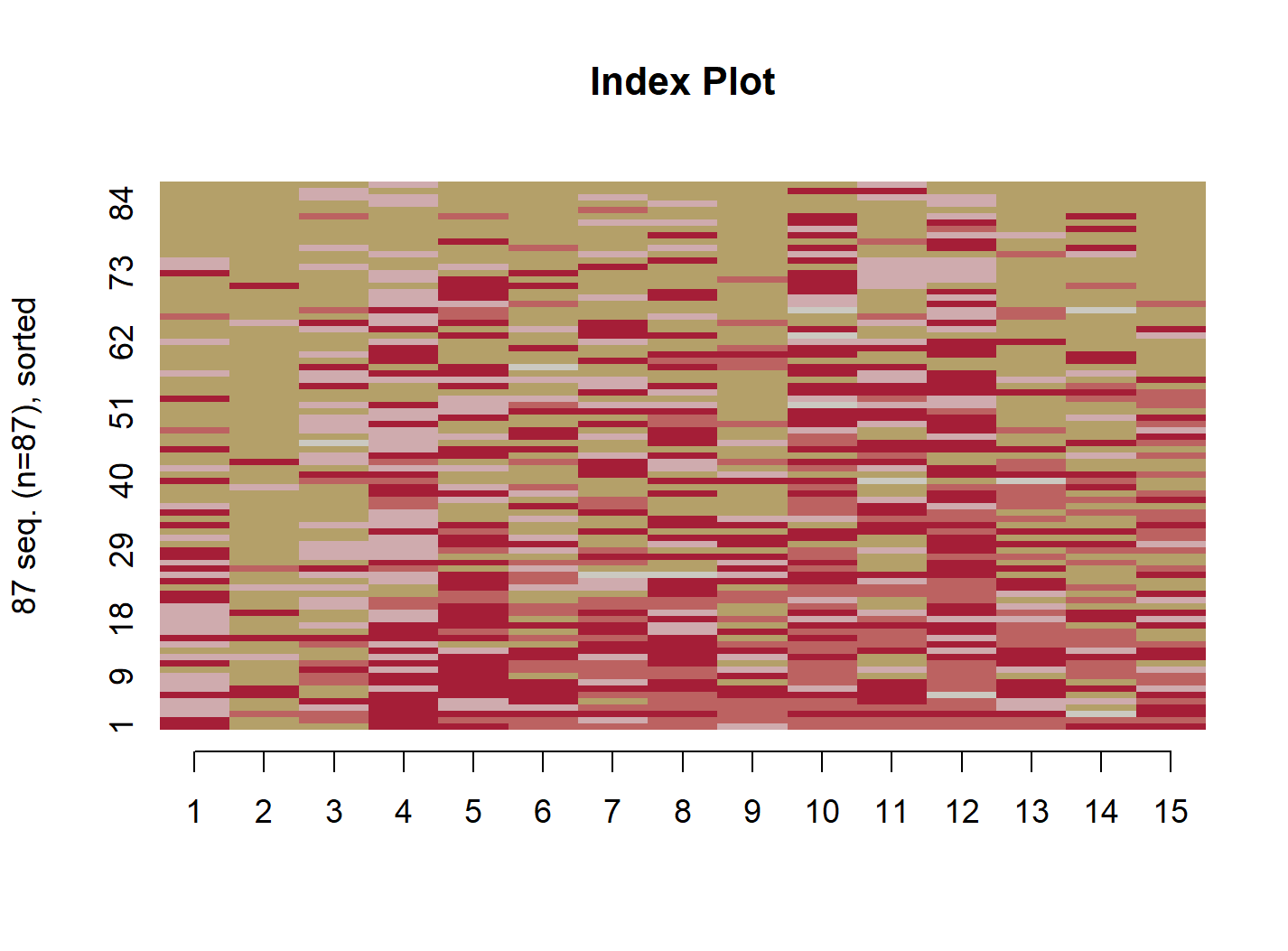}
    \includegraphics[width=0.5 \textwidth]{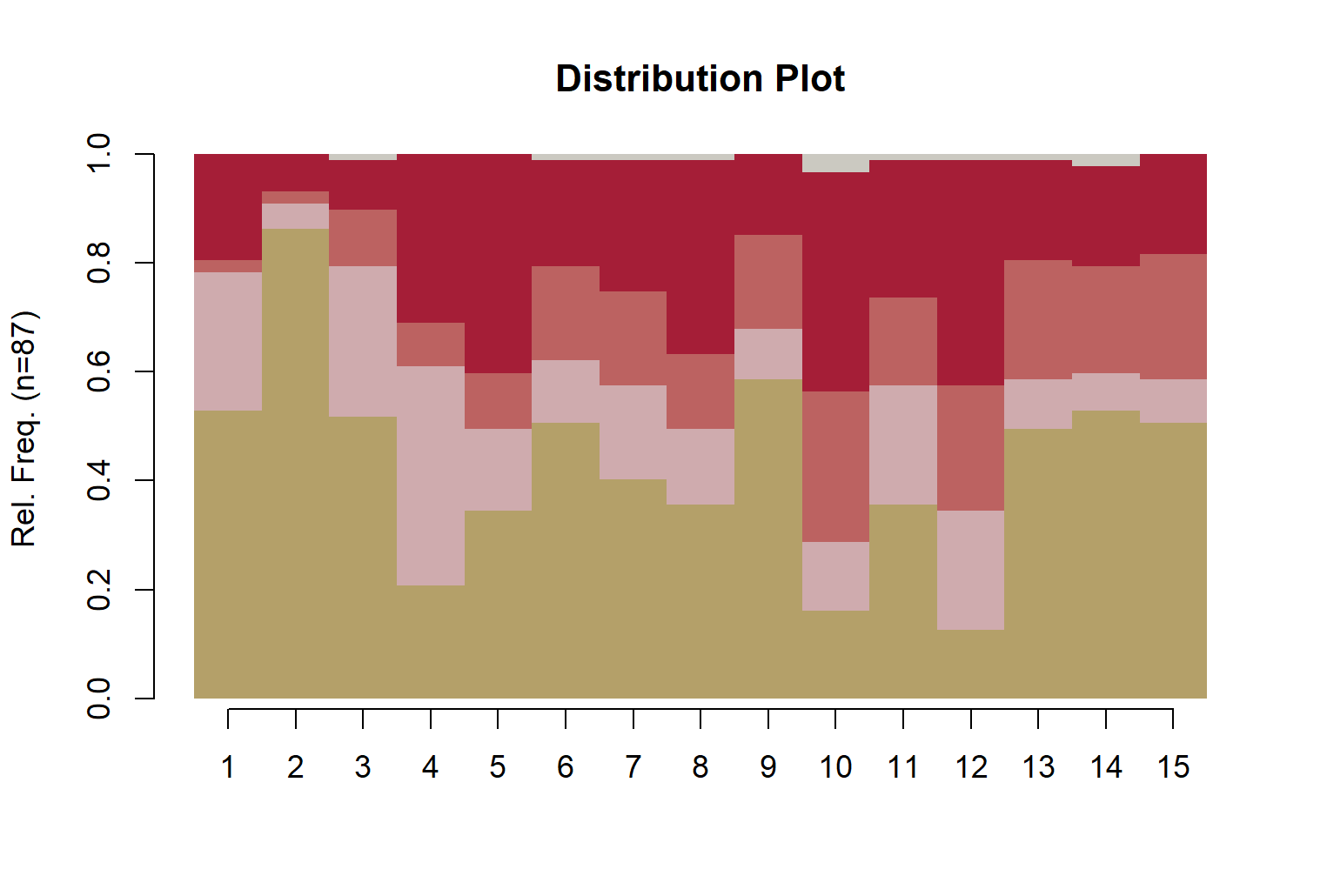}
    \begin{center}
    \includegraphics[width=0.2\textwidth]{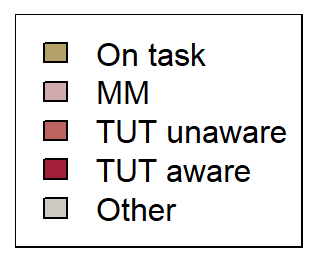}
    \end{center}
    \label{fig:seq}
\end{figure}

Figure \ref{fig:seq} shows the self-reported thought category sequences by person (left) and the category distribution (right) over time. The visualization of individual sequences shows the dynamic transition between different thought and mind-wandering categories of participants over the course of the lecture video. In the state distribution plot, it can be seen that TUTs begin to increase around Probes 4 and 5, which were administered around 10 to 15 min into the video lecture. It is possible to observe a slight increase in overall zone-outs over time, whereas the tune-out rates fluctuated slightly. There was a brief interruption after Probe 8 (approximately 30 min into the video) to recalibrate the eye trackers, potentially reflected in the slightly lower TUT rates at Probe 9.

\begin{figure}[h!]
    \caption{Sequence Index Plots by Cluster.}
    \includegraphics[width=\textwidth]{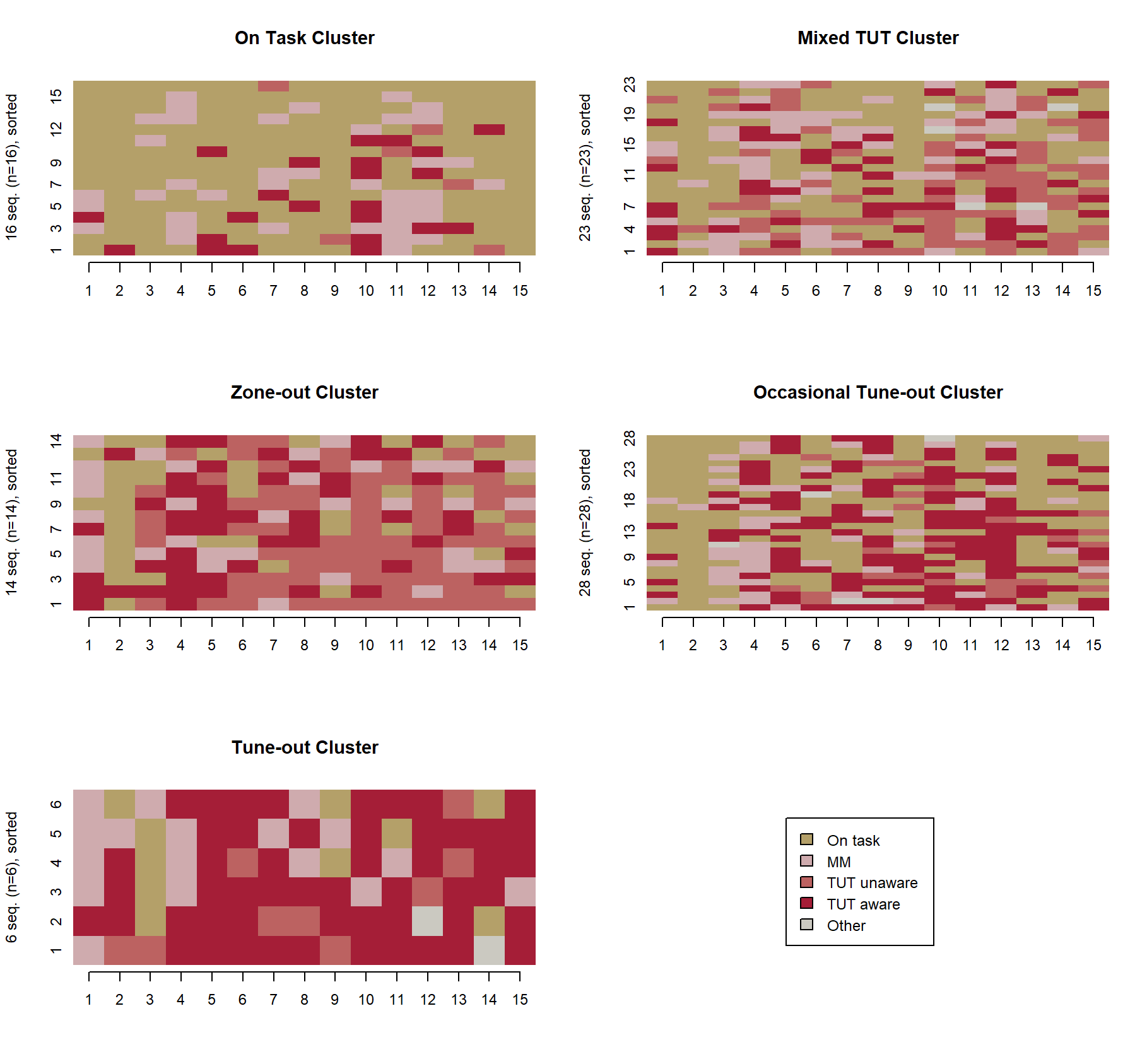}
    \label{fig:clu_ind}
\end{figure}

\begin{figure}[h!]
    \caption{State Distribution Plots by Cluster.}
    \includegraphics[width=\textwidth]{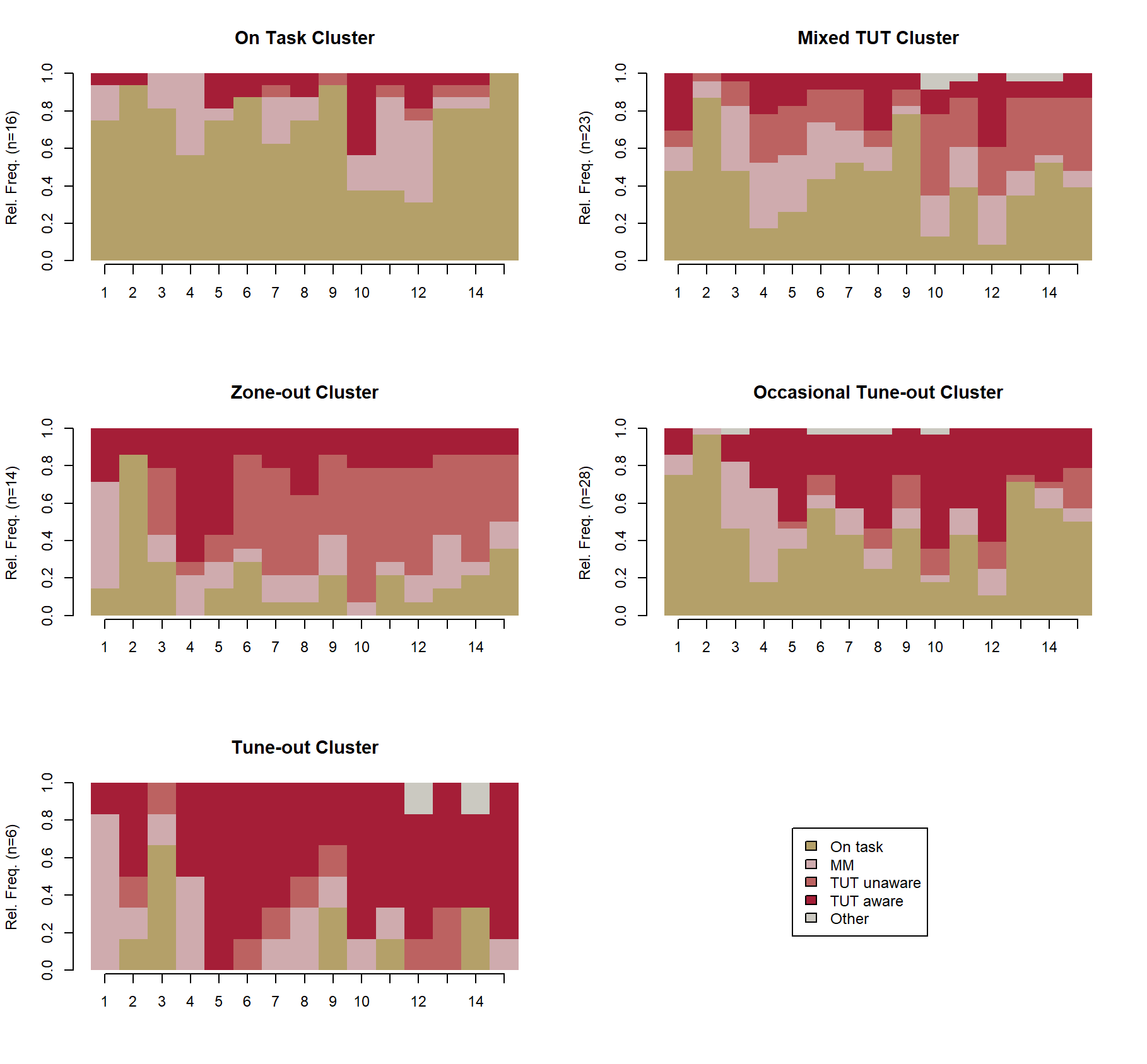}
    \label{fig:clu_st}
\end{figure}

\begin{figure}
    \centering
    \caption{Average Reports per Category Plots by Cluster.}
    \includegraphics[ width=\textwidth]{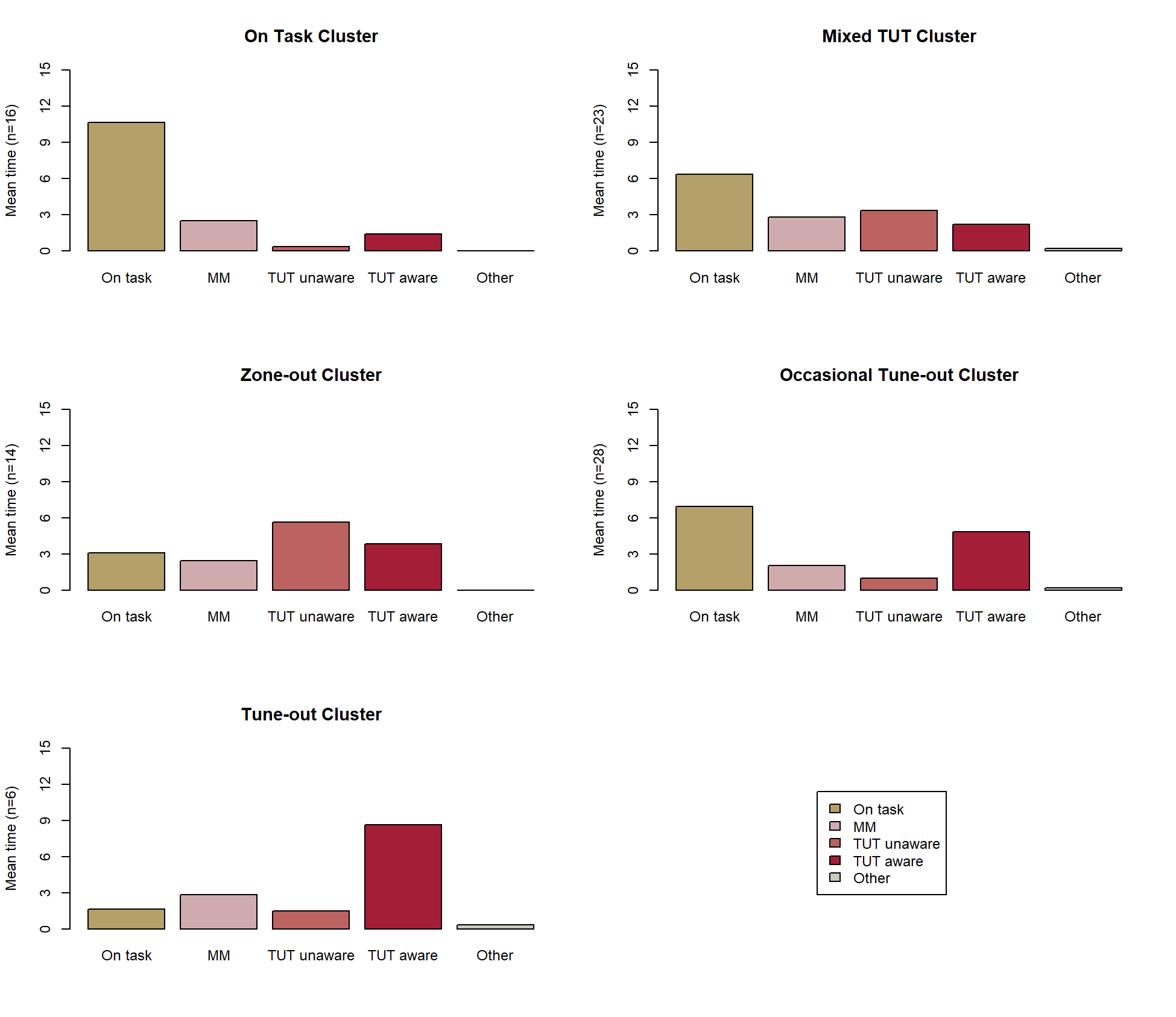}
    \label{fig:clu_mean}
\end{figure}

Utilizing OM distances to apply hierarchical clustering to cluster thought probes on the basis of their temporal progression yielded the identification of five distinct clusters that represented different thought patterns. The individual thought sequences and the distribution of thought categories within each cluster are graphically presented in Figures \ref{fig:clu_ind} and \ref{fig:clu_st}, respectively. The clusters were assigned titles that correspond to their predominant thought structures. 
The first cluster, referred to as the on-task cluster, consisted primarily of participants who remained focused on the task at hand throughout the lecture. On average, they indicated being engaged in the task during 11 of the 15 thought probes (refer to Figure \ref{fig:clu_mean}). These individuals reported thoughts related to understanding, particularly in the latter part of the lecture, along with sporadic instances of aware TUTs. Overall, this cluster demonstrated a high level of sustained attention toward the lecture and exhibited the lowest rate of mind wandering.
The second cluster, the mixed-TUT cluster, displayed an increasing occurrence of unaware mind wandering over time compared with the first cluster. On average, participants in this cluster reported an equal number of on-task instances and instances of TUTs, with the majority of TUTs being unaware. Participants reported metacognitive monitoring throughout the lecture, particularly in the first half. 

By contrast, the zone-out cluster was primarily characterized by a prevalence of unaware TUTs. This cluster displayed a brief peak in aware TUTs followed by a relatively consistent occurrence of zone-outs throughout the last two-thirds of the lecture. Participants in this cluster reported experiencing zone-outs in six of the 15 probes on average, followed by four instances of tune-outs and only three instances of being on task.
Conversely, the occasional tune-out cluster demonstrated minimal unaware TUTs. On average, participants in this cluster reported being on task approximately half of the time. However, after a transitional phase characterized by increased reports of metacognitive monitoring interspersed with indications of tune-outs, participants in this cluster reported being on task an average of five times per person. 
The tune-out cluster exhibited a more extreme form of aware mind wandering and was also the smallest cluster we identified (\textit{n} = 6). Participants in this cluster reported tune-outs in nine of the 15 probes on average, with only two instances of being on task and minimal zone-outs. After Probes 3 to 4, which occurred around 10 to 15 min into the lecture, participants in this cluster consistently engaged in aware TUTs.
An interesting observation is that the overall rates of metacognitive monitoring were similar across all five clusters, suggesting that self-reflection on understanding was a common element within each distinct TUT pattern.

\begin{figure}[h!]
    \caption{Thought Content Rates by Cluster and MANOVA-Based Games-Howell Post Hoc Pairwise Comparisons.}
    \includegraphics[ height=5.6in]{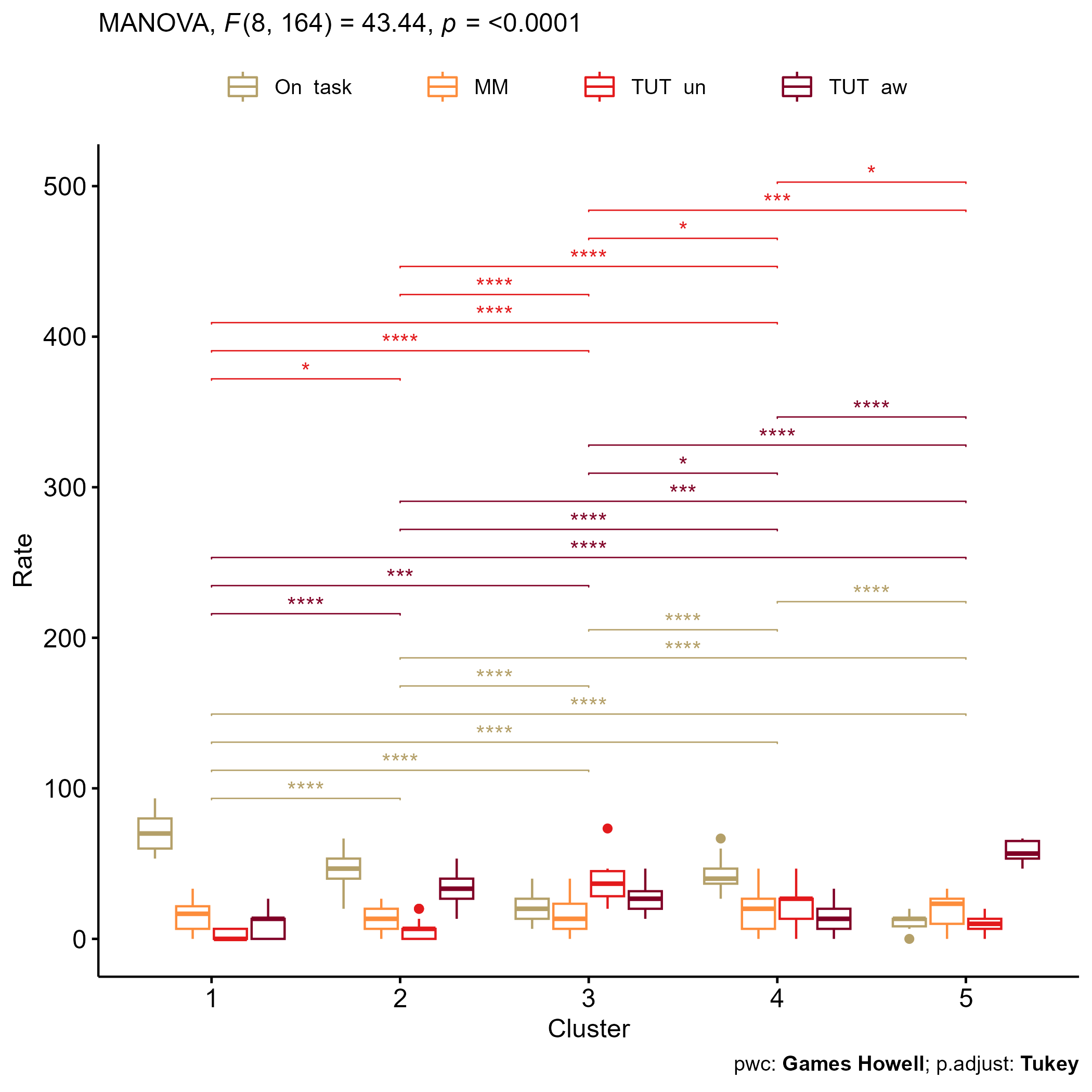}
    \label{fig:mano}
\end{figure}

To investigate the differences between the single clusters, we conducted a one-way multivariate analysis of variance (MANOVA) to determine the differences in the proportion of overall on-task, MM, unaware, and aware TUT reports per person. There was a statistically significant difference between the clusters on the combined dependent variables (on-task rate, MM rate, unaware TUT rate, aware TUT rate), \textit{F}(8, 164) = 43.441, \textit{p} < .001. Follow-up univariate Welch ANOVAs, using a Bonferroni-adjusted alpha level of .0125, showed that there was a statistically significant difference in task rate, \textit{F}(4, 27.9) = 69, \textit{p} < .001, unaware TUT rate, \textit{F}(4, 25.9) = 47.5, \textit{p} < .001, and aware TUT rate, \textit{F}(4, 24.7) = 35.5, \textit{p} < .001, between clusters. No significant difference was found for MM rates. Pairwise comparisons, employing a Games-Howell post hoc test between clusters by outcome variables, can be seen in Figure \ref{fig:mano}. On-task rates were significantly different between all clusters except the mixed-TUT and occasional-TUT clusters, as well as between the zone-out and tune-out clusters. Aware TUT rates differed significantly between all clusters except for the on-task and occasional-TUT clusters, as well as the mixed-TUT and zone-out clusters. For unaware TUT rates, no significant differences were found between on-task and tune-outs and mixed TUTs and tune-outs, but significant differences were found for all other cluster pairs.

\subsubsection{Mind-Wandering Patterns and Learning Outcomes}

To investigate the potential associations between specific thought patterns and varying levels of learning, a comprehensive linear regression analysis was conducted to assess the relationships between the thought clusters and the different test scores. The results are presented in Table \ref{tab:clu_reg}.
The findings from the linear regression analysis indicated that membership in the zone-out cluster, characterized by thought patterns predominantly defined by increasing rates of unaware mind wandering over time, showed a significant negative association with fact-based memory learning and deep-level inference learning compared with the on-task cluster. Conversely, belonging to the tune-out cluster, which entailed consistent engagement in aware-task-unrelated thoughts throughout the lecture, was significantly associated with lower performance solely in inference learning. Notably, both the zone-out and tune-out clusters demonstrated significantly poorer performance in terms of the overall test score when compared with the on-task cluster. The two clusters, namely, the mixed-TUT and occasional tune-out clusters, which exhibited less extreme patterns of TUTs, did not show statistically significant relationships with any of the test scores compared with the on-task cluster.
Furthermore, higher levels of self-concept in statistics were positively related to learning outcomes across all three test scores. Additionally, prior knowledge exhibited a positive relationship with the test scores, indicating that possessing a greater knowledge base in the subject matter contributes to improved performance. A small significant gender effect was found for the inference-based test score such that male participants displayed slightly inferior performance compared with their female counterparts.

To validate these results, we additionally conducted a regression analysis to investigate the relationship between thought category rates per person and posttest scores (see Table \ref{tab:rate_reg}). Similar overall patterns were identified: The overall zone-out rate had negative associations with fact- and inference-based scores, whereas aware-TUT rates had a significant negative impact only on inference-based learning. Notably, the MM rate and deep-level understanding had a significant negative relationship.

\begin{sidewaystable}[h!]

\caption{Linear Regression Analyses Predicting Fact-Based, Inference-Based, and Total Test Scores From the Mind-Wandering Clusters}

\begin{tabular}{lcccccc}
\hline
                        & \multicolumn{2}{l}{Fact-Based Test Score} & \multicolumn{2}{l}{Inference-Based Test Score} & \multicolumn{2}{l}{Total Test Score} \\
Predictors              & Estimates         & CI                    & Estimates            & CI                      & Estimates       & CI                 \\ \hline
Intercept               & 4.58 **           & 1.75 – 7.41           & 4.73 ***             & 2.27 – 7.20             & 9.31 ***        & 4.97 – 13.66       \\
Mixed TUT Cluster       & 0.04              & -0.82 – 0.90          & -0.41                & -1.16 – 0.34            & -0.37           & -1.70 – 0.95       \\
Zone-out Cluster        & -1.17 *           & -2.18 – -0.17         & -1.31 **             & -2.18 – -0.43           & -2.48 **        & -4.02 – -0.94      \\
Occasional Tune-out Cl. & -0.62             & -1.52 – 0.29          & -0.48                & -1.27 – 0.31            & -1.09           & -2.49 – 0.30       \\
Tune-out Cluster        & -0.92             & -2.27 – 0.43          & -1.61 **             & -2.79 – -0.44           & -2.54 *         & -4.61 – -0.46      \\
Previous Knowledge      & 0.36 *            & 0.00 – 0.72           & 0.39 *               & 0.08 – 0.71             & 0.76 **         & 0.20 – 1.31        \\
Self-Concept& 0.83 ***          & 0.43 – 1.23           & 0.62 ***             & 0.27 – 0.97             & 1.45 ***        & 0.84 – 2.07        \\
Male                    & -0.36             & -1.09 – 0.37          & -0.68 *              & -1.31 – -0.04           & -1.04           & -2.15 – 0.08       \\
Age                     & -0.05             & -0.16 – 0.07          & -0.06                & -0.16 – 0.04            & -0.10           & -0.28 – 0.08       \\ \hline
Observations            & \multicolumn{2}{l}{87}                    & \multicolumn{2}{l}{87}                         & \multicolumn{2}{l}{87}               \\
R2 / R2 adjusted        & \multicolumn{2}{l}{0.333 / 0.264}         & \multicolumn{2}{l}{0.390 / 0.327}              & \multicolumn{2}{l}{0.445 / 0.389}    \\ \hline
\end{tabular}
\begin{tablenotes}[para,flushleft]
    {\small
        * \textit{p} \textless{  } .05.   ** \textit{p} \textless{ } .01.   *** \textit{p} \textless{ } .001. \\
        \textit{Note.} Reference category is the on-task cluster. 
     }
\end{tablenotes}
\label{tab:clu_reg}
\end{sidewaystable}

\subsubsection{Mind-Wandering Meta-Awareness Prediction Using Machine Learning}

Based on the hypothesis that meta-awareness in mind wandering is reflected in gaze behavior, we employed a predictive modeling approach based on eye tracking features. Descriptive comparisons of mean levels of extracted gaze features across distinct self-reported thought groups can be found in the appendix in Table \ref{tab:etsum}. Using the self-reported thought probes, we employed machine-learning classifiers to predict instances of aware and unaware mind wandering in distinction to a combined on-task category, which encompassed reports of being on task, TRT, and MM. The results of the trained classifiers are presented in Table \ref{tab:mlclass}. The first three columns provide F1 scores, precision, and recall for each prediction category, whereas the last column reports macro scores across all categories. The last row represents the baseline performance levels of predicting by chance.
The findings indicate that the models achieved predictions slightly above chance for unaware TUTs, with the Random Forest model performing the best, yielding an F1 score of 0.215 (approximately 9\% above chance). For aware TUTs, the SVM and MLP classifiers achieved F1 scores of 0.332 (approximately 12\% above chance) and 0.282 (approximately 5\% above chance), respectively, resulting in Macro F1 scores of 0.377 (approximately 5\% above chance) and 0.361 (approximately 3\% above chance). A Random Forest classifier achieved the best overall results with macro F1 scores of 0.387 (approximately 6\% above chance).
Comparatively, when applying binary classification to distinguish between on-task and combined TUT instances using the same data, the F1 scores for TUTs reached 0.529 when an MLP classifier was used (see Table \ref{tab:mlbin}), indicating a detection accuracy of approximately 24\% above the chance level.

\begin{sidewaystable}[h!]
    {\setlength{\tabcolsep}{1.2pt} 
    \caption{Mind Wandering by Meta-Awareness Classification Results.}
    \label{tab:mlclass}
    \begin{tabular}{l ccc ccc ccc cccl}\toprule
     & \multicolumn{3}{c}{On Task} & \multicolumn{3}{c}{Aware TUT} & \multicolumn{3}{c}{Unaware TUT} & \multicolumn{3}{c}{Macro Scores}
    \\\cmidrule(lr){2-4}\cmidrule(lr){5-7}\cmidrule(lr){8-10}\cmidrule(lr){11-13}
                Model    & F$_{1}$ Score    & Precision  & Recall & F$_{1}$ Score      & Precision & Recall & F$_{1}$ Score  & Precision & Recall & F$_{1}$ Score  & Precision & Recall \\\midrule 
        Random forest   & 0.710         & 0.680     & 0.743       &  0.236   & 0.280 & 0.217      & \textbf{0.215}    & 0.228         & \textbf{0.231}    & \textbf{0.387} & 0.396 & \textbf{0.397} \\
        XGBoost         &  \textbf{0.721} & 0.664   & \textbf{0.788}    & 0.150         & 0.213     & 0.129         & 0.190 & \textbf{0.245}& 0.204        & 0.353         & 0.374 & 0.373  \\ 
        SVC             & 0.638         & 0.663     & 0.629   & \textbf{0.332} & \textbf{0.317} & \textbf{0.357} & 0.161     & 0.218         & 0.156  & 0.377         &  \textbf{0.399} & 0.381 \\
        MLP             & 0.592         &  \textbf{0.687}    & 0.521   & 0.282 & 0.243 & 0.341              & 0.210    & 0.179     & 0.272               & 0.361        & 0.370 & 0.378      \\
    
    \midrule
         Chance Level  & 0.615       & 0.615 &0.618 & 0.244& 0.246& 0.246 & 0.134     &0.135 & 0.133        & 0.343 & 0.344 & 0.344\\
    \bottomrule
    \end{tabular}}

\end{sidewaystable} 

\begin{figure}
    \centering
    \caption{SHAP Analysis by Prediction Class.}
    \includegraphics[width=0.65\textwidth]{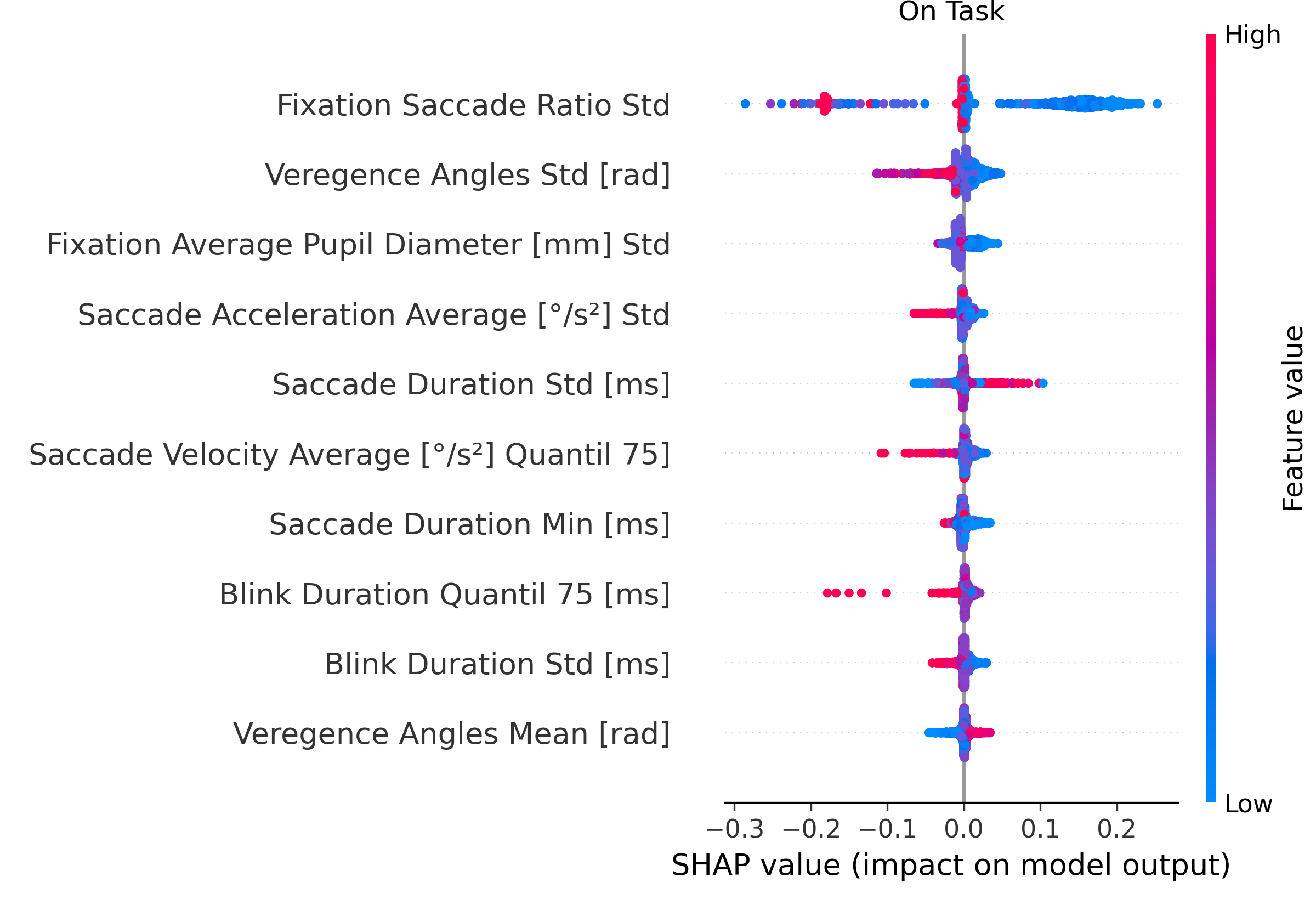}
    \includegraphics[width=0.65\textwidth]{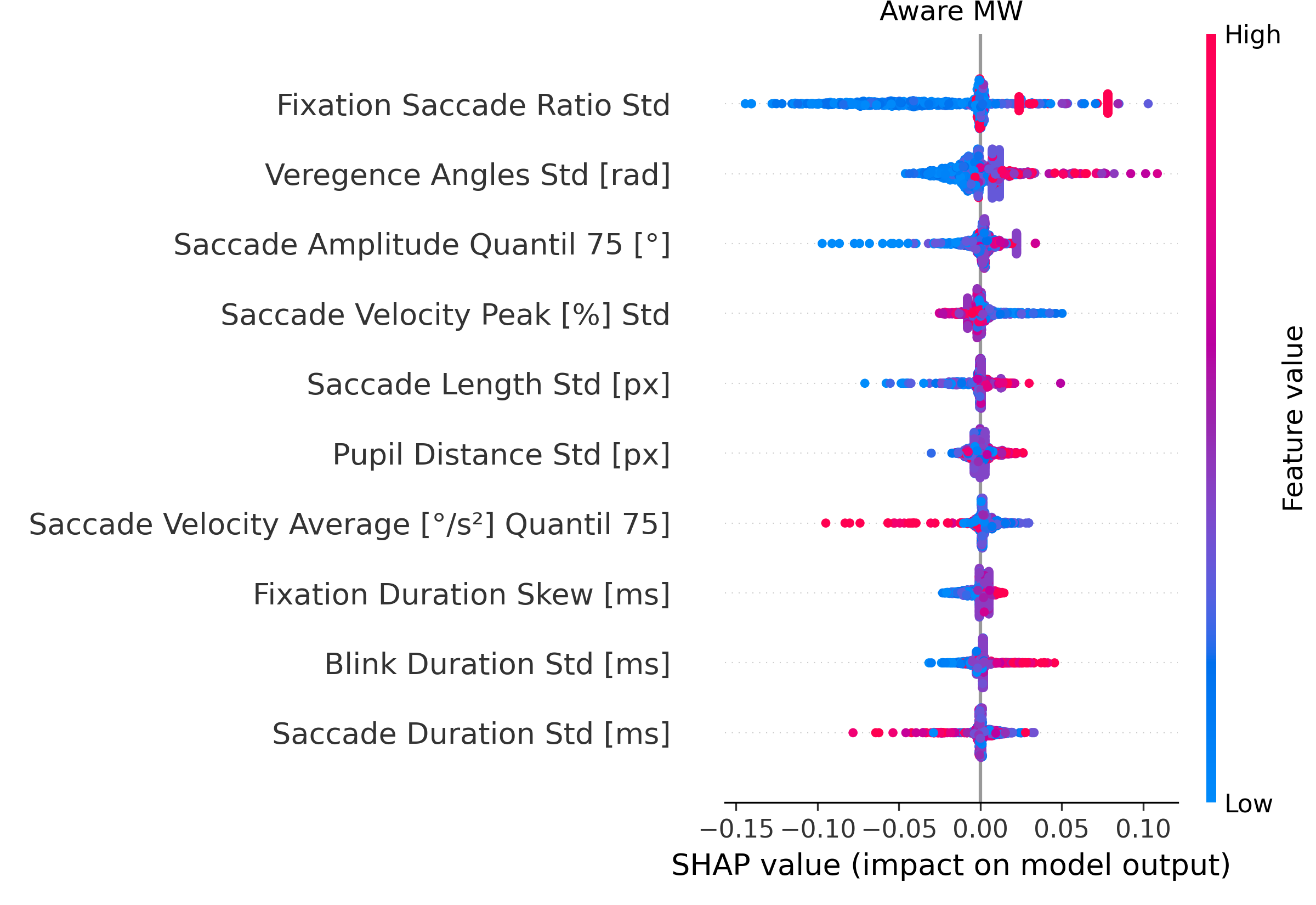}
    \includegraphics[width=0.65\textwidth]{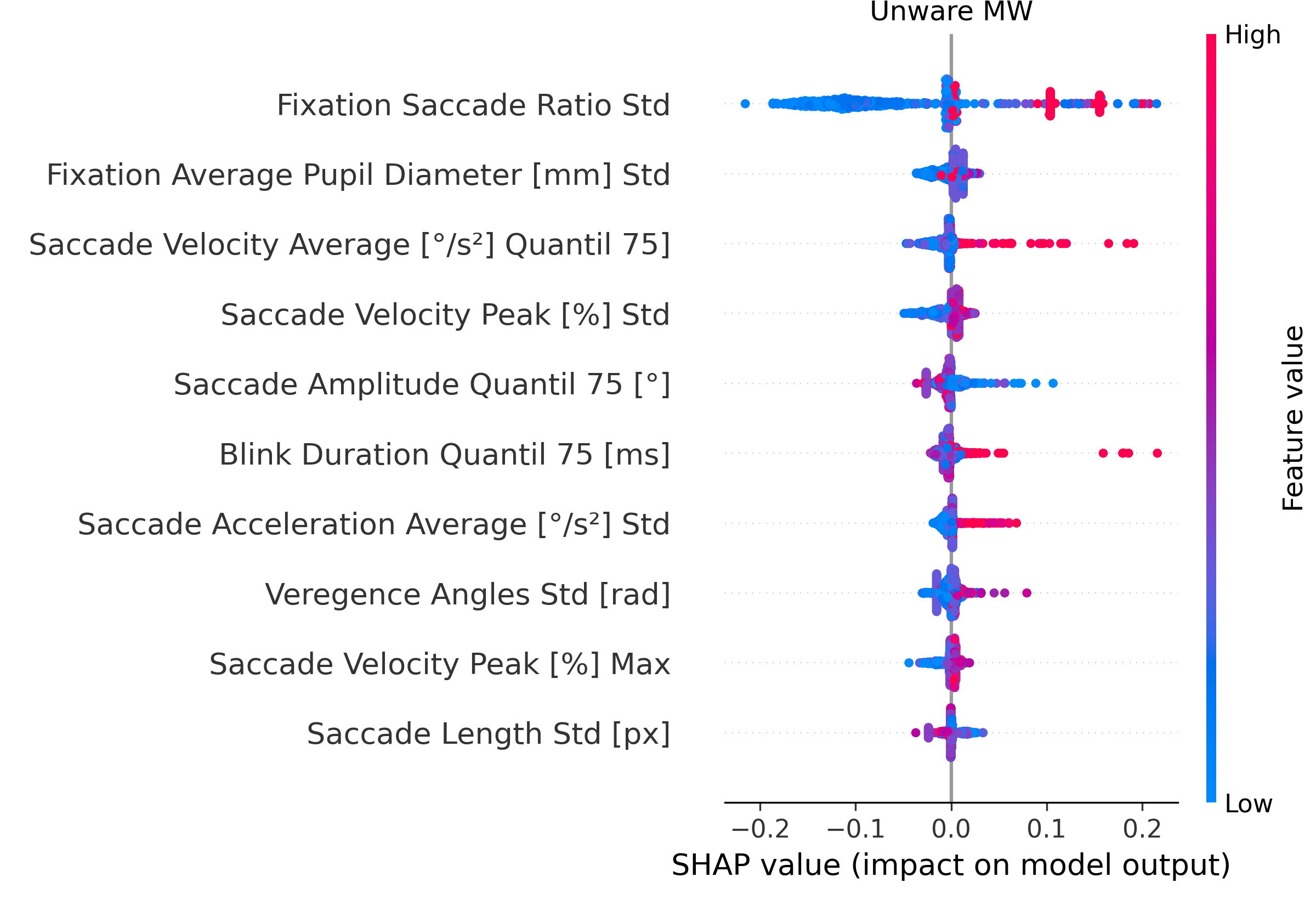}
    \label{fig:shap}
\end{figure}

To delve deeper into the connection between gaze features and the meta-awareness of TUTs, as modeled with our machine-learning approach, we computed SHAP values \parencite{shap2017} for the Random Forest classifier, which achieved the highest macro F1 score. 
SHAP is a local explainability approach for machine learning models, quantifying how each feature contributes to an individual prediction made. The SHAP values for the ten most important features by predicted class are depicted in Figure \ref{fig:shap}. 
The most important features for predicting aware and unaware TUTs include saccade-, blink-, vergence- and fixation-derived duration features.

The preeminent feature for classifying all three classes is the ratio of the standard deviations of fixation and saccade durations in the 10 s before the probe. A low value in this variable, indicating that the variability in the duration of fixations relative to saccades is smaller, contributes to on-task prediction, whereas the relationship with both mind-wandering classes is reversed. 

Aggregated data on the two vergence features we utilized, namely, vergence angles and pupil distance, which characterize eye rotations and can capture the phenomenon of staring into nothingness, constituted another vital feature group. The figure reveals the importance of variation in gaze vergence, depicted by the standard deviation of vergence angles and standard deviation of pupil distance, for predicting aware TUTs. A high variance in vergence, which might reflect the occurrence of staring into nothingness, increases the likelihood of an aware mind wandering prediction. On the contrary, low values for variance in vergence angles, which might indicate a stable focus on the computer screen assuming a constant distance, increase the probability of on-task predictions. This is further supported by the positive impact of high average vergence angles on on-task predictions. 

Ranking third in importance for on-task and second for unaware mind-wandering predictions is the standard deviation of average pupil diameter during fixations, elucidating the fluctuation in pupil diameters between fixations. It appears to have a diverse effect on the prediction of both classes with the plot depicting both positive and negative influences, suggesting that this feature interacts with other factors in complex ways.

Further, the 75\% quantile of the average saccade velocity feature heavily impacts the classification, with higher average velocities often leading to an unaware mind-wandering classification, whereas lower velocities are associated with on-task and aware mind-wandering classifications. Several important features like the saccade velocity peak maximum value, the standard deviation of saccade velocity peak, the standard deviation in saccade acceleration, and average high saccade velocity indicate the importance of high saccade velocity and high-velocity variance for unaware mind wandering predictions. This becomes especially visible in comparison to aware mind wandering predictions, which are positively impacted by low variance of saccade velocity peaks and lower values for average saccade velocity. Further, also the distance the eye travels during a saccade impacts the prediction. Lower values in the 75\% quantile of saccade amplitudes and lower variability in saccade lengths increase the likelihood of predicting unaware mind wandering whereas the effects of both features seem to be reversed for predicting aware mind wandering.

Another very important gaze indicator impacting the predictions is the duration of blinks. A higher variance in blink durations tilts the prediction towards aware mind wandering, whereas lower values slightly push the prediction to on task. Higher values in the blink duration 75\% quantile lead to unaware mind wandering classification, in contrast, to lower blink durations increasing the likelihood of on-task predictions. In general, the aggregation form of standard deviations is the most prominent one in these groups of most important features. This indicates that especially the amount of variance in certain gaze features over the observed time windows yield significant information.
 Summary statistics for the most important features identified by SHAP analysis can be found in the appendix Table \ref{tab:SHAPetsum}.

In terms more accessible to general education researchers and practitioners, learners were predicted to be mind wandering and unaware of it when their eye movements showed overall higher speed, high variability in how quickly their eyes moved, longer blink durations, and shorter distances moved by the eyes. Learners were predicted to be mind wandering and aware of it when their eye movements were slower, their blink durations were more variable, the distances their eyes moved were more variable, and they showed signs of not focusing on the computer screen, such as staring into space.

\subsection{Discussion}
 
This study examined the meta-awareness of mind wandering during video lecture viewing by employing probes that asked participants' about the content and their awareness of their thoughts. The findings revealed distinct patterns of thought sequences, shedding light on the temporal dynamics of aware and unaware mind wandering. These patterns exhibited varying associations with learning outcomes, such that zone-out patterns were negatively related to fact-based and inference-based learning, whereas tune-out patterns were negatively associated with deep-level understanding. Moreover, an initial exploration into the predictive modeling of mind wandering--in which we distinguished TUTs by meta-awareness and utilized gaze data--appears to have the potential to successfully distinguish between these two forms of mind wandering, offering prospects for future research.

\subsubsection{A Better Understanding of Mind-Wandering Meta-Awareness}

The findings from the current study provide several significant observations that can be applied to improve the understanding of meta-awareness in mind wandering during learning. First, approximately 41\% of the probes administered during Zoom video lecture viewing uncovered instances of mind wandering, which predominantly focused on participants' current state and personal concerns. This number is slightly higher than the average mind-wandering rates reported by \textcite{wong2022} of about 30\% during educational activities. By contrast, participants reported being on task in our study only 35.4\% of the time, with the rest of the time engaging in elaborations and meta-cognitive monitoring. Interestingly, the majority (60\%) of the TUTs that were reported occurred with meta-awareness, indicating that participants were conscious of their mind wandering before they were interrupted by the probe. 

In these instances, participants appear to allow themselves to continue mind wandering after becoming aware of it, thus not fully attempting to focus on the lecture and to ignore distractors \parencite[]{seli2017}, as such distractors might be more personally important to them \parencite[]{scheiter2014}. This aware mind wandering might also regulate affective states, as it allows learners to cope with a lack of motivation or increased frustration \parencite[]{Risko.2012}. This hypothesis was supported by the negative association between positive emotions experienced during the lecture and the number of reported aware TUTs in this study. Unaware TUTs, namely, engaging in non-lecture-related thought without realizing that the mind has wandered off, constituted the smaller part of the observed mind-wandering instances in this study (30\%). Specifically, this particular form of mind wandering exhibited a negative correlation with participants' metacognitive self-regulation skills. This finding reinforces the supposition that from an SRL perspective, the incidence of unaware TUTs during learning represents students' temporal inability to effectively monitor and regulate cognitive activities, whereas aware mind wandering may be indicative of what would be referred to in an SRL framework as failures in regulating motivational or affective states. These findings stress the importance of differentiating between the different manifestations of mind wandering during learning activities. 

Second, similar to observations made in previous research \parencite[]{Risko.2012}, the examination of person-specific thought sequences revealed an increase in TUTs as a function of time for the first half of the video lecture. This pattern was evident for both aware and unaware instances of TUTs. Nevertheless, during the latter 30 min of the lecture, a relatively consistent rate of TUTs emerged without a discernible continued upward trajectory, implying the presence of a potential ceiling effect. An in-depth examination of individual sequences by clustering participants' thoughts into distinct thought patterns unveiled nuanced insights, revealing disparate temporal effects among participants and characteristic thought trajectories.
Through cluster analysis, five distinct patterns of thought sequences were identified: an on-task cluster, a mixed-TUT cluster, a zone-out cluster, an occasional tune-out cluster, and a tune-out cluster. 

The clustering of temporal sequences of meta-awareness in TUTs offers valuable insights into the temporal unfolding of both aware and unaware mind wandering that revealed how distinct groups of participants exhibit analogous patterns. This distinction thereby provides a way to identify characteristic thought trajectories. Whereas a small group (\textit{n} = 16) of participants in the on-task cluster was able to maintain their attention throughout the whole lecture, participants in the mixed-TUT cluster (\textit{n} = 23) showed a slight increase over time in unaware TUTs interspersed with aware TUTs and on-task episodes, a pattern that probably shows the process of catching one's mind wandering and refocusing on the lecture. The smaller zone-out cluster (\textit{n} = 14), on the other hand, shows increasing and then fairly constant unaware TUTs over the last two-thirds of the lecture.
To explain these patterns robustly, it is necessary to explore whether the observed increase in unaware TUT rates over time within the zone-out clusters can be attributed to failures in executive control \parencite[]{kane2012}, possibly linked to individual differences in working memory capacities \parencite[]{mcvay2012wmc}. By contrast, the persistent rates of aware mind wandering evident in the tune-out cluster (\textit{n} = 6) almost from the start suggest that some participants may deliberately choose to allow their minds to wander throughout the video. This awareness brings up the intriguing possibility that such behavior might overlap with intentionality, reflecting a conscious decision to engage in TUTs \parencite[]{Risko.2012} and might be related to a lack of motivation \parencite[]{Seli.2016}. Further research is warranted to comprehensively address these nuanced aspects of mind wandering and its underlying mechanisms, especially with regard to potential strategies for reducing mind wandering. For instance, a learner categorized within the occasional tune-out cluster, who briefly engages in aware TUTs, as they are already familiar with the current lecture content, but reverts to attentive listening later on, requires different support than an individual who, for instance, due to a lack of motivation, consistently tunes-out for the entire duration.

Third, the temporal mind-wandering patterns we identified exhibited varying associations with learning outcomes. The zone-out pattern was significantly correlated with lower fact-based and inference-based learning outcomes, whereas the tune-out pattern had a statistically significant relationship solely with deep-level understanding when compared with on-task thought sequences. The other two clusters were not significantly related to the learning outcomes. These findings are consistent with previous studies that showed a negative association between higher TUT rates and learning \parencite[]{wong2022}: Clusters characterized by higher proportions of TUTs and lower on-task rates were more strongly negatively related to the learning outcomes. Furthermore, akin to previous studies that highlighted the heightened influence of zone-outs on attentional task performance \parencite[]{Smallwood.2007} and mental model construction during reading \parencite[]{smallwood2008}, our results that pertained to lecture retention indicated a stronger statistically significant negative relationship between zone-outs and fact-based learning. However, concerning deep-level understanding, both types of mind wandering exhibited comparable and significant negative associations in the present study.

Lastly, 
an initial exploration into the predictive modeling of the meta-awareness of mind-wandering employing gaze data achieved prediction accuracies of 12\% and 9\% above the chance level for aware and unaware TUTs, respectively. Whereas the predictive modeling of a summarized mind-wandering state, combining aware and unaware TUTs resulted in slightly higher prediction accuracies (24\% above the chance level; see Table \ref{tab:mlbin}), similar to previous research on eye-tracking-based automated mind-wandering detection during lecture viewing, which ranged from 11\% to 24\% above the chance level \parencite[]{Hutt.2017,Zhang.2020, bixler2021crossed}, the more fine-grained approach appears to be more challenging. This finding may have resulted primarily from the limited data available for this specific classification in the current study, notably the smallest class representing unaware mind wandering comprising only 147 instances. This assumption gained support from the observed significant disparities in various gaze features between unaware TUTs and instances of being on task during binary group analysis. However, there may be upper limits to the extent to which the meta-awareness of mind wandering is distinctly reflected in observable eye-tracking data, considering that they represent inherently internal cognitive processes.

Employing SHAP explainability methods, the most influential feature we identified was the ratio between fixation and saccade duration variance. Another very important feature group impacting the prediction of aware TUT was vergence features, indicating that the occurrence of staring into nothingness or the fixation on something aside from the computer screen could indicate aware mind wandering. Gaze behavior increasing the likelihood of predicting unaware mind wandering indicated higher saccade velocities, while the reverse relationships are found for aware TUT prediction. Other important features include blink duration aggregations, the standard deviation of pupil diameter, and duration and amplitude features from saccades, shedding light on predictive factors for both aware and unaware TUTs.

This study tested complex theoretical assumptions about the temporal dynamics of mind wandering and its meta-awareness through sophisticated analytical techniques, including cluster analysis, machine learning algorithms, and SHAP explainability methods.
The outcomes derived from this computational modeling approach demonstrate the discernibility between meta-awareness in mind wandering and on-task behavior through the objective behavioral indicator of eye gaze.
Utilizing Machine Learning allowed us to explore the nuanced and often non-linear relationships between those cognitive processes and gaze behavior. Unlike traditional statistical methods, these techniques can extract insights from datasets with high dimensionality of features, allowing for a comprehensive analysis of numerous gaze indicators. We unraveled the complex interplay between these indicators and different types of mind wandering using SHAP explanatory methods.
This discernibility delineates two forms of mind wandering previously reported by learners and theoretically postulated, thereby contributing to advancing further theory development. 
Moreover, these findings highlight the promising potential of utilizing gaze data for continuous, non-interruptive measurement in future endeavors, as accuracies might be improved by substantively increasing the sample size or including additional data sensors, such as video or physiological data \parencite[]{Brishtel.2020}.
This opens up new opportunities for a deeper understanding of their impact on learning. Consequently, our results contribute to refining the conceptualization of mind wandering meta-awareness in video learning and illustrate the practical implications of leveraging predictive modeling to discern between aware and unaware mind-wandering patterns.

\subsubsection{Implications for Educational Practice}

The present study highlights that mind wandering in an educational context includes several facets with respect to the unfolding of meta-awareness that need to be considered in order to support learners in practice. It is plausible that addressing these diverse manifestations of mind wandering may necessitate distinct and tailored solutions concerning the role of teachers and the design of learning materials. Furthermore, the proposed fine-grained automated detection, based on low threshold sensors, has the potential to allow for the personalized adaptivity of learning content and targeted interventions in the development of intelligent learning systems to provide suitable support for self-regulated learning. 

Examples of possible interventions consist of providing feedback, suggesting rewatching, asking intermediate questions, and adapting the presented content when a user loses focus. Most intervention studies on mind wandering \parencite[e.g.,][]{Kane.2017, Szpunar.2013b} have not been geared to be specific to the mental state of the learner and have therefore been limited to general support. An attention-aware system, implementing eye tracking, could predict the learner's attentional state in real time by analyzing gaze data using the pre-trained mind-wandering detection models. When mind wandering is detected, the system can attempt to redirect the learner's attention to the learning material, for instance, by asking the learner a question on the content. This intervention might cause the learner to mentally re-engage with the content and possibly realize that some information has been missed. The impact of the respective intervention on attention, learning, and performance can be evaluated.

On account of the presumably different explanations for the occurrence of aware and unaware TUTs, consequently requiring distinct solutions, the refined fine-grained recognition of aware and unaware TUTs enables even more precise interventions for learners by taking into account different underlying processes. For example, in the case of frequent zone-outs, which might be caused by failures in executive control, one might aim to support learners by letting them take a break or giving them feedback to improve their metacognitive self-regulation skills and strategies. Meanwhile, in the case of persistent tune-outs, which might be caused by frustration or boredom, one might want to adapt learning content to make it easier to understand and more appealing.

 Initial research on real-time interventions during automatically detected mind-wandering instances (e.g., prompting rereading \parencite[]{Mills.2020} or repeating content and asking questions in an ITS \parencite[]{hutt2021}) have shown promising results for long-term retention and comprehension, even when they were based on imperfect prediction accuracies. However, given the current limitations in predicting the meta-awareness of mind wandering and the potential for false positives, interventions should prioritize methods that do not disrupt the learning process. Nonintrusive interventions, such as follow-up prompts for revisiting critical learning content, focus on fostering self-regulated learning while minimizing distractions. Additionally, implementing thresholds that are based on confidence in predictions or duration can help maintain intervention quality.

The effectiveness of these and other interventions requires extensive empirical research, the premise of which we established in this study by refining the underlying theory and automated recognition approach. Recent studies on AI support and student agency have indicated a trend where students tend to depend on the support that is provided rather than actively learning from it \parencite[]{darvishi2023}. This highlights the need to ensure learners' agency and responsibility when designing interventions \parencite[]{winnie1998}.

Further, this potential integration of real-time interventions raises points that should be considered regarding ethics, learner privacy, and data use. Obtaining informed consent becomes crucial when implementing any form of automated detection techniques. Students should be aware of the data that are being collected and how that data will be used and should have the right to opt-out if they desire to guarantee full autonomy. Further, ensuring that learners' personal information is ethically managed and that their data are safeguarded is paramount, as intelligent learning systems advance in personalized adaptivity. Further, steps must be taken to avoid potential biases that may disproportionately impact certain groups of learners.

\subsubsection{Limitations and Future Work}
A substantial proportion of participants exhibited awareness of TUTs during the lecture-viewing task. We posit that this finding can be attributed in part to the lab setting, where participants might not be completely aligned with the primary goal of comprehending the video lecture \parencite[]{Risko.2012}. Given the absence of personal gain and potentially limited intrinsic motivation to sustain attention, participants may exhibit a tendency to engage in distracting thoughts that hold greater personal significance \parencite[]{scheiter2014}. Thus, this tendency might impact the ecological validity of the reported results. Even if we assume that aware mind wandering is lower in more naturalistic learning environments, competing personal thoughts with heightened short-term importance may persist, particularly in lectures in real study programs, where exams may be scheduled several months ahead. Hence, conducting analogous investigations in regular university lecture settings will be of considerable interest for further exploration. Further, it would be interesting to assess the overlap with the intentionality of mind-wandering episodes to gain deeper insights into whether specific episodes of aware mind wandering can be regarded as a deliberate allocation of cognitive resources \parencite[]{Risko.2012}. Moreover, distractions from the environment and the participants' reflections on study-related prompts reported in the open-answer option of thought prompts suggest that learners were cognizant of their participation in a study. This awareness might imply increased self-monitoring among the learners during the study session and might have had an impact on the observed outcomes.

Although we emphasized the importance of investigating thought patterns of mind wandering with respect to the meta-awareness of TUTs, the current analysis was based on discrete measurement points. As a result, it cannot provide the granularity required for a comprehensive analysis of the temporal unfolding and switching between states of aware and unaware mind wandering during learning, highlighting the importance of using a continuous measurement approach in future research.
Furthermore, due to the limited amount of data available, the clusters identified in our analysis became relatively small and might not be distinct enough to fully capture the heterogeneity inherent in thought patterns. This study was conducted with university students, and thus, the results might not be generalizable to other age or educational groups. Increasing the sample size and obtaining more diverse data sets would enhance the robustness and generalizability of the findings. Another limitation lies in the measurement of learning outcomes. Although we assessed both surface and deep-level understanding immediately after the lecture, the evaluation focused on short-term memory and comprehension. Long-term memory and understanding were not addressed but should be considered in future investigations.

Regarding the automated approach using eye tracking and machine learning, the classification results obtained in this study did not show high predictive performance. Whereas they showed the potential of the approach to distinguish between aware and unaware TUTs, they have not yet obtained the performance required for reliable continuous measurement and automated detection. 
A significant constraint in this study was using a relatively small data set, especially for implementing complex machine-learning algorithms. Subsequent research should investigate the opportunities to improve the prediction of the meta-awareness of TUTs using larger-scale data sets. Further, more advanced eye-tracking features, such as scanpath features and temporal algorithms, could be employed to enhance prediction accuracies. The use of more elaborate methods, for instance temporal models and further optimization of the hyperparameters tested, could also contribute to this. Additionally, the Overall, the outcomes of this study underscore the significance of future research in making finer distinctions regarding the meta-awareness of mind wandering throughout the learning process.

\subsubsection{Conclusion}

In summary, this study sheds light on the prevalence and characteristics of mind wandering and meta-awareness during Zoom video-lecture viewing. The findings underscore the significance of different mind-wandering patterns and their implications for metacognitive self-regulation. The identified patterns revealed nuanced associations with learning, showing a negative relation between patterns dominated by unaware mind wandering with fact-based and deep-level understanding, whereas persistent aware mind-wandering patterns specifically deteriorated deep-level understanding. Through gaze tracking, specific eye-movement indicators were linked to different mind-wandering states. Notably, predictive modeling using gaze data displayed great promise for distinguishing between aware and unaware mind wandering, albeit with constraints due to limited data set size and the inherently internal nature of cognitive processes. 
This study underscores the complexity of mind wandering in educational settings, suggesting the need for tailored interventions based on the nuanced understanding of meta-awareness and its manifestations. It lays the groundwork for future investigations into that fine-grained, automated detection of mind wandering through low-threshold sensors. Such detection could enable personalized learning adaptations and interventions, potentially enhancing self-regulated learning through strategies like feedback, content adaptation, and intermediate questioning while highlighting ethical considerations in implementing such technologies.

\subsection*{Supplemental Material}

\subsubsection{Eye-Gaze Features by Mind Wandering Meta-Awareness} \label{etsum}

The investigation of how meta-awareness in mind wandering manifests in eye movements involved comparing the average gaze features extracted during aware and unaware mind wandering instances with those that occurred during on-task instances. To achieve this comparison, we computed \textit{t} tests to compare the mean values of each eye-gaze feature among groups that reported being on task, aware of mind wandering, and unaware of mind wandering.

We investigated how meta-awareness in mind wandering is reflected in gaze behavior by initially comparing mean levels of extracted gaze features across distinct self-reported thought groups. Table \ref{tab:etsum} delineates all features' mean values and standard deviations from 10-s time windows categorized by self-reported thought type. For simplicity, we only report summary statistics of eye-tracking features resulting from mean aggregations over time in this table omit other statistical aggregations computed during data pre-processing. Employing \textit{t} tests presented in the final three columns of Table \ref{tab:etsum}, we scrutinized the statistical differences across the groups. Notably, the \textit{t} tests revealed that instances of unaware mind wandering exhibited a significantly smaller number of fixations ($p < .01$), averaging 16.43 fixations per 10 s compared with 18.73 fixations during on-task instances. Additionally, the ratio between average fixation and saccade duration was notably higher ($p < .05$) during episodes of TUTs without awareness compared to attentive lecture viewing. Discrepancies in eye movement speed were also apparent. Specifically, the mean and peak saccade velocity and acceleration were notably higher ($p < .05$), whereas the peak deceleration was lower during instances of unaware TUTs. No significant differences were found in any extracted features between on-task and aware mind-wandering instances. Nevertheless, it is intriguing to note that the general trend paralleled unaware mind wandering across most features, except for fixation count, with differences being less pronounced. Notably, modestly higher values for fixation saccade ratio, saccade velocity, and acceleration features were observed in comparison with the on-task group, positioning these values between the mean values of the on-task and unaware TUT instances across these dimensions. When comparing aware and unaware TUT instances, a significant difference ($p < .05$) in the number of fixations was noted, indicating a higher fixation count on the screen when participants engaged in mind wandering with awareness, comparable to on-task instances. This analysis focused on a subset of the features utilized for subsequent computational modeling, specifically mean values over time. This approach was adopted to enhance clarity and facilitate interpretability purposes. In the predictive modeling stage, various other statistical aggregation methods, including minimum and maximum values along with standard deviations, were also incorporated, as detailed in the Instruments section.

\begin{sidewaystable}[]
    {\small\setlength{\tabcolsep}{1.4pt} 
    \caption{Average eye-tracking feature and \textit{t} test values by mind-wandering self-report categories.}
    \label{tab:etsum}
    \begin{adjustbox}{scale=0.95,center}
    \begin{tabular}{l@{\hskip 7.5pt}lll@{\hskip 9.5pt}lll}
    \hline
     & \multicolumn{3}{c}{\textit{M (SD)}}                                  & \multicolumn{3}{c}{\textit{t values}}                                                                                                                                                          \\
    \multicolumn{1}{l}{Gaze Feature}                              & On Task              & Aware TUTs& Unaware TUTs& \begin{tabular}[c]{@{}l@{}}On Task vs. \\ Aware\end{tabular} & \begin{tabular}[c]{@{}l@{}}On Task vs. \\ Unaware\end{tabular} & \begin{tabular}[c]{@{}l@{}}Aware vs. \\ Unaware\end{tabular} \\ \hline
    Fixation Count                                    & 18.73 (7.33)        & 18.32 (7.07)        & 16.44 (7.54)        & 0.65                                                      & 2.97 **                                                         & 2.14*                                                          \\
    Fixation Duration Mean {[}ms{]}                   & 499.77 (377.51)    & 518.136 (469.83)     & 563.09 (648.16)     & -0.53                                                     & -1.39                                                           & -0.678                                                         \\
    Fixation Dispersion X Mean {[}px{]}               & 44.54 (29.90)       & 44.03 (17.902)       & 44.29 (15.40)       & 0.21                                                      & 0.09                                                            & -0.125                                                         \\
    Fixation Dispersion Y Mean {[}px{]}               & 49.52 (24.55)      & 50.60 (16.55)       & 52.17 (16.14)       & -0.55                                                     & -1.08                                                           & -0.787                                                         \\
    Fixation Saccade Ratio Mean                       & 10.73 (7.61)       & 11.43 (11.01)       & 13.42 (21.79)       & -0.95                                                     & -2.28 *                                                         & -1.023                                                         \\
    Saccade Duration Mean {[}ms{]}                    & 46.77 (7.26)        & 47.20 (8.36)        & 45.64 (7.18)        & -0.67                                                     & 1.49                                                             & 1.621                                                          \\
    Saccade Amplitude Mean {[}°{]}                    & 4.96 (2.44)        & 5.20 (2.18)          & 5.25 (2.80)         & -1.17                                                     & -1.09                                                           & -0.157                                                         \\
    Saccade Length Mean {[}px{]}                      & 316.49 (127.83)    & 326.95 (124.22)     & 315.02 (143.25)      & -0.96                                                     & 0.11                                                            & 0.745                                                          \\
    Saccade Velocity Average Mean {[}°/s²{]}          & 101.62 (56.03)      & 109.22 (60.12)      & 115.09 (72.45)      & -1.55                                                     & -2.18 *                                                         & -0.742                                                         \\
    Saccade Velocity Peak Mean {[}°/s²{]}            & 238.76 (146.02)    & 257.00 (147.28)     & 279.83 (191.39)     & -1.45                                                     & -2.54 *                                                         & -1.137                                                         \\
    Saccade Velocity Peak Mean {[}\%{]}               & 40.74 (3.85)        & 41.17 (4.09)        & 41.25 (3.50)        & -1.26                                                     & -1.28                                                           & -0.178                                                         \\
    Saccade Acceleration Average  Mean {[}°/s²{]}     & 5934.17 (3379.08)  & 6490.86 (3972.81)   & 6886.43 (4507.41)   & -1.83                                                     & -2.53 *                                                         & -0.778                                                         \\
    Saccade Acceleration Peak Mean {[}°/s²{]}         & 10096.78 (5288.92) & 10866.85 (6365.60)   & 11349.70 (7074.60)  & -1.61                                                     & -2.13 *                                                         & -0.598                                                         \\
    Saccade Deceleration Peak Mean {[}°/s²{]}         & -9322.59 (5910.94) & -10129.84 (6075.57) & -11061.46 (7656.50) & 1.58                                                      & 2.66 **                                                         & 0.255                                                          \\
    Blink Count                                       & 4.11 (3.14)        & 4.33 (3.15)         & 4.15 (2.87)          & -0.76                                                     & -0.12                                                           & 0.454                                                          \\
    Blink Duration Mean {[}ms{]}                      & 224.36 (48.66)     & 224.44 (50.41)       & 232.13 (56.84)      & -0.02                                                     & -1.41                                                           & -1.141                                                         \\
    Vergence Angles  Mean {[}rad{]}                  & 0.07 (0.04)        & 0.07 (0.044)         & 0.06 (0.04)          & 0.60                                                      & 1.89                                                            & 1.186                                                          \\
    Vergence Pupil Distance Mean {[}px{]}                       & 270.23 (91.196)     & 264.06 (79.34)      & 281.25 (127.74)     & 0.81                                                      & -1.07                                                           & -1.409                                                         \\
    Average Pupil Diameter Mean {[}mm{]}             & -0.46 (0.35)        & -0.42 (0.32)         & -0.44 (0.31)        & -1.42                                                     & -0.46                                                           & 0.65                                                           \\ \hline
    \end{tabular}
    \end{adjustbox}
    \begin{tablenotes}[para,flushleft]
        {\small
            * \textit{p} \textless{} .05.   ** \textit{p} \textless{} .01.   *** \textit{p} \textless{} .001. \\
         }
    \end{tablenotes}
    }   
\end{sidewaystable}

This investigation into differences in gaze indicators by the meta-awareness of mind wandering showed that differences in single features were more pronounced and were statistically significant only during unaware TUTs compared with on-task episodes. Specifically, a higher number of fixations, a higher fixation saccade duration ratio, and higher values for saccade velocity and acceleration were found for mind wandering without meta-awareness. Whereas for most of these gaze characteristics, a similar, less strong trend was found for aware TUTs, we found a significantly lower number of fixations compared with unaware TUTs, which was very similar to the average fixation count while being on task. These results confirm the previously identified high task and stimulus dependency of gaze features \parencite{Faber.2020}.

\subsubsection*{Evaluation Metrics}\label{evalmet}
The F$_{1}$ scores were calculated as follows: 

\[
F_{1}=2\times\frac{Precision \times Recall}{Precision + Recall}
\]

where the precision (i.e., the proportion of correctly predicted positives of all predicted positives) and recall (i.e., the proportion of correctly predicted positives of all positive instances; sensitivity) are defined as: 
\[
Precision = \frac{TP}{TP+FP},\qquad Recall = \frac{TP}{TP+FN}. 
\]
TP, FP, TN, and FN represent true positive, false positive, true negative, and false negative, respectively.

\subsubsection*{Additional Figures and Tables}

\begin{sidewaystable}[h]
{\setlength{\tabcolsep}{1.7pt} 
\caption{Means, Standard Deviations, and Correlations of all Variables.} 
\label{tab:corr_tab}

\begin{adjustbox}{scale=0.93,center}
\begin{tabular}{lllccccccccccccccc}
\hline
Variable                            & M     & SD    & 1      & 2     & 3      & 4      & 5     & 6     & 7     & 8    & 9     & 10    & 11     & 12   & 13    & 14    & 15    \\ \hline
1. On task rate                     & 43.22 & 20.46 &        &       &        &        &       &       &       &      &       &       &        &      &       &       &       \\
2. MM rate                          & 16.32 & 10.94 & -.23*  &       &        &        &       &       &       &      &       &       &        &      &       &       &       \\
3. Aware TUT rate                   & 24.21 & 15.56 & -.57** & -.25* &        &        &       &       &       &      &       &       &        &      &       &       &       \\
4. Unaware TUT  rate                & 2.30  & 2.25  & -.58** & -.13  & -.10   &        &       &       &       &      &       &       &        &      &       &       &       \\
5. Fact-based score                 & 3.06  & 1.59  & .38**  & -.07  & -.19   & -.26*  &       &       &       &      &       &       &        &      &       &       &       \\
6. Inference-based score            & 2.48  & 1.45  & .43**  & -.20  & -.28** & -.16   & .55** &       &       &      &       &       &        &      &       &       &       \\
7. Total test score                 & 5.54  & 2.69  & .46**  & -.15  & -.27*  & -.24*  & .89** & .87** &       &      &       &       &        &      &       &       &       \\
8. Pretest score                    & 0.61  & 0.85  & .14    & -.06  & -.16   & -.03   & .20   & .25*  & .25*  &      &       &       &        &      &       &       &       \\
9. Self concept statistics           & 2.82  & 0.78  & .28**  & -.25* & -.17   & -.03   & .45** & .42** & .49** & .07  &       &       &        &      &       &       &       \\
10. Dispositional interest          & 1.66  & 0.72  & .26*   & -.24* & -.08   & -.13   & .37** & .34** & .40** & .11  & .57** &       &        &      &       &       &       \\
11. Dispositional MW    & 3.81  & 0.74  & -.20   & -.05  & .12    & .20    & -.02  & -.09  & -.06  & -.03 & .08   & .20   &        &      &       &       &       \\
12. Meta-cognitive SR & 3.82  & 0.54  & .34**  & -.17  & -.09   & -.24*  & .15   & .34** & .27** & .09  & .12   & .13   & -.33** &      &       &       &       \\
13. Situational interest            & 2.63  & 0.80  & .47**  & .01   & -.30** & -.34** & .31** & .39** & .39** & .26* & .29** & .28** & -.20   & .16  &       &       &       \\
14. Involvement                     & 3.13  & 0.77  & .36**  & .01   & -.29** & -.22*  & .22*  & .28** & .28** & .26* & .03   & .20   & -.28** & .12  & .46** &       &       \\
15. Cognitive engagement            & 2.60  & 0.53  & .46**  & .05   & -.25*  & -.42** & .29** & .36** & .37** & .17  & .15   & .20   & -.19   & .26* & .60** & .67** &       \\
16. Positive emotions               & 2.18  & 0.73  & .33**  & -.07  & -.29** & -.10   & .20   & .23*  & .24*  & .09  & .22*  & .10   & -.15   & .08  & .68** & .27*  & .31** \\ \hline
\end{tabular}
\end{adjustbox}}
\begin{tablenotes}[para,flushleft]
    {\small
        
        \textit{Note.} M and SD are used to represent mean and standard deviation, respectively. * indicates p < .05. ** indicates p < .01. The abbreviations MW and SR stand for mind wandering, and self regulation respectively.
     }
\end{tablenotes}

\end{sidewaystable}

\begin{table}[h]
\centering
\caption{Content Probes.}
\begin{tabular}{lll}
\hline
Content Probe                       & Count & Proportion \\ \hline
On task                             & 462   & 0.354   \\
Ideas about lecture (TRT)           & 102   & 0.078   \\
Lecture Comprehension (MM)        & 213   & 0.163   \\
Current state (TUT)                 & 244   & 0.187   \\
Personal matters (TUT)              & 157   & 0.120   \\
Daydreaming (TUT)                   & 61    & 0.047   \\
Video stimulus (TUT)                & 20    & 0.015   \\
Lecture triggered (TUT)             & 16    & 0.012   \\
Experiment (TUT)                    & 11    & 0.008   \\
Blank (TUT)                         & 7     & 0.005   \\
Other                               & 12    & 0.009   \\ \hline
Total                               & 1305  & 1          \\ \hline
\end{tabular}
\label{tab:con}
\end{table}

\begin{table}[h]
\centering
\caption{Content Probe TUTs by Meta-Awareness.}
\begin{tabular}{llll}
\hline
Content   Probe (TUT)   & Awareness Probe & Count & Proportion \\ \hline
Current state (TUT)     & aware           & 171   & 0.713     \\
                        & unaware         & 69    & 0.288    \\
Personal matters (TUT)  & aware           & 80    & 0.513   \\
                        & unaware         & 76    & 0.487   \\
Daydreaming (TUT)       & aware           & 27    & 0.458   \\
                        & unaware         & 32    & 0.542  \\
Lecture triggered (TUT) & aware           & 13    & 0.867   \\
                        & unaware         & 2     & 0.133  \\
Video stimulus (TUT)    & aware           & 14    & 0.737   \\
                        & unaware         & 5     & 0.263   \\
Experiment (TUT)        & aware           & 9     & 0.818   \\
                        & unaware         & 2     & 0.182  \\
Blank (TUT)             & aware           & 2     & 0.333   \\
                        & unaware         & 4     & 0.667   \\\hline
Total                   & aware           & 316   & 0.625   \\ 
                        & unaware         & 190   & 0.375   \\ \hline
\end{tabular}
\label{tab:conaw}
\end{table}

\begin{figure}
    \caption{Modal Category by Probe Plots by Cluster.}
    \includegraphics[width=\textwidth]{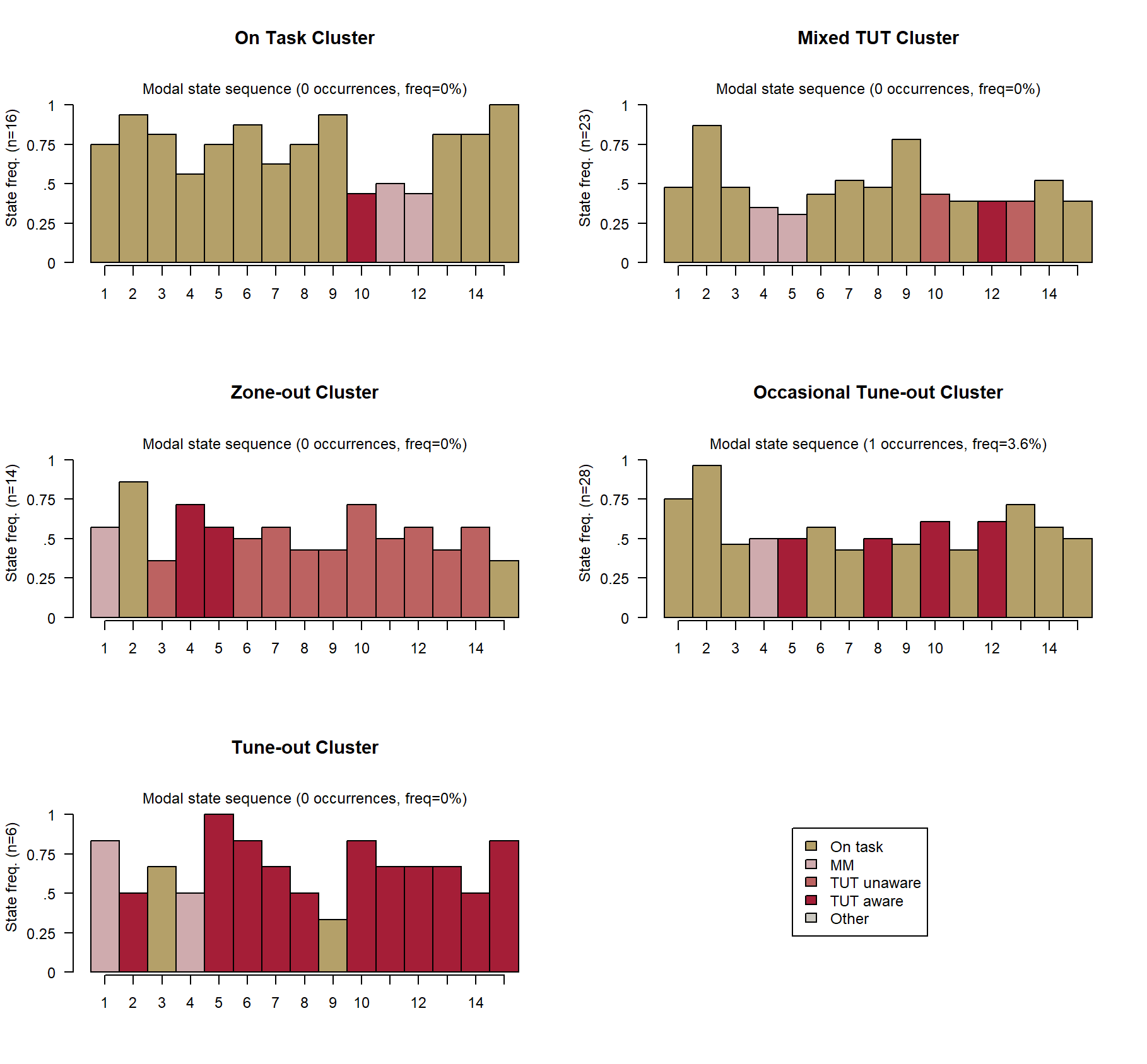}
    \label{fig:clu_modal}
\end{figure}

\begin{sidewaystable}[h]
\caption{Linear Regression Analysis of Thought Category Rates on Fact-Based, Inference-Based, and Total Test Scores}
\begin{tabular}{lllllll}
\hline
                   & \multicolumn{2}{l}{Fact-Based Test Score} & \multicolumn{2}{l}{Inference-Based Test Score} & \multicolumn{2}{l}{Total Test Score} \\
Predictors         & Estimates         & CI                    & Estimates            & CI                      & Estimates       & CI                 \\ \hline
Intercept          & 5.00 **           & 2.00 – 8.00           & 5.77 ***             & 3.14 – 8.39             & 10.76 ***       & 6.10 – 15.43       \\
MM Rate            & -0.01             & -0.04 – 0.02          & -0.03 *              & -0.06 – -0.00           & -0.04           & -0.08 – 0.01       \\
Aware TUT Rate     & -0.01             & -0.04 – 0.01          & -0.03 **             & -0.04 – -0.01           & -0.04 *         & -0.07 – -0.01      \\
Unaware TUT Rate    & -0.03 **          & -0.05 – -0.01         & -0.02 *              & -0.04 – -0.00           & -0.05 **        & -0.08 – -0.01      \\
Previous Knowledge & 0.28              & -0.08 – 0.63          & 0.34 *               & 0.02 – 0.65             & 0.61 *          & 0.06 – 1.17        \\
Self Concept       & 0.79 ***          & 0.37 – 1.21           & 0.49 **              & 0.12 – 0.86             & 1.28 ***        & 0.62 – 1.93        \\
Male               & -0.25             & -0.97 – 0.48          & -0.71 *              & -1.34 – -0.07           & -0.95           & -2.08 – 0.18       \\
Age                & -0.05             & -0.16 – 0.07          & -0.07                & -0.17 – 0.04            & -0.11           & -0.29 – 0.07       \\ \hline
Observations       & \multicolumn{2}{l}{87}                    & \multicolumn{2}{l}{87}                         & \multicolumn{2}{l}{87}               \\
R2 / R2 adjusted   & \multicolumn{2}{l}{0.320 / 0.260}         & \multicolumn{2}{l}{0.374 / 0.319}              & \multicolumn{2}{l}{0.422 / 0.371}    \\ \hline
\end{tabular}
\begin{tablenotes}[para,flushleft]
    {\small
        \textit{Note.} Reference Category is the On Task Rate. * p\textless{}0.05   ** p\textless{}0.01   *** p\textless{}0.001 \\
     }
\end{tablenotes}
\label{tab:rate_reg}
\end{sidewaystable}

\begin{table}
\centering
\caption{Hyperparameter Grids by Classification Model. }
\label{tab:hyper}

\begin{tabular}{lll}
\hline
Model                          & Hyperparameter       & Grid                    \\ \hline
Random Forest & balancing technique  & SMOTE, Upsampling, None \\
                               & bootstrap            & true, false             \\
                               & max depth            & 6, 10, 20, 50, None     \\
                               & max features         & sqrt, None             \\
                               & min samples leaf     & 1, 2, 4                   \\
                               & min samples split    & 2, 5, 10                  \\
                               & number of estimators & 50 100, 200            \\
XGBoost      & balancing technique  & SMOTE, Upsampling, None \\
                               & subsample            & 0.6, 0.8, 1             \\
                               & max depth            & 3, 6, 10                  \\
                               & number of estimators & 100, 500                \\
                               & colsample by tree    & 0.1, 0.5,  1.0            \\
SVM          & balancing technique  & SMOTE, Upsampling, None  \\
                               & C                    & 0.1, 5, 10              \\
                               & gamma                & 1, 0.1, 0.01            \\
                               & kernel               & linear, rbf             \\
MLP           & balancing technique  & SMOTE, Upsampling, None \\
                               & activation           & tanh, relu              \\
                               & solver               & sdg, adam, lbfgs        \\
                               & alpha                & 0.06, 0.05, 0.004       \\
                               & learning rate        & constant, adaptive      \\ \hline
\end{tabular}

\end{table}

\begin{table}
\centering
 \caption{Binary Mind Wandering Classification Results. }
\begin{tabular}{lrrrr} 
\toprule
        Model    & F$_{1}$ Score & Precision & Recall & AUC-PR\\
\midrule 
    Random forest & 0.474       & 0.424         & 0.538         & 0.399 \\
    XGBoost     & 0.488         & \textbf{0.479} & 0.497        & 0.453\\
    SVC         &  0.474        & 0.435         & 0.525         & \textbf{0.471} \\
    MLP         & \textbf{0.529} & 0.460        & \textbf{0.624}& 0.463\\
     \midrule
    Chance Level        & 0.384 & 0.387 & 0.383 & 0.401\\
\bottomrule
\end{tabular}

    \label{tab:mlbin}
\end{table}

\begin{sidewaystable}
{\small\setlength{\tabcolsep}{1.4pt} 
\caption{SHAP most important eye-tracking features and \textit{t} test values by mind-wandering self-report categories.}
\label{tab:SHAPetsum}
\begin{adjustbox}{scale=0.92,center}

\begin{tabular}{l@{\hskip 7.5pt}lll@{\hskip 7.5pt}lll}
\hline
                     & \multicolumn{3}{c}{\textit{M (SD)}}                                                                      & \multicolumn{3}{c}{\textit{t values} }                                                                                                                                                                                                                            \\
      Gaze Feature                                            & \multicolumn{1}{l}{On Task} & \multicolumn{1}{l}{Aware TUTs} & \multicolumn{1}{l}{Unaware TUTs} & \multicolumn{1}{l}{\begin{tabular}[l]{@{}l@{}}On Task vs. \\ Aware\end{tabular}} & \multicolumn{1}{l}{\begin{tabular}[l]{@{}l@{}}On Task vs. \\ Unaware\end{tabular}} & \multicolumn{1}{l}{\begin{tabular}[l]{@{}l@{}}Aware vs. \\ Unaware\end{tabular}} \\ \hline
Blink Duration Quantil 75 {[}ms{]}                & 242.6   (55.573)            & 245.271   (59.53)              & 256.761   (67.793)               & -0.522                                                                           & -2.24*                                                                             & -1.437                                                                           \\
Blink Duration Std {[}ms{]}                       & 51.292   (36.536)           & 59.175   (37.585)              & 67.371   (48.358)                & -2.17*                                                                           & -3.48***                                                                           & -1.403                                                                           \\
Fixation Average Pupil Diameter {[}mm{]} Std      & 0.171   (0.138)             & 0.186   (0.156)                & 0.206   (0.152)                  & -1.25                                                                            & -2.34*                                                                             & -1.03                                                                            \\
Fixation Duration Skew {[}ms{]}                   & 1.8   (0.844)               & 1.911   (0.891)                & 1.76   (0.841)                   & -1.516                                                                           & 0.443                                                                              & 1.417                                                                            \\
Fixation Saccade Ratio Std                        & 155.452   (2110.786)        & 1786.707   (23428.101)         & 40.329   (45.546)                & -1.615                                                                           & 0.564                                                                              & 0.771                                                                            \\
Pupil Distance Std {[}px{]}                       & 105.647   (78.612)          & 115.272   (80.409)             & 126.023   (88.192)               & -1.416                                                                           & -2.423*                                                                            & -1.061                                                                           \\
Saccade Acceleration Average {[}°/s²{]} Std       & 4957.889   (4352.558)       & 5688.987   (5140.366)          & 6395.516   (5484.498)            & -1.864                                                                           & -2.987**                                                                           & -1.097                                                                           \\
Saccade Amplitude Quantil 75 {[}°{]}              & 6.657   (3.753)             & 6.909   (3.401)                & 6.711   (4.364)                  & -0.798                                                                           & -0.134                                                                             & 0.429                                                                            \\
Saccade Duration Min {[}ms{]}                     & 27.111   (4.493)            & 27.577   (4.882)               & 27.606   (4.471)                 & -1.179                                                                           & 0.777                                                                              & 0.234                                                                            \\
Saccade Duration Std {[}ms{]}                     & 16.024   (4.996)            & 15.755   (5.015)               & 15.605   (5.646)                 & 0.625                                                                            & -1.051                                                                             & -0.051                                                                           \\
Saccade Length Std {[}px{]}                       & 276.852   (126.916)         & 298.825   (151.096)            & 291.664   (153.37)               & -1.916                                                                           & -1.066                                                                             & 0.386                                                                            \\
Saccade Velocity Average {[}°/s²{]} Quantil 75{]} & 127.856   (87.189)          & 138.276   (96.36)              & 149.074   (121.978)              & -1.354                                                                           & -2.159*                                                                            & -0.833                                                                           \\
Saccade Velocity Peak {[}\%{]} Max                & 61.059   (8.315)            & 61.622   (8.407)               & 62.495   (9.154)                 & -0.785                                                                           & -1.62                                                                              & -0.826                                                                           \\
Saccade Velocity Peak {[}\%{]} Std                & 10.815   (2.544)            & 11.018   (2.662)               & 11.826   (2.741)                 & -0.918                                                                           & -3.713***                                                                          & -2.456*                                                                          \\
Veregence Angles Mean {[}rad{]}                   & 0.068   (0.042)             & 0.066   (0.044)                & 0.06   (0.038)                   & 0.603                                                                            & 1.885                                                                              & 1.186                                                                            \\
Veregence Angles Std {[}rad{]}                    & 0.011   (0.007)             & 0.013   (0.01)                 & 0.012   (0.007)                  & -3.378***                                                                        & -2.175*                                                                            & 0.604                                                                            \\ \hline
\end{tabular}
\end{adjustbox}}
\begin{tablenotes}[para,flushleft]
    {\small
         * p\textless{}0.05   ** p\textless{}0.01   *** p\textless{}0.001 \\
     }
\end{tablenotes}

\end{sidewaystable}

\newpage

\begin{table} 
\centering
\caption{The 10 Most Important Gaze Features in Random Forest Classification of Meta-Awareness in Mind Wandering.}
\label{tab:featimp}
\begin{tabular}{lc}
\hline
Gaze Feature                                        & Gini importance \\ \hline
Fixation Saccade Ratio Std                     & 0.048                  \\
Saccade Velocity Average Quantile 75 {[}°/s²{]}  & 0.017                  \\
Fixation Average Pupil Diameter Std {[}mm{]}   & 0.016                  \\
Saccade Duration Min {[}ms{]}                  & 0.014                  \\
Saccade Velocity Peak Std {[}\%{]}             & 0.014                  \\
Veregence Angles Mean {[}rad{]}                & 0.012                  \\
Blink Duration Quantile 75 {[}ms{]}             & 0.012                  \\
Fixation Duration Min {[}ms{]}                 & 0.011                  \\
Pupil Distance Std {[}px{]}                    & 0.011                  \\
Saccade Velocity Peak Median {[}\%{]}          & 0.010                  \\ \hline
\end{tabular}
\end{table}

\FloatBarrier

%% file: 2_Multimidal_MW.tex
\section[Detecting Aware and Unaware Mind Wandering: A Multimodal Machine Learning Approach]{Detecting Aware and Unaware Mind Wandering During Lecture Viewing: A Multimodal Machine Learning Approach Using Eye Tracking, Facial Videos and Physiological Data}

\subsection{Abstract}

Learners often experience aware and unaware mind wandering during educational tasks, both negatively impacting learning outcomes. Differentiating these types of task-unrelated thoughts is crucial, as they stem from different cognitive processes and warrant tailored support that addresses the specific nature of mind wandering. Automated detection of these episodes could help mitigate their adverse effects, for example, by developing adaptive, attention-aware learning environments. In this study (N = 87), we explored a novel multimodal approach, combining eye tracking, facial videos, and physiological wristbands (i.e., electrodermal activity and heart rate), to predict aware and unaware mind wandering during lecture video watching. In addition, to allow comparison to previous research, we also predicted an integrated mind-wandering category. Mind wandering was assessed using 15 two-stage thought probes to determine task-unrelated thoughts and the participants' awareness of their mind wandering.
Our findings indicate that a multimodal approach, utilizing the top 100 features from the fused data, outperforms unimodal methods. Specifically, aware mind wandering was detected at 20\% above chance (AUC-PR = 0.396), unaware mind wandering at 14\% above chance (AUC-PR = 0.267), and the combined category at 40\% above chance (AUC-PR = 0.637). Eye tracking and video features proved more predictive than physiological measures when used as standalone modalities. SHAP analysis, employed to explain the results, highlighted the significance of integrating features from all three modalities for effective detection, particularly emphasizing the role of video-based facial expressions in identifying unaware mind wandering. Going beyond the current state of the art, this study demonstrates the potential of leveraging multimodal data to enhance the precision of aware and unaware mind-wandering detection and differentiation, setting a foundation for advancing educational technologies that respond dynamically to learners' cognitive states.

\subsection{Introduction}

Learners' minds wander about 30\% of the time spent in learning activities, preventing them from absorbing learning content, which in turn has a negative effect on learning outcomes \cite{wong2022}. Mind wandering is defined as a shift of attention away from the current task to task-unrelated thoughts (TUTs) \cite{Smallwood.2006}. Specifically in remote learning and self-study settings, like video lecture viewing, learners experience mind-wandering episodes \cite{Wammes.2017, Risko.2012, Szpunar.2013, Hollis.2016, Pan.2020}, presumably as they offer less external support to sustain attention on a given learning task. 
One way to promote learners' self-regulation in such situations is through attention-aware learning technologies \cite{hutt2021, Mills.2020, dmello2017zone}that can provide support measures or adapt learning content based on the learners' needs. The prerequisite for such supportive measures is the automated detection of mind wandering using observable indicators. Several studies have explored this automated detection \cite{mills2021, hutt2016eyes} and demonstrated the suitability of eye tracking \cite{hutt2016eyes}, physiological signals, such as skin conductance \cite{Brishtel.2020} and heart rate \cite{Pham.2015}, and facial video recordings \cite{Bosch.2021, Lee.2022} as input modalities for mind-wandering detection. Beyond unimodal approaches, several studies combined two or more modalities \cite{bixler2015gazeeda, Brishtel.2020, Hutt.2019, khosravi2024}.

The goals of combining different modalities are to enhance performance and increase robustness by compensating for temporarily noisy data, resolving ambiguities, and leveraging correlations across modalities \cite{baltruvsaitis2018multimodal}. A recent meta-analysis by \citet{kuvar2023} found that multimodal approaches combining two or more modalities for mind-wandering detection consistently outperformed unimodal approaches when directly compared, showing great opportunities when combining complementary information gathered from different modalities. However, the review also highlights the modest size of improvements and between-study differences, showing partially superior detection performance in unimodal studies. This suggests that feature performance may not be additive and vary across tasks, warranting further exploration of feature combinations \cite{kuvar2023}. Motivated by these findings and going beyond the current state of the art, we employ and compare a novel combination of modalities in this study, fusing features extracted from eye tracking, facial videos, and physiological sensors. 

In addition, previous research on automated detection has predominantly treated mind wandering as a unitary state. However, mind wandering is not a homogeneous construct \cite{Seli.2018family}. Individuals may engage in TUTs without noticing them, without meta-awareness, also called ``zone-out'' \cite[]{Smallwood.2006, Smallwood.2007, Schooler.2011}. However, over time, a person usually catches their mind wandering and gains awareness of the content of thoughts. The person then may redirect their attention to the task. However, people also report engaging in TUTs with meta-awareness \cite[]{seli2017, Smallwood.2007}, so-called ``tune-outs.'' In such a case, the individual is not able or willing to direct their attention back to the learning task. Those two types of mind wandering--aware and unaware--are neurologically dissociable \cite{Christoff.2009}, therefore are hypothesized to have distinct underlying cognitive processes and appear to have distinct effects on learning \cite{smallwood2008, buhler2024edpsych}. During the task of lecture video watching, persistent zone-out patterns are associated with lower fact- and inference-based learning, whereas persistent tune-outs mostly had a negative impact on deep-level understanding \cite{buhler2024edpsych}. This suggests that different interventions or forms of support are needed in case of the prevalence of either mind-wandering type, like suggesting a short break or taking notes in contrast to adapting the content difficulty. This highlights the importance of distinguishing mind wandering along the lines of meta-awareness for automated detection. An initial study using gaze data to model the awareness levels of mind-wandering showed the potential to automatically differentiate between the two types of mind-wandering \cite{buhler2024edpsych}.


\subsubsection{Novelty and Contribution}
The present research explores, for the first time, the automated detection of aware and unaware mind wandering, employing a novel combination of modalities, namely eye tracking, video, and physiology, namely heart rate and electrodermal activity (EDA). This study makes several important contributions to research on mind wandering and learning. First, it compares and contrasts prediction performances across three individual and combined modalities for different categories of mind wandering observed during lecture viewing. We present an in-depth analysis of which features, derived from which modalities, most significantly influence the recognition of aware, unaware, and combined forms of mind wandering. Furthermore, this research contributes to a deeper understanding of the temporal dynamics of mind-wandering meta-awareness and its impact on learning outcomes. Demonstrating automated detection and differentiation of mind wandering by meta-awareness levels will enable continuous and fine-grained assessments in future research into the causes, effects, and mitigation strategies for mind wandering during learning. Consequently, this study lays a crucial foundation for further developing effective adaptive learning technologies and targeted interventions to support learners in remote learning environments.

\subsection{Related Work}


Automatically detecting mind wandering during learning is challenging, as mind wandering is an internal cognitive process that even human observers cannot detect \cite{bosch2022}. Therefore, the ground truth typically used to train detection models is self-reports gained by experience sampling. These self-reports are either gathered in a probe-caught or self-caught fashion \cite{Weinstein.2018}. In the first case, participants are interrupted at specific time points during a task and asked where their attention was directed; in the latter, participants report, for instance, via a key press, when they catch themselves mind wandering (i.e., become aware of it). However, there is evidence that mind wandering is also reflected in observable indicators. Various sensors, in conjunction with machine learning algorithms, have been employed to automatically assess mind wandering non-intrusively during various learning tasks \cite{kuvar2023}. We focus on eye tracking, video, and physiological EDA and heartrate modalitites employed in this study. Other modalities used encompass, for instance, log data \cite{dasilva2020mouse} and EEG  \cite{dhindsa2019, Conrad.2021}.

\subsubsection{Unimodal Mind Wandering Detection Approaches}
\paragraph{Eye-Tracking-Based Detection}
The most popular and most successfully employed modality is eye tracking \cite{kuvar2023}. Due to the mind-eye link, cognitive processes are reflected in eye movements \cite{just1976eye, rayner1998eye, reichle2012using}, and task-demand-specific gaze indices have been linked to mind wandering \cite{Faber.2020}. \citet{dmello2013} used eye-tracking derived global and local gaze features to detect TUTs during reading, achieving a Kappa of 0.23. Similarly, \citet{Faber.2018read} predicted mind wandering during reading, employing global gaze features and achieving AUROCs of 0.64 (28\% above chance). Developing mind wandering detection during narrative film watching \citet{mills2016} employed local and global gaze features, achieving an F$_{1}$ score of 0.49 (29\% above chance). For detection during video lecture viewing \citet{Hutt.2017} used global and local eye-tracking features, achieving an F$_{1}$ score of 0.47, which was 24\% above the chance level (0.30), using 30-second windows. \citet{Zhao.2017} compared webcam-based gaze estimation to eye-tracking for mind wandering detection during massive open online courses (\gls{mooc}s), finding similar SVM classifier performance with an F$_{1}$ score of 0.41 (16\% above chance) for eye tracking and F$_{1}$ score of 0.40 (11\% above chance) for webcam features. \citet{bixler2021crossed} examined cross-domain predictions of TUTs, achieving a 21\% improvement over chance (F$_{1}$ = 0.57) for within-dataset predictions on a video lecture task using global gaze features from 40-second windows. 
Several studies combined eye-tracking-derived features with some contextual features, mostly derived from the stimulus, which could already be considered a form of multimodality. \citet{bixler2016automatic} employed a combination of gaze features derived by eye tracking and context features (e.g., reading time, words skipped) to predict TUTs during reading, achieving a Kappa of 0.31. \citet{hutt2016eyes} also fused global gaze and contextual features (i.e., pretest score, time into session) during intelligent tutoring system (ITS) usage, achieving F$_{1}$-scores of 0.49 (37\% above chance). In a follow-up study on the same ITS, \citet{hutt2017out} brought their study from the lab to a classroom setting and added local gaze features, achieving F$_{1}$-scores of 0.59 (46\% above chance). 
In a recent study \citet{buhler2024edpsych} attempted to automatically differentiate meta-awareness of mind wandering during lecture video watching, employing global gaze features and achieved prediction accuracies of 12\% (F$_{1}$= 0.33) and 9\% (F$_{1}$= 0.22) above chance level for aware and unaware TUTs.

\paragraph{Video-Based Detection}
Another strand of literature focused on the more scalable use of facial videos. A simple webcam, built-in in most personal computers, can enable low-cost, in-the-wild assessment. Features like facial expressions, head pose, and appearance-based gaze estimation can be extracted from those videos. 
\citet{stewart2017} initiated webcam-based mind wandering detection during film viewing in a lab, achieving an F$_{1}$ score of 0.39 (13\% above chance) using SVM on aggregated facial features from 45-second windows. Subsequently, \citet[][]{stewart2017generalizability} explored cross-task classification, predicting TUTs in reading from film-watching data with a C4.5 decision tree, achieving F$_{1}$ scores of 0.41 and 0.44 (21\% and 22\% above chance) for film to reading and vice versa. 
\citet{Bosch.2021} conducted mind wandering detection using facial features in lab and classroom settings during reading tasks and ITS usage, respectively, achieving F$_{1}$ scores of 0.478 and 0.41 (25\% and 20\% above chance). Adopting a similar approach in in-the-wild settings, \citet{Lee.2022} detected mind wandering during lecture viewing using webcam videos. The authors expanded the set of video-derived features to extract gaze dynamics, emotion prediction, and head movement, achieving an F$_{1}$ of 0.36 (15\% above chance) with XGBoost. \citet{hutt2023webcam} conducted in-the-wild studies using webcam-based eye tracking for reading tasks, achieving an F$_{1}$ score of 0.25 (9\% above chance).

\paragraph{Physiology-Based Detection}
The third set of modalities employed in mind wandering detection research is physiological sensors, which are meaningful due to the relationship between physiological arousal and attention states \cite{smallwood2004phys}. \citet{Blanchard.2014} used a combination of physiology, namely skin conductance and skin temperature, with context features, for instance, text difficulty and time elapsed, as input for their detector during a reading task, achieving Kappa of 0.22. Additionally, episodes of mind wandering were associated with increased heart rate due to heightened arousal, ~\cite[][]{Smallwood.2007}; thus, this physiological marker has also been utilized for automated detection \cite{Pham.2015}. Heart rate measures and lecture content features were merged in a study by \citet{Pham.2015} on learning with MOOCs, obtaining a Kappa of 0.22. Heart rates were extracted from image frames via mobile cameras, capturing participants' fingertips, so-called photoplethysmography sensing.

\subsubsection{Multimodal Mind Wandering Detection Approaches}
All those modalities capture different observable indicators of mind wandering. Therefore, several researchers have aimed to combine modalities to improve the precision and robustness of predictions. \citet{bixler2015gazeeda} combined gaze, physiological (i.e., skin conductance and temperature), and context (e.g., time on task, difficulty) features for mind wandering detection during reading. The authors achieved a Kappa of 0.19 with feature-level fusion, representing an 11\% improvement over the best-performing unimodal approach. Similarly, \citet{Brishtel.2020} employed eye tracking, EDA, and questionnaire (i.e., interest) features during a reading task. The best models obtained prediction accuracies of Kappa 0.41. A recent study by \citet{khosravi2024} explored merging physiology, galvanic skin response, and heart rate with eye tracking during video watching, reporting accuracies of 0.90.  
A study that fused video and eye tracking features was presented by \citet{Hutt.2019}. The fusion (F$_{1}$= 0.45; 29\% above-chance accuracy) could not improve over the unimodal approach only employing eye-tracking features (Global gaze features F$_{1}$= 0.45, 29\% above-chance accuracy; local gaze features F$_{1}$: 0.49, 34\% above-chance accuracy), which also outperformed the facial-feature based classifier (AUs F$_{1}$: 0.31, 10\% above-chance accuracy; co-occurring AUs F$_{1}$: 0.3, 9\% above-chance accuracy). However, a fusion of both features could increase robustness by accounting for missing values in one of the two modalities. 

All of these studies, with the exception of \citet{buhler2024edpsych}, have in common that they predicted a unitary mind-wandering state. In this study, we compare more fine-granular mind-wandering definitions to a combined mind-wandering category and dissect how different modalities contribute to predicting these categories. For this, we employ a novel combination of previously successfully utilized modalities: Eye tracking, video, and physiology (EDA and heart rate) to assess TUTs at a more fine-granular level, distinguishing between aware and unaware mind wandering. 

\subsection{Methods}

\subsubsection{Data}

\paragraph{Participants}
In this study, we collected data from 96 university students. Due to technical issues during data collection, six participants had to be excluded, and an additional three were excluded for not having native-level language proficiency, which was a prerequisite to understanding the lecture, resulting in a final sample of 87 participants for analysis. The age range of participants was 19 to 33 years (\textit{M} = 23.44, \textit{SD} = 2.6), with 19\% of the sample being male. 

\paragraph{Study Procedure}
Participants watched a 60-minute prerecorded Zoom lecture video on introduction to statistics. The lecture video comprised both lecture slides and the lecturer’s webcam image. The participant's eye gaze, facial videos, and physiological data were recorded. Remote SMI eye trackers with a sampling frequency of 250 Hz were used for eye tracking. Facial videos of the participant were recorded using off-the-shelf webcams (1090p, 30 frames per second). E4 Empatica wristbands were employed to collect physiological data, recording EDA (4Hz) and blood volume pressure (BVP; 64Hz). To gather mind-wandering self-reports, 15 thought probes were interspersed at 3- to 5-minute intervals. The session included a mid-point recalibration of the eye trackers. Including pre- and post-questionnaires, the entire session lasted approximately 120 minutes. Participants received 20€ for their participation. The ethics committee of \textit{-- blinded for review --} 
approved our study procedures, and all participants gave written consent to the data collection. Parts of this dataset, specifically self-reports and gaze data, have previously been used for analysis in \citet{buhler2024edpsych}.

\paragraph{Mind Wandering Experience Sampling}

Mind wandering among participants was assessed using the probe-caught method, which intermittently asked for self-reports on their thoughts during the lecture. The lecture featured 15 such probes at 3- to 5-minute intervals. Participants responded to a two-stage probe (depicted in Figure \ref{fig:probec} in the Appendix) during each interruption. Initially, they were prompted to classify their thought content into one of six predefined categories or provide an open-ended response. These categories, based on previous research \cite{Kane.2017}, included several that reflected TUTs, hence mind wandering (i.e., everyday personal concerns, daydreaming, current physical or emotional state). The first three categories, describing following the lecture, lecture-related thoughts, or understanding of the lecture, were combined due to their close relation to an ``On-Task'' category. 
In the second stage of the probe, participants reported their meta-awareness of TUTs, which allowed for categorizing TUTs into aware and unaware mind wandering (MW). This led to a final classification of thoughts into three categories: ``On-Task'' (\textit{n}=779), ``Aware MW'' (\textit{n}=313), and ``Unaware MW'' (\textit{n}=192). These self-reports then served as labels for the machine-learning approach.  

As each participant answered 15 thought probes, the total dataset comprised 1305 instances. Unfortunately, the eye trackers had occasional recording failures. A known deficiency of eye trackers is that tracking failures are sometimes recorded as unusually long blinks \cite{castner2020}. Therefore, blinks longer than 500 ms (i.e., those exceeding an expected blink duration range between 100 to 400 ms; \cite[]{schiffman2001}) were excluded. These recording failures led to entirely missing eye-tracking data for 11 instances. We excluded those from the whole analysis to ensure comparability between modalities, resulting in a final dataset of 1284 instances, as depicted in Table \ref{tab:data}.

\begin{table}[]
\centering
\caption{Data by Mind Wandering Type}
\label{tab:data}
\resizebox{0.7\textwidth}{!}{\begin{tabular}{lllllll}
\hline
\multicolumn{1}{c}{}      & \multicolumn{2}{c}{Aware MW}                               & \multicolumn{2}{c}{Unaware MW}                             & \multicolumn{2}{c}{Combined MW}                            \\
\multicolumn{1}{c}{Class} & \multicolumn{1}{c}{Count} & \multicolumn{1}{c}{Percentage} & \multicolumn{1}{c}{Count} & \multicolumn{1}{c}{Percentage} & \multicolumn{1}{c}{Count} & \multicolumn{1}{c}{Percentage} \\ \hline
0                         & 971                       & 75.62\%                        & 1092                      & 85.05\%                        & 779                       & 60.67\%                        \\
1                         & 313                       & 24.38\%                        & 192                       & 14.95\%                        & 505                       & 39.33\%                        \\ \hline
Total                     & 1284                      & 100\%                          & 1284                      & 100\%                          & 1284                      & 100\%                          \\ \hline
\end{tabular}}
\end{table}

\subsubsection{Feature Engineering}

We extracted information from all collected modalities for 30-second intervals right before each thought probe. Window sizes were chosen based on results of previous research, finding that 30-second windows were the most suitable for mind wandering detection during lecture video watching \citet{Hutt.2017}.

\begin{table*}[]

\centering
\caption{Feature Groups and Number of Features Extracted from Eye Tracking, Video and Physiology Modalities.}

\resizebox{\textwidth}{!}{\begin{tabular}{lllcc}
\hline
\multicolumn{1}{c}{Modality}  & \multicolumn{1}{c}{Feature Group} & \multicolumn{1}{c}{Example Features}                                                  & \# & \# Agg. \\ \hline
\multirow{5}{*}{Eye Tracking} & Fixations                         & Fixation number, duration, dispersion, fixation-saccade ratio                         & 5           & 37                     \\
                              & Saccades                          & Saccade duration, amplitude, acceleration, velocity, length, regression proportion    & 9           & 81                     \\
                              & Blinks                            & Blink number, duration                                                                & 2           & 10                     \\
                              & Pupil Diameter                    & Fixation average pupil diameter                                                       & 1           & 9                      \\
                              & Vergence                          & Vergence angle radians, pupil distance                                                & 2           & 18                     \\ \hline
\multirow{4}{*}{Video}        & Facial Action Units               & Lid tightener, lips part, jaw drop, blink presence (0,1) and intensities (1-5)        & 35          & 315                    \\
                              & Face Shape Parameters             & Face shape location, scale and rotation, non-rigid face deformation parameters (0-33) & 40          & 360                    \\
                              & Head Pose                         & Roll, pitch and yaw, head location                                                    & 6           & 54                     \\
                              & Gaze                              & Left and right eye gaze direction vectors (x,y,z), gaze angels in radians             & 8           & 75                     \\ \hline
\multirow{2}{*}{Physiology}   & EDA                               & Cleaned signal, tonic, phasic, peak amplitude, peak rise and recovery time           & 8           & 72                     \\
                              & BVP                               & Cleaned signal, rate, peak                                                            & 4           & 36                     \\ \hline
Total                         &                                   &                                                                                       & 120         & 1064                   \\ \hline
\end{tabular}}

\begin{tablenotes}
    {\footnotesize
    \item \textit{Note.} All features that are not already aggregated (i.e., number of fixations) were aggregated over 30 seconds using nine summary statistics (e.g., mean, median, std, etc.).
     }
\end{tablenotes}
\label{tab:feat}
\end{table*}

\paragraph{Eye Tracking Data}
Eye tracking events (i.e., fixations, saccades, blinks) were extracted from the raw data employing the proprietary SMI BeGaze Software. This processing provided details such as event durations, fixation locations, saccade velocities, and pupil diameters during fixations. We also calculated additional features like the total number of fixations and blinks, fixation dispersion, saccade length, and eye vergence (see Table \ref{tab:feat} for a list of features). We focused on global gaze features, capturing broad eye movement patterns across a visual scene, because they are less task-dependent than local features, which focus on specific areas and, therefore, more generalizable \cite{kuvar2023} and showed exceeding performance in previous research \cite{hutt2016eyes}. To summarize this data over 30-second intervals, we computed various statistics for each feature, including minimum, maximum, mean, median, standard deviation, 25\% and 75\% percentiles, skewness, and kurtosis. After aggregation, the set comprised 155 global gaze features.

\paragraph{Video Data}

We extracted facial expressions, head poses, and gaze features from the recorded facial videos using the OpenFace toolbox \cite{openface2016}. Features included facial action units (AUs; presence and intensity information) and face shape parameters, describing the rigid face shape (location, scale, rotation) and non-rigid face shape (deformation due to expression and identity). The head pose was depicted by pitch, yaw, and roll of the head and the distance to the camera. Additionally, it included appearance-based gaze estimation and deriving gaze vectors and angles. Those features were extracted frame-wise and summarized using the same aggregation statistics as eye-tracking features, leading to 804 features. 

\paragraph{Physiological Data}
We extracted EDA and blood-volume pressure (BVP), an indicator for heart rate, from the Empatica E4 wristbands. Both signals were z-standardized within participants to account for interindividual differences and cleaned using the NeuroKit2 \cite{Makowski2021neurokit} package in Python. We decomposed the cleaned EDA signal (4 Hz) into phasic (i.e., phasic skin conductance response; SCR) and tonic (i.e., tonic skin conductance level; SCL) components using NeuroKit2's standard method. The phasic component reflects stimulus-dependent fast-changing signal parts, i.e., it describes peaks in the signal; more specifically, their amplitude, height, and time taken to rise (rise time) and fall again (recovery time). The tonic component describes the slowly changing, continuous EDA baseline signal. 
A PPG sensor with a 64 Hz sampling frequency measured the BVP signal, providing information on its frequency and shape. We preprocessed the BVP signal using the NeuroKit2 package, resulting in a cleaned BVP signal, the rate (i.e., heart rate baseline), and peaks (i.e., systolic (local maxima) and diastolic (local minima) peak points in the signal). Feature aggregation over time, as described above, resulted in 72 EDA and 36 BVP features and, consequently, 108 physiological features. After combining all modalities by feature concatenation, the complete dataset entailed 1064 features.

\subsubsection{Supervised Classification}

\begin{figure}
    \centering
  \includegraphics[width=0.7\textwidth]{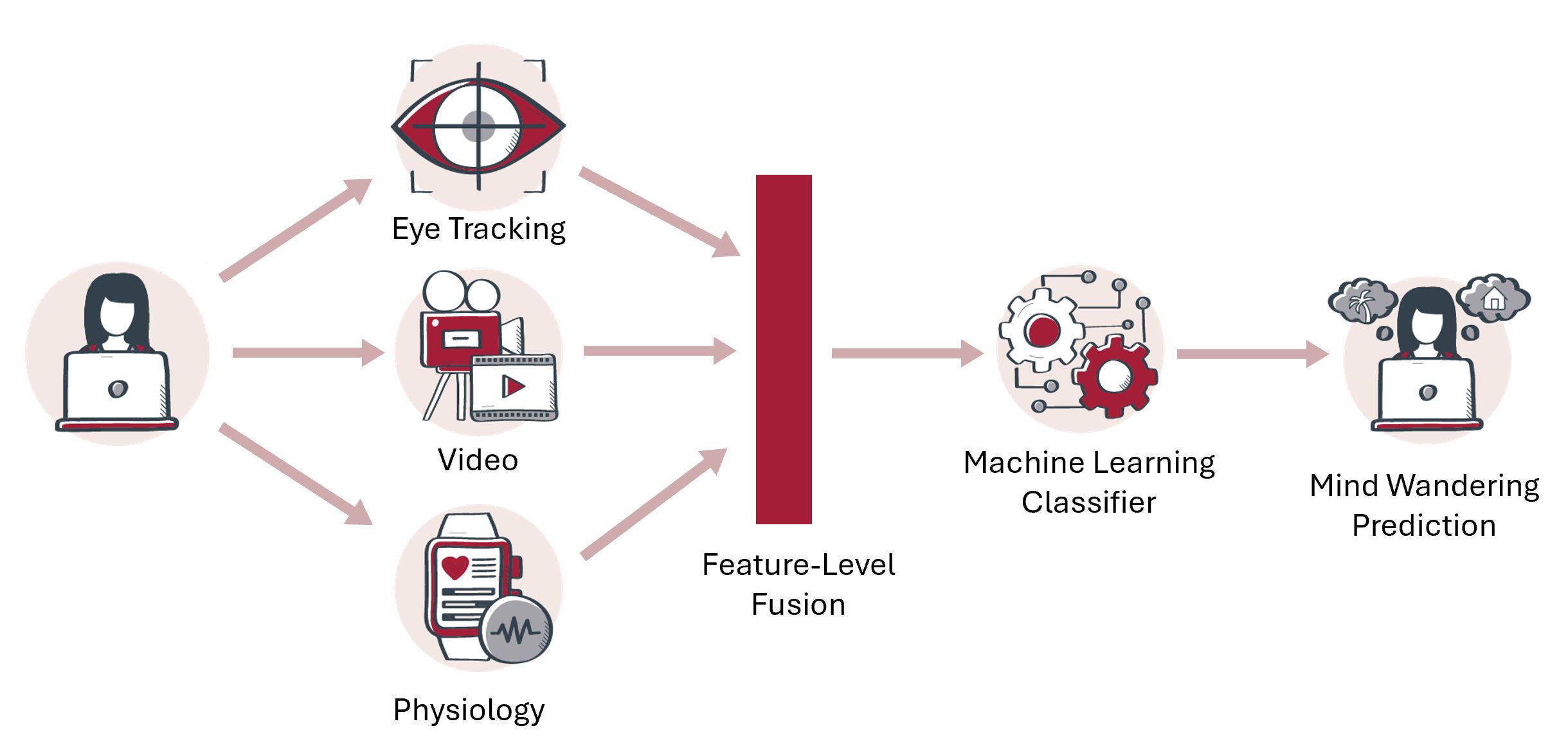}
  \caption{Mind Wandering Detection Pipeline.}
  \label{fig:pipe}
\end{figure}

\paragraph{Multimodal Fusion}
We conducted three binary classification tasks to predict both types of mind wandering—aware and unaware—separately and combined. We utilized all available modality feature sets separately and in combination for each task. Following an early fusion approach, we integrated the modalities at the feature level by concatenating aggregated feature arrays. An overview of the classification pipeline is depicted in Figure \ref{fig:pipe}. Feature level fusion was also found to outperform decision level fusion in previous research \cite{bixler2015gazeeda}. We implemented mean imputation for missing values and scaled the data. Due to the temporal aggregations resulting in constant features (i.e., min values of AUs were often 0), a variance threshold of 0 was applied to remove these features.

\paragraph{Model Building}

To address the highly imbalanced nature of the data, we applied SMOTE \cite{smote2002} or random oversampling to balance the classes during training. Using the sklearn \cite{scikit-learn} package in Python, we trained Random Forest (RF), Support Vector (SVC), and Multi-layer Perceptron (MLP) classifiers, as well as XGBoost (XGB) utilizing the XGBoost \cite{xgb} package. We primarily relied on default parameters, making adjustments to prevent overfitting. For the RF classifier, we set the maximum depth to 50. For the MLP classifier, we implemented early stopping, and for XGB, we tested learning rates of 0.3 and 0.1. 

We employed the explainability (XAI) method SHAP (SHapley Additive exPlanations) \cite{shap2017} to generate explanations of the trained multimodal models for each of the mind-wandering types. SHAP generates post-hoc, local explanations by calculating Shapley values. These Shapley values gave us insights into how different features contributed to the overall prediction of the different mind-wandering types by quantifying the influence per variable on the outcome's difference from the base value (i.e., the average prediction). Positive Shapley values indicate an increase and negative values indicate a decrease in the model's prediction \cite{shap2017}. The feature fusion increased the size of feature vectors to over a thousand, and large feature vectors can lead to overfitting and exacerbate the curse of dimensionality, which increases the data sparsity and requires more data. Therefore, according to the SHAP analysis, we retrained the models only on the 100 most important features. 

\paragraph{Validation}
To maximize training data while ensuring individual independence, we employed Leave-One-Person-Out Cross Validation, testing up to 15 instances stemming from one individual and training on all other instances. To evaluate the performance of the classifiers on our highly imbalanced dataset, we focused on the area under the precision-recall curve (AUC-PR) of the minority class' mind wandering. We focus on this metric over commonly used threshold metrics as F$_{1}$ scores because they can be driven by high recall and low prediction precision \cite{jeni13}. Such a discrepancy results in a high number of false positives in mind-wandering detection, making it problematic to apply these classification algorithms, for instance, in delivering interventions. The rank method AUC-PR, in contrast, shows precision as a function of recall; therefore, it helps to balance the two measures and is threshold-independent. For a better comparison of performances across mind-wandering categories with different base rates, we also report the improvement of the model above the chance level (refer to Appendix \ref{app} for details on the computation).

\subsection{Results}

\subsubsection{Aware, Unaware, and Combined Mind Wandering Detection}

\begin{figure}
    \centering
    \includegraphics[width=0.7\textwidth]{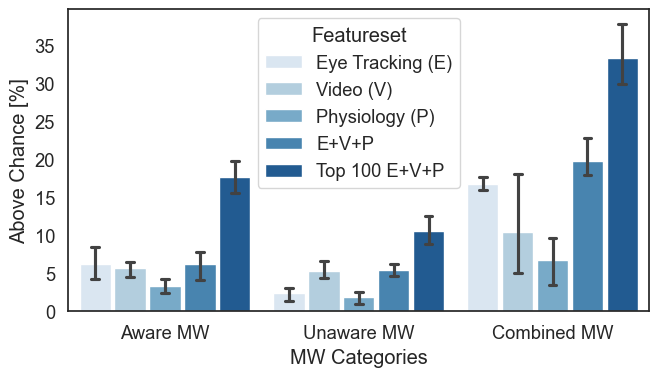}
    \caption{Avergage Classification Results (Above Chance Based on AUC-PR) by Mind Wandering Category and Modality Feature Set.}
    \label{fig:res}
\end{figure}

We conducted three separate binary classification experiments to predict aware, unaware, and a combined mind-wandering class. Classifiers were trained on single-modality-derived feature sets and fused, multimodal feature sets. 
To manage the large size of the multimodal feature vectors, which exceeded a thousand due to feature concatenation, we narrowed the features down to the top 100 most important ones. This selection was based on the SHAP analysis from the best-performing classifier for each category of mind wandering. Notably, each of the 100 feature subsets contained features of all three modalities. After selecting these features, we re-trained all classifiers using these refined feature sets. The classification results are in Table \ref{tab:res}. 
For all three predicted mind-wandering categories, we found that multimodal classifiers trained on the top 100 features outperformed unimodal approaches. Multimodal approaches trained on the whole feature set not or only slightly (combined MW) outperform unimodal approaches. However, when focusing on only about 10\% of the best features, the improvement over unimodal approaches becomes more distinct, as depicted in Figure \ref{fig:res}.

For aware mind wandering, an AUC-PR of 0.396, representing a prediction accuracy of 20.21\% above chance, was obtained when training on the 100 most important multimodal features. For aware mind wandering, the most predictive modality in unimodal approaches is eye tracking, identifying these instances 9\% above chance. 
Turning to unaware mind wandering, the highest AUC-PR of 0.267, approximately 14\% above chance (chance level being 0.151), can be achieved using the 100 most important features from the multimodal dataset. In contrast to aware mind wandering, the most successful unimodal approach was based on video-derived features, predicting unaware mind wandering 7\% above chance. 
We formulated the detection tasks as binary classifications, necessitating combining one of the two distinct mind-wandering categories with on-task instances into a single negative category. One could assume that the two mind-wandering categories are more similar to one another and, therefore, harder to distinguish. To explore this, we analyzed misclassified examples to determine the composition of false positives. Our findings reveal that for aware mind wandering, only 15\% of false positives were actually unaware mind wandering instances, while 85\% were on-task instances. This means only 8\% of unaware mind wandering instances were classified as aware mind wandering. Conversely, for unaware mind wandering classification, 25\% of false positives were aware mind wandering instances, and only 12\% of all aware mind wandering instances in the data were classified as unaware mind wandering. This analysis suggests that the two mind-wandering categories are distinguishable within a binary classification framework.
When integrating aware and unaware mind wandering into one category, models trained on a set entailing eye tracking, video, and physiological features obtained an AUC-PR of 0.637, a prediction of about 40\% above chance (chance level being 0.393). For this broader mind-wandering definition, overall eye tracking-based classifiers (16-18\% above chance) perform best when comparing unimodal approaches, except for the video-based XGBoost classifier (21\% above chance).

In all three experiments, physiological features were the least predictive of mind wandering when used as a standalone modality. However, physiological features were included in the 100 most essential features of multimodal classifiers, highlighting their importance in the prediction, potentially in correlation with other modalities. There is no clear trend regarding which of the tested classifiers is the best suited for which classification problem or feature set.


\begin{sidewaystable}[]
\caption{Classification Results by Mind Wandering Category and Employed Modality Feature Sets.}
\label{tab:res}
\resizebox{\textwidth}{!}{
\begin{tabular}{lllllll|lllll|lllll}
\hline
\multicolumn{1}{c}{}           & \multicolumn{1}{c}{}      & \multicolumn{5}{c|}{Aware Mind Wandering}                                                                                                                        & \multicolumn{5}{c|}{Unaware Mind Wandering}                                                                                                                       & \multicolumn{5}{c}{Combined Mind Wandering}                                                                                                                   \\
\multicolumn{1}{c}{Featureset} & \multicolumn{1}{c}{Model} & \multicolumn{1}{c}{AUC-PR} & \multicolumn{1}{c}{\% AC} & \multicolumn{1}{c}{F$_{1}$} & \multicolumn{1}{c}{Precision} & \multicolumn{1}{c|}{Recall} & \multicolumn{1}{c}{AUC-PR} & \multicolumn{1}{c}{\% AC} & \multicolumn{1}{c}{F$_{1}$} & \multicolumn{1}{c}{Precision} & \multicolumn{1}{c|}{Recall} & \multicolumn{1}{c}{AUC-PR} & \multicolumn{1}{c}{\% AC} & \multicolumn{1}{c}{F$_{1}$} & \multicolumn{1}{c}{Precision} & \multicolumn{1}{c}{Recall} \\ \hline
Eye Tracking (E)                             & RF             & 0.306                    & 9.08                                & 0.226                  & 0.354                         & 0.166                       & 0.157                    & 0.71                                 & 0.073                  & 0.167                         & 0.046                       & 0.492                    & 16.31                               & 0.449                  & 0.507                         & 0.402                      \\
                               & XGB                       & 0.288                    & 6.32                                & 0.261                  & 0.342                         & 0.211                       & 0.177                    & 3.06                                 & 0.118                  & 0.205                         & 0.082                       & 0.488                    & 15.65                               & 0.453                  & 0.481                         & 0.428                      \\
                               & SVC                       & 0.285                    & 5.87                                & 0.337                  & 0.268                         & 0.454                       & 0.172                    & 2.47                                 & 0.257                  & 0.184                         & 0.428                       & 0.503                    & 18.12                               & 0.507                  & 0.483                         & 0.535                      \\
                               & MLP                       & 0.269                    & 3.56                                & 0.333                  & 0.268                         & 0.441                       & 0.178                    & 3.18                                 & 0.27                   & 0.196                         & 0.433                       & 0.496                    & 16.97                               & 0.496                  & 0.472                         & 0.523                      \\ \hline 
Video (V)                             & RF             & 0.272                    & 3.98                                & 0.132                  & 0.284                         & 0.086                       & 0.196                    & 5.3                                  & 0.081                  & 0.321                         & 0.046                       & 0.432                    & 6.43                                & 0.39                   & 0.416                         & 0.367                      \\
                               & XGB                       & 0.286                    & 6.02                                & 0.204                  & 0.352                         & 0.144                       & 0.192                    & 4.83                                 & 0.084                  & 0.222                         & 0.052                       & 0.526                    & 21.91                               & 0.443                  & 0.512                         & 0.391                      \\
                               & SVC                       & 0.29                     & 6.62                                & 0.328                  & 0.281                         & 0.393                       & 0.212                    & 7.18                                 & 0.247                  & 0.225                         & 0.273                       & 0.415                    & 3.62                                & 0.472                  & 0.443                         & 0.505                      \\
                               & MLP                       & 0.285                    & 5.87                                & 0.33                   & 0.28                          & 0.403                       & 0.185                    & 4                                    & 0.228                  & 0.22                          & 0.237                       & 0.451                    & 9.56                                & 0.493                  & 0.446                         & 0.55                       \\ \hline 
Physiology (P)                             & RF             & 0.277                    & 4.7                                 & 0.208                  & 0.299                        & 0.16                        & 0.155                    & 0.47                                 & 0.105                  & 0.161                         & 0.077                       & 0.403                    & 1.65                                & 0.372                  & 0.41                          & 0.341                      \\
                               & XGB                       & 0.259                    & 2.16                                & 0.203                  & 0.245                         & 0.173                       & 0.173                    & 2.59                                 & 0.152                  & 0.229                         & 0.113                       & 0.428                    & 5.77                                & 0.4                    & 0.42                          & 0.383                      \\
                               & SVC                       & 0.269                    & 3.56                                & 0.315                  & 0.251                         & 0.422                       & 0.169                    & 2.12                                 & 0.235                  & 0.166                         & 0.402                       & 0.446                    & 8.73                                & 0.459                  & 0.43                          & 0.491                      \\
                               & MLP                       & 0.263                    & 2.71                                & 0.292                  & 0.274                         & 0.313                       & 0.172                    & 2.47                                 & 0.222                  & 0.158                         & 0.376                       & 0.457                    & 10.54                               & 0.438                  & 0.423                         & 0.454                      \\  \hline
E + V + P                     & RF             & 0.295                    & 7.38                                & 0.115                  & 0.31                          & 0.07                        & 0.189                    & 4.48                                 & 0.057                  & 0.353                         & 0.031                       & 0.501                    & 17.79                               & 0.433                  & 0.516                         & 0.373                      \\
                               & XGB                       & 0.301                    & 8.3                                 & 0.196                  & 0.344                         & 0.137                       & 0.202                    & 6.01                                 & 0.083                  & 0.375                         & 0.046                       & 0.541                    & 24.38                               & 0.496                  & 0.549                         & 0.452                      \\
                               & SVC                       & 0.288                    & 6.32                                & 0.327                  & 0.295                         & 0.367                       & 0.205                    & 6.36                                 & 0.261                  & 0.25                          & 0.273                       & 0.503                    & 18.12                               & 0.492                  & 0.47                          & 0.517                      \\
                               & MLP                       & 0.265                    & 2.99                                & 0.332                  & 0.285                         & 0.399                       & 0.191                    & 4.71                                 & 0.237                  & 0.237                         & 0.237                       & 0.508                    & 18.95                               & 0.469                  & 0.468                         & 0.469                      \\ \hline  
Top 100 E + V + P               & RF             & 0.389                    & 19.29                               & 0.259                  & \textbf{0.449}                & 0.182                       & 0.228                    & 9.07                                 & 0.162                  & 0.318                         & 0.108                       & 0.593                    & 32.95                               & 0.507                  & 0.584                         & 0.448                      \\
                               & XGB                       & \textbf{0.396}           & \textbf{20.21}                      & 0.299                  & 0.426                         & 0.23                        & 0.225                    & 8.72                                 & 0.143                  & 0.316                         & 0.093                       & \textbf{0.637}           & \textbf{40.2}                       & \textbf{0.58}          & \textbf{0.616}                & 0.548                      \\
                               & SVC                       & 0.367                    & 16.38                               & 0.396                  & 0.341                         & 0.473                       & \textbf{0.267}           & \textbf{13.66}                       & 0.27                   & \textbf{0.296}                & 0.247                       & 0.579                    & 30.64                               & 0.569                  & 0.542                         & \textbf{0.6}               \\
                               & MLP                       & 0.354                    & 14.66                               & \textbf{0.399}         & 0.325                         & \textbf{0.514}              & 0.242                    & 10.72                                & \textbf{0.306}         & 0.275                         & \textbf{0.345}              & 0.57                     & 29.16                               & 0.55                   & 0.527                         & 0.574                      \\ \hline
\end{tabular}}
\begin{tablenotes}
    {\footnotesize
        \item \textit{Note.} Chance level is 0.243 for aware MW, 0.151 for unaware MW, and 0.393 for combined MW. AC stands for above chance, E for eye tracking, V for video, and P for physiology.
     }
\end{tablenotes}
\end{sidewaystable}

\subsubsection{Feature Importance Analysis}

\begin{figure*}
    \centering
    \includegraphics[width=0.43\textwidth]{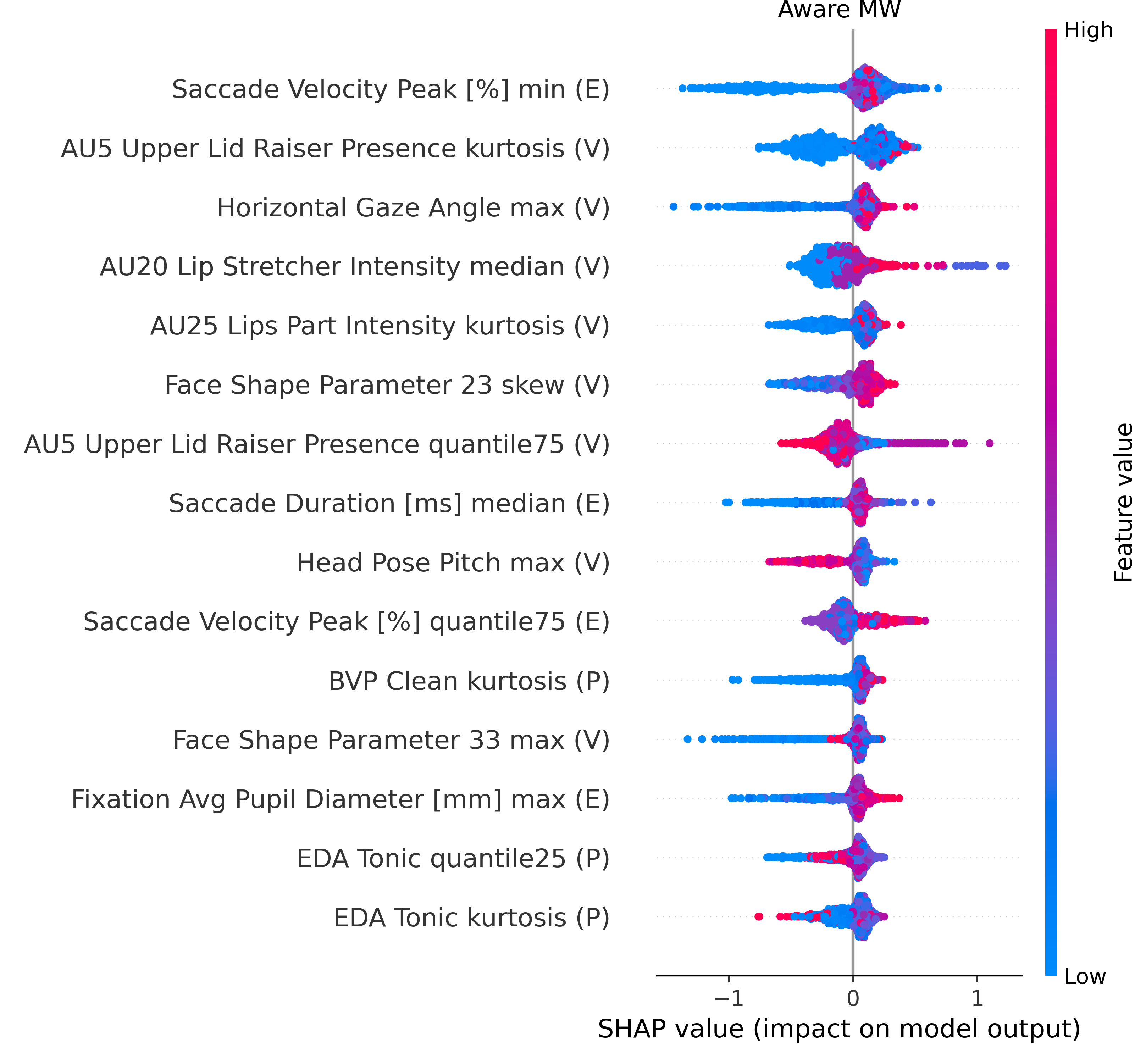}
    
    \includegraphics[width=0.43\textwidth]{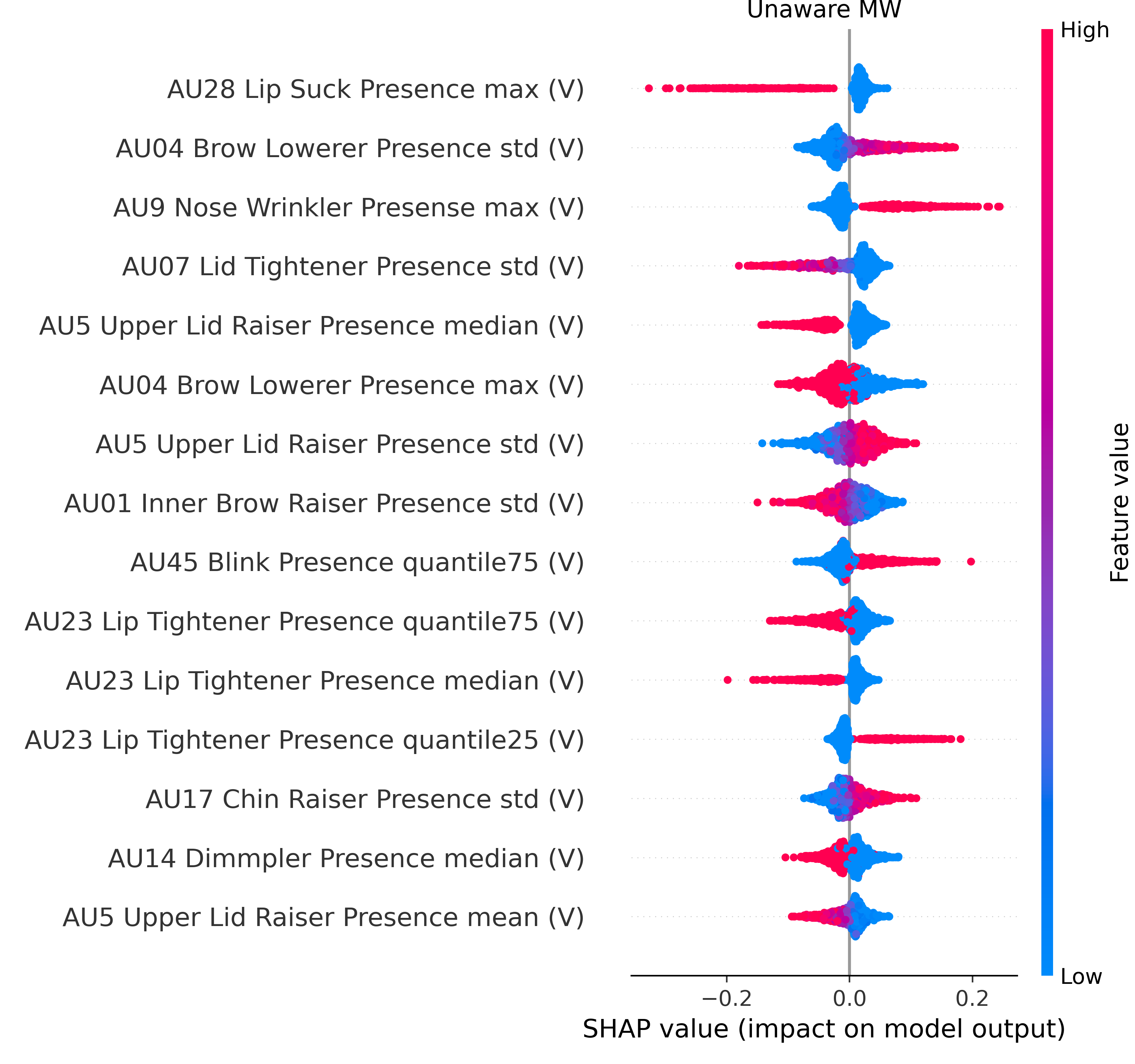}
    
    \includegraphics[width=0.43\textwidth]{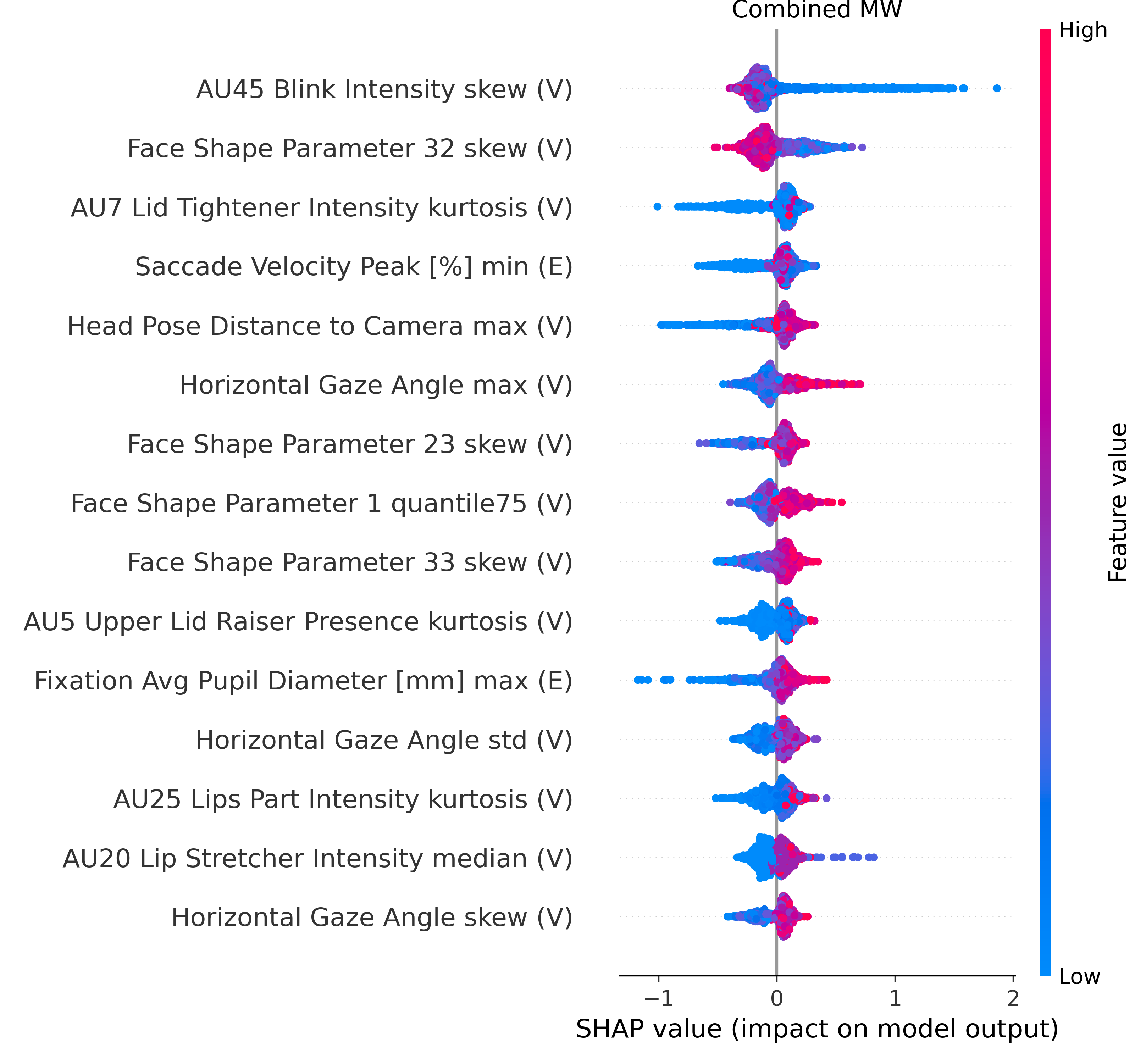}
    \caption{SHAP Summary Plots of Best Performing Classifiers by Mind Wandering Type. \footnotesize Characters in parenthesis indicate the corresponding feature modality (E = Eye Tracking, V = Video, P = Physiology).}
    \label{fig:shapc}
\end{figure*}

Using the XAI method SHAP \cite{shap2017}, we generated explanations for the test set predictions based on the best-performing classifier for each classification problem. Figure \ref{fig:shapc} depicts SHAP summary plots -- one for each mind-wandering type -- giving information on the top 15 most influential features (on the y-axis) for the model's prediction. Each dot in the plot represents one instance in the test set, and the color represents the feature value from low (= blue) to high (= red). Positive Shapley values indicate an increase and negative values indicate a decrease in the model's prediction \cite{shap2017}.

For aware mind wandering detection, the model used features stemming from all three modalities, as seen in the first plot. Taking a look at the most important feature, minimum saccade velocity peak, which describes the lowest peak velocity recorded for eye movements between two fixation points, we observed a mixed pattern in the SHAP values, i.e., low saccade velocity peak minima were related to positive and negative Shapley values. This hints that potential interactions with other features, as well as complex non-linearities, have an impact. A similar relationship was found for the eye tracking-based feature median durations of saccades, where shorter saccades seem to have both positive and negative impacts on predictions.
In contrast, peaks in the 75\% quantile of saccade velocity seemed to have a linear effect on the prediction, with higher peaks in the 75\% quantile of saccade velocity peaks being associated with positive predictions. In addition, we found a negative impact of smaller maximum pupil diameters during fixations on aware mind wandering predictions. 
Key features obtained from video data include facial expressions like AUs encoding upper eyelid raising and lip stretching, as well as face shape parameters. Additionally, maximum horizontal gaze angle and head pose (pitch) are important features.
Regarding the physiological features, the kurtosis of the cleaned BVP signal and the 25\% quantile and kurtosis of the tonic signal were important, also showing a mixed impact on the outcome, hinting again at non-linearities and higher-order interactions between the input variables.

From the SHAP summary plot for the unaware mind wandering classifier (center in Figure \ref{fig:shapc}), the model evidently used predominantly facial expression features derived from the video recordings for prediction. The top 15 features are exclusively summary statistics of the presence of certain AUs during the regarded time frame. More specifically, the presence of facial action units describing a lip suck, lid tightener, and upper lid raiser contributed negatively to the prediction, as higher values of this feature were associated with negative mind-wandering model outputs. This might be typical facial expressions associated with concentration. In contrast, higher values in aggregations of AUs depicted brow lowering, nose wrinkling, lip tightener, or more blinks, tilting the prediction towards unaware mind wandering. 

The top 15 features for integrated aware and unaware mind-wandering predictions consisted of a mixture of video and eye-tracking features. We found blink intensities extracted as AU from video data to be especially important. For example, lower skew values, indicating a skew towards more frequent, higher blink intensities, push the model towards mind-wandering predictions, similar to the effects of more frequent blinks observed for unaware mind wandering. Further gaze-related features extracted from facial videos, like the maximum, standard deviation, and skew of horizontal gaze angles, strongly impact the model. Higher values in maximum horizontal gaze, indicating fixations of further right areas on the screen, where the instructor image was located, push the predictions towards mind wandering. Lower standard deviations in gaze angle, indicating less horizontal dispersion of fixations, are associated with negative mind-wandering predictions. In contrast to the aware mind wandering predictions, we found a mixed impact of facial expression features, like different face shape parameters, upper lid raiser lips part, and lip stretcher AUs. Similar to the aware mind wandering predictions, we found a mixed relation of minimum saccade velocity peak and a negative impact of small maximum pupil diameters on combined mind wandering predictions.

\subsection{Discussion}

\subsubsection{Main Findings}

In this study, we explored a novel combination of modalities to predict aware and unaware mind wandering, as well as combined mind wandering during video lecture watching in the lab. We found that combining the three modalities, eye tracking, facial videos, and physiology, improves the accuracy of mind wandering prediction over single modalities. This finding was consistent over the three different mind-wandering categories we predicted. The highest accuracies could be achieved when subsetting the fused multimodal feature set to the 100 most important features, based on Shapley values. Notably, the top 100 feature subsets of each mind-wandering classification task contained features of all three modalities, highlighting the importance of multimodality while simultaneously showcasing opportunities for minimal data use. As hypothesized by \citet{kuvar2023}, performances of modality feature sets are not additive; however, in contrast to previous research, the improvements over unimodal approaches are considerable in this study. 

When comparing prediction accuracies over the three different tasks, 14\% above chance for unaware, 20\% for aware, and 40\% for integrated mind wandering, it appears that unaware mind wandering is the most difficult to detect. However, there seems to be a correlation between the number of instances and detection accuracies, as typically observed in machine learning tasks, suggesting that more data could improve the prediction accuracy. It remains uncertain whether the low detection accuracy of unaware mind wandering is truly due to its inherent difficulty to detect or simply because of the limited number of cases available (192), compared to unaware (313) and combined (505) mind wandering, restricts the model's ability to learn effectively.
Our analysis of unimodal approaches revealed differences in performance between individual modalities for detecting aware and unaware mind wandering. Eye tracking features were, on average, most predictive for aware mind wandering, while for unaware mind wandering, the models based on features derived from the facial videos performed best. Physiology, specifically EDA and heart rate features, displayed the lowest predictive power when used as a standalone modality for all mind-wandering types.


This modality gains significance when combined with others, as SHAP analysis of best-performing multimodal models revealed. Key features for predicting aware mind wandering include saccade velocities, pupil diameter, facial expressions, head pitch, and physiological signals such as BVP and tonic EDA. For unaware mind wandering, the most critical features are facial action units, particularly the absence of expressions like lip suck and the presence of movements like nose wrinkling. These findings suggest a higher predictive power of facial expressions for this category. However, frequent blinking, a feature associated with video-based gaze information, also significantly contributes to detecting unaware mind wandering. For simultaneous prediction of both aware and unaware types, the most significant features encompass blinks, facial expressions, head distance, horizontal gaze angle from video analysis, and eye-tracking metrics like saccade velocity peaks and pupil diameter

The observation that gaze-related features from facial videos had more influence on predictions than those from eye-tracking may relate to the moderate eye-tracking data quality in this study, with an average tracking ratio of 82\% (\textit{std} = 22\%). Despite these issues, we opted not to apply strict exclusion criteria, which is common in eye-tracking research, to maximize data utilization and capitalize on a multimodal approach \cite{baltruvsaitis2018multimodal}. We excluded only time windows where complete eye-tracking data were missing, reflecting a more naturalistic approach that acknowledges tracking issues in real-world applications. This approach might explain the lesser influence of eye-tracking features in this study compared to others \cite{Hutt.2019, Brishtel.2020, bixler2015gazeeda}. It's important to note that this reduced influence does not diminish the overall significance of gaze in detecting mind wandering; rather, the importance of video-based gaze features underscores its pivotal role. These findings illustrate the challenges of using remote eye trackers in realistic, prolonged learning scenarios like lectures and the increased robustness that additional reliable modalities provide

\subsubsection{Applications}

This study highlights the complexity of mind wandering in educational contexts, arguing for the need for differentiated support that considers distinct aspects of meta-awareness. Utilizing automated, fine-grained detection across multiple modalities could facilitate personalized learning adjustments and targeted interventions, thereby supporting self-regulated learning. Potential interventions might include feedback provision, content review prompts, in-lesson questioning, and material adaptation to enhance learner engagement. However, most intervention research currently offers only general support, such as intermediate testing \cite{Szpunar.2013} or note-taking \cite{Kane.2017}, without addressing specific attentional states. Initial studies employing real-time interventions during mind-wandering episodes detected through automated methods in ITS learning \cite{hutt2021breaking} and reading contexts \cite{mills2021} have shown promise in enhancing long-term retention and comprehension, despite the challenges posed by moderate prediction accuracies.

This study underscores the opportunity to adress specific types of mind wandering, such as prompting breaks for learners during unaware episodes or adapting content to mitigate boredom or frustration during aware episodes. However, given the moderate detection accuracies, the importance of non-intrusive, confidence-based interventions is emphasized. Such interventions set a minimum certainty threshold for predictions before implementing actions, minimizing the risk of inappropriate interventions. The effectiveness of facial expression features and gaze data, particularly for detecting unaware mind wandering, supports the viability of scalable detection in naturalistic settings with high-quality video, as demonstrated in prior research \cite{Lee.2022}. Furthermore, webcam-based eye tracking has shown potential for mind-wandering detection \cite{hutt2023webcam, zhao2017}. As intelligent learning systems become increasingly personalized, ethical considerations, privacy concerns, and potential biases also highlight the need for informed consent, transparency, and careful data management. The implications of interventions on educational practices and outcomes warrant a thorough investigation to prevent adverse effects like students becoming overly reliant on automated support.

\subsubsection{Limitations and Future Work}
This study was conducted in a lab setting, allowing for comparatively high data quality, whereas limiting the ecological validity of our prediction results. Future research aiming at employing detection for attention-aware learning technologies should test the approach in more naturalistic learning settings. The limited sample size and the homogeneity of our participant pool, restricted to university students, might limit the generalizability of our results to other demographics. Consequently, expanding the sample size and including more diverse participant groups would enhance the robustness of the findings. Further, the more accurate detection, with a larger number of mind-wandering instances, indicates that with a larger dataset size, there is also the potential for more accurate predictions. Future research should explore the opportunities of employing more sophisticated machine learning algorithms, such as temporal models, on larger datasets. 

\subsubsection{Conclusion}

Our study demonstrates the improved accuracy of multimodal mind-wandering detection, fusing eye tracking, video, and physiology, outperforming unimodal approaches during lecture video watching. It highlights the potential to detect mind-wandering types at a more fine granular level, specifically to classify aware (20\% above chance) and unaware (14\% above chance) mind wandering when employing those modalities. The consideration of meta-awareness of mind wandering becomes highly relevant when employing automated detection for attention-aware learning technologies, as the two forms may require tailored support.

\subsection*{Acknowledgements}
We thank Sidney D'Mello for his invaluable advice on study design and analysis. This research was supported by the LEAD Graduate School \& Research Network, which is funded by the Ministry of Science, Research and the Arts of the state of Baden- W{\"u}rttemberg within the framework of the sustainability funding for the projects of the Excellence Initiative II. Babette B{\"u}hler and Hannah Deininger are doctoral candidates supported by the LEAD Graduate School and Research Network. This research was also partly funded by the Deutsche Forschungsgemeinschaft (DFG, German Research Foundation) under Germany's Excellence Strategy - EXC number 2064/1 - Project number 390727645. ChatGPT was utilized for language editing in this work.

\renewcommand\thefigure{\thesection.\arabic{figure}}    
\setcounter{figure}{0}   

\subsection*{Research Methods}
\label{app}
\begin{figure}[h]
    \centering
    \includegraphics[width=0.9\textwidth]{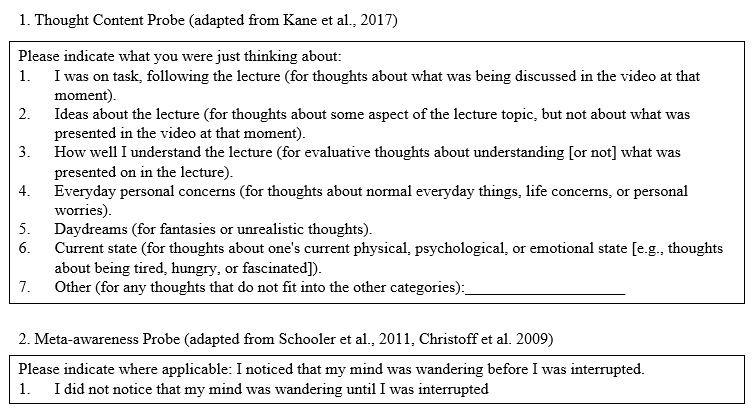}
    \caption{Two-Stage Mind Wandering Thought Probe}
    \label{fig:probec}
\end{figure}

The full two-stage thought probe employed to assess mind wandering ground truth is depicted in Figure \ref{fig:probec}.
 
The improvement of classification accuracy, defined by AUC-PR, above chance was computed as follows: 
\[
Above Chance Level = \frac{Actual Performance - Chance}{Perfect Performance - Chance}
\]

\FloatBarrier

%% file: 3_LabToWild.tex
\section[Examining Generalizability of Video-based Mind Wandering Detection]{From the Lab to the Wild: Examining Generalizability of Video-based Mind Wandering Detection}

\subsection{Abstract}
Student's shift of attention away from a current learning task to task-unrelated thought, also called mind wandering, occurs about 30\% of the time spent on education-related activities. Its frequent occurrence has a negative effect on learning outcomes across learning tasks. Automated detection of mind wandering might offer an opportunity to assess the attentional state continuously and non-intrusively over time and hence enable large-scale research on learning materials and responding to inattention with targeted interventions. To achieve this, an accessible detection approach that performs well for various systems and settings is required. In this work, we explore a new, generalizable approach to video-based mind wandering detection that can be transferred to naturalistic settings across learning tasks. Therefore, we leverage two datasets, consisting of facial videos during reading in the lab (N = 135) and lecture viewing in-the-wild (N = 15). When predicting mind wandering, deep neural networks (\gls{dnn}) and long short-term memory networks (LSTMs) achieve F$_{1}$ scores of 0.44 (AUC-PR = 0.40) and 0.459 (AUC-PR = 0.39), above chance level, with latent features based on transfer-learning on the lab data. When exploring generalizability by training on the lab dataset and predicting on the in-the-wild dataset, BiLSTMs on latent features perform comparably to the state-of-the-art with an F$_{1}$ score of 0.352 (AUC-PR = 0.26). Moreover, we investigate the fairness of predictive models across gender and show based on post-hoc explainability methods that employed latent features mainly encode information on eye and mouth areas. We discuss the benefits of generalizability and possible applications.

\subsection{Introduction}\label{intro}

Attention plays a central role in learning and knowledge construction \cite[][]{levine1990}. However, a recent meta-analysis by \cite{wong2022} showed that about 30\% of the time learners spend in educational activities, their thoughts are elsewhere. This shift of attention away from the current task to task-unrelated thought is called mind wandering~\cite[][]{Smallwood.2006}. \cite{wong2022} further demonstrated that frequent occurrence of task-unrelated thoughts during learning is significantly associated with lower test performance and explains about 7\% of the variability in learning outcomes. This negative relationship holds equally for surface and inference-level learning and is consistent across tasks. For instance, mind wandering has shown to have a negative effect on reading comprehension~\cite[][]{Smallwood.2011, Feng.2013, dmello2021, bonifacci2022, caruso2023} and  lecture retention~\cite[][]{Risko.2012, Szpunar.2013, Hollis.2016, Pan.2020}.

This evident effect of mind wandering on learning should not be neglected. Therefore, learning environments - physical as well as online - aim to create appealing conditions that allow students to focus their attention on the relevant content and support successful learning. The Covid19 pandemic has greatly accelerated the use \cite[][]{lemay2021} and development \cite[][]{dhawan2020} of online learning tools  at all levels of education. These include intelligent tutoring systems (\gls{its}s), massive open online courses (MOOCs), as well as online lecture portals. To support learners in online learning settings, one can either improve the presented learning materials in a way that decreases mind wandering, for instance, by making texts more interesting \cite[][]{bonifacci2022}, or one can try to direct attention back to the learning task, for instance, through targeted interventions \cite[e.g.,][]{mills2021}.  

One step that is foundational to those two approaches to support learners is the automated detection of mind wandering, as it allows for continuous and unobtrusive measurement of the state of attention over time. It can be used to test and optimize learning materials and to conduct further research on the conditions under which mind wandering occurs and its effects on learning outcomes. At the same time, it offers online learning systems the possibility to implement targeted interventions to respond to the learners' shift of attention with adaptive and supporting actions. It has been demonstrated that such automated interventions can reduce mind wandering and thus support learning. For example, feedback following eye-tracking-based mind wandering detection mitigated its negative effect on reading comprehension during computerized reading \citep[][]{dmello2017zone, mills2021}. Furthermore, repetition and questioning interventions based on automatically detected mind wandering reduced mind wandering and improved retention of students with low prior knowledge in an ITS in certain cases~\cite[][]{hutt2021breaking}.

Mind wandering detectors using supervised machine learning mostly rely on data from modalities such as eye trackers~\cite[][]{Hutt.2017, Faber.2018, Hutt.2019, Faber.2020, Zhang.2020, mills2021} or physiological sensors such as EEG~\cite[][]{jin2019eeg, dong2021eeg}. While these modalities provide very useful process information, the use of such sensors requires a well-controlled environment, is quite costly, and is difficult to scale. However, another strand of recent research has shown that mind wandering can also be detected above chance level using video recordings of the face obtained from consumer-grade webcams, such as those found in almost all laptops~\cite[][]{Bosch.2021, Lee.2022}. The use of video recordings enables the detection of mind wandering on a large scale in natural environments where online learning systems are commonly used, such as classrooms or homes. 

Video-based mind wandering detectors have the potential to be used in many different systems and environments. Intelligent user interfaces for learning can combine multiple stimuli, such as text and video, and may be used globally, i.e., by culturally diverse target groups. However, to train suited machine-learning models, labeled ground truth data is needed, i.e., collecting learner self-reports makes the whole process time-consuming, effortful, and costly. Thus, approaches that generalize well across different settings, learning tasks, and target groups are required in order to ensure the applicability of such solutions. 
 
To obtain generalizability, facial features rather than gaze features, which are highly predictive but also highly stimulus dependent~\cite[][]{Faber.2020}, are suitable as they can achieve greater transferability between tasks. In this study, we use transfer-learning-based features trained on a dataset of facial expressions in the wild~\cite[][]{affectnet2017}, i.e., on a highly diverse set of facial images. Previous research has shown that affective features such as facial action units (\gls{au}s) and predictions of emotional state are informative for predicting mind wandering~\citep{stewart2017, Bosch.2021}. However, this could pose a challenge when thinking about generalizing models across subject groups with different cultural backgrounds, as there is an ongoing scientific debate about the universality of facial expressions of emotion across different cultures~\cite[][]{russell1994, ekman1994, jack2012facial}. This could have implications for cross-cultural generalizability when using features derived from such classification tasks. This highlights one of the major limitations of previous approaches—namely, the lack of sample diversity and the unexplored effects on algorithms, which may bias results and impact generalizability~\cite[][]{kuvar2023}.

The goal of the present work is to examine the generalizability of video-based mind wandering detection. Towards this goal, we investigate whether (1) a new feature set based on transfer learning of facial-expression recognition can be used in combination with temporal models exploiting temporal relationships in video data to improve model performance compared to explicit facial features. Furthermore, (2) we explore the potential of its generalization across two datasets that differ with regard to the environment (lab vs. in the wild), task (reading vs. watching a video lecture), and cultural background of the target groups (American vs. Korean students).  We then (3) examine the fairness of our models across genders and (4) use explainable AI tools to investigate the information encoded in latent features. Accurate detection of mind wandering is the first critical step towards large-scale research and adaptive learning technologies that aim to enhance engagement and learning outcomes.

\subsection{Related Work}

The automated detection of mind wandering episodes allows measuring this state non-obtrusively and continuously over time. This is achieved by employing machine learning methods and self-reports from learners as ground truth. Here, self-reports are used because an objective, reliable measurement by neurophysiological or behavioral markers is not possible so far~\cite[][]{Smallwood.2015}. When collecting self-reports with the probe-caught method, subjects are repeatedly interrupted by a probe and explicitly asked about the direction of their attention~\cite[][]{Smallwood.2006}, while in the self-caught method, participants are instructed to report whenever they become aware of their own shift of attention~\cite[][]{Schooler.2011}. Research suggests that both approaches allow for reliable measurement of mind wandering~\cite[][]{schubert2020validity, VaraoSousa.2019}. Self-reports are associated with physiological signals~\cite[][]{franklin2013, Blanchard.2014, Christoff.2009} and consistently correlate with objective performance measures~\cite[][]{randall2014}, demonstrating predictive validity. 
Moreover, the datasets on which mind wandering is examined are all very unbalanced with approximately 25-30\% mind wandering rates. Therefore, the results of the prediction are reported using the F$_1$ measure, which represents a harmonized mean of precision (proportion of predicted mind wandering instances that are truly mind wandering) and recall (proportion of true mind wandering instances predicted as mind wandering). The reported improvement over chance level is the proportion of above chance performance of the perfect above chance prediction. For more details on the evaluation metrics, we refer the reader to section \ref{eval}.

Most studies on automated mind wandering can be divided into two main strands according to the sensing modalities used for the detection: Eye tracking and physiological measures, and video recordings. Therefore, in the following, we review this research focusing on these areas and describe the novelty of our study. 

\subsubsection{Eye-tracking and Physiological Sensor-based Approaches}

Most research on mind wandering detection has focused on eye movement data obtained by eye trackers. According to the so-called mind-eye link, cognitive processes are reflected in eye movements~\cite[][]{just1976eye, rayner1998eye, reichle2012using} thus making eye tracking suitable to identify mind wandering. Global gaze features, such as fixations and saccades, as well as locality features describing the spatial properties of gaze have widely been used in research to estimate attentional states during a variety of learning-related tasks such as reading ~\cite[][]{bixler2014, dmello2016, Faber.2018b, mills2021}, watching video lectures~\cite[][]{Hutt.2017,Jang.2020, Zhang.2020} or using an ITS~\cite[][]{Hutt.2019}. Also cross-task prediction of such features was examined~\cite[][]{Faber.2020,2021crossed}. Additionally, pupil size and blink rates have been shown to be meaningful features to off-task thought detection~\cite[][]{smilek2010, Brishtel.2020}. 

Further, several physiological sensors are utilized for mind wandering detection. Electrodermal activity (EDA) was used as a standalone modality~\cite[][]{Blanchard.2014} and in combination with eye tracking~\cite[][]{Brishtel.2020} for mind wandering detection during reading. Furthermore, an increased heart rate was detected in mind wandering episodes due to greater arousal~\cite[][]{Smallwood.2007}, thus being deployed for automated detection as well~\cite[][]{Pham.2015}. Another way of assessing mind wandering is using EEG ~\cite[][]{jin2019eeg, dong2021eeg, dhindsa2019, Conrad.2021}, which has also been employed for learning related mind wandering during the watching of online lectures ~\cite[][]{dhindsa2019, Conrad.2021}.

The collection of data employing eye trackers or physiological sensors requires highly controlled settings (i.e., laboratory). Consequently, most of the studies were conducted in a lab setting, with the exception of Hutt et al.~\cite[][]{Hutt.2019}; who used commercial eye trackers in a classroom setting. Furthermore, respective modalities are often very expensive, which again limits scalability. The alternative of video-based detection is more cost-effective. Consumer-grade webcams can be employed and thus detection can be upscaled easily in naturalistic settings (e.g., at home or in the classroom). Therefore, in this study, we focus on mind wandering detection based only on facial videos.

\subsubsection{Video-based Approaches}

The first approach to detect mind wandering based on facial videos from webcams was provided back in 2017 by ~\cite{stewart2017}. The authors predicted self-reported mind wandering during narrative film watching in a laboratory setting, based on features such as AUs, head pose, face position, face size, and gross body movement. Employing support vector machine (SVM) models, they achieved an F$_{1}$ score of 0.39, which is an improvement of 13\% above chance level, on aggregated features from 45-second windows. 
In a following study, the potential of cross-task classification in laboratory settings was shown, by predicting mind wandering on a reading task from a model trained on mind wandering from a film watching task and vice versa~\cite[][]{stewart2017generalizability}. Employing the same feature set as in the previous study and a decision tree based C4.5 classifier, their models generalized well and almost maintained within-dataset prediction performance when training on film watching data and predicting reading data (F$_{1}$: 0.407; 21\% above chance level) and also after adjusting the classification threshold the other way around (F$_{1}$: 0.441; 22\% above chance level).

In a laboratory experiment to detect mind wandering during MOOCs, ~\cite{Zhao.2017} implemented webcam-based gaze estimation and compared it to predictions with specialized eye-tracking data. With probe-caught mind wandering reports as ground truth they concluded that SVM classifiers on both data sources perform equally well with the webcam-based approach achieving an F$_{1}$ score of 0.405, a 16\% above-chance improvement.

Another recent study by ~\cite{Bosch.2021} examined face-based mind wandering detection in the laboratory during a reading task with self-caught mind wandering reports and in the classroom during the usage of an ITS based on probe-caught mind wandering reports. In addition to AUs, head pose, and body movement, they hand-crafted new features depicting co-occurring AUs, temporal dynamics of AUs, and facial texture. In the classroom setting, those features were extracted in real-time, avoiding the recording of children due to privacy concerns. With SVM and deep neural networks on aggregated features sets over 10-second windows, they achieved F$_{1}$ scores of 0.478 in the lab and 0.414 in the classroom setting, which represents 25\% and 20\% above-chance improvements respectively.
Although eye-tracking features (Global gaze features F$_{1}$: 0.45, 29\% above-chance accuracy; Locality gaze features F$_{1}$: 0.49, 34\% above-chance accuracy)  outperform a facial-feature based approach (AUs F$_{1}$: 0.31, 10\% above-chance accuracy; Co-occurring AUs F$_{1}$: 0.3, 9\% above-chance accuracy) for mind wandering detection while using an intelligent tutor system in the same classroom setting~\cite[][]{Hutt.2019}, a fusion of both features could increase robustness by accounting for missing values in one of the two modalities. 

In a recent paper by ~\cite{Lee.2022} mind wandering detection based on facial webcam videos during lecture viewing in the wild, for example at home, was examined. Gaze-related features (i.e., speed, dispersion, horizontal movement ratio), Eye Aspect Ratio (\gls{ear}) features, head movement, as well as emotion predictions, were extracted from facial videos. They employed eXtreme Gradient Boosting (XGBoost), Deep Neural Network (DNN), and SVM classifiers on different time windows, achieving the best results with XGBoost with an F$_{1}$ score of 0.36 (15\% over chance level improvement) utilizing 10-second windows.

Adopting webcam-based eye tracking for reading tasks, \cite{hutt2023webcam} executed two in-the-wild studies: the first recruited participants through a university, and the second utilized Prolific for participant recruitment. They predicted probe-caught mind wandering using global and local gaze features, achieving an F$_{1}$-score of 0.25 on a combined dataset (9\% above chance). In cross-dataset prediction between two data collections, training on one dataset and predicting on the other, they achieved Kappa values of 0.09 and 0.15, respectively.

The overwhelming majority of these studies, except one cross-task laboratory-based generalization study~\cite[][]{stewart2017generalizability} and a very recent webcam-eye-tracking study \cite[][]{hutt2023webcam}, consider the performance of their models only on the labeled data on which they were trained and therefore do not test the generalizability of their models. As mentioned above, collecting labeled data for this specific task is costly and may not be possible for every use case, implying that generalizable models are necessary.

\subsubsection{Novelty of this Work}

While a large variety of machine learning and computer vision methods are used for feature extraction, they are not directly targeted to generalizability. Hence, there are still unexploited potentials of machine learning techniques, which we investigate in this study. These techniques include utilizing features from pre-trained networks for similar but more general tasks, employing temporal models, and an evaluation of generalizability, which are discussed as follows.

\paragraph{Deep Learning-based Features}
To our knowledge, all previous research in this field used explicit features, such as AUs~\cite[][]{stewart2017, Bosch.2021} and gaze features~\cite[][]{Lee.2022} for predictions. In some cases, additional hand-manufactured features, such as gaze dispersion~\cite[][]{Lee.2022}, have been created. 
While those appear to be informative in the data at hand, there is a risk that some of those features, especially gaze-related features, might be very stimulus-specific. Gaze in a reading task has a highly specific signature, whose features may be difficult to transfer to different tasks \cite[][]{2021crossed, Faber.2020}.  While there have been advances with webcam-based eye tracking in recent years, enabling the detection of fixations of specific areas of interest (AOIs) on the computer screen, there are still severe limitations compared to specialized eye trackers. Especially in terms of mind wandering detection, more fine granular eye movement indicators such as saccade duration are predictive and most likely more generalizable. Those cannot be observed based on consumer-grade webcams at this time, underlining the advantages of facial expression features. Moreover, recent studies have showcased the utility of advanced Facial Emotion Recognition (FER) methods on webcam videos in adjacent domains, particularly for evaluating emotion regulation in remote collaborative learning settings~\cite[][]{nguyen2022, ngo2024fer}. Further, the creation of hand-manufactured features requires domain knowledge and might be tailored to the setting at hand.

In other image-based classification tasks such as facial expression recognition, deep learning methods have been used successfully~\cite[][]{rawat2017deep, Li2020deepfer}. Due to the limited sample sizes of data for mind wandering detection, we use transfer learning, where the feature extraction part of networks pre-trained on a similar task with available large datasets, is applied to a new problem.
Based on the previous successful use of facial expression features as facial texture patches and AUs~\cite[][]{Bosch.2021} and emotional state features~\cite[][]{Lee.2022} for mind wandering detection, we argue that latent features learned by a CNN trained on the related task of FER can be informative to the mind wandering classification problem at hand. Furthermore, as they are trained on an in-the-wild data set, which assures large data variability, and thus utilizing these features may contribute to making the models generalizable to in-the-wild settings, i.e. in terms of image quality. To increase the confidence in these latent features, which cannot be directly interpreted by humans, we use the Explainable AI tool LIME \cite[][]{lime} to illustrate which parts of the face are used for classification and are encoded respectively in our latent representation.

\paragraph{Temporal Models}

Previous research indicates that temporal dynamics can be informative for mind wandering detection. Recent studies showed the importance of hand-crafted features such as co-occurring AU pairs and temporal dynamics of AUs~\cite[][]{Bosch.2021}, as well as body~\cite[][]{stewart2017, Bosch.2021} and head movement~\cite[][]{Lee.2022}, which are explicitly designed to depict these dynamics. If not represented by such manually generated features, they might be blurred by aggregation over time. For this reason, we employ models that are able to take time-series data such as the video data at hand as input and directly learn temporal relations between features. The use of these temporal models allows us to additionally train an end-to-end model, in which we can fine-tune the pre-trained Convolutional Neural Network (\gls{cnn}) used for frame-wise feature extraction to our specific classification task.

\paragraph{Generalizability}

In previous face-based mind wandering detection research, only cross-task prediction in laboratory settings was explored~\cite[][]{stewart2017generalizability}, but the generalizability of lab settings to other more naturalistic settings is still to be examined. 
We trained a model on data collected during a reading task in the lab~\cite[][]{Bosch.2021} and then applied it to in-the-wild data of students watching a lecture video at home~\cite[][]{Lee.2022}, resulting in a cross-context and cross-task prediction. However, the two data sets differ additionally in the cultural backgrounds of their subjects, as one was collected in the U.S. and the other in Korea. With regard to the discussion about the cross-cultural universality of facial expressions in the literature~\cite[][]{russell1994, ekman1994, Benitez2017}, this is a further challenge for our model, which is based on such features to generalize across culturally diverse user groups. Although the in-the-wild dataset is relatively modest in size, it represents to our knowledge the only publicly available dataset of its kind, enabling a crucial initial stride towards achieving generalizability in naturalistic environments. For comparability, we used the reading-task data to carry out within-context and within-task evaluation. We further compare the classification results for both across gender to investigate potential biases.

\subsection{Methodology}
In this section, we discuss the details of the employed data, features that are utilized for mind wandering detection, as well as training and inference processes. 

\subsubsection{Data}
We employ two different datasets in our work, one~\cite[][]{Bosch.2021} for within-dataset evaluation and the other~\cite[][]{Lee.2022} for cross-dataset evaluation. Table~\ref{data_table} provides a general overview of the datasets, and they are described as follows.  

\begin{figure}[h]
  \centering
  \includegraphics[width = 220pt]{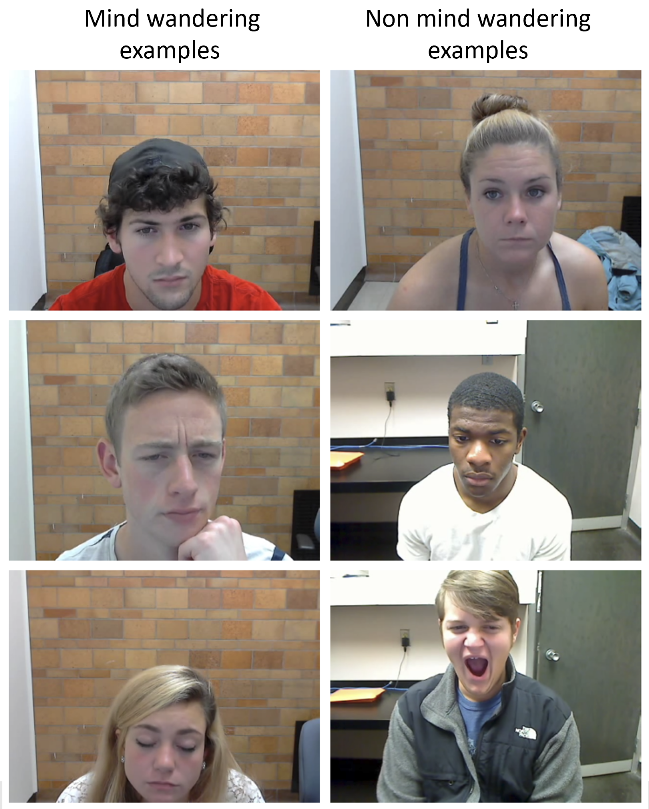}
  \caption{Example images of mind wandering (left) and non mind wandering (right) instances in the lab data \cite[][]{Bosch.2021}.}
  \label{fig_ex}
\end{figure}

\paragraph{Lab Data by ~\citet{Bosch.2021}} This dataset is from a lab study, containing facial videos, recorded using a Logitech C270 webcam, of $N = 135$ university students from the U.S. reading a scientific text and their self-caught reports on mind wandering. Mind wandering instances and on-task moments of 10 seconds each were cut from the original videos~\cite[][]{Bosch.2021}. The mind wandering instances are the time windows right before a self-report, with a 4 second buffer before the self-report, to ensure  the exclusion of the key-press movement. The buffer length was validated in a pilot study. The on-task examples are 10-second clips taken from the time in-between, that do not include a page turn or fall into 30 seconds before a mind wandering self-report. The resulting dataset contains $N = 1031$ mind wandering instances and $N = 2406$ non-mind wandering instances. We use this dataset for training our models in both within- and cross-task evaluation scenarios as it contains more subjects and samples.

\paragraph{In-the-wild Data by ~\citet{Lee.2022}} This open-source dataset available for research purposes contains facial videos of $N = 15$ university students in Korea and probe-based mind wandering reports. The participants watched a one-hour-long lecture video at home, were probed for mind wandering in 40-second intervals, and were filmed by their webcams. In contrast to previously employed datasets, the data was collected in the wild (i.e., at students' homes), which is an increasingly realistic learning setting for MOOCs and other online learning tools. In total, it contains $N = 205$ mind wandering and $N = 1009$ non-mind wandering instances with 30 FPS. Due to the smaller dataset size, we use this dataset solely for evaluation purposes to detect mind wandering in the wild.

\begin{table}[h]
\begin{center}
\begin{minipage}{\textwidth}
\caption{Dataset comparison.}
\label{data_table}
\resizebox{\textwidth}{!}{\begin{tabular}{lll}\toprule
\multirow{2}{*}{Specification} & \multicolumn{2}{c}{Datasets}    \\
                               &  Lab data  &  In-the-wild data \\
\midrule
Study                       &\cite{Bosch.2021} &\cite{Lee.2022} \\
Task                           & Reading scientific text     & Watching lecture video \\
Setting                        & Laboratory  & In the wild   \\
Country                        & USA         & Korea         \\
Mind wandering self reports     & Self-caught & Probe-caught  \\
Participants                   & 135         & 15            \\
Total instances                & 3,437       & 1,220         \\
\hspace*{0.3cm}Mind wandering        & 1,031       & 206           \\
\hspace*{0.3cm}Non mind wandering        & 2,406      & 1,014         \\
Video FPS                       & 12.5        & 30            \\
\bottomrule
\end{tabular}}
\end{minipage}
\end{center}
\end{table}

\subsubsection{Features} \label{features}

\begin{figure}[h]
  \centering
  \includegraphics[width=0.8\textwidth]{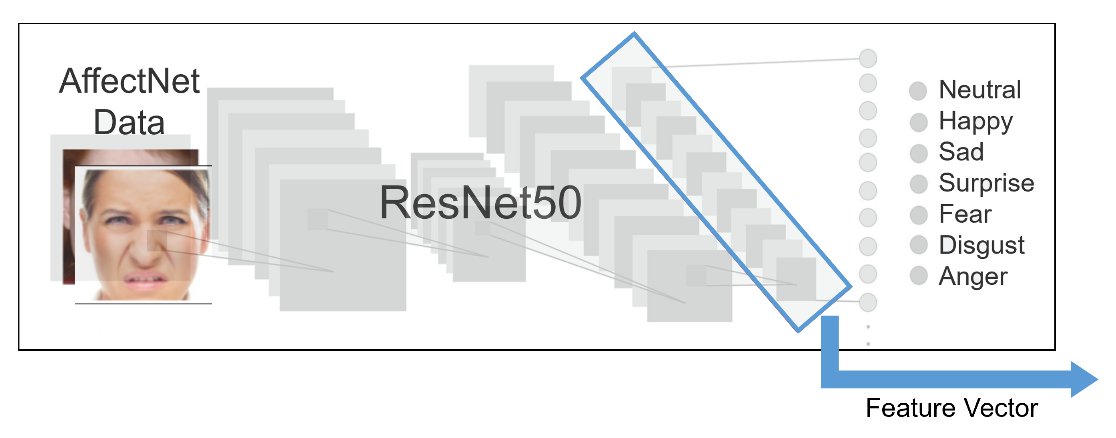}
  \caption{Pre-trained Resnet50 as feature extractor.}
  \label{fig_Resnet50features}
\end{figure}

To extract deep learning-based facial expression features, we use a CNN with a ResNet50~\cite[][]{He_2016_CVPR_resnet50} architecture pre-trained on the AffectNet dataset~\cite[][]{affectnet2017} containing 23,901 images classified as belonging to seven discrete facial expressions (neutral, happy, sad, surprise, fear, disgust, anger), as a feature extractor. The process is depicted exemplarily in Fig.~\ref{fig_Resnet50features}. This model achieves an accuracy of $58\%$ on the AffectNet validation set (see~\cite{sumer2021} for details on model training). To extract latent features from our video clips, we apply frame-wise face detection by employing RetinaFace~\cite[][]{retinaface2020} to our videos. In this step, we had to remove 74 instances, 31 of them mind wandering, because no face could be detected across all frames. We pre-process the resulting face images, aligning them based on five facial key points extracted by the face detector, then they are cropped to size $224 \times 224$ and normalized. We then insert them into the pre-trained ResNet50 model, from which the FER classification layer was removed. The extracted latent feature vector is 2048 digits long and can be fed into a downstream classifier. We provide insights into the most important image areas encoded in the feature vectors by applying the Explainable AI tool LIME \cite[][]{lime} to the feature extraction model.

To compare our latent deep learning features to more explicit features, similar to those used in previous research, we extract AUs, facial landmark locations, head location, pose and rotation, as well as face shape parameters. Additionally, we extract gaze direction vectors for both eyes and gaze angles, as well as 2D and 3D eye region landmarks, consisting of 55 landmark points for each. All features were extracted using the OpenFace toolkit~\cite[][]{openface2016} from each video frame respectively.

\subsubsection{Training and Inference}
In our analysis approach for binary mind wandering classification, we aim to leverage pre-trained features from deep neural networks by employing transfer learning from the related task of facial expression recognition and compare this approach to employing explicit features as AUs and gaze vectors. Further, we examine whether these pre-trained features enable our model to generalize to a new in-the-wild dataset. 

\paragraph{Handling Temporal Data}
We aim to leverage temporal dynamics information which may be lost by aggregation over time in non-temporal models, by employing recurrent neural network models for supervised classification. In particular, we train long short-term memory (LSTM)~\cite[][]{hochreiter1997} and bidirectional long short-term memory (\gls{bilstm})~\cite[][]{baldi1999, schuster1997} recurrent neural networks. For both models, we employ an architecture consisting of three recurrent layers with 100 neurons respectively, taking frame-wise extracted feature vectors with 125-time steps as input. A dense layer stacked on top, implementing a sigmoid activation function, outputs the binary mind wandering predictions.

The aforementioned self-reports serve as ground truth for our supervised machine learning approach. In order to compare our results in a within-task and within-context evaluation scenario, we employ the exact same person-independent 4-fold validation splits as in~\cite[][]{Bosch.2021}. For each fold, a model is trained separately and results are averaged over all test folds. To emphasize the minority class mind wandering during training we employ class weighting. Furthermore, we employ early stopping of model training to avoid overfitting with a patience of 5 epochs. 

To compare the performance of temporal models with non-temporal models, we additionally train SVMs~\cite[][]{Cortes.1995} with a radial basis function (RBF) kernel, which showed to be the best performing models in previous research \cite[][]{Bosch.2021}, as well as XGBoost~\cite[][]{Chen.2016} models and simple DNNs with one hidden layer. To find the optimal parameter settings for each model, we used person-independent, nested 4-fold cross-validation to apply grid-search hyperparameter tuning. The best performing settings determined in inner 4-fold cross-validation were used for prediction in each outer fold. An overview of the parameters tested can be found in Table \ref{tab:hyper_grid} in the appendix. To generate suitable input, we aggregate frame-wise extracted explicit features, computing the mean, median, minimum, maximum, and standard deviation values for each feature over the whole clip. For our latent features, we create statistical aggregations of the 2048-numbers long feature vector over all 125 frames. Since this procedure results in a large number of features with a lot of redundancy, we apply mutual information-feature selection, a univariate feature selection method based on the dependency between variables. Based on the training data, the 100 most meaningful features were selected in each fold. To account for the imbalanced data, we employ weighting or up-sampling of the training split using Synthetic Minority Over-sampling Technique (\gls{smote})~\cite[][]{Chawla.2002}. We report the best performing combinations of balancing and hyper-parameters.

\begin{figure}[h]
  \centering
  \includegraphics[width=0.9\textwidth]{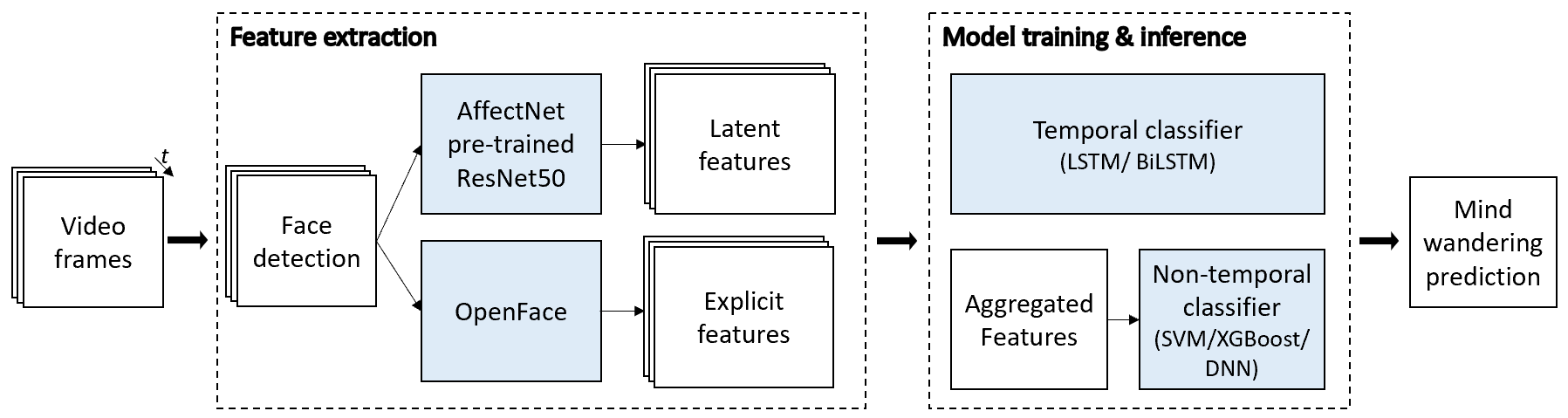}
  \caption{Mind wandering detection pipeline.}
  \label{fig_pipeline}
\end{figure}

\paragraph{Fine-tuned CNN-LSTM}

Employing temporal models allows us to not only employ the pre-trained AffectNet-CNN as a feature extractor but also train and fine-tune an end-to-end CNN-LSTM model. This allows not only to train the temporal inference and classification parts of the model for our present problem of mind wandering detection but also to adapt the image feature extraction part of the model more precisely to the particularities of our problem. As a first step, the pre-trained ResNet50 without classification layer is included in the LSTM architecture to allow for video input and frame-wise feature extraction. Then, the previously described RNN architectures are stacked on top. In a first training run, the pre-trained weights in the CNN part of the model are frozen and only the top part of the model is allowed to adapt with a learning rate of 0.001. In a second training run, the last convolutional block of the ResNet50 model is set trainable and thus adapted to the mind wandering classification task with a smaller learning rate of 0.00001.

\paragraph{Cross-dataset Prediction}

To examine the generalizability of the proposed approach over different settings, tasks, and target groups, we perform cross-dataset prediction employing our new approach. This means the aforementioned in-the-wild data is used to predict mind wandering instances. To this end, we use the video data from our lab reading task and train a model on the entire data. Using this model, we then predict mind wandering instances in the in-the-wild lecture viewing data from~\cite[][]{Lee.2022}. We pre-process the in-the-wild dataset in the same way as the lab data. Based on the provided mind wandering probes, we cut 10-second windows before each probe and extracted all features described in Section~\ref{features}. For temporal models, we downsampled the 300 frames resulting from a higher recording frame rate to 125 frames to match the sequence length.

\paragraph{Evaluation Metrics} \label{eval}

Due to the unbalanced nature of both our datasets with lab data having 30\% and in-the-wild data having 25\% mind wandering instances, reporting the accuracies of mind wandering prediction models could be misleading. A classifier always predicting non-mind wandering would achieve 70-75\% accuracy without recognizing a single mind wandering instance. 
Also commonly used threshold metrics focusing on the minority class as F$_{1}$ scores are highly influenced by the skew in the data \cite[][]{jeni13}. Such measures can be driven by high recall at low detection precision. For this reason, we report area under Precision-Recall curve (AUC-PR) values, which show the precision as function of the recall. This rank metric helps to balance precision and recall of the minority class.

Another measure that helps to evaluate the performance, especially when comparing performances of datasets with differing class distributions, is the improvement of the model above chance level, which is calculated as follows:
\[
Above Chance Level = \frac{Actual Performance - Chance}{Perfect Performance - Chance}
\]

To allow comparability to previous research, we additionally report the F$_{1}$ scores for the minority-class mind wandering, which is calculated as follows:
\[
F_{1}=2\times\frac{Precision \times Recall}{Precision + Recall}
\]
where the Precision (i.e., the proportion of correct mind wandering predictions of all mind wandering predictions) and Recall (i.e., share of correctly predicted mind wandering instances of all mind wandering instances) are defined as: 
\[
Precision = \frac{TP}{TP+FP},\qquad Recall = \frac{TP}{TP+FN}. 
\]
TP, FP, TN, and FN represent "true positive", "false positive", "true negative", and "false negative", respectively.

\subsection{Results}

In this section, we report within- and cross-dataset mind wandering detection results and investigate potential model biases by comparing our results across gender and explain the employed latent features.

\subsubsection{Mind Wandering Detection}

The results for within-task mind wandering detection, training, and predicting on the lab data, are depicted in Table~\ref{tab:results_within_task}. In the within-task prediction setting, models trained on latent features outperform the random baseline (i.e., $F_{1} = 0.3$~\cite[][]{Bosch.2021} and $AUC-PR = 0.3$) to a significant extent. 

The results are comparable to the state-of-the-art on this data~\cite[][]{Bosch.2021}, which achieved similar performance between $F_{1}$ scores of 0.414 and 0.478, mostly with hand-crafted features and SVMs. However, the observed $F_{1}$ scores are clearly driven by high recall values, while precision values are rather low. While a high recall means that most mind wandering instances are detected, simultaneously low precision also means that many instances are falsely classified as mind wandering. Therefore, we introduce the $AUC-PR$ score, which is a rank metric that also reflects the ratio of precision and recall. Thus, the above chance level values reported in the table are calculated on the basis of $AUC-PR$ scores. 

\begin{table}[h]
\begin{center}
\begin{minipage}{\textwidth}
\caption{Results of mind-wandering detection for the within-lab-data prediction.} \label{tab:results_within_task}
\resizebox{\textwidth}{!}{\begin{tabular}{lllcccccc}

\toprule
Model      & Feature Set                & Method    & AUC-PR & Above Chance-Level & F$_{1}$  & Precision & Recall            & ROC-AUC\\
\midrule
SVM        & Explicit   features        & OpenFace  & 0.288         & -1.70\%            & 0.391        & 0.286     & 0.636  & 0.482    \\
           & Latent   features          & AffectNet & 0.365         & 9.30\%             & 0.443        & 0.347     & 0.63   &  0.587    \\
           &                            &           &               &                    &              &           &        &     \\
XG Boost   & Explicit   features        & OpenFace  & 0.346         & 6.60\%             & 0.332        & 0.348     & 0.321  &  0.543       \\
           & Latent   features          & AffectNet & 0.372         & 10.30\%            & 0.358        & 0.356     & 0.368  &  0.576  \\
           &                            &           &               &                    &              &           &        &       \\
DNN        & Explicit   features        & OpenFace  & 0.331         & 4.40\%             & 0.367        & 0.333     & 0.439  & 0.507\\
           & Latent   features          & AffectNet & \textbf{0.398} & \textbf{14.00\%}   & 0.44         & 0.361     & 0.576 & 0.601\\
           &                            &           &               &                    &              &           &        & \\
LSTM       & Explicit   features        & OpenFace  & 0.323         & 3.23\%             & 0.375        & 0.323     & 0.477  & 0.524 \\
           & Latent   features          & AffectNet & 0.391         & 13.00\%            & \textbf{0.459} & 0.362     & 0.636 &  \textbf{0.612}\\
           &                            &           &               &                    &              &           &        & \\
BiLSTM     & Explicit   features        & OpenFace  & 0.303         & 0.40\%             & 0.335        & 0.288     & 0.462  & 0.483\\
           & Latent   features          & AffectNet & 0.383         & 11.90\%            & 0.453        & 0.348     & \textbf{0.658} & 0.602\\
           &                            &           &               &                    &              &           &        & \\
CNN-LSTM   & Fine-tuned latent features & AffectNet & 0.394         & 13.40\%            & 0.41         & \textbf{0.389} & 0.445&  0.605\\
CNN-BiLSTM & Fine-tuned latent features & AffectNet & 0.384         & 12.00\%            & 0.377        & 0.388     & 0.368 &   0.602
\\ \bottomrule 
\end{tabular}}
\footnotetext{\footnotesize \textit{ Random baseline lab data: F$_{1}$ = 0.3; Base rate = 0.3}}
\end{minipage}
\end{center}
\end{table}

The results for cross-task mind wandering detection using in-the-wild data for evaluations are reported in Table~\ref{tab:results_cross_task}. Similar to the within-task setting, our results outperform the random baseline (i.e., $F_{1} = 0.25$~\cite[][]{Lee.2022}, $AUC-PR = 0.17$) despite the cross-task prediction and in-the-wild setting. 
State-of-the-art on this dataset~\cite[][]{Lee.2022} achieved $F_{1}$ scores between 0.25 and 0.36 using SVMs, XGBoost, and DNNs for mind wandering prediction in a within-dataset setting. We achieve a very comparable performance to the best performance of the state-of-the-art~\cite[][]{Lee.2022} despite cross-task evaluation, a culturally different target group, and an in-the-wild setting. 

\begin{table}[]
\begin{center}
\begin{minipage}{\textwidth}
\caption{Results of mind-wandering detection for cross-dataset prediction. Training on lab data, testing on in-the-wild data.}
\label{tab:results_cross_task}
\resizebox{\textwidth}{!}{\begin{tabular}{@{}lllcccccc@{}}
\toprule
Model      & Feature Set                & Method    & AUC-PR         & Above Chance-Level & F$_{1}$             & Precision     & Recall        & ROC-AUC \\
\midrule
SVM        & Explicit   features        & OpenFace  & 0.214          & 5.30\%             & 0.330          & 0.241         & 0.524          & 0.6       \\
           & Latent   features          & AffectNet & 0.242          & 8.70\%             & 0.301          & 0.261         & 0.355          & 0.596             \\
           &                            &           &                &                    &                &               &                &                   \\
XGBoost    & Explicit   features        & OpenFace  & 0.18           & 1.20\%             & 0.169          & 0.193         & 0.15           & 0.534             \\
           & Latent   features          & AffectNet & 0.196          & 3.10\%             & 0.193          & 0.194         & 0.192          & 0.574             \\
           &                            &           &                &                    &                &               &                &                   \\
DNN        & Explicit   features        & OpenFace  & 0.218          & 5.80\%             & 0.291          & 0.23          & 0.398          & 0.589             \\
           & Latent   features          & AffectNet & 0.257          & 10.50\%            & 0.296          & \textbf{0.32} & 0.276          & 0.646             \\
           &                            &           &                &                    &                &               &                &                   \\
LSTM       & Explicit   features        & OpenFace  & 0.194          & 2.90\%             & 0.283          & 0.217         & 0.408          & 0.551             \\
           & Latent   features          & AffectNet & 0.252          & 9.90\%             & 0.323          & 0.234         & 0.524          & 0.634             \\
           &                            &           &                &                    &                &               &                &                   \\
BiLSTM     & Explicit   features        & OpenFace  & 0.168          & -0.20\%            & 0.251          & 0.16          & 0.583          & 0.47\\
           & Latent   features          & AffectNet & 0.261          & 11.00\%            & \textbf{0.352} & 0.224         & \textbf{0.825} & \textbf{0.658}    \\
           &                            &           &                &                    &                &               &                &                   \\
CNN-LSTM   & Fine-tuned latent features & AffectNet & 0.263          & 11.00\%            & 0.238          & 0.308         & 0.195          & 0.652        \\
CNN-BiLSTM & Fine-tuned latent features & AffectNet & \textbf{0.265} & \textbf{11.40\%}   & 0.254          & 0.285         & 0.229          & \textbf{0.658}             \\
\bottomrule
\end{tabular}}
\footnotetext{\footnotesize \textit{Random baseline in-the-wild data: F$_{1}$ = 0.25; Base rate = 0.17}}
\end{minipage}
\end{center}
\end{table}

\begin{figure}[h]
  \centering
  \includegraphics[width=0.46\textwidth]{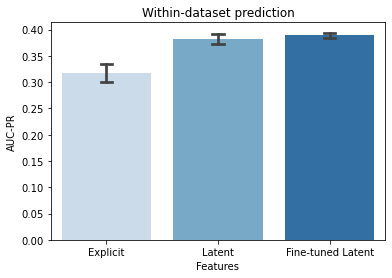}
  \includegraphics[width=0.46\textwidth]{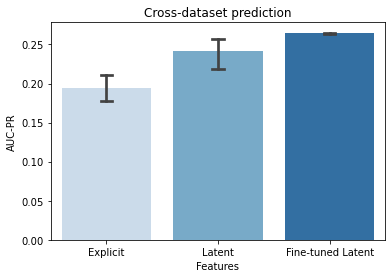}
  \includegraphics[width=0.46\textwidth]{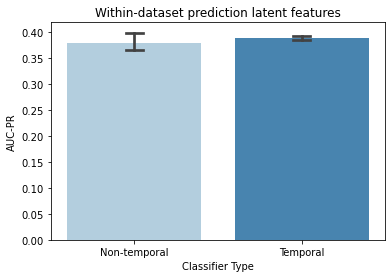}
  \includegraphics[width=0.46\textwidth]{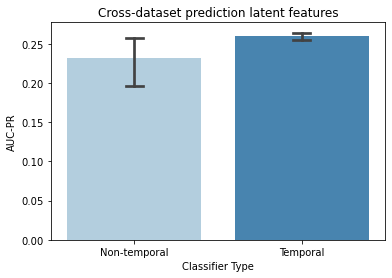}
  
  \caption{Average Performance by Feature Sets and Classifier Type.}

  \label{feature_comp}
\end{figure}

A comparison of performance along the employed feature sets over all classifiers and prediction scenarios, depicted in Figure \ref{feature_comp}, indicates that overall the latent features allow a better mind wandering detection. Especially in cross-dataset prediction, the fine-tuning of the latent features in an end-to-end CNN-LSTM leads to further improvement. This could be due to the fact that more data, i.e., the complete lab data set, were available for the training, hence fine-tuning, than in the case of the within the prediction. When employing latent features, the use of temporal (i.e., LSTM, BiLSTM, CNN-LSTM, CNN-BiLSTM) models, allows for a small improvement of prediction performance compared to non-temporal classifiers (i.e., SVM, XGB, DNN) (Figure \ref{feature_comp}, 2nd row), especially in the cross-dataset prediction.

\begin{figure}[h!]
  \centering
  \includegraphics[width=0.36\textwidth]{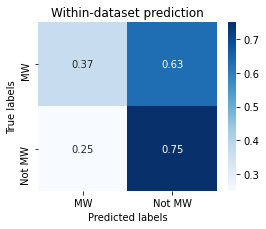}
  \includegraphics[width=0.36\textwidth]{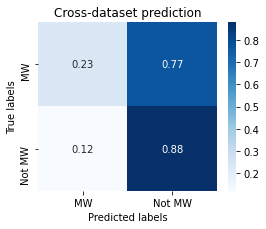}
  \includegraphics[width=0.46\textwidth]{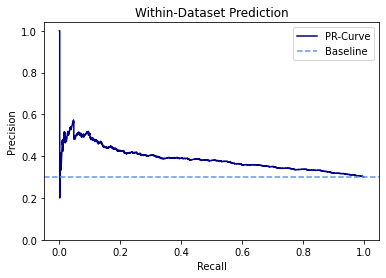}
  \includegraphics[width=0.46\textwidth]{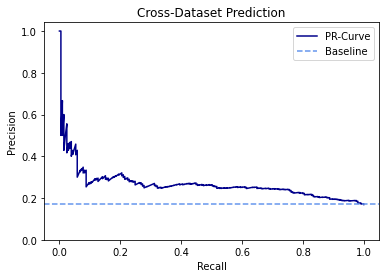}
  \caption{Confusion Matrices and Precision-recall Curves for Within- and Cross-dataset Prediction with CNN-BiLSTM.}

  \label{confusion_mat}
\end{figure}

Figure \ref{confusion_mat} shows the confusion matrices for within- and cross-dataset mind wandering prediction of the fine-tuned CNN-BiLSTM model, as well as precision-recall curves for both prediction scenarios. The confusion matrices reveal that the models are able to detect 37\% and 23\% of mind wandering instances in the two scenarios. Depending on the detection use case high recall might be favored over precision, which can be achieved by lowering the prediction threshold. 

When aggregating instance-by-instance predictions over time per participant, to assess performance over longer time windows, we obtain a Pearson's correlation coefficient of 0.578 ($P < 0.001$) between true and predicted mind wandering proportion per person in the within-dataset prediction. For cross-dataset predictions, we found a correlation of 0.48 ($P = 0.07$) on a person-level. However, when adapting the prediction threshold from 0.5 to 0.3 we achieve a correlation of 0.8 ($P < 0.001$).
This finding underscores the potential to enhance mind wandering detection by fine-tuning the prediction threshold for specific use cases, particularly in cross-dataset predictions that exhibit varying overall proportions of mind wandering. Consequently, we conducted systematic experimentation to optimize thresholds for enhanced detection accuracy and applicability.

\subsubsection{Threshold Optimization}

In alignment with previous research \cite[][]{stewart2017generalizability, Faber.2018b} we employ threshold optimization to further improve the detection of mind wandering instances. Threshold optimization is a technique used to enhance the performance of a classification model by selecting the optimal cutoff value for distinguishing between the two classes. In binary classification, models often output a probability score that indicates the likelihood of an instance belonging to the positive class. The default threshold is typically set at 0.5. However, the default threshold of 0.5 may not always be the best choice for all datasets or objectives, especially in cases where there is class imbalance. Particularly in the cross-dataset scenario, where the two different datasets yielded varying mind wandering rates of 30\% and 17\%, threshold optimization emerges as a critical approach, enabling the adjustment of classification cutoffs to better reflect the disparate prevalence rates across datasets \cite[][]{stewart2017generalizability}.

\begin{figure}[h]
  \centering

  \includegraphics[width=0.48\textwidth]{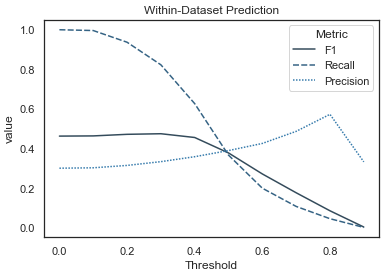}
  \includegraphics[width=0.48\textwidth]{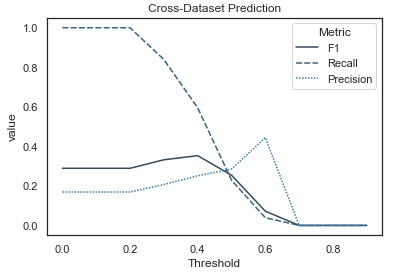}

  \caption{Evaluation Metrics by Prediction Thresholds for Within- and Cross-dataset Prediction with CNN-BiLSTM.}
  \label{th_fig}
\end{figure}

We systematically tested thresholds from 0.1 to 0.9 in increments of 0.1 for all employed models. Tables \ref{tab:th_results_within_task} and \ref{tab:th_results_cross_task} show the results for optimization according to $F_{1}$ values for within- and cross-dataset predictions. Optimal thresholds are smaller than 0.5 for most classifiers. Those lower thresholds lead to notable improvements in $F_{1}$ values. Figure \ref{th_fig} shows evaluation metrics by prediction thresholds for CNN-BiLSTM models in both prediction scenarios. We can see that $F_{1}$ values drop at the 0.3 and 0.4 thresholds, which were identified as optimal for those models.

\begin{table}[h]
\caption{Results of threshold optimization for within-lab-data mind wandering prediction.}
\label{tab:th_results_within_task}
\resizebox{\textwidth}{!}{\begin{tabular}{llllccccc}
\hline
Model      & Threshold & Feature Set                & Methods   & AUC-PR & F$_{1}$    & Precision & Recall & ROC-AUC   \\ \hline
SVM        & 0.5       & Explicit features          & OpenFace  & 0.288  & 0.391 & 0.286     & 0.636  &  0.482      \\
           & 0.5       & Latent features            & AffectNet & 0.365  & 0.443 & 0.347     & 0.63   &  0.587 \\
           &           &                            &           &        &       &           &        &       \\
XGBoost    & 0.1       & Explicit features          & OpenFace  & 0.346  & 0.434 & 0.313     & 0.703  & 0.543 \\
           & 0.1       & Latent features            & AffectNet & 0.372  & 0.456 & 0.334     & 0.717  & 0.576 \\
           &           &                            &           &        &       &           &        &       \\
DNN        & 0.1       & Explicit features          & OpenFace  & 0.331  & 0.461 & 0.3       & \textbf{0.992}  & 0.507  \\
           & 0.4       & Latent features            & AffectNet & \textbf{0.398}  & 0.467 & 0.338    &  0.752 & 0.601 \\
           &           &                            &           &        &       &           &        &       \\
LSTM       & 0.1       & Explicit features          & OpenFace  & 0.323  & 0.459 & 0.301     & 0.959  & 0.524 \\
           & 0.4       & Latent features            & AffectNet & 0.391  & \textbf{0.48}  & \textbf{0.345}     & 0.789  & \textbf{0.612} \\
           &           &                            &           &        &       &           &        &       \\
BiLSTM     & 0.1       & Explicit features          & OpenFace  & 0.303  & 0.459 & 0.302     & 0.954  & 0.483 \\
           & 0.4       & Latent features            & AffectNet & 0.383  & 0.478 & 0.34      & 0.806  & 0.602 \\
           &           &                            &           &        &       &           &        &       \\
CNN-LSTM   & 0.3       & Fine-tuned latent features & AffectNet & 0.394  & 0.472 & 0.334     & 0.806  & 0.605 \\
CNN-BiLSTM & 0.3       & Fine-tuned latent features & AffectNet & 0.384  & 0.475 & 0.334     & 0.823  & 0.602 \\ \hline
\end{tabular}}
\footnotetext{\footnotesize \textit{ Random baseline lab data: F$_{1}$ = 0.3; Base rate = 0.3}}
\end{table}

\begin{table}[h]
\caption{Results of threshold optimization for cross-dataset mind wandering prediction. Training on lab data, testing on in-the-wild data.}
\label{tab:th_results_cross_task}
\resizebox{\textwidth}{!}{\begin{tabular}{llllccccc} 
\hline 
Model      & Threshold & Feature Set                & Methods   & AUC-PR & F$_{1}$    & Precision & Recall & ROC-AUC   \\ \hline
SVM        & 0.5       & Explicit features          & OpenFace  & 0.214  & 0.324 & 0.231     & 0.539  & 0.6   \\
           & 0.2       & Latent features            & AffectNet & 0.242  & 0.329 & 0.225     & 0.611   & 0.596   \\
           &&&&&&&&\\
XGBoost    & 0.1       & Explicit features          & OpenFace  & 0.18   & 0.286 & 0.188     & 0.597  & 0.534 \\
           & 0.1          & Latent features         & AffectNet & 0.196  & 0.297 & 0.182     & 0.793  & 0.574     \\
           &&&&&&&&\\
DNN        & 0.4       & Explicit features          & OpenFace  & 0.218  & 0.316 & 0.216     & 0.592  & 0.589 \\
           & 0.2       & Latent features            & AffectNet & 0.257  & 0.349 & 0.248     & 0.586  & 0.646 \\
           &&&&&&&&\\
LSTM       & 0.2       & Explicit features          & OpenFace  & 0.194  & 0.289 & 0.171     & 0.927  & 0.551 \\
           & 0.5       & Latent features            & AffectNet & 0.252  & 0.323 & 0.234     & 0.624  & 0.634 \\
           &&&&&&&&\\
BiLSTM     & 0.1       & Explicit features          & OpenFace  & 0.168  & 0.3   & 0.177     & \textbf{0.99}   & 0.47 \\
           & 0.5       & Latent features            & AffectNet & 0.261  & 0.352 & 0.224     & 0.825  & \textbf{0.658} \\
           &&&&&&&&\\
CNN-LSTM   & 0.3       & Fine-tuned latent features & AffectNet & 0.263  & \textbf{0.356} & 0.235     & 0.737  & 0.652 \\
CNN-BiLSTM & 0.4       & Fine-tuned latent features & AffectNet & \textbf{0.265}  & 0.353 & \textbf{0.251}     & 0.595  & \textbf{0.658} \\ \hline
\end{tabular}}
\footnotetext{\footnotesize \textit{Random baseline in-the-wild data: F$_{1}$ = 0.25; Base rate = 0.17}}
\end{table}

\subsubsection{Comparison across Gender}

To detect potential bias in the classification models and provide transparency with regard to the fairness of prediction across gender, we compare the detection performance of mind wandering by gender for the within- and cross-dataset predictions. We compare the results of all models trained on latent features for both prediction scenarios. The lab data employed for within-dataset predictions consists of 40.77\% male and 59.23\% female participants, with females reporting an overall higher mind wandering rate of 32.64\% and males a rate of 26.41\%.

The in-the-wild data, serving as an evaluation set for our cross-dataset predictions, contains 53.33\% female and 46.67\% male participants, reporting mind wandering in 18\% and 16\% of the mind wandering probes on average, respectively.  
The prediction results by gender are displayed in Table \ref{gender_table}, with above chance values based on different base rates by gender and dataset. In general, mind wandering instances are predicted more accurately for females than for males. These results are most likely rooted in the imbalance of the underlying datasets, both regarding overall gender rates as well as gender-specific mind wandering rates, leading to overall fewer male mind wandering instances in the data. However, this difference becomes larger when employing temporal models, especially for within-dataset prediction scenarios. We can assume that this is due to the increased number of parameters in the temporal models, requiring large training data. While the overall performance increases, only females seem to benefit from improved detection, as there is more training data available.    

\begin{table}[]
\caption{Results of mind wandering detection using latent features by gender.}
\label{gender_table}

\resizebox{\textwidth}{!}{\begin{tabular}{llcccc}
\toprule
\multirow{2}{*}{Model} & \multirow{2}{*}{Gender} & \multicolumn{2}{l}{Within-dataset   prediction} & \multicolumn{2}{l}{Cross-dataset   prediction} \\
                       &                         & AUC-PR              & Above chance              & AUC-PR              & Above chance             \\
                       \midrule
SVM                    & Female                  & 0.38                & 8\%                       & 0.283               & 12.6\%                   \\
                       & Male                    & 0.326               & 8.4\%                     & 0.189               & 3.5\%                    \\
                       \\
XGB                    & Female                  & 0.365               & 5.7\%                     & 0.21                & 3.7\%                    \\
                       & Male                    & 0.301               & 5.0\%                     & 0.186               & 3.1\%                    \\
                       \\
DNN                    & Female                  & 0.419               & 13.7\%                    & 0.351               & 20.9\%                   \\
                       & Male                    & 0.355               & 12.4\%                    & 0.186               & 3.1\%                    \\
                       \\
LSTM                   & Female                  & 0.458               & 19.5\%                    & 0.335               & 18.9\%                   \\
                       & Male                    & 0.328               & 8.7\%                     & 0.196               & 4.3\%                    \\
                       \\
BiLSTM                 & Female                  & 0.445               & 17.6\%                    & 0.335               & 18.9\%                   \\
                       & Male                    & 0.306               & 5.7\%                     & 0.186               & 0.7\%                    \\
                       \\
CNN-LSTM               & Female                  & 0.429               & 15.0\%                    & 0.29                & 13.4\%                   \\
                       & Male                    & 0.334               & 9.5\%                     & 0.207               & 5.6\%                    \\
                       \\
CNN-BiLSTM             & Female                  & 0.439               & 16.7\%                    & 0.333               & 18.7\%                   \\
                       & Male                    & 0.316               & 7.1\%                     & 0.199               & 4.6\%                   
 \\
               \bottomrule
\end{tabular}}
\end{table}

\subsubsection{Latent Feature Explanation}

In order to gain deeper insights into the employed latent features, we apply the explainability algorithm LIME \cite[][]{lime} on our feature extraction CNN model, pre-trained on facial expression recognition task, and fine-tuned to our mind wandering detection task. LIME is an Explainable AI tool that helps to understand the decision-making of an algorithm, as it allows highlighting areas of interest in the image, the so-called super-pixels, that contribute positively or negatively to the model’s prediction. This helps to get an intuition on why the model thinks this image belongs to a certain class and which part of a given image the decision is based on. It is important to take this step to uncover any unintended correlations that the classifier may have learned, such as those resulting from artifacts produced in the data collection process~\cite[][]{lime}.

\begin{figure}[h!]
  \centering
  \includegraphics[width=0.5\textwidth]{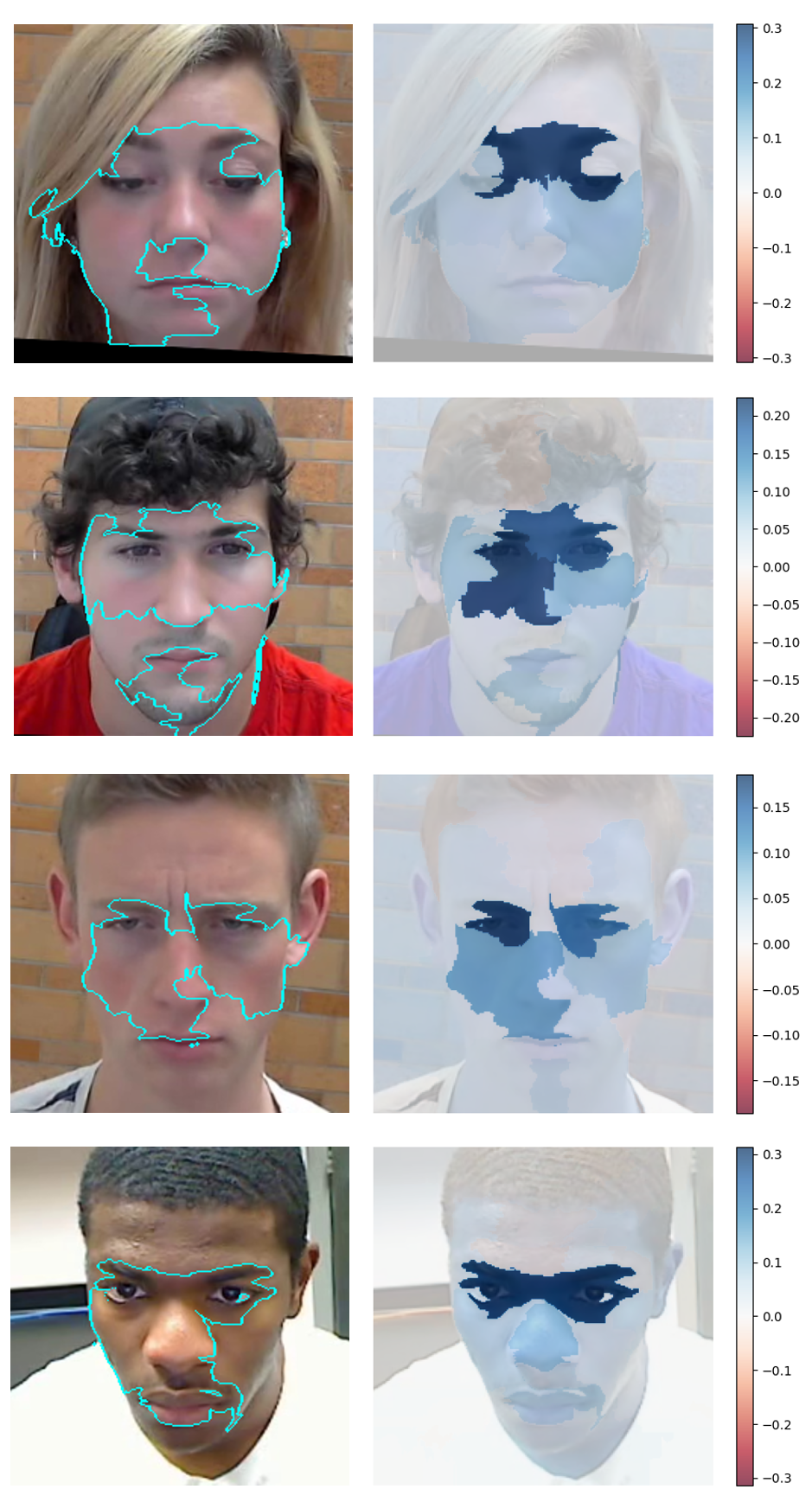}
  \caption{Most important super-pixel boundaries (left) and super-pixel importance heatmaps (right) of pre-trained ResNet50 FER model, calculated with LIME for a true positive, true negative, false positive, and false negative samples in the final mind wandering classification (top to bottom).}
  \label{LIME}
\end{figure}

In our study, these regions of interest let us draw conclusions about the information encoded in the latent features that are fed into the temporal mind wandering classifier. Therefore, it ensures the quality of feature extraction, by validating that the learned information aligns with established theoretical frameworks.  Example images from the lab data including super-pixel boundaries and heatmaps, with dark blue encoding the most important areas, are provided in Figure \ref{LIME}. The super-pixel boundaries include the 5 most important features  positively contributing to the obtained prediction. The heatmaps depict super-pixels by importance, by coloring the most important areas in dark blue. We see that the CNN part of our classification model, employed for feature extraction, mainly relies on information from the eye and mouth areas of the participants, which are consequently inherently encoded in our feature vectors as well. We included example images based on our mind wandering prediction results including a true positive, true negative, false positive, and a false negative. These results are in line with previous studies suggesting facial textures and AUs are meaningful features for video-based mind wandering detection \cite[][]{Bosch.2021}.

\begin{figure}[h!]
  \centering
  \includegraphics[width=0.97\textwidth]{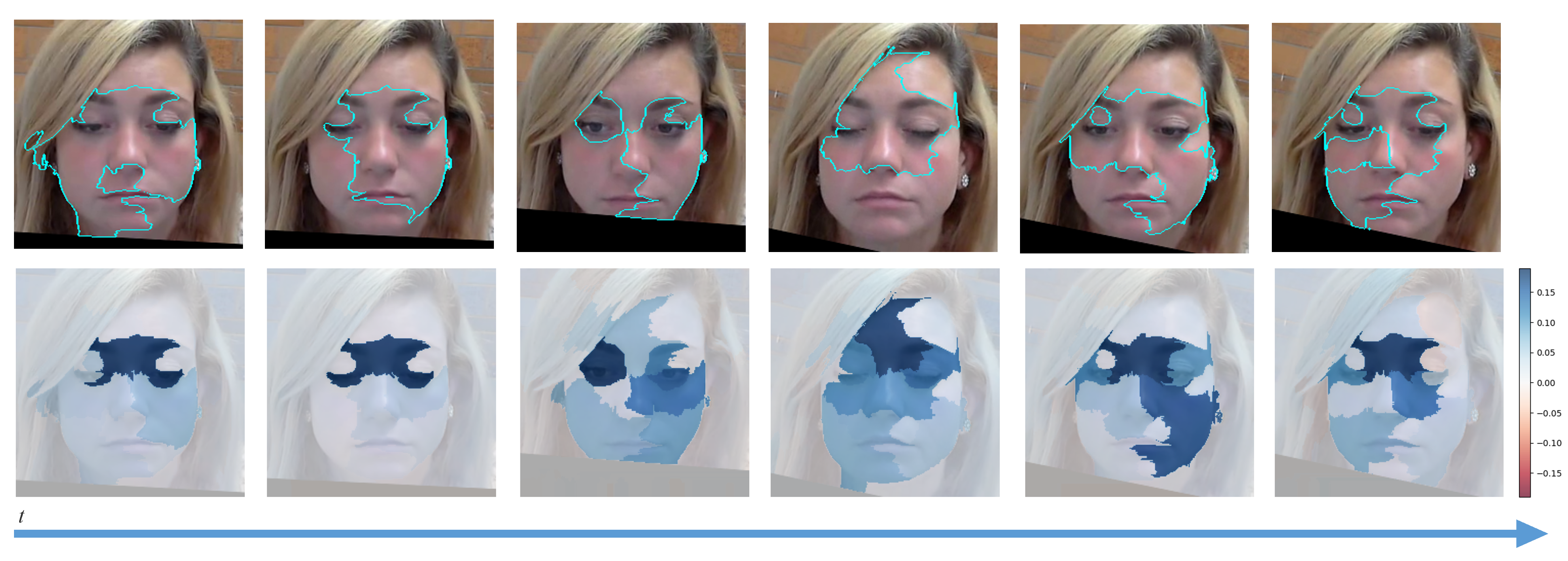}
  \caption{Temporal dynamics of extracted latent feature super pixels. Example depiction of every 25th frame from one correctly predicted mind wandering instance.}

  \label{LIME_temp}
\end{figure}

Since our analysis is based on video clips, the features shown are extracted from each frame individually and then transferred to the LSTM module as a time series. In this way, temporal dynamics can be displayed. To visualize this process, we have mapped the time course of the most important features for a correctly predicted mind wandering instance in Figure \ref{LIME_temp}. Every 25th frame from a 10 second clip is depicted. It is evident that the focused regions undergo minimal changes based on facial expression and eye opening, but the primary focus on the eye region remains consistent.

\subsection{Discussion}
In this section, we discuss our findings, also focusing on application scenarios and ethical issues.

\subsubsection{Main Findings}

We showed that features extracted from a pre-trained CNN on the AffectNet facial expression recognition dataset are informative for predicting mind wandering, even to a higher degree when they are utilized with temporal models that use frame-wise extracted features as input. The best results for within-dataset prediction were obtained using deep learning models such as DNN, LSTM, and BiLSTM, which allowed us to detect mind wandering $\approx14\%$ above the chance level. This indicates that the latent representations provided by a model based on the recognition of six basic facial expressions of emotion contain information at the same level for the task of mind wandering recognition as a set of explicit out-of-the-box features such as AUs, facial landmarks, and gaze vectors extracted by the OpenFace toolbox. Further, visualizations of areas of interest from the pre-trained FER model, fine-tuned on the mind wandering task showed that mainly eye and mouth areas were encoded by the model and, therefore, most likely encoded in the latent representation. these findings show that the features used represent meaningful information and can increase confidence in the latent representations. 
The results are comparable to those previously obtained by ~\cite{Bosch.2021} using a CART fusion of five explicit and hand-crafted feature groups on this data ($F_{1}$ = 0.478; $\approx25\%$ above chance level). By using transfer learning, the problem of insufficient data to train deep learning for this task can be overcome. Fine-tuning these features in an end-to-end model on the task at hand did slightly improve performance for the respective models but depends strongly on the available amount of data. 

When we evaluated these models with a new dataset that differed in the task, setting, and target group, we showed that latent features generalized very well and achieved a prediction performance ($\approx11.4\%$ above chance level) that was comparable to the best performance of a within-task prediction in the state-of-the-art~\cite[][]{Lee.2022}. Also for the cross-dataset setting, the models on the latent features performed better overall than those on the OpenFace features. This indicates that the transfer-learned features generalize better than the explicit out-of-the-box features without manual feature engineering. This may be due to the fact that they are based purely on facial expression, which is less stimulus-specific than, for example, gaze. Beyond the only existing generalization study of video-based mind wandering recognition~\cite[][]{stewart2017generalizability}, which was based on AUs, we were able to show that latent facial features are also predictive beyond the lab in naturalistic settings, such as the home. These features were trained on an in-the-wild dataset, which might favor the transferability of the features to an in-the-wild setting.

Nevertheless, the generalization results are remarkable because not only the setting and task differ in our cross-dataset prediction, but also the culture of the target group. As mentioned in section \ref{intro}, there is an ongoing debate about the universality of facial expressions of emotion \cite[][]{russell1994, ekman1994, jack2012facial}. One position taken is that the facial expressions of the six basic emotions relate to Western interpretations of these, but are not representative for instance, of East Asian culture~\cite[][]{jack2012facial}. While often universality of facial expressions of the six basic emotions is assumed, several studies tackled cross-cultural FER ~\cite[][]{dailey2010evidence, Benitez2017, ali2020artificial} and highlighted culture-specific differences in automated detection. The features used in this work are based on a model that was pre-trained on AffectNet data without accounting for culture-specific differences. Nevertheless, they generalize well from U.S. to Korean subjects in detecting mind wandering.

When comparing the prediction performances across gender, for both scenarios, we achieve better classification results for the female group. This difference could partly be the result of a combination of a greater proportion of women in both datasets, and the generally higher mind wandering rate for females in the training sample. However, gender differences are higher for temporal models with a higher number of parameters, in need of large training data.  Another challenge is choosing the best prediction model. We demonstrated that adjusting the threshold for decision-making, depending on its application, can improve the effectiveness of these predictions, especially in cross-dataset prediction. However, focusing on F$_1$ scores can lead to an imbalance between recall and precision of the mind wandering class, therefore we focus on the rank metric AUC-PR. Depending on the application scenario, the need for optimization towards one or the other can be advantageous. While it is important to capture all mind wandering instances when used for research and testing of materials, it may be more useful to have high precision when using interventions not to disturb learners in attentive moments.

\subsubsection{Applications}

The ability of a machine learning model to generalize is of primary importance in terms of its potential applications. On the one hand, it enables finding the optimal design of learning materials and systems in a way that less mind wandering occurs.  On the other hand, automated detection of mind wandering has the potential to help learners in online learning contexts such as ITSs and MOOCs to focus their attention on the task at hand, which often consists of tasks such as the ones we explored in this study: reading or video lecture viewing. This can be done through interventions such as providing feedback, suggesting re-watching or re-reading, asking intermediate questions, or adapting the presented content when a user loses focus. Such interventions can reduce user's mind wandering~\cite[][]{dmello2017zone,hutt2021breaking, mills2021} and therefore, based on the observed negative relationship between mind wandering episodes and learning success~\cite[][]{Smallwood.2011, Feng.2013,Szpunar.2013, Hollis.2016}, improve learning outcomes. However, since all of these studies examine short-term effects, there is a need to validate the long-term impact of mind wandering interventions.

The automatic detection of learners' off-task thoughts presented here can be used in future research to provide learners with different types of interventions and feedback. An experimental examination with corresponding questionnaires and knowledge tests could examine whether these are suitable to help learners sustain their attention and whether they actually lead to improved learning success without being distracting or disruptive. While the efficacy of some interventions based on eye tracking has been investigated in both lab and classroom context~\cite[][]{dmello2017zone, hutt2021breaking, mills2021}, a generalizable video-based recognition approach proposed in this study will allow future research to investigate the effectiveness of interventions outside the laboratory in naturalistic settings on a large scale. Furthermore, it will allow to deploy mind wandering detection modules in interfaces used in settings, in which eye tracking is not feasible.  

While the current precision in predicting isolated occurrences of mind wandering might evoke uncertainties, our research shows that by consolidating predictions at the participant level correlations with the ground truth, ranging from moderate to high, can be achieved. This underscores that when aggregated over longer periods of time, analogous to the temporal scope utilized in interventions, models can yield improved outcomes. However, given the current performance of state-of-the-art video-based mind wandering prediction and the potential for false positive predictions, it is imperative to consider interventions that do not impede the learning process. Among the viable options, two seem most suitable: Non-interruptive interventions, like follow-up prompts to re-read or re-watch critical parts of learning materials, prioritize fostering self-regulated learning while minimizing disruptions and distractions. By seamlessly integrating these interventions into the learning environment, a more favorable and effective learning experience can be achieved. Furthermore, the implementation of thresholds, based on criteria such as prediction confidence or time durations, ensures the maintenance of intervention quality. This approach mitigates the risk of learner irritation or negative experiences resulting from inadequately controlled interventions. In doing so, even if video-based mind wandering detection is less performant than comparable tasks, such as emotion recognition, its use in online learning systems can still lead to an improvement in attention. It is crucial to determine the required level of precision for targeted interventions to attain their desired effect without impeding learning. In general, the goal should be to support learners in self-regulated learning \cite[][]{jarvela2023} rather than to monitor or assess them. Therefore, it is important to consider user and data privacy in any potential application.

\subsubsection{Ethical Considerations}

While the application of video-based mind wandering detection holds promise for supporting learning in various applications, it also raises concerns regarding privacy, fairness, and inclusiveness. Since videos capture the identifiable faces of students during learning tasks, responsible handling of this data becomes imperative. While we work with raw videos, for model development, one of the advantages of our approaches is that for real world applications both explicit and latent features can be extracted on the fly in real time without the need of saving videos in the first place. A similar approach was employed in the classroom study of~\cite{Bosch.2021} to preserve privacy. This aspect is very important as particularly vulnerable target groups such as school children (i.e., minors in dependent relationships), as those can also benefit from MOOCs and ITSs that utilize mind wandering detection. The HCI community has been addressing data privacy-related issues in different tasks such as face recognition~\cite[][]{Erkin.2009}, gaze estimation~\cite[][]{Bozkir.2020}, or even in the classroom context~\cite[][]{Suemer.2020} and similar approaches that preserve privacy should be utilized if one uses raw videos in an end-to-end fashion when mind wandering detection modules are deployed in real-world systems. Another convenient approach would also include utilizing federated learning systems~\cite[][]{Li.2020} where the training of machine learning models is carried out locally in user devices by keeping the sensitive data away from other parties and only sending the trained models to the computation party for aggregation. In brief, while the performance of mind wandering detection is an important aspect, privacy issues are of paramount importance for real-world use. In any case the process of collecting and analyzing any form of data should be transparent to the user with informed consent and clarity of the purposes the data will be used for~\cite[][]{nguyen2022ethical}.
Another important aspect of deploying automated mind wandering detection in educational settings is the inclusiveness in data and algorithms \cite[][]{nguyen2022ethical}. To ensure fairness and equality across user groups, it is crucial that these algorithms are trained and tested on diverse, unbiased data. This research underscores the significance of these considerations by aiming to develop a culturally generalizable algorithm and conduct a transparent analysis of gender fairness. While the results are promising with regard to generalization, especially the found gender differences stemming from unbalanced data, highlight the necessity for diverse and representative data that facilitate even better adherence to these principles in future research.

\subsubsection{Limitations and Future Work}

Despite its novelty, the presented work exhibits several limitations. First of all, we examined cross-setting predictions from the lab dataset to the in-the-wild dataset. This is partly due to the fact that lab-based data collection is more controlled, and therefore, allows for higher data quality. As reported in the original study~\cite[][]{Lee.2022} of the in-the-wild dataset, a large amount of data had to be excluded during data preparation due to insufficient data quality, for example, due to low luminance. From a practical standpoint, the generalizability from labeled lab data to in-the-wild data is a likely setting for future applications. Related to this point, the present in-the-wild dataset is rather small to train our proposed models on it. The need for a comparatively large labeled dataset to enable us to use the proposed temporal methods is a further limitation of this work. Currently, to our best knowledge, no larger in-the-wild mind wandering dataset is available. Despite the limited size of the in-the-wild dataset, this study examines the potential generalizability of state-of-the art mind wandering detection to diverse datasets. However, it is important to exercise caution when interpreting these findings due to the dataset's size. Future research should focus on collecting larger datasets to provide a more robust evaluation of the model's performance and enhance our understanding of its broader applicability. In general, further advances in automated detection of mind wandering should be evaluated based on their generalizability to in-the-wild settings to avoid the optimization of performance on single datasets at the expense of generalizability, as was recently discussed in the related field of affect detection ~\cite[][]{dmello2023affect}.

While our cross-dataset prediction examines the models’ ability to generalize two culturally different target groups, the sample of university and college students is somewhat homogeneous regarding other demographics, such as age. Future research should investigate, how well the model can be transferred to other target groups, for example, to school children. Furthermore, differences in predictive power by gender were revealed. A balanced sample by gender, as well as further investigation of gender differences in mind wandering, should be an important consideration for future studies to avoid bias.

In general, the presented results underscore the complexity of the recognition of mind wandering solely through facial video analysis. The approach presented in this paper, alongside previous approaches, can predict off-task thoughts above chance level, for dataset-specific class distributions, albeit without achieving exceptionally high prediction performances. While the achieved prediction accuracy may be considered moderate, and will potentially lead to false positive predictions when employed in educational systems, it is crucial to acknowledge the inherent difficulties associated with accurately identifying such a nuanced cognitive process. A prior study hinting at upper bounds for appearance-based detection showed that human observers only could identify mind wandering episodes to a similar extent ($F_{1}$= 0.406) as machine-learning based detection algorithms on the lab data ~\cite[][]{bosch2022}. These results strengthen the assumption that there is a limit to the accuracy with which the covert cognitive state of mind wandering can be detected based only on appearance. 
However, the required detection accuracy heavily depends on the desired application. Initial studies exploring interventions, like reiterating content, and asking questions, based on similar prediction performances show the potential success of those~\cite[][]{dmello2017zone, mills2021, hutt2021breaking}. Future investigations should deploy the presented automated mind wandering detection approach to deliver interventions while assessing learning outcomes to determine the extent to which interventions improve attention and learning outcomes, whether frequent or incorrectly delivered mind wandering interventions hinder or interfere with learning and whether predictions at the current precision level are sufficient to be integrated into systems. As mentioned above, especially the use of non-intrusive interventions and the implementation of quality thresholds should be considered in the context of moderate prediction accuracies.

Additionally, the limited performance of these elaborate models can be attributed to the size constraints of the employed training dataset. Unfortunately, to our best knowledge, no other video-based mind wandering detection dataset is publicly available at this moment. To further advance the field and improve prediction outcomes, future research should place emphasis on collecting large-scale datasets, particularly in in-the-wild scenarios. By incorporating such diverse and extensive data, researchers can enhance the models' performance and pave the way for more accurate and robust mind wandering detection systems. A commendable approach was employed by \cite{hutt2023webcam}, who gathered webcam eye tracking data both in university environments and via the platform Prolific, thereby ensuring the inclusion of diverse target groups and settings.
Further potential strategies for enhancing prediction performance may include the implementation of personalization, such as the use of personalized prediction thresholds. This approach could partially compensate for the presumed interpersonal variations.
Another way to improve prediction accuracy is to add other signals, like behavioural trace data or physiological signals. In this work, we focused on webcam videos due to the goal of scalability and low-threshold application in naturalistic settings. One potential for future research in this realm could be the further improvement of webcam-based eye tracking, since eye movements have been shown to be highly predictive for mind wandering~\cite[][]{hutt2023webcam, bixler2014,dmello2016, Faber.2018b, mills2021}.

\subsection{Conclusion}

In this work, we proposed a novel and generalizable approach for mind wandering detection utilizing facial features based on transfer learning from videos. Our results show the meaningfulness and transferability of those features with within- and cross-dataset prediction tasks on two challenging datasets. In particular, the results of the cross-dataset setting that differed with regard to the task, target group, and environment show the generalizability of our approach, which is key to the deployment of such models in intelligent learning systems to support learners to keep their attention on a given task.

\subsection*{Declarations}

\subsubsection*{Funding}

Babette B{\"u}hler is a doctoral candidate and supported by the LEAD Graduate School and Research Network, which is funded by the Ministry of Science, Research and the Arts of the state of Baden-W{\"u}rttemberg within the framework of the sustainability funding for the projects of the Excellence Initiative II. This research was partly funded by the Deutsche Forschungsgemeinschaft (DFG, German Research Foundation) under Germany's Excellence Strategy - EXC number 2064/1 - Project number 390727645. Sidney D'Mello was supported by the National Science Foundation (DRL 1920510). The opinions expressed are those of the authors and do not represent the views of the funding agencies.

\subsubsection*{Competing Interests}

The authors have no relevant financial or non-financial interests to disclose.

\subsubsection*{Author Contributions}

All authors contributed to the study conception and design. Data analysis was performed by Babette B{\"u}hler. The first draft of the manuscript was written by Babette B{\"u}hler, Efe Bozkir and all authors commented on previous versions of the manuscript. All authors read and approved the final manuscript.

\subsection*{Additional Tables}\label{secA1}

    \begin{table}[H]
        \begin{center}
            \caption{Hyperparameter grids for employed non-temporal classification models.}
            \label{tab:hyper_grid}
        
            \begin{tabular}[t]{ll}
                \toprule
                Model & Hyperparameter   Grid                                                                                                                                                                                             \\
                \midrule
                SVM   & C (1, 10,   1000, 1000)\\    
                      &  Gamma (0.001,   0.01, 0.1) \\
                DNN   & Hidden layer size (100, 64, 32)\\    
                      & Learning rate (0.001, 0.0001)\\    
                      & Alpha (0.0001, 0.001, 0.005)\\    
                      & Early stopping (True, False)                      \\
                XGB   & Gamma (0.5, 1, 2,5)\\    
                      & Subsample (0.6, 0.8, 1)\\    
                      & Column sample by tree (0.6, 1)\\    
                      & Max depth (3, 5, 9)\\    
                      & Learning rate (0.01, 0.1, 0.3)
                                        \\
                \bottomrule
            \end{tabular}
        \end{center}
    \end{table}

    \begin{table}[H]
    \begin{center}
    \begin{minipage}{\textwidth}
    \caption{Results of mind-wandering detection for within- and cross-dataset prediction with feature level fusion.}
    \label{tab:results_fusion}
    \resizebox{\textwidth}{!}{\begin{tabular}{@{}lllccccc@{}}
    \toprule
    Prediction & Model   & Method           & F$_{1}$ & Precision & Recall & AUC-PR  & Above Chance-Level \\
    \midrule

    Within-dataset & SVM                        & OpenFace, AffectNet                    & 0.405        & 0.331        & 0.525     & 0.336  & 5.143\%\\
    
      &  XG Boost                   & OpenFace, AffectNet                    & 0.338        & 0.382        & 0.307     & 0.364  & 9.143\%\\
      & DNN                      & OpenFace, AffectNet                   & 0.401         & 0.382        & 0.693     & 0.397  & 13.857\%\\
      &  LSTM                      & OpenFace, AffectNet                    & \textbf{0.461}	    & 0.334	       & \textbf{0.761}	  & 0.373 & 10.429\%\\
      &  BiLSTM                    & OpenFace, AffectNet                    & 0.450	    & 0.346	       & 0.697	    & 0.377 & 11.000\% \\
                          &&&&&\\
    Cross-dataset & SVM                   & OpenFace, AffectNet & 0.299 & 0.218          & 0.473     & 0.207 & 4.5\%\\
     & XGBoost                & OpenFace, AffectNet & 0.202 & 0.198          & 0.207     & 0.199 & 3.5\%\\
     & DNN                   & OpenFace, AffectNet & 0.311 & 0.238          & 0.448     & 0.220 & 6.0\%\\
     & LSTM                  & OpenFace, AffectNet & 0.315 & 0.192       & \textbf{0.864}     & 0.250 & 9.6\% \\
     & BiLSTM                 & OpenFace, AffectNet & 0.335 & 0.214          & 0.772     & 0.229 & 7.1\%\\ 
    \bottomrule
    \end{tabular}}
    \end{minipage}
    \end{center}
    \end{table}

\FloatBarrier

%% file: 4_OnTaskInSync.tex
\section[Examining the Relationship between Gaze Synchrony and Self-Reported Attention]{On Task and in Sync: Examining the Relationship between Gaze Synchrony and Self-Reported Attention During Video Lecture Learning}

\subsection{Abstract}
Successful learning depends on learners’ ability to sustain attention, which is particularly challenging in online education due to limited teacher interaction. A potential indicator for attention is gaze synchrony, demonstrating predictive power for learning achievements in video-based learning in controlled experiments focusing on manipulating attention. This study (\textit{N}=84) examines the relationship between gaze synchronization and self-reported attention of learners, using experience sampling, during realistic online video learning. Gaze synchrony was assessed through Kullback-Leibler Divergence of gaze density maps and MultiMatch algorithm scanpath comparisons. Results indicated significantly higher gaze synchronization in attentive participants for both measures and self-reported attention significantly predicted post-test scores. In contrast, synchrony measures did not correlate with learning outcomes. While supporting the hypothesis that attentive learners exhibit similar eye movements, the direct use of synchrony as an attention indicator poses challenges, requiring further research on the interplay of attention, gaze synchrony, and video content type.

\subsection{Introduction}

The increasing transition from traditional classroom settings to digital learning environments changes student-teacher interactions, which are crucial for learning. Without a physically present instructor to tailor content and provide support, learners need to self-regulate to sustain focus on educational tasks \cite{schacter2015}. This presents a significantly greater challenge for students in a video lecture than in a traditional face-to-face lecture \cite{wammes2017}. Although online teaching can take place in real-time, it proves more difficult for lecturers to monitor and manage learners' attention, causing a desire for teachers to receive feedback \cite{sun2019}. This has led to extensive research on assessing attention in online learning \cite{schacter2015, wammes2019, wisiecka2022}. 
Eye gaze is one of the most crucial sensing modalities for measuring human attention. As such it has been extensively studied also in the educational context~\cite{goldberg2021attentive, sumer2018teachers, castner2018scanpath, gao2021digital, bozkir2021exploiting}. Recent studies showed that visualizations of gaze data enable instructors to estimate the level of attention of learners \cite{sauter2023} and their comprehension of learning content \cite{Kok.2023}. In a pioneering study, \cite{Madsen.2021} hypothesized that attentive learners follow instructional videos similarly with their eyes. Measuring synchrony with inter-subject correlation (ISC) of eye gaze within experimental groups of attentive and distracted students, they showed that gaze synchrony levels were predictive of test scores. Additionally, they expanded their approach to webcam-based eye-tracking data, suggesting the potential for real-time assessment of attention and adaption of learning content. In a later study, \cite{Sauter.2022} could not reproduce these findings with unreliably sampled data, highlighting the challenges of webcam-based eye-tracking for educational contexts. In another study, \cite{Liu.2023} only found weak correlations between gaze synchrony and test scores and decreased synchrony when using eye gaze models, whereas confirming previous findings that eye gaze models foster learning \cite{jarodzka2013}.

A major limitation of previous research is that participants' attention level was experimentally controlled by introducing a secondary distraction task in the inattentive condition~\cite{Madsen.2021, Liu.2023}. It remains uncertain whether these experimental manipulations accurately reflect naturally occurring distractions \cite{Liu.2023}. The underlying mechanisms of inattention during video learning are multifaceted and range from overt distraction by the environment, for example, an incoming email \cite{wammes2019}, to hidden cognitive processes such as mind wandering, i.e., the engagement in task-unrelated thought, for example, daydreaming \cite{lindquist2011}. 
In realistic learning scenarios, the level of attention can be assumed to exhibit more gradation. Distraction and cognitive disengagement are dynamic processes that evolve and fluctuate over time \cite{farley2013}. Intruding thoughts may be less persistently inhibitory, potentially less evident in eye movements than the artificially induced counting task, yet still detrimental to learning. Further, only the very time-sensitive frame-wise ISC has been employed as a synchrony measure, suitable for very dynamic stimuli, which are not necessarily given in the educational context. A widespread video lecture setup presents an instructor and slides.

In this paper, we investigate the hidden link between gaze synchrony, attention, and learning outcomes. We examine the relationship between gaze synchrony and self-reported attention in a realistic learning scenario, specifically during learning with a pre-recorded Zoom video lecture. Gaze synchrony is assessed with two different measures: the Kullback-Leibler divergence between gaze density maps and the MultiMatch scanpath comparison method. Further, we examine the suitability of gaze synchrony as an indicator of attention by investigating the relationship between gaze synchrony, self-reported attention, and learning outcomes in the form of post-test scores. Our primary contributions include investigating naturally occurring self-reported (in)attention and its association with gaze synchrony, employing novel measures for gaze synchronization in video learning, and examining whether gaze synchrony predicts learning outcomes in this setting.

\subsection{Related Work}
\subsubsection{Synchrony during Learning}

In their study, \cite{Madsen.2021} proposed that learners synchronize their gaze during cognitive processing of lecture videos, indicating attentiveness. To investigate this, participants in two experimental conditions, attentive and distracted, viewed short informal instructional videos. In the distracted condition, participants were instructed to do the serial subtraction task (counting back in their minds from a randomly chosen prime number in decrements of 7) while watching the video. The attentive condition did not get any extra instructions. The synchronicity of eye movements was significantly higher in the attentive than in the distracted condition. They found a significant correlation between the level of gaze synchronicity and test scores, and the results were robust across factual and comprehension questions, as well as different video styles (animation vs. drawing figures). 

In replicating the prior study, including various lecture video styles and using webcam-based eye tracking, without extensive exclusion of low sampling rates, \cite{Sauter.2022} did not observe a predictive correlation between gaze ISC and test scores in a comprehensive quiz. This underscores the constraints of current webcam-based eye-tracking methods and their limitations in employing a synchrony measure in real-time remote settings. Additional studies with high-end eye trackers confirmed the correlation between experimentally induced distraction and ISC \cite{Liu.2023}. In a second experiment, however, \cite{Liu.2023} showed that the use of gaze modeling decreased the level of gaze synchrony compared to a normal viewing condition while still showing higher post-test results. A very weak correlation between total gaze synchrony and test scores was found.

These studies strongly suggest the significance of gaze synchronization during video learning, correlating with learning success, and suggest an interrelation to attention. Previous research investigating gaze patterns in live online lectures reinforces these observations, highlighting a positive correlation between students' focal attention and their ability to retain lecture content \cite{wisiecka2022}. However, it remains unclear if gaze (dis-)synchrony can indeed be used as an indicator for naturally occurring (in-)attention and employed for attention detection in a real-world learning scenario. 

\subsubsection{Measuring Gaze Synchrony}

The only measure employed to date to assess gaze synchrony during learning is the \gls{isc} of vertical and horizontal gaze positions and pupil diameter per video frame \cite{Madsen.2021, sauter2023, Liu.2023}. It is computed within the experimentally induced attentive and distracted groups. In the webcam-based eye tracking setting, pre-computed median values and eye movement velocity instead of pupil diameter were used for correlation computation~\cite{Madsen.2021,sauter2023}. The ISC is a very time-sensitive measure that assumes synchronization with minimal spatial-temporal distances, which appears to be suitable for a very dynamic video stimulus, which more strongly drives eye movements \cite{Dorr.2010}. This raises the question of the applicability of ISC for more static settings in the educational domain. A frequently selected format, particularly in live teaching settings, is the presentation of lecture slides. This represents a relatively static stimulus, in which the gaze is mainly guided by the lecturer's verbal description of the content and potential pointers like cursors \cite{sauter2023}.  

Various other measures have been proposed in different contexts to capture the synchronicity of eye movements in dynamic scenes. This includes clustering-based approaches, measuring the percentage of gaze falling into the main cluster \cite{goldstein2007}.
Another set of proposed measures works with fixation maps or probability distributions created by the additive superposition of Gaussians, assessing the differences between those maps employing the sum of squared pointwise subtraction \cite{wooding2002} or computing the Kullback-Leibler divergence (\gls{kld}) \cite{rajashekar2004, tatler2005}. 
Other studies computed the entropy of gaze density based on Gaussian Mixture Modeling \cite{sawahata2008, smith2013} and temporal Normalized Scanpath Saliency (\gls{nss}) \cite{Dorr.2010}, demonstrating a high correlation with KLD to quantify gaze similarity.
 A multidimensional scanpath comparison approach including gaze characteristics alongside spatial and temporal properties is the vector-based MultiMatch \cite{jarodzka2010, dewhurst2012, foulsham2012} method. The derived similarity measure combines sub-measures assessing similarity in scanpath shape, saccadic direction, length, fixation position, and duration. MultiMatch's inherent alignment of scanpaths reduces sensitivity for minor temporal shifts and manages variations in the lengths of scanpaths \cite{le2013methods}. However, a limitation of this metric is its comparison between only two scanpaths at a time, while the objective in synchrony analysis often involves comparing entire groups of subjects.

Each of the proposed methods captures slightly different characteristics of gaze behavior. For this study, we chose to compute and compare two established gaze synchrony measures that account for the properties of our relatively static video stimulus and incorporate gaze characteristics beyond spatio-temporal properties: KLD of gaze density maps and MultiMatch scanpath comparison.

\subsection{Methods}

\subsubsection{Experiment}

The ethics committee of the Faculty of Economics and Social Sciences, University of T\"ubingen (Date of approval 13 January 2022, approval \#A2.5.4-210\textunderscore ns) 
approved our study procedures, and all participants have given written consent to the data collection. 

\paragraph{Participants}

The data employed for this study was collected from \textit{N} = 96 university students. Five participants had to be excluded from further data analysis due to technical errors during the experiment, such as malfunction of the eye tracker or crashing of experiment computers. 
Additionally, three participants were excluded from the further analysis because they were not fluent in the language used for the instructions, questionnaires, and video content. Consequently, the study was completed with a total of \textit{N} = 88 participants (Ages 19-33, \textit{M} = 23.44, \textit{SD} = 2.6), of which 19\% were male.

\paragraph{Study Procedure and Setup}

The data was collected in the laboratory using the SMI Red remote eye tracker with a sampling rate of 250 Hz. We refrained from using chin rests, as we wanted to ensure an ecologically valid setting. Further, research has shown that even without a chin rest acceptable levels of accuracy for purposes not relying on small eye movements are achieved \cite{carter2020}. After completing a short questionnaire and a test on previous knowledge on the session topic of statistics, participants performed a nine-point pulsating calibration of the eye tracker. Participants then watched a pre-recorded Zoom lecture on introductory statistics. The video lecture's total duration was approximately 60 minutes, which required a re-calibration of the eye tracker after about 30 minutes. Participants were instructed to focus on the lecture and were not allowed to take notes or use electronic devices, including phones, during the study. After the video was completed, a comprehensive post-test of 14 questions targeting factual knowledge and deep-level understanding was conducted. 
Including the time allocated for general instructions and filling out questionnaires, the overall duration per participant averaged approximately 120 minutes. Participants received a compensation of \texteuro20.

\paragraph{Video Stimulus}

\begin{figure}[h]
  \centering
  \includegraphics[width=0.5\linewidth]{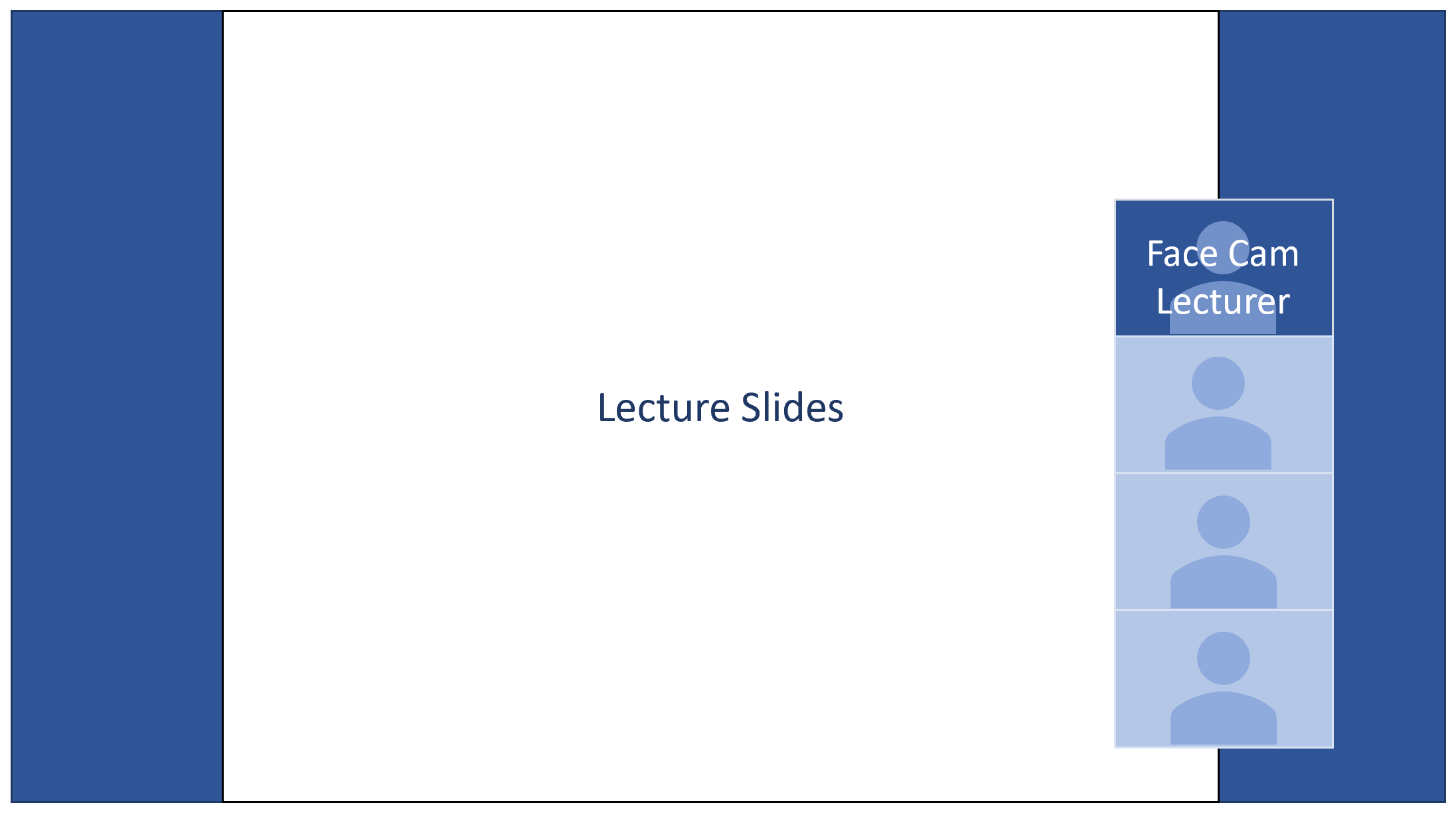}
  \caption{Zoom lecture video layout.}
  \label{fig:vid}
\end{figure}

The video stimulus represented a typical Zoom layout, depicted in Figure \ref{fig:vid}, including lecture slides and a webcam display of the lecturer's face on the top right. Other participants in the Zoom lecture turned off their cameras so only participant tiles were visible. The slides were primarily static, yet the instructor used the cursor to point at specific locations on the slides.

\begin{figure}[h]
  \centering
  \includegraphics[width=\linewidth]{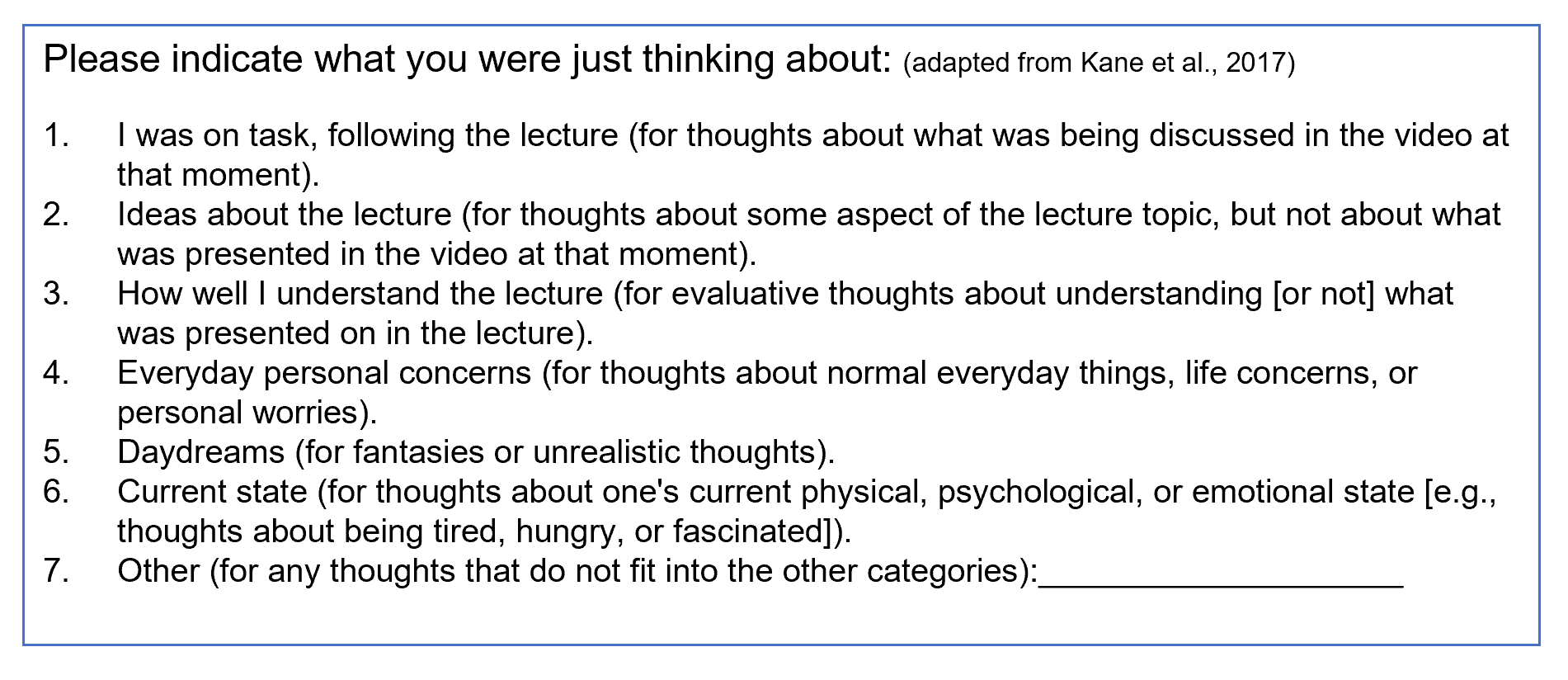}
  \caption{Attention thought probes, adapted from \cite{Kane.2017}.}
  \label{fig:probed}
\end{figure}

\paragraph{Attention Experience Sampling}
The most common method for directly assessing the internal state of learners is through self-reports \cite{weinstein2018}. Given the potential inaccuracies associated with retrospective reports, experience sampling, also known as probe-caught method, is typically employed. This involves intermittently stopping participants during a task and asking them to indicate where their attention was focused at that very moment. Although these self-reports are subjective, previous research has revealed their correlation with more objective but indirect measures of attention, such as physiological indicators, response times, and task performance \cite{weinstein2018}.
The study incorporated 15 quasi-randomized thought probes presented at fixed three- to five-minute intervals to assess the participants' attentional state throughout the lecture. All participants received the probes at identical moments during the lecture. A probe was administered by displaying a screen with a multiple-choice question asking what the participants thought about just now. Six answer categories adapted from \cite{Kane.2017}, ranging from ``I was on task, following the lecture'' to ``Everyday personal concerns'' (See figure \ref{fig:probed} for all categories), and an open response option was provided. Responses within the open-ended category underwent manual coding by two independent raters. The process followed an iterative method involving assigning responses to existing and establishing new off-task categories, like ``External Distraction.'' As this study focuses on the difference between attentive and inattentive learners, we dichotomized the answers accordingly. The ``I was on task, following the lecture'' answer option was coded as attentive. In contrast, answers to all other categories encompassing meta-cognitive monitoring, elaborations, distractions, and mind wandering were coded as inattentive. Although elaborations or meta-cognitive monitoring mechanisms are not inherently disruptive but are considered an essential part of the learning process, they still impede learners from following the lecture's content at that very moment.

\begin{figure}[h]
  \centering
  \includegraphics[width=0.6\linewidth]{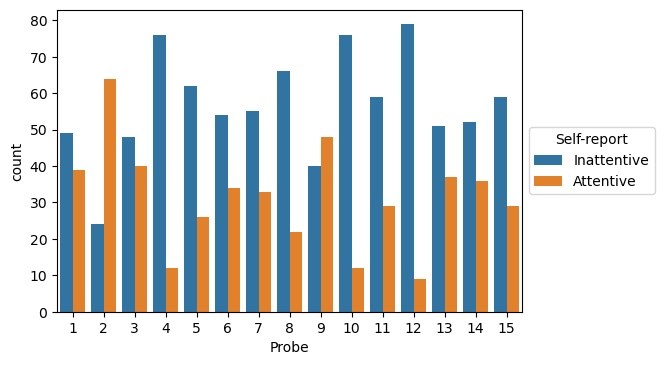}
  \caption{Absolute frequency of attention self-reports by experience sampling thought probe.}
  \label{fig:attention}
\end{figure}

For $36\%$ of all thought probes during the 60-minute lecture, participants reported being on task and following the lecture. Most of the time, however, their thoughts were preoccupied with elaborations about the lecture topic, whether they understood the lecture or something else entirely, such as personal concerns or their current state, for example, tiredness or hunger ($64\%$). The difference between attentive and inattentive self-reports across all probes is significantly different from the uniform distribution ($\chi^2 = 98.337$, $p < 0.001$). Figure \ref{fig:attention} shows the absolute frequency of the participant's attention self-reports for the attentive and inattentive categories by thought probe. It becomes visible how the level of attention fluctuates over time. The peak in attention occurs at the second probe, approximately ten minutes into the lecture. Another local peak can be observed at probe nine after the eye tracker re-calibrated, which allowed for a short break.

\paragraph{Learning Outcomes}

\begin{figure}[h]
  \centering
  \includegraphics[width=0.9\linewidth]{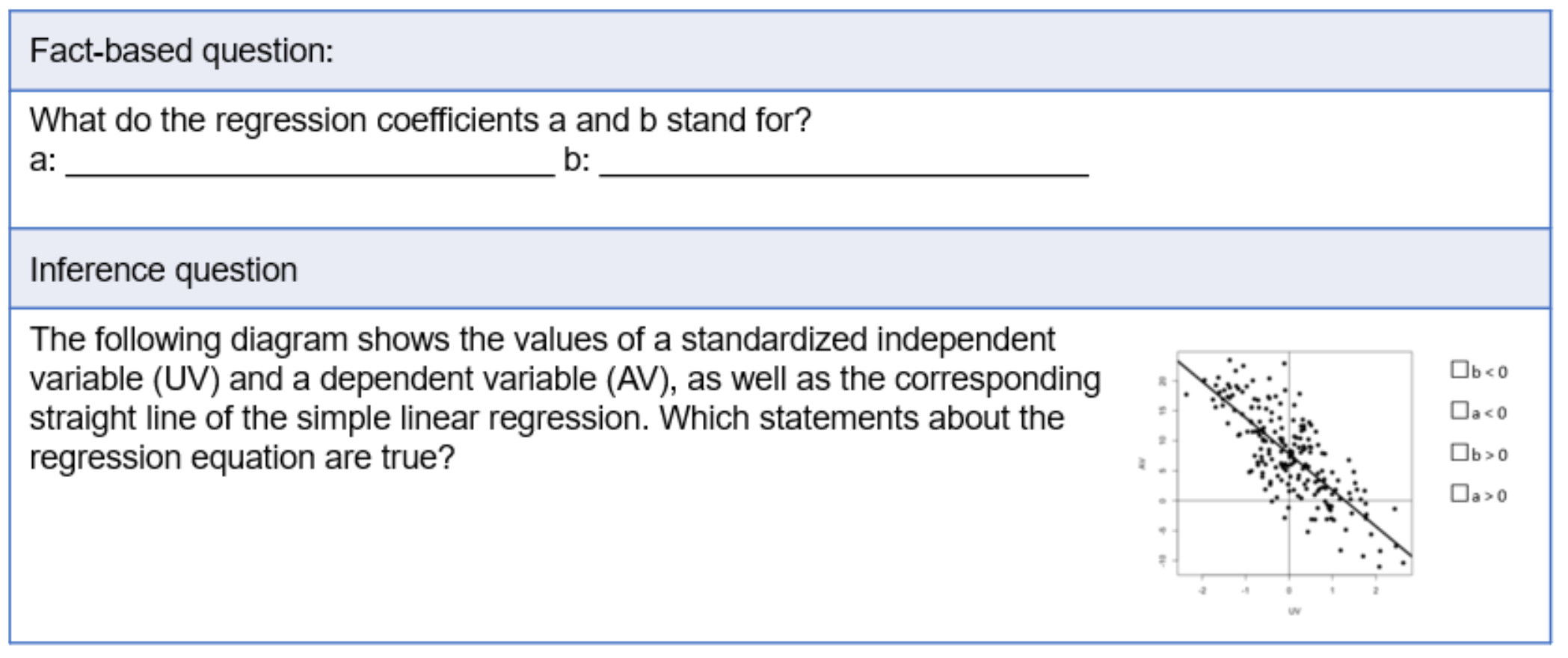}
  \caption{Example Posttest Questions.}
  \label{fig:testc}
\end{figure}

To assess learning outcomes, we administered a post-video knowledge test comprising 14 multiple-choice and open-ended questions to assess participants' comprehension levels related to the video lecture content. This assessment included seven fact-based memory questions and seven questions targeting deep-level understanding (see Figure \ref{fig:testc} for an example of both types). Specifically tailored for the prerecorded lecture on linear regression analysis, the questions covered topics such as empirical covariance, method of local averaging, and least squares estimation. We ensured to include all topics covered right before the thought probes in the test. Examples of both fact-based and inference questions are illustrated in Figure \ref{fig:testc}. The summative scores, ranging from 0 to 14, were derived by assigning 1 point for each correctly answered question. Participants achieved an average score of 5.63 ($\textit{sd} = 2.663$; see Figure \ref{fig:posttest} in the appendix for score distribution). To control for previous knowledge, a pre-test on the general topic of linear regression analysis was conducted before the video lecture, including eight multiple-choice and open-ended questions.

\subsubsection{Eye Gaze Data Pre-processing}

The average calibration error for our nine-point calibration procedure was 0.31°. We used the analysis software BeGaze by SMI to extract eye movement events, including fixations (determined by dispersion-based threshold), saccades (determined by velocity-based threshold), and blinks from the raw gaze data \cite{begaze}. We regard the time windows right before the experience sampling probes for our gaze synchrony analysis. Therefore, we cut 10-second windows before each of the 15 probes for each participant, resulting in a set of 1335 instances with corresponding self-reported attention information. Furthermore, a deficiency of the employed eye trackers that have been reported before is that tracking failures are recorded as unusually long blinks \cite{castner2020}. Consequently, blinks longer than 500 ms, exceeding an expected blink duration range between 100 to 400 ms \cite[]{schiffman2001}, were excluded. We excluded gaze sequences with less than a 75\% gaze tracking ratio to ensure high data quality. For four participants, applying this criterion resulted in excluding all 15 sequences due to insufficient data quality with less than 75\% tracking ratio across the board. Overall, this exclusion threshold led to 785 examples from 84 participants and an average tracking ratio of 92.87\%.  

\subsubsection{Gaze Synchrony Assessment}

We employed two established measures to assess gaze synchrony. First, we computed the KLD between gaze density maps \cite{rajashekar2004, tatler2005}. Gaze density maps per person and regarded video sequence were created by superposing Gaussian probability density functions on fixation counts and durations of the ten-second time windows before a thought probe. When applied to gaze density maps, the KLD measures the discrepancy between two distributions of gaze points, quantifying how much one density map diverges from another \cite{le2013methods}. This involves comparing the probability distribution of gaze points across the density map generated by one set of gaze data to the reference distribution from another set, thereby indicating the degree of similarity or difference in visual attention patterns between the two. For each point on the map, the KLD quantifies the difference by calculating the logarithm of the ratio of the gaze density at that point in the first map to the gaze density at the corresponding point in the second map, then weighting this by the gaze density of the first map, and summing these values across all points. Equal gaze distributions would result in a KLD of zero, while higher values would signify a more considerable dissimilarity and, consequently, less synchronous gaze. We argue that gaze density maps, only capturing the spatial distribution of the gaze, are suitable due to the relatively static nature of the video stimulus presenting PowerPoint slides.  

Additionally, to incorporate temporal synchrony dimensions, we employed the MultiMatch scanpath comparison algorithm \cite{jarodzka2010, dewhurst2012, foulsham2012}. MultiMatch consists of five separate measures to compare scanpaths, capturing a range of characteristics: shape, direction, length, position, and duration. The shape similarity is derived by vector differences between aligned pairs of saccades and averaged over the whole scanpath. The direction subdimension is computed by the angular differences between saccades, whereas length similarity is defined as the absolute difference in amplitude of aligned saccades. These measures are insensitive to fixation locations or durations. The position similarity is computed as the Euclidean distance between aligned fixations. The duration measure is defined by the absolute difference in fixation duration of aligned fixations and is insensitive toward fixation locations and saccade information. An overall similarity score can be computed by averaging all subdimension scores. We did not simplify the scanpaths, as small changes can already be meaningful in our video setting. We used the MultiMatch\_gaze python implementation\cite{wagner2019multimatch} to compute similarities. To obtain an aggregated scanpath similarity measure, all five sub-measures were averaged. 

For a baseline comparison, we calculated the ISC, previously utilized in research \cite{Madsen.2021, sauter2023, Liu.2023}. This involved computing correlations for vertical and horizontal gaze positions and pupil diameter, which were then aggregated into a single ISC metric.
All synchrony measures were computed probe-wise for all 15 10-second video sequences preceding the attention self-reports separately. 
The similarity score of each participant for a given sequence was computed by comparing their gaze data with that of all other participants in the same peer group, grouped as either attentive or inattentive based on their corresponding self-reports of attention. This comparison was conducted in a pairwise fashion, and the resultant similarity scores were averaged. To compare the synchrony between the two groups across the video sequences, the synchrony values per video sequence were z-standardized.

\subsubsection{Analysis}

\paragraph{Gaze Synchrony and Attention}
In the next step, we analyzed the relationship between attention self-reports and gaze synchrony. This analysis was conducted at the probe level, focusing on the relationship between each attention self-report and the gaze synchrony values computed for the 10 seconds immediately preceding the report. Since we had up to 15 probes and thus multiple measurement time points per person, we employed a multi-level analysis approach. Specifically, we utilized mixed linear regression with gaze synchrony as the dependent variable. Within this model, participant ID was treated as a random effect,  and self-reported attention was incorporated as a fixed effect. Furthermore, to account for potential variability across the video segments before each probe under analysis, we included the probe number as a categorical variable in the model. 

\paragraph{Predicting Learning Outcomes}
To explore the potential of using gaze synchrony measures as indicators for attention, we investigated the relationship between gaze synchrony measures and learning outcomes. To this end, we aggregated gaze synchrony scores to the participant level by averaging the obtained synchrony values over all considered time windows for each participant. Similarly, to aggregate the attention self-reports, we calculated the proportion of on-task self-reports from the total number of reports for each participant, effectively determining the share of self-reported attention. With these aggregated values, we then performed linear regression analyses to compare the relation of self-reported attention, KLD, and MultiMatch similarities to post-test scores, incorporating pre-test scores into all models to adjust for prior knowledge.

\subsection{Results}


\subsubsection{Gaze Synchrony and Attention}

We calculated ISC as a baseline synchrony measure. When z-standar-dizing the measure at the probe level, we found average ISC values of 0.104 ($\textit{SD} = 1.019$) for the attentive group and -0.069 ($\textit{SD} = 0.983$) for the inattentive group, indicating a slightly higher gaze synchronization for attentive learners. Results of Linear mixed-effects modeling of self-reported attention on ISC can be found in the first two columns of Table \ref{tab:syncreg}. Specifically, our results indicated that participants who self-reported as attentive demonstrated significantly higher inter-subject correlation of gaze position and pupil diameter ($\textit{Estimate} = 0.21$, $\textit{p} < 0.01$) compared to those who reported being inattentive.

\begin{table*}[h]
\centering
\caption{Linear mixed effects model of self-reported attention on Inter Subject Correlation (ISC), Kullback-Leibler divergence (KLD) and MultiMatch similarity (MM).}
\label{tab:syncreg}
\begin{adjustbox}{max width = 0.9\linewidth}
\begin{tabular}{lcccccccc}
\hline
                    & \multicolumn{2}{c}{\textbf{ISC}} & & \multicolumn{2}{c}{\textbf{KLD}}       &         & \multicolumn{2}{c}{\textbf{MM}} \\
Predictors          & \multicolumn{1}{c}{\textit{Estimates}} & \multicolumn{1}{c}{\textit{CI}}  && \multicolumn{1}{c}{\textit{Estimates}}               & \multicolumn{1}{c}{\textit{CI}}           & & \multicolumn{1}{c}{\textit{Estimates}}       & \multicolumn{1}{c}{\textit{CI}}            \\ \hline
(Intercept)         & -0.09              & -0.34 – 0.15 && 0.17                    & -0.07 – 0.41  && -0.21           & -0.45 – 0.03  \\
Attentive           & 0.21 **            & 0.06 – 0.37  && -0.36 ***               & -0.51 – -0.22 && 0.38 ***        & 0.23 – 0.53   \\
Stimulus 2          & -0.06              & -0.39 – 0.27 && 0.13                    & -0.17 – 0.43  && -0.08           & -0.39 – 0.22  \\
Stimulus 3          & -0.02              & -0.37 – 0.33 && 0.12                    & -0.20 – 0.43  && -0.04           & -0.36 – 0.28  \\
Stimulus 4          & 0.06               & -0.28 – 0.39 && -0.08                   & -0.38 – 0.21  && 0.10            & -0.20 – 0.41  \\
Stimulus 5          & 0.04               & -0.31 – 0.39 && -0.03                   & -0.35 – 0.29  && 0.09            & -0.24 – 0.41  \\
Stimulus 6          & -0.01              & -0.38 – 0.35 && 0.08                    & -0.25 – 0.40  && -0.02           & -0.36 – 0.31  \\
Stimulus 7          & 0.01               & -0.34 – 0.37 && 0.09                    & -0.23 – 0.41  && 0.01            & -0.32 – 0.34  \\
Stimulus 8          & 0.04               & -0.32 – 0.40 && 0.02                    & -0.30 – 0.35  && 0.04            & -0.30 – 0.37  \\
Stimulus 9          & -0.04              & -0.40 – 0.32 && 0.13                    & -0.19 – 0.45  && -0.09           & -0.42 – 0.24  \\
Stimulus 10         & 0.07               & -0.31 – 0.45 && -0.08                   & -0.42 – 0.26  && 0.09            & -0.26 – 0.44  \\
Stimulus 11         & 0.02               & -0.35 – 0.38 && 0.06                    & -0.27 – 0.39  && -0.01           & -0.35 – 0.33  \\
Stimulus 12         & 0.07               & -0.29 – 0.44 && -0.18                   & -0.51 – 0.15  && 0.19            & -0.15 – 0.54  \\
Stimulus 13         & -0.02              & -0.40 – 0.36 && 0.05                    & -0.29 – 0.39  && -0.04           & -0.39 – 0.31  \\
Stimulus 14         & -0.01              & -0.40 – 0.37 && 0.04                    & -0.30 – 0.39  && -0.02           & -0.38 – 0.34  \\
Stimulus 15         & 0.02               & -0.38 – 0.42 && -0.02                   & -0.38 – 0.34  && 0.01            & -0.36 – 0.38  \\
Random Effects      &                    &              &&                         &               &&                 &               \\
$\sigma^2$          & 1                  &              && 0.79                    &               && 0.84            &               \\
$\tau_{00id}$              & 0.02               &              && 0.21                    &               && 0.17            &               \\
ICC                 & 0.02               &              && 0.21                    &               && 0.17            &               \\
$N_{id}$                 & 84                 &              && 84                      &               && 84              &               \\ \hline
Observations        & 785                &              && 785                     &               && 785             &               \\
Marg. R2 / Cond. R2 & \multicolumn{2}{l}{0.009   / 0.025} & &  \multicolumn{2}{l}{0.028 / 0.234} && \multicolumn{2}{l}{0.029 / 0.196 } \\ \hline
\multicolumn{9}{r}{* p\textless{}0.05   ** p\textless{}0.01   *** p\textless{}0.001}                                               
\end{tabular}
\end{adjustbox}
\end{table*}

\paragraph{Kullback-Leibler Divergence}

\begin{figure}[]
  \centering
  \includegraphics[width=0.6\linewidth]{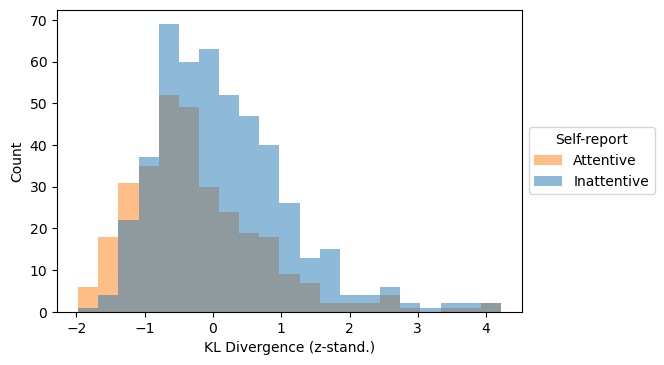}
  \caption{Kullback-Leibler divergence of gaze density maps by self-reported attention, z-standardized by video sequence.}
  \label{fig:kldiv}
\end{figure}

The analysis of gaze density maps, utilizing KLD to quantify synchrony in gaze patterns, was conducted probe-wise. Each video sequence, spanning 10 seconds and leading up to a self-report probe, was individually z-standardized to ensure comparability across different segments. The standardization process modifies the KLD scale, allowing us to interpret smaller or negative values as indicative of reduced differences between gaze maps, thereby signifying greater gaze synchrony. The distribution of KLD, divided into attentive ($\textit{M} = -0.219$, $\textit{SD} = 1.009$) and inattentive ($\textit{M} = 0.142$, $\textit{SD} = 0.968$) instances based on participant's self-reports are depicted in Figure \ref{fig:kldiv}. The visible delineation suggests that participants who self-reported as inattentive exhibited a marginally higher divergence in gaze patterns within their peer group, indicating a reduced level of gaze synchrony compared to their attentive counterparts. This higher divergence signifies smaller gaze synchrony within the distracted group. Although the trend is visible, the two distributions largely overlap, illustrating that the differences are not particularly large.

Further exploration through multi-level analysis, employing linear mixed effects models, accentuated these findings. Specifically, our results indicated that participants who self-reported as attentive demonstrated significantly lower gaze divergence ($\textit{Estimate} = -0.36$, $\textit{p} < 0.001$) compared to those who reported being inattentive. This statistical significance underscores a greater degree of gaze synchrony among participants who were on task and focused on the lecture, as per their self-reports, albeit by a small effect size. Such a relationship between self-reported attention states and gaze synchrony metrics, detailed in Table \ref{tab:syncreg}, provides empirical evidence supporting the assumption that attention levels, as self-reported by participants, are intricately associated with measurable gaze behaviors during video lecture viewing.

\begin{figure*}[h]
  \centering
  \includegraphics[width=\linewidth]{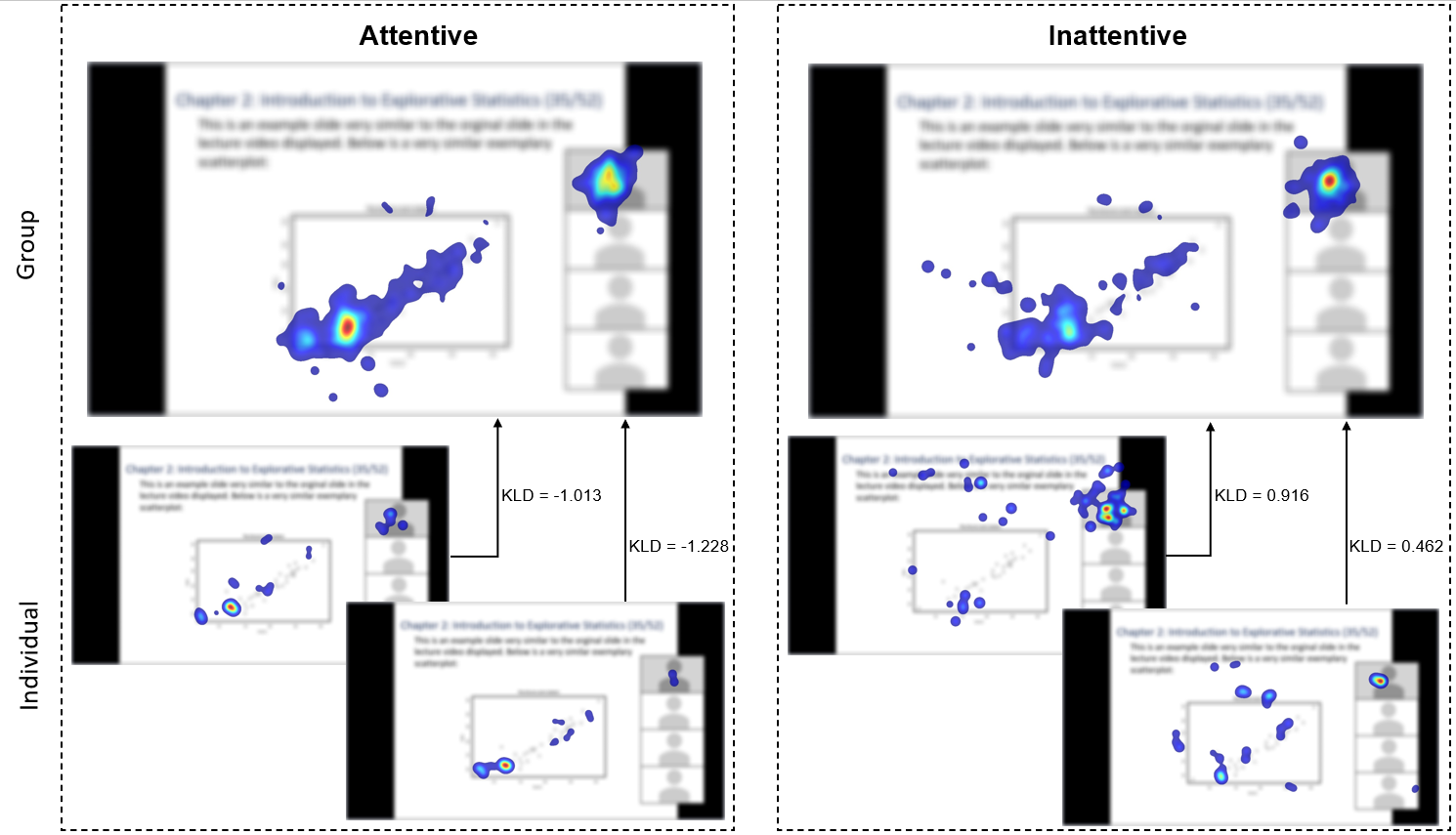}

  \caption{Example gaze density heatmap visualizations by attentive and inattentive self-reports in a video sequence. The top row shows the gaze density at the group level, while images on the bottom show examples of individual participants' gazes and corresponding average Kullback-Leibler Divergences (KLD).}
  \label{fig:hm}
\end{figure*}

Figure \ref{fig:hm} shows visualizations of gaze density maps for attentive and inattentive learners, according to their self-reports, for one video sequence of 10 seconds before a preceding probe. The group-level depiction of gaze density illustrates that the attentive group focused clearly on a specific point in the graphic on the slide shown. In contrast, the gaze of the distracted group tended to be scattered across the slide, and a high gaze density can only be seen directly by the lecturer. When looking at examples of single participants, attentive heatmaps appeared more similar to each other and to the group depiction, which is supported by the lower KLD values of $-1.013$ and $-1.228$. Contrarily, the inattentive participants' gaze patterns appeared less focused and more random, also reflected in higher KLD Values of $0.916$ and $0.462$. Additionally, a correlation analysis between the baseline ISC similarity scores and KLD distance values revealed a weak negative significant correlation ($\textit{r} = -0,12$).

\paragraph{Scanpath Comparisons with MultiMatch}

\begin{figure}[h]
  \centering
  \includegraphics[width=0.6\linewidth]{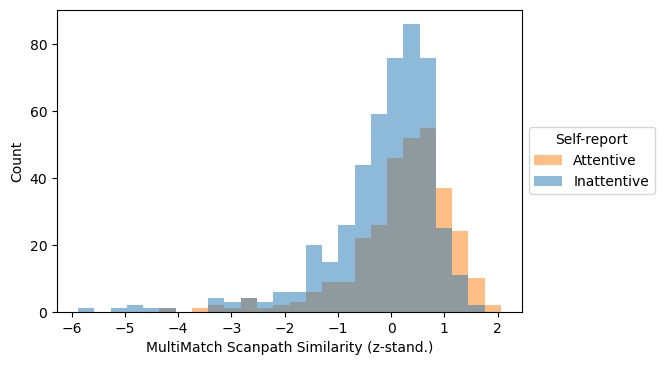}
  \caption{MultiMatch scanpath similarity scores by self-reported attention, z-standardized by video sequence.}

  \label{fig:mmsim}
\end{figure}

As a second measure of synchrony, we employed the MultiMatch method to assess scanpath similarities computed at the probe level for individual instances. 
The calculated similarities are presented in Figure \ref{fig:mmsim}, where we observed that attentive participants ($\textit{M} = 0.207$, $\textit{SD} = 0.946$) exhibited marginally more similar scanpaths compared to their inattentive counterparts ($\textit{M} = -0.137$, $\textit{SD} = 1.013$). This suggests a higher level of gaze synchronization among participants who reported being on task during the respective video lecture sequences.

To delve deeper into the differences between groups, we applied mixed linear models to analyze the data at the probe level. Table \ref{tab:syncreg} displays the models for MM and its subdimensions. This analysis revealed that attentive learners demonstrated higher MM scanpath similarities ($\textit{Estimate} = 0.38, \textit{p} < 0.001$), indicating a greater degree of gaze synchronization though the effect size suggests these differences, while statistically significant, are modest in magnitude. This finding was consistent across all subdimensions of the MM analysis, as displayed in Table \ref{tab:MMreg}, with the exception of saccade length similarity ($\textit{Estimate} = 0.02, \textit{p} = 0.825$), which did not show a significant difference between attentive and inattentive groups. Notably, the most pronounced effect of attention was observed in the similarity of gaze positions ($\textit{Estimate} = 0.42, \textit{p} < 0.001$), underscoring the impact of the attentional state on visual engagement with the video content.
When comparing KLD scores to our ISC baseline, we found a weak, significantly positive correlation ($\textit{r} = 0,17$). Further, we conducted a comparative analysis of MM similarities and KLD scores. This comparison revealed a modest yet significant correlation ($\textit{r} = -0.45$) between MultiMatch and KLD, suggesting that both metrics, although distinct in their computational approaches, provide complementary insights into the nature of gaze behavior and its association with attention.

\begin{figure*}[h]
  \centering
  \includegraphics[width=0.9\linewidth]{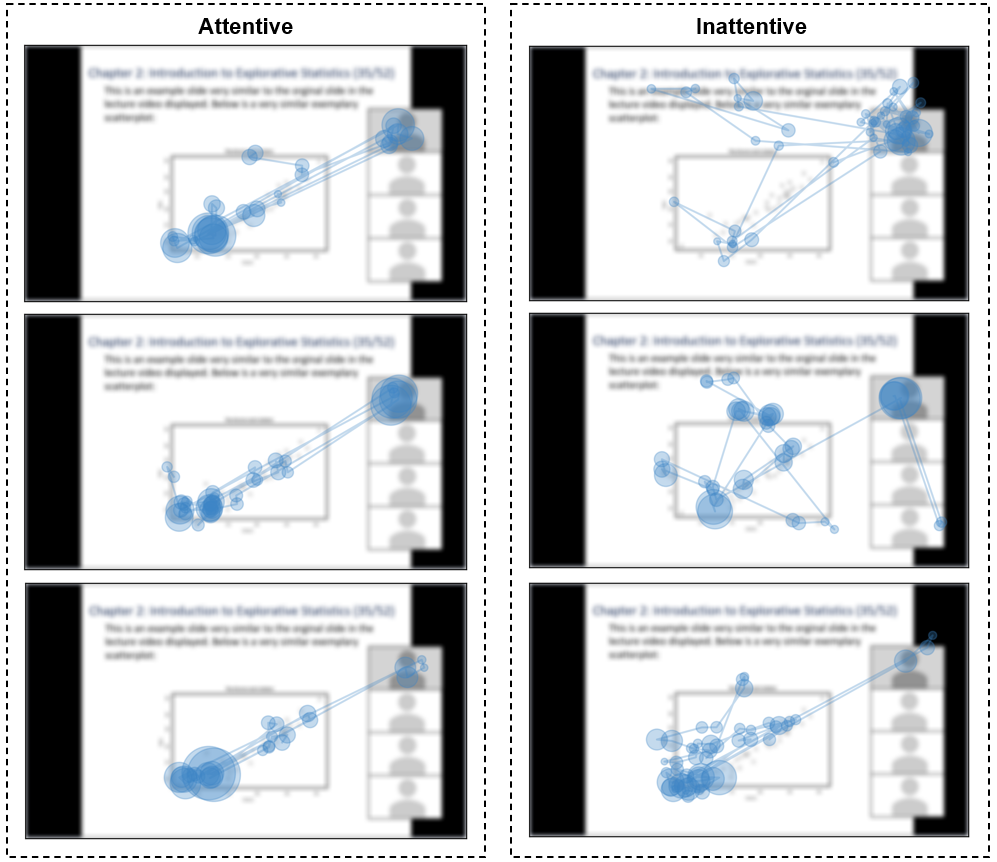}
  \caption{Example scannpath visualizations of one 10-second video sequence by self-reported attention.}
  \label{fig:sp}
\end{figure*}

A visualization of example scanpaths can be seen in Figure \ref{fig:sp}. The three participants depicted, attentive according to their self-reports, show very similar fixation patterns that move back and forth between the presenter and the very few specific relevant points on the slides. The scanpaths of distracted learners exhibit significantly greater diversity. In these cases, fixations are dispersed across a broader area of the slide and do not seem to follow a distinct pattern. While their gaze also briefly touches the relevant areas of the graphic, they occasionally fixate on empty areas on the slides.


\subsubsection{Predicting Learning Outcomes}

We explored the relationship between gaze synchrony as an indicator of attention and learning outcomes by aggregating self-reported attention and gaze synchrony metrics to the participant level and then conducting linear regression analysis on post-test scores. This approach facilitates a detailed exploration and comparison of the relationships between self-reported attention and gaze synchrony with learning outcomes, examining how each correlates with educational success over the entire session. 
The results of linear regression analysis are detailed in Table \ref{tab:regpt}. Our findings indicate a significant positive relationship between the overall proportion of time participants reported being on task and their post-test scores, even after adjusting for their prior knowledge of the session topic. This suggests that self-reported attention strongly predicts learning success, underscoring the importance of maintaining focus during educational sessions.
In contrast, we could not find a significant relationship between the computed gaze (a-)synchrony measures KLD and MultiMatch similarity and learning outcomes. However, in these models, previous knowledge becomes significant. This lack of significant correlation suggests that while these measures provide valuable insights into participants' engagement and attention alignment, they may not directly predict learning effectiveness as measured by post-test scores. Interestingly, in the models incorporating gaze synchrony metrics, prior knowledge emerged as a significant predictor. Models that do not account for previous knowledge are detailed in the appendix, specifically in Table \ref{tab:singlept_reg}.

\begin{table*}[h]
\caption{Linear regression of share of attentive self-reports, average Kullback-Leibler divergence (KLD), and average MultiMatch scanpath similarity (MM) on posttest scores, controlling for pretest scores.}
\label{tab:regpt}
\begin{adjustbox}{max width =\linewidth}
\begin{tabular}{lllllll}
\hline
                                                                      & \multicolumn{2}{l}{}              & \multicolumn{2}{c}{Post-Test Score} & \multicolumn{2}{l}{}              \\
Predictors                                                            & Estimates      & CI               & Estimates       & CI                & Estimates      & CI               \\ \hline
Intercept                                                             & $2.89^{\ast\ast\ast}$      & 1.63 – 4.15      & $5.16^{\ast\ast\ast}$       & 4.44 – 5.88       & $5.17^{\ast\ast\ast}$      & 4.46 – 5.88      \\
Pre-Test Score                      & 0.55           & -0.06 – 1.15     & $0.72^{\ast}$         & 0.05 – 1.39       & $0.70^{\ast}$        & 0.03 – 1.37      \\
Attentive Share & $0.05^{\ast\ast\ast}$     & 0.03 – 0.08      &                 &                   &                &                  \\
Average KLD                                                           &                &                  & 0.14            & -0.74 – 1.02      &                &                  \\
Average MM                                                            &                &                  &                 &                   & -0.17          & -1.06 – 0.72     \\ \hline
Observations                                                          & \multicolumn{2}{l}{84}            & \multicolumn{2}{l}{84}              & \multicolumn{2}{l}{84}            \\
R2 / R2 adjusted                                                      & \multicolumn{2}{l}{0.224 / 0.205} & \multicolumn{2}{l}{0.054 / 0.030}   & \multicolumn{2}{l}{0.054 / 0.031} \\ \hline
\multicolumn{7}{r}{ $^{\ast}$ p\textless{}0.05    $^{\ast\ast}$ p\textless{}0.01    $^{\ast\ast\ast}$ p\textless{}0.001}                                                                                               
\end{tabular}
\end{adjustbox}
\end{table*}

\subsection{Discussion}

Our study fills a crucial gap by exploring the relationship between gaze synchrony and self-reported attention during lecture video watching. We identified significant differences in gaze synchrony by self-reported attention, indicating higher synchronization when students report attentiveness. However, these differences were observed to be relatively small in magnitude. As a first study, this work establishes a connection between gaze synchrony and experience-sampled attention reports, reinforcing the hypothesis that gaze synchrony, beyond experimental conditions, is related to naturally occurring (in)attentiveness during video lecture viewing.

When comparing and contrasting the two measures employed to assess gaze synchrony, namely the Kullback-Leibler divergence and the MultiMatch Scanpath comparison, the small significant correlation reveals a common trend but shows that the two measures still depict distinct characteristics of the eye movements. While KLD focuses mainly on the spatial distribution of fixations, MultiMatch incorporates the temporal dimension by considering the sequence and a range of other multidimensional gaze properties, such as overall scanpath shape. Interestingly, the sub-dimension of MultiMatch that is most strongly linked to self-reported attention is the one that assesses how similar fixation positions are. This underscores the significance of where the eyes focus and the visual engagement with specific content.
On the contrary, the only subdimension that did not show a significant relation was saccade length similarity. This discrepancy may indicate that the amplitude of aligned eye movements does not exhibit increased synchronization in the same way as other aspects of gaze behavior when learners are attentively engaged in a task. Location may be more directly influenced by where attention is focused, reflecting the cognitive engagement with specific content areas.
Consequently, both measures appear to be suitable for assessing gaze synchrony in the setting of video lecture learning while providing complementary insights. When synchrony is assessed with these measures, it demonstrates a stronger connection to self-reported attention than the previously used ISC measure. This likely stems from their lower time sensitivity, which is more suitable for relatively static stimuli like lecture slides. The exemplary visualization of gaze density heatmaps and scanpaths shows that in many cases, the discrepancy in gaze movement patterns associated with different attentional states can be readily discerned through visual inspection, as described in previous studies \cite{Sauter.2022b}.

Our study did not replicate the previously suggested finding \cite{Madsen.2021, Liu.2023} that the average level of gaze synchrony significantly predicts post-test scores. Conversely, as anticipated, self-reported attention demonstrated a significant association with learning outcomes. A similar finding was reported by \cite{Sauter.2022}, who attributed the lack of association between gaze synchrony and test scores primarily to the poorer webcam-based eye-tracking sampling rates. However, these diverging findings may also partially be attributed to distinctions between the employed video stimuli. \cite{Madsen.2021} used short, informal instructional videos, i.e., animations, while \cite{Sauter.2022} and this study displayed more traditional lecture videos featuring slides and a lecturer. These differences in video content might influence the observed gaze synchrony patterns, highlighting the potential impact of video format on eye movement behaviors. The presence of a presenter in the video changes the gaze distribution and potentially how eye movements synchronize during attention. This suggestion is also supported by findings of \cite{Sauter.2022}, that the amount of gaze on the presenter interacted with ISC in the regression on post-test scores. Other studies revealed longer fixations on the instructor during mind-wandering episodes \cite{zhang2020}.

A limitation of this study is the observation of relatively short time windows for calculating gaze synchrony instead of the continuous observation over the entire lecture video. These brief intervals were specifically chosen to accommodate the inherently fluctuating nature of attention, recognizing that it can vary significantly within short periods. However, this methodological choice means that only a small proportion of the total gaze data is utilized for synchrony computation and, consequently, to predict learning outcomes. Observing eye movements over longer periods could increase the synchrony measures' robustness and potentially reveal a clearer relation to learning outcomes. 
Further, our study sample consists of university students within a narrow age range, potentially affecting the generalizability to other age groups such as school children. Moreover, the data is unbalanced in terms of gender, with a smaller proportion of male participants. Additionally, the chosen video lecture, which is not part of participants' regular study programs, may have contributed to lower intrinsic motivation, influencing their attention. While our findings provide important insights into gaze synchrony and attention, they may not directly apply to live online lectures, where participants might be visible through a webcam. Prior research has identified a negative correlation between the time spent actively looking at one's own and other students' webcam images and learning outcomes \cite{wisiecka2022}. This suggests that the mere visibility of these images could act as a distractor in live online educational contexts. Another limitation of the current study is that in our effort to increase ecological validity, we refrained from using chin rests. This decision, aimed at creating a more naturalistic setting for participants, might have affected the accuracy of our eye-tracking measurements.



The observations in this study underline the complexity of directly using synchrony as an attention indicator, suggesting that the findings are not as robust as previously thought. Future research should investigate how gaze synchrony is influenced by the educational video type, especially by the presence of a presenter. The choice of a synchrony measure in future research should be informed by the dynamic nature of the video or learning task at hand. Considering that KLD and MM appear to perform comparably for static, slide-based stimuli, their selection might hinge on the computational demands for real-time applications. KLD might be more computationally efficient, especially for group-based comparisons. Nonetheless, for dynamic stimuli or in scenarios involving longer time windows, MM might be preferable due to its capacity to incorporate the temporal dimension. Additionally, the moderate correlation observed between KLD and MM suggests they could be complementary. To leverage the unique insights provided by each, employing a hybrid approach, such as Normalized Scanpath Saliency, may offer significant benefits \cite{le2013methods, Dorr.2010}. Future research should systematically test and compare various synchrony measures and how they relate to video types to find a more precise measure of synchrony that can eventually serve as a robust attention indicator during online learning. 

This is particularly relevant for further research aimed at a better understanding of the learning process in online environments and, for example, improving the quality of learning materials. Moreover, this advancement carries significant implications for practical applications. Prior research proposed the potential of employing webcam-based eye tracking to enable real-time adaptability in online education based on attention \cite{Madsen.2021}. The finding that gaze synchrony correlates with momentary, naturally occurring attention presents the prospect that an overall degree of gaze synchrony could be a meaningful metric for lecturers and online educators, providing insights into the attentiveness of learners in live online settings. This could potentially guide instructional strategies to enhance student engagement and learning outcomes. However, ethical considerations take center stage when contemplating the application of eye tracking in real-world settings. Researchers must prioritize students' privacy \cite{suemer_2020_automated_anon} by implementing robust measures to secure and anonymize eye movement data~\cite{diff_privacy_eye_tracking_2021, brendan_streaming_et_data_tvcg_2021}, as using eye movement data, it is possible to infer various sensitive user attributes~\cite{Liebling_preibusch_2014, Kroeger_etal_2020}. Therefore, it is also essential to carefully consider issues of consent, data security, and the responsible use of technology. Furthermore, researchers should be mindful of the potential impact on learners, and steps should be taken to avoid potential biases that can disadvantage learner groups. 

\subsection{Conclusion}

In conclusion, this study investigated the relationship between gaze synchrony and self-reported attention in a realistic video lecture setting. While we found attentive participants exhibited higher synchronization of eye movements, our results did not show a significant association between gaze synchrony and learning outcomes. The findings underscore the complexity of using gaze synchrony as a reliable indicator of attention. Further research is required to explore the interplay between attention, gaze synchrony, and the educational video type to better understand their relationship.

\subsection*{Acknowledgements}

This research was supported by the LEAD Graduate School \& Research Network, which is funded by the Ministry of Science, Research and the Arts of the state of Baden- W{\"u}rttemberg within the framework of the sustainability funding for the projects of the Excellence Initiative II. Babette B{\"u}hler and Hannah Deininger are doctoral candidates supported by the LEAD Graduate School and Research Network. This research was also partly funded by the Deutsche Forschungsgemeinschaft (DFG, German Research Foundation) under Germany's Excellence Strategy - EXC number 2064/1 - Project number 390727645.

\subsection*{Appendices}

\begin{figure}[h]
  \centering
  \includegraphics[width=0.5\linewidth]{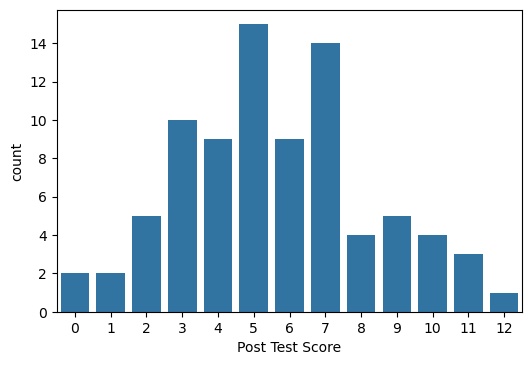}
  \caption{Post test scores.}
  \label{fig:posttest}
\end{figure}

\begin{table*}[]
\caption{Linear mixed effects models of self-reported attention on MultiMatch scanpath similarity (MM) sub-dimensions.}
\label{tab:MMreg}
\begin{adjustbox}{max width = \linewidth}
\begin{tabular}{lcccccccccc}
\cline{1-11}
                                                                                     & \multicolumn{2}{c}{\textbf{MM Shape}}      & \multicolumn{2}{c}{\textbf{MM Length}}     & \multicolumn{2}{c}{\textbf{MM Direction}}  & \multicolumn{2}{c}{\textbf{MM Position}}   & \multicolumn{2}{c}{\textbf{MM Duration}}    \\
Predictors                                                                           & \textit{Estimates}      \textit{CI}               & \textit{Estimates}      & \textit{CI}               & \textit{Estimates}      & \textit{CI}               & \textit{Estimates}      & \textit{CI}               & \textit{Estimates}      & \textit{CI}                 \\ \hline
(Intercept)                                                                          & -0.28 *        & -0.54 – -0.02    & -0.03          & -0.27 – 0.21     & -0.30 *        & -0.57 – -0.04    & -0.20          & -0.44 – 0.04     & -0.16          & -0.40 – 0.08       \\
Attentive                                                                            & 0.30 ***       & 0.17 – 0.42      & 0.02           & -0.14 – 0.17     & 0.29 ***       & 0.17 – 0.42      & 0.42 ***       & 0.27 – 0.58      & 0.30 ***       & 0.15 – 0.45        \\
Stimulus 2                                                                           & -0.05          & -0.30 – 0.19     & 0.01           & -0.31 – 0.33     & -0.05          & -0.30 – 0.20     & -0.12          & -0.44 – 0.21     & -0.06          & -0.37 – 0.25       \\
Stimulus 3                                                                           & -0.02          & -0.27 – 0.24     & 0.01           & -0.32 – 0.35     & -0.02          & -0.29 – 0.24     & -0.05          & -0.39 – 0.28     & -0.03          & -0.35 – 0.30       \\
Stimulus 4                                                                           & 0.05           & -0.20 – 0.29     & 0.02           & -0.30 – 0.34     & 0.06           & -0.20 – 0.31     & 0.12           & -0.21 – 0.44     & 0.09           & -0.23 – 0.40       \\
Stimulus 5                                                                           & 0.06           & -0.20 – 0.32     & 0.04           & -0.30 – 0.38     & 0.06           & -0.21 – 0.33     & 0.08           & -0.27 – 0.42     & 0.07           & -0.27 – 0.40       \\
Stimulus 6                                                                           & -0.04          & -0.30 – 0.23     & 0.01           & -0.34 – 0.36     & -0.03          & -0.30 – 0.25     & -0.03          & -0.38 – 0.32     & -0.01          & -0.35 – 0.33       \\
Stimulus 7                                                                           & -0.04          & -0.30 – 0.23     & 0.01           & -0.33 – 0.36     & -0.02          & -0.29 – 0.25     & -0.00          & -0.35 – 0.35     & 0.03           & -0.30 – 0.37       \\
Stimulus 8                                                                           & -0.01          & -0.28 – 0.25     & 0.00           & -0.34 – 0.35     & -0.01          & -0.29 – 0.26     & 0.06           & -0.29 – 0.41     & 0.04           & -0.30 – 0.38       \\
Stimulus 9                                                                           & -0.02          & -0.29 – 0.24     & -0.01          & -0.36 – 0.33     & -0.03          & -0.30 – 0.25     & -0.07          & -0.42 – 0.27     & -0.07          & -0.41 – 0.27       \\
Stimulus 10                                                                          & 0.17           & -0.11 – 0.45     & -0.02          & -0.38 – 0.34     & 0.16           & -0.13 – 0.45     & 0.14           & -0.23 – 0.50     & 0.07           & -0.29 – 0.42       \\
Stimulus 11                                                                          & 0.05           & -0.22 – 0.33     & -0.02          & -0.37 – 0.33     & 0.05           & -0.23 – 0.33     & 0.02           & -0.34 – 0.37     & 0.00           & -0.34 – 0.35       \\
Stimulus 12                                                                          & 0.26           & -0.02 – 0.53     & 0.03           & -0.33 – 0.38     & 0.25           & -0.03 – 0.53     & 0.18           & -0.18 – 0.54     & 0.13           & -0.22 – 0.48       \\
Stimulus 13                                                                          & 0.13           & -0.15 – 0.41     & -0.04          & -0.40 – 0.33     & 0.11           & -0.18 – 0.40     & -0.05          & -0.41 – 0.32     & -0.04          & -0.40 – 0.32       \\
Stimulus 14                                                                          & 0.02           & -0.26 – 0.31     & 0.02           & -0.35 – 0.39     & 0.04           & -0.25 – 0.33     & -0.02          & -0.39 – 0.35     & -0.02          & -0.38 – 0.34       \\
Stimulus 15                                                                          & 0.09           & -0.21 – 0.38     & 0.00           & -0.38 – 0.39     & 0.08           & -0.22 – 0.39     & 0.03           & -0.36 – 0.41     & 0.00           & -0.38 – 0.38       \\
Random Effects                                                                       &                &                  &                &                  &                &                  &                &                  &                &                    \\
$\sigma^2$                                                                                     & \multicolumn{2}{l}{0.52}          & \multicolumn{2}{l}{0.90}          & \multicolumn{2}{l}{0.56}          & \multicolumn{2}{l}{0.93}          & \multicolumn{2}{l}{0.87}            \\
$\tau_{00id}$                                                                                  & \multicolumn{2}{l}{0.77}       & \multicolumn{2}{l}{0.12}       & \multicolumn{2}{l}{0.78}       & \multicolumn{2}{l}{0.05}       & \multicolumn{2}{l}{0.14}         \\
ICC                                                                                  & \multicolumn{2}{l}{0.60}          & \multicolumn{2}{l}{0.12}          & \multicolumn{2}{l}{0.58}          & \multicolumn{2}{l}{0.05}          & \multicolumn{2}{l}{0.14}            \\
$N_{id}$                                                                                    & \multicolumn{2}{l}{84}         & \multicolumn{2}{l}{84}         & \multicolumn{2}{l}{84}         & \multicolumn{2}{l}{84}         & \multicolumn{2}{l}{84}           \\ \hline
Observations                                                                         & \multicolumn{2}{l}{785}           & \multicolumn{2}{l}{785}           & \multicolumn{2}{l}{785}           & \multicolumn{2}{l}{785}           & \multicolumn{2}{l}{785}             \\
Marg. R2 / Cond. R2                                                         & \multicolumn{2}{l}{0.017 / 0.603} & \multicolumn{2}{l}{0.000 / 0.122} & \multicolumn{2}{l}{0.015 / 0.589} & \multicolumn{2}{l}{0.036 / 0.087} & \multicolumn{2}{l}{0.018 / 0.153}   \\ \hline \multicolumn{11}{r}{* p\textless{}0.05   ** p\textless{}0.01   *** p\textless{}0.001} 
\end{tabular}
\end{adjustbox}
\end{table*}

\begin{table*}[h]
\caption{Linear regression of pre-test Score, attentive self-report share, average Kullback-Leibler divergence (KLD) and average MultiMatch scanpath similarity (MM) on posttest scores.}
\label{tab:singlept_reg}
\begin{adjustbox}{max width = \linewidth}
\begin{tabular}{lllllllll}
\hline
                            & \multicolumn{8}{c}{\textbf{Post-Test Score} }                                                                                                                                                                                   \\
Predictors                  & Estimates & CI                                 & Estimates & CI                                  & Estimates & CI                                  & Estimates & CI                                    \\ \hline
(Intercept)                 & $5.18^{\ast\ast\ast}$    & 4.48 – 5.89 &  $3.10^{\ast\ast\ast}$      & 1.85 – 4.36 &  $5.62^{\ast\ast\ast}$       & 5.03 – 6.21   & $5.60^{\ast\ast\ast}$      & 5.01 – 6.20  \\
Pre-Test Score              & $0.71^{\ast}$      & 0.04 – 1.37        &           &                                                  &              &                                     &              &                           \\
Attentive Self-Report Share &           &                                      & $0.06^{\ast\ast\ast}$      & 0.03 – 0.08 &           &                                        &           &                                         \\
Average KLD                 &           &                                       &           &                                       & 0.05      & -0.84 – 0.95                     &           &                                        \\
Average MM Similarity       &           &                                       &           &                                      &           &                                        & -0.22     & -1.13 – 0.69                     \\ \hline
Observations                & \multicolumn{2}{l}{84}                              & \multicolumn{2}{l}{84}                              & \multicolumn{2}{l}{84}                               & \multicolumn{2}{l}{84}                               \\
R2 / R2 adjusted            & \multicolumn{2}{l}{0.053 / 0.041}                   & \multicolumn{2}{l}{0.193 / 0.183}                   & \multicolumn{2}{l}{0.000 / -0.012}                   & \multicolumn{2}{l}{0.003 / -0.009}                   \\ \hline
\multicolumn{9}{r}{$^{\ast}$$ p\textless{}0.05    $$^{\ast\ast}$ p\textless{}0.01   $^{\ast\ast\ast}$ p\textless{}0.001}   
\end{tabular}
\end{adjustbox}
\end{table*}

%% file: 5_Hand-raising.tex
\section[Automated hand-raising detection in classroom videos]{Automated hand-raising detection in classroom videos: A view-invariant and occlusion-robust machine learning approach}

\subsection{Abstract}
Hand-raising signals students’ willingness to participate actively in the classroom discourse. It has been linked to academic achievement and cognitive engagement of students and constitutes an observable indicator of behavioral engagement. However, due to the large amount of effort involved in manual hand-raising annotation by human observers, research on this phenomenon, enabling teachers to understand and foster active classroom participation, is still scarce. An automated detection approach of hand-raising events in classroom videos can offer a time- and cost-effective substitute for manual coding. From a technical perspective, the main challenges for automated detection in the classroom setting are diverse camera angles and student occlusions. In this work, we propose utilizing and further extending a novel view-invariant, occlusion-robust machine learning approach with long short-term memory networks for hand-raising detection in classroom videos based on body pose estimation. We employed a dataset stemming from 36 real-world classroom videos, capturing 127 students from grades 5 to 12 and 2442 manually annotated authentic hand-raising events. Our temporal model trained on body pose embeddings achieved an $F_{1}$ score of 0.76. When employing this approach for the automated annotation of hand-raising instances, a mean absolute error of 3.76 for the number of detected hand-raisings per student, per lesson was achieved. We demonstrate its application by investigating the relationship between hand-raising events and self-reported cognitive engagement, situational interest, and involvement using manually annotated and automatically detected hand-raising instances. Furthermore, we discuss the potential of our approach to enable future large-scale research on student participation, as well as privacy-preserving data collection in the classroom context.

\subsection{Introduction}

Students’ active participation in classroom discourse contributes to academic achievement in the school context \cite{sedova2019}. To contribute verbally to the classroom discourse, students are required to raise their hands. Therefore, hand-raising in the classroom is an indicator of active participation and behavioral engagement, which is associated with achievement, cognitive engagement, perceived teacher emotional support \cite{boheim2020engagement}, and motivation \cite{Boheim.2020}. Further, a significant variation in the hand-raising frequency of eighth graders was due to situational interest in language art classes, and  the self-concept in maths classes \cite{Boheim.2020}. Results of such pioneering studies show the importance of hand-raising research to enable educators to understand and foster student engagement and active classroom participation, as well as the potential of employing hand-raising as an objective, low-inference behavioral engagement indicator.
To study active participation, human observers often rate student behavior manually, which is time- and cost-intensive. Crowdsourcing strategies are often not applicable due to data protection regulations. This is part of the reason why studies of hand-raising have small sample sizes limited to certain grades, age groups, and school subjects \cite{Boheim.2020,boheim2020engagement}, resulting in a lack of generalizability of results. Advances in computer vision and machine learning offer alternatives through automated recognition. This study aims to develop a robust approach to detect hand-raising events in classroom videos in an automated fashion and thus, develop a time- and cost-effective hand-raising assessment tool, replacing manual annotations to enable future large-scale research. 

Previous research tackled automated hand-raising detection by either aiming to detect image patches of raised hands \cite{Si.2019} or employing body pose estimation \cite{YuTe.2019}. This research mostly focused on the real-time assessment of hand-raisings on the classroom level, i.e., as part of classroom monitoring systems \cite{Ahuja.2019}. However, research on hand-raising and its role in individual students' learning processes is still scarce. Further, previous approaches, mapping hand-raisings to individual students, are often not evaluated on real-world classroom videos containing authentic hand-raisings. Therefore, we propose a hand-raising detection approach, built and evaluated on diverse real-world classroom data, to identify individual students' hand-raisings for post hoc analysis in education research. To investigate the relation between hand-raising and learning activities, (1) we conduct a correlation analysis of manually annotated hand-raisings to cognitive engagement, involvement, and situational interest reported after each lesson. For enabling such research with automated action recognition in the classroom, one of the biggest challenges is that students are often filmed from diverse angles and might be partially occluded by classmates or learning materials. Therefore, (2) we propose a novel hand-raising gesture recognition approach based on view-invariant and occlusion-robust embeddings of body pose estimations and temporal classification. Since we are not directly working on the image stream, this approach allows student privacy to be protected by directly extracting poses in real-time and eliminating the need to store sensitive video data. Since, in addition to recognizing the gesture itself, identifying who and how often someone raised their hand is of particular interest for education research, (3) we apply and evaluate our classification approach for the automated annotation of hand-raising instances for individual students. We then conduct correlation analysis to learning-related activities comparing manually and automatically annotated hand-raisings.

\subsection{Related Work}

Initial work addressing the automatic recognition of hand-raising gestures formulated it as an object detection task, localizing raised hands and arms frame-by-frame. It investigated hand-raising in simple and staged settings with few people in the frame, focusing on techniques such as temporal and spatial segmentation, skin color identification, shape and edge feature analysis~\cite{JieYao.2002} or the geometric structure of edges~\cite{Bo.2011}. 
 
However, such methods reach their limits when applied in a real classroom where a large number of students are visible, they occlude each other, and image resolution becomes lower.

Therefore, other works~\cite{Si.2019,TaoLiu.2020,nguyen2022new} aimed to overcome challenges of various gestures and low resolution, by introducing architectural adaptions of deep learning models. They achieved reliable results for detecting raised-hand image patches in realistic classroom scenes, with mean average precision (\gls{map}) ranging between 85.2\% \cite{nguyen2022new} and 91.4\% \cite{TaoLiu.2020}.

Object detection approaches are useful for measuring hand-raising rates at the class level, but it is challenging to analyze individual student behavior because raised hands cannot be easily attributed to specific students. To this end, a two-step approach combining object detection and pose estimation was performed by \cite{HuayiZhou.2018}, heuristically matching the detected hand bounding box to a student based on body keypoints. On six real-world classroom videos, 83.1\% detection accuracy was achieved.
\cite{Liao.2019} chose the reverse approach, using pose estimation to obtain arm candidate areas and then classifying the corresponding image patch, achieving a $F_{1}$-score of 0.922 on a test video of college students. 

Another strand of work directly employed pose estimation algorithms to detect hand-raising in the classroom. A classroom monitoring system by \cite{Ahuja.2019} used OpenPose~\cite{cao2019openpose} to extract eight upper body keypoints, for which direction unit vectors and distance between points were computed as geometric features. On scripted classroom scenes, a hand-raising prediction accuracy of 94.6\% with a multi-layer perceptron was reported. The evaluation on real-world videos, only containing six hand-raising instances, yielded a recall of 50\%. 
Furthermore, a classroom atmosphere management system for physical and online settings implemented rule-based hand-raising detection employing Kinect pose estimations \cite{YuTe.2019}. The detection accuracy was however not evaluated on real-world classroom videos. Likewise, geometric features (i.e., normalized coordinates, joint distances, and bone angles) extracted from pose estimations were used by \cite{lin2021student} for student behavior classification. Their approach to detecting four behaviors was evaluated on six staged classroom videos with a precision of 77.9\% in crowded scenes.

As described above, utilizing pose estimation offers the advantage that hand-raisings can be attributed to a specific student, which is important when studying participation on the student level. One common limitation of previous research relying on pose estimation is that either the performance of models has only been extensively evaluated on scripted videos and small real-world datasets \cite{Ahuja.2019,lin2021student}, or it implemented rule-based detection that was not evaluated on classroom videos at all \cite{YuTe.2019}. How these approaches perform in a real-world classroom scenario, where hand-raisings are likely to be expressed with various, sometimes subtle gestures, is unknown. 
 In conclusion, automated annotation of hand-raisings needs to be student specific and methodologically robust for real-world classroom scenarios. Therefore, we propose a recognition approach based on pose estimation trained and tested on a challenging dataset containing real-world school lessons with realistic hand-raising behaviors, representing a wide range of age groups (grades 5 to 12) and school subjects. Furthermore, our approach copes with the main technical challenges of such videos, such as camera angles and occlusions of students. To tackle those, we adapted and extended state-of-the-art view-invariant, occlusion-robust pose embeddings. Going beyond related works which detected hand-raising in a frame-by-frame fashion, we developed a classification model which integrates temporal information and is thus able to capture the large variety of hand-raisings expressed in their dynamic process.

\subsection{Methodology}
\subsubsection{Data}
\paragraph{Data Collection} 
The classroom videos utilized in the study were recorded in real-world lessons at a German secondary school, approved by the ethics committee of the University of Tübingen (Approval \verb|#|A2.5.4-097\_aa).
A total of 127 students, 56.3\% male, from grades 5 to 12 were videotaped during 36 lessons across a wide variety of subjects (see Table \ref{tab:overview_videos}). 
All recordings were captured by cameras (24 frames per second) mounted in the front left or right corner of the classroom. After each lesson, students completed a questionnaire on learning activities in the lesson, including self-reported involvement \cite{frank2014presence}, cognitive engagement \cite{rimm2015extent}, and situational interest \cite{knogler2015situational}. The employed scales have been shown to be related to engagement observer ratings and learning outcomes~\cite{goldberg2021}. This resulted in video and questionnaire data for 323 student-lesson instances. Due to missing questionnaire information (i.e., item non-response), 18 instances had to be excluded for correlation analysis with learning-related activities.

\paragraph{Manual Annotation}
\label{sec:annot}
Two human raters manually annotated hand-raisings in all 36 videos. The intra-class correlation coefficient of the two raters for the number of hand-raisings per student and lesson was 0.96, indicating very high inter-rater reliability. The number of hand-raisings averaged across the two observers was employed to analyze associations between student hand-raising and self-reported learning activities. A total of 2442 hand-raising events were annotated. Summary statistics on hand-raisings by students and lessons across grades are shown in Table \ref{tab:overview_videos}. On average, one student raised their hand 7.4 times per lesson, while an average total amount of 64.6 hand-raisings occurred per lesson.

Half of the data was used to build the automated hand-raising detection model, requiring more fine-granular annotations. To this end, for 18 videos, hand-raising was annotated in a spatio-temporal manner using the VIA software~\cite{dutta2019via}, including the start and end time of hand-raisings as well as bounding boxes, with a joint agreement probability of 83.36\%. To increase reliability, we combined the two observers' annotations by temporal intersection, 
resulting in 1584 hand-raising instances with an average duration of 15.6 seconds. 
The remaining 18 videos, only coded with regard to the number of hand-raisings of each student, in addition to four videos employed as a test set for the hand-raising classifier, then served to validate our developed model for automated annotation.

\begin{table} []
\centering
\caption{Summary statistics of hand-raisings (N = 305).}
\label{tab:overview_videos}
\resizebox{\textwidth}{!}{\begin{tabular}{lllllllllllll}
\toprule
\multirow{2}{*}{Grade} & \multirow{2}{*}{Subjects} & \multirow{2}{*}{Lessons} & \multicolumn{5}{l}{Hand-raisings per student} & \multicolumn{5}{l}{Hand-raisings per lessons} \\ \cline{4-13} 
                       &                           &                                    & M         & SD        & Md    & Min   & Max   & M         & SD         & Md   & Min    & Max  \\ \midrule
5                      &    B      & 2                                  & 15.093    & 6.091     & 15    & 5     & 26    & 203.750   & 6.010   & 204  & 199.5  & 208  \\
6                      &   B, L               & 3                                  & 8.625     & 7.934     & 6     & 0     & 33    & 126.500   & 89.705  & 85   & 66     & 230  \\
7                      &   E, I    & 2                                  & 9.420     & 12.991    & 5     & 0     & 60    & 117.750   & 64.700  & 118  & 72     & 164  \\
8                      & E, F, G, H, IMP, L  & 8                                  & 5.079     & 4.860     & 4     & 0     & 23    & 44.438    & 20.711  & 40   & 8      & 74   \\
9                      &  E, G  & 2                                  & 6.053     & 5.550     & 5     & 0     & 24    & 57.500    & 23.335  & 58   & 41     & 74   \\
10                     &  G, P.    & 3                                  & 8.912     & 7.997     & 5     & 0     & 26    & 50.500    & 14.309  & 58   & 34     & 60   \\
11                     &  A, ET, G, P, PS & 6                                  & 7.813     & 7.203     & 6     & 0     & 24    & 52.083    & 17.019  & 53   & 33     & 83   \\
12                     & C, H, M, P & 10                                 & 5.169     & 5.061     & 4     & 0     & 31    & 36.700    & 22.671  & 40   & 7      & 89   \\ \hline
Total                  &                           & 36                                 & 7.425     & 7.451     & 5     & 0     & 60    & 64.556    & 53.155     & 48   & 7      & 230  \\ \hline
\end{tabular}}
\scriptsize{\textit{Subject abbreviations:} B Biology, L Latin, E English, I Informatics, F French, G German, H History, IMP Informatics Math \& Physics, P Physics, A Arts, ET Ethics, PS Psychology, C Chemistry, M Math}
\end{table}

\subsubsection{Skeleton-based Hand-raising Detection}
\label{sec:hand-raising_detection}
\begin{figure}[h]
\centering
\includegraphics[width=\textwidth]{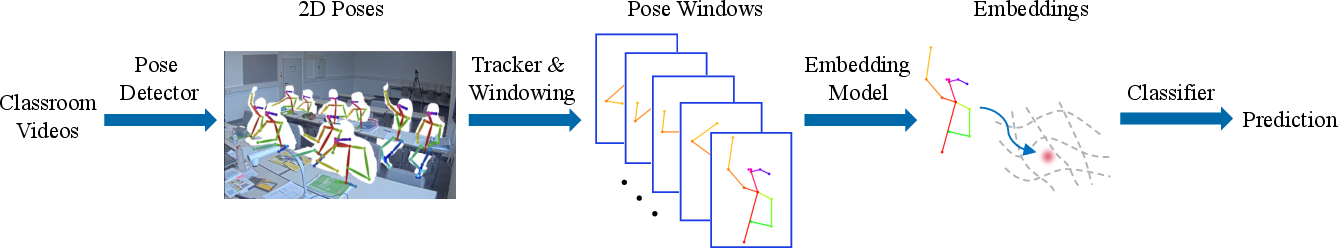}
\caption{Hand-raising detection pipeline.} \label{fig:pipeline}
\end{figure}

This section presents our machine learning-based approach to automated hand-raising gesture detection. Fig.~\ref{fig:pipeline} depicts the processing pipeline: generating sequential poses, extracting embeddings, and performing binary classification.

\paragraph{Preprocessing}
To extract 2D human poses, we used the OpenPose library~\cite{cao2019openpose}, estimating 25 body keypoints per student in each frame. Since students' lower body parts are often invisible due to occlusion, we focused on the 13 upper body keypoints representing the head, torso, and arms. 
To generate one skeleton tracklet for each student over time, we implemented an intersection-over-union tracker. 
Then, we used the 18 videos with spatio-temporal annotations (Sect.~\ref{sec:annot}) to create a dataset for classifier development.
Based on the annotations, tracklets were divided into subsequences labeled as "hand-raising" and "non-hand-raising" and then split into 2-second windows (48 frames) without overlap. This resulted in a highly imbalanced large-scale dataset of 243,069 instances, of which 12,839 (ca. 5\%) represented hand-raisings. 

\paragraph{Pose Embeddings}
To handle issues of viewpoint change and partial visibility, we adopted a recent approach by Liu et al.~\cite{liu2022prvipe}, attempting to extract pose embeddings. First, the embedding space is view-invariant, where 2D poses projected from different views of the same 3D poses are embedded close and those from different 3D poses are pushed apart. Second, occlusion robustness is achieved by using synthetic keypoint occlusion patterns to generate partially visible poses for training. Third, the pose embeddings are probabilistic, composed of a mean and a variance vector, defining thus a Gaussian distribution that takes ambiguities of 2D poses during projection from 3D space into consideration.
We tailored the training procedure to our classroom setup, training the embedding model from scratch on Human3.6M~\cite{h36m_pami}, a human pose dataset containing a large number of recordings from four camera views. We executed OpenPose on all images and neglected lower body keypoints to focus on learning similarity in the upper body poses. We employed the neck keypoint as the origin coordinate and normalized the skeletons by the neck-hip distance.
To find a trade-off between performance and computational cost, the embedding dimension was selected as 32. This training process was independent of the classroom data, ensuring the generalization capability of our approach. 
To leverage the probabilistic embeddings as input to a downstream classifier, we concatenated the two embedding outputs (mean and variance) into a 64D feature vector. Moreover, we extracted the following two types of geometric features as baselines to compare their recognition performance against pose embeddings:
First, Lin et al.~\cite{lin2021student} distinguished between four student behaviors, by using eight keypoints and designing a 26D feature vector that concatenates normalized keypoint coordinates, keypoint distances, and bone angles. 
Second, Zhang et al.~\cite{zhang2017geometric} proposed using joint-line distances for skeleton-based action recognition.
We computed the corresponding distances on the basis of the 13 upper body keypoints, resulting in a 297D feature vector.

\paragraph{Classification}
To achieve binary classification based on sequential inputs, we employed long short-term memory (LSTM) models.
Similar to~\cite{zhang2017geometric}, we constructed three LSTM layers with 128 hidden units, followed by a fully connected layer applying a sigmoid activation function. Thus, the multi-layer model directly takes a 2-second sequence of frame-wise feature vectors as input, encodes temporal information, and estimates hand-raising probability. 
We trained the model using an empirically derived batch size of 512, binary cross-entropy loss, and Adam optimizer
with a fixed learning rate of 0.001. To avoid overfitting, we held out 10\% of training examples as validation data and set up an early stopping callback with the patience of 10 epochs according to validation loss.
Additionally, we trained non-temporal random forest (RF)
models for baseline comparisons. For model input, we generated an aggregated feature vector for each time window by calculating and concatenating the mean and standard deviation of all features.
To handle imbalanced data, we compared the performance with and without class weighting while training both models. Class weights were set inversely proportional to class frequencies in the training set.
We evaluated model performance in a video-independent manner, using 14 videos for training and 4 videos for testing. Our videos, stemming from un-staged classroom lessons, contain highly varying numbers of hand-raisings, making video-independent cross-validation infeasible due to unequal class distributions in each fold. 

\subsubsection{Automated Hand-raising Annotation}
\label{sec:automated_annotation}
Afterwards, we applied this hand-raising detection technique to estimate the number of hand-raisings of a student in class. To take full advantage of our data, we utilized 22 videos to evaluate the automated annotation performance, consisting of the second half of our videos and the four videos used to test classifiers. 
Following the pipeline in Fig.~\ref{fig:pipeline}, we generated one tracklet for each student in each video. To achieve a more robust temporal detection, a tracklet was sliced into 48-frame sliding windows with a stride of 8 frames. Then, we extracted pose embeddings and applied the trained classifier to estimate the hand-raising probability for each window. When the average probability exceeded 0.5, a frame was assigned to hand-raising, and consecutive frames were combined into one hand-raising instance.
We merged any two adjacent hand-raisings with an interval of fewer than 4 seconds and discarded those with a duration of less than 1 second, which only occur in less than 5\% of the cases respectively according to the annotations of the training videos. 
This resulted in the number of estimated hand-raisings for each student in each video.

\subsection{Results}

\subsubsection{Relation between hand-raising and self-reported learning activities} 

To demonstrate the importance of hand-raising analysis in classroom research, we examined the association between manually annotated hand-raisings and self-reported learning activities. The number of hand-raising instances of a student per lesson, annotated by human raters, is significantly positively correlated to self-reports of cognitive engagement ($r = 0.288$, $p < 0.001$), situational interest ($r = 0.379$, $p < 0.001$), and involvement ($r = 0.295$, $p < 0.001$) of students.


\subsubsection{Hand-raising Classification}
To assess the performance of hand-raising classifiers, we utilized the $F_{1}$-score which is the harmonic mean of precision (i.e., the proportion of correct hand-raising predictions of all hand-raising predictions) and recall (i.e., the share of correctly predicted hand-raising instances of all hand-raising instances). 
We first tested different classification models with and without class weighting, using pose embeddings as input features. The results are shown in the upper part of Table~\ref{tab:classification_results}. The greater penalty for misclassifying any hand-raising example when using weighting results in fewer false negatives and more false positives, i.e., a higher recall but lower precision.
The best performance was obtained by the temporal LSTM model without class weighting, achieving a $F_{1}$-score of 0.76.


In the second step, we compared three pose representations, namely pose embeddings and two types of geometric features, employing LSTM classifiers. As shown in the lower part of Table~\ref{tab:classification_results}, the pose embeddings generally yield better performance than the geometric features with respect to all three metrics, revealing the effectiveness of the data-driven approach. Notably, the pose embeddings achieve a noticeable increase in $F_{1}$-score over the features used in~\cite{lin2021student}. Moreover, despite the trivial improvement over the features used in~\cite{zhang2017geometric}, the pose embeddings benefit from less computational effort for inference, in comparison to the distance calculation over 297 joint-line combinations.

\begin{figure}[] 
    \centering
    \begin{subfigure}[t]{0.24\textwidth}
        \centering
        \includegraphics[width = 0.4\textwidth]{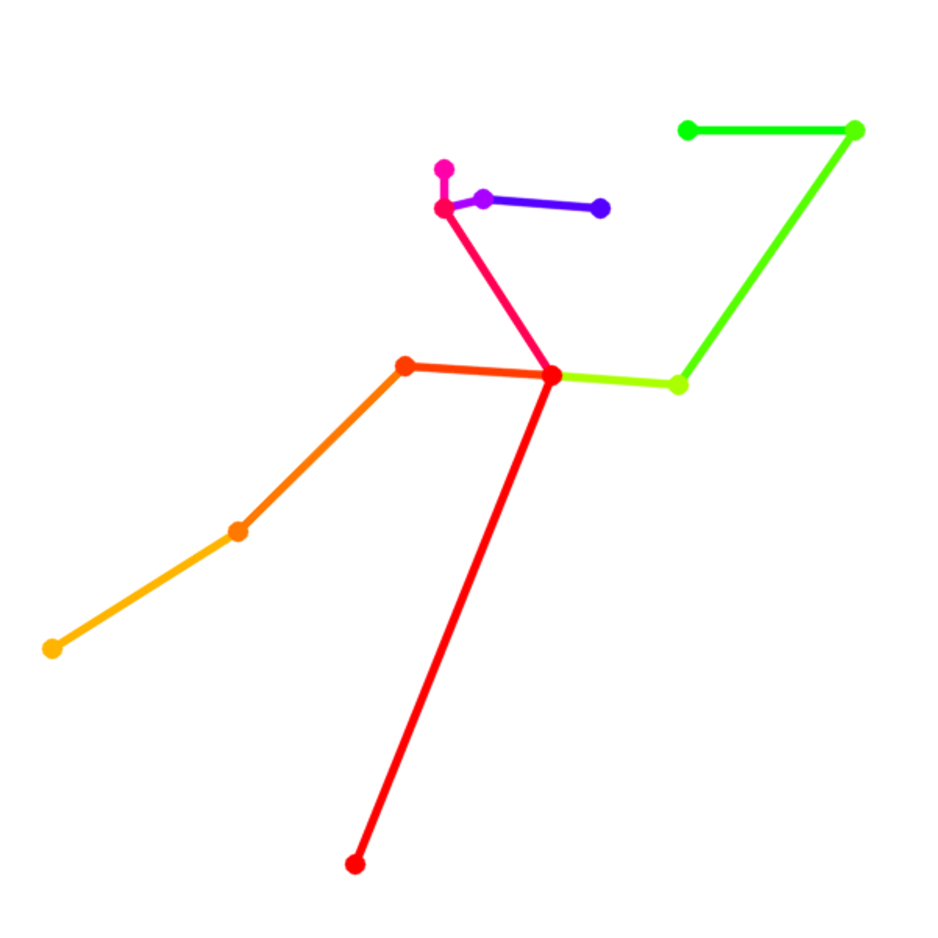}
        \caption{}
        \label{fig:misclassified_non_hand-raising_3}
    \end{subfigure}
    \begin{subfigure}[t]{0.24\textwidth}
        \centering
        \includegraphics[width = 0.4\textwidth]{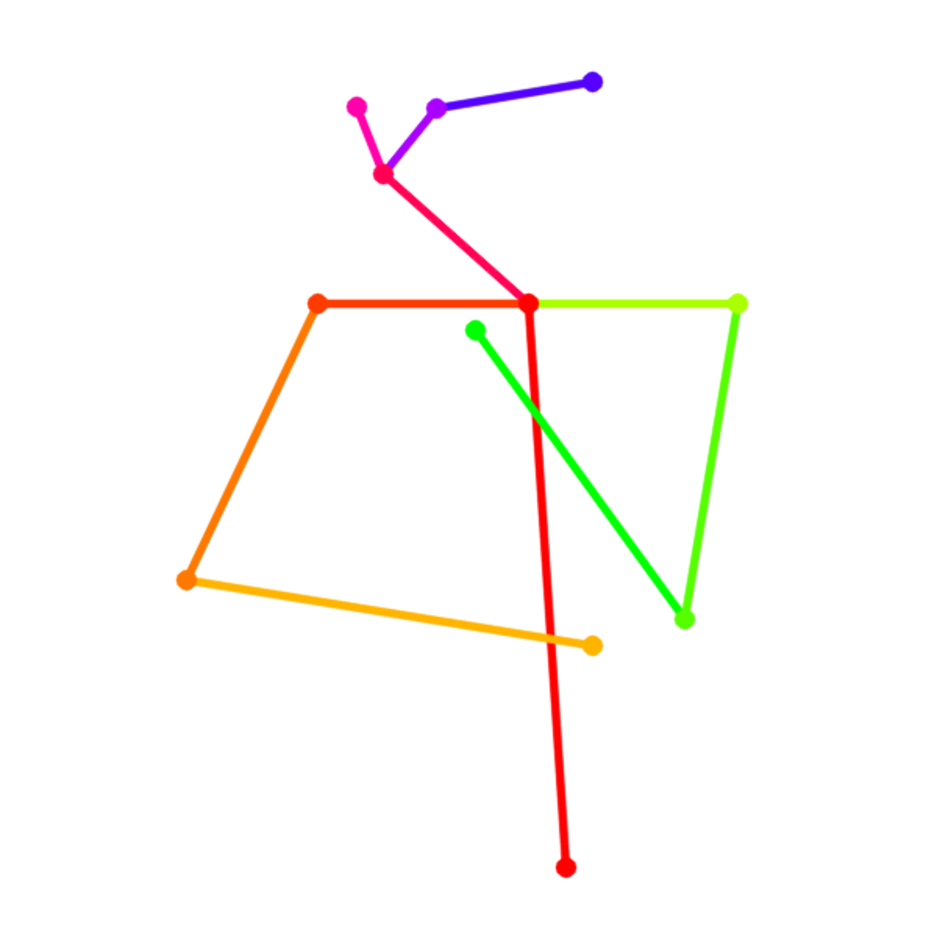}
        \caption{}
        \label{fig:misclassified_non_hand-raising_2}
    \end{subfigure}
    \begin{subfigure}[t]{0.24\textwidth}
        \centering
        \includegraphics[width = 0.4\textwidth]{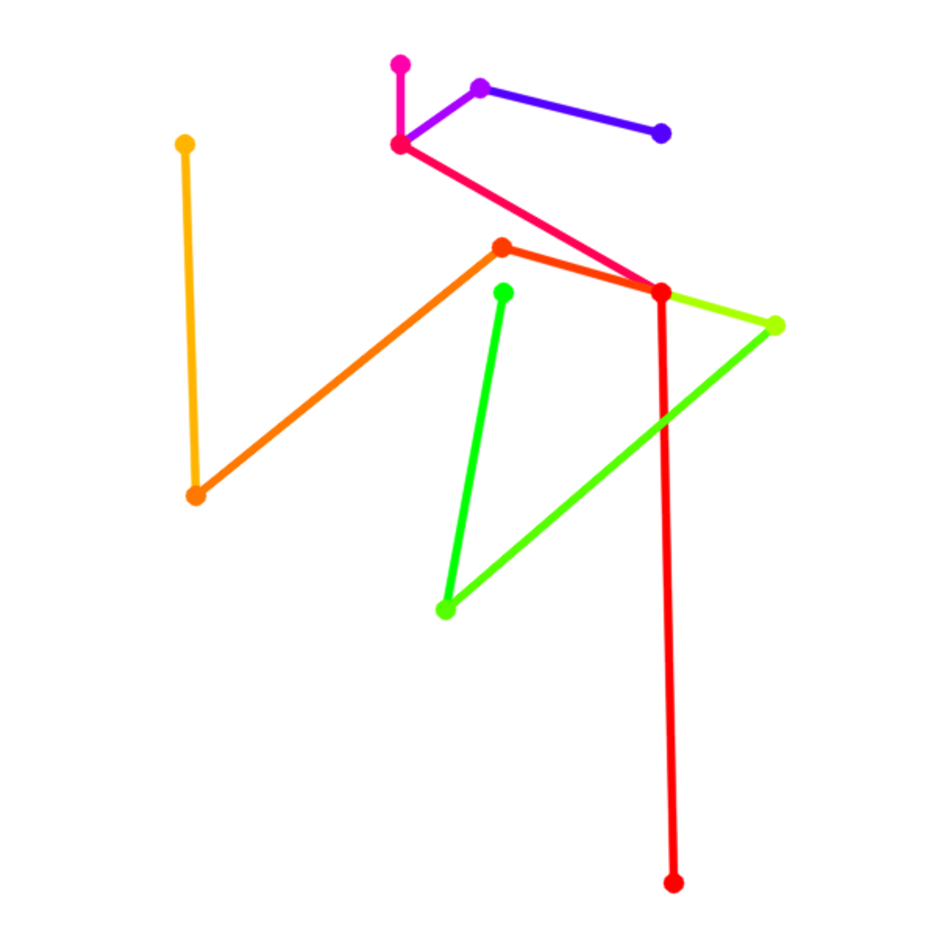}
        \caption{}
        \label{fig:misclassified_hand-raining_1}
    \end{subfigure}
    \begin{subfigure}[t]{0.24\textwidth}
        \centering
        \includegraphics[width = 0.4\textwidth]{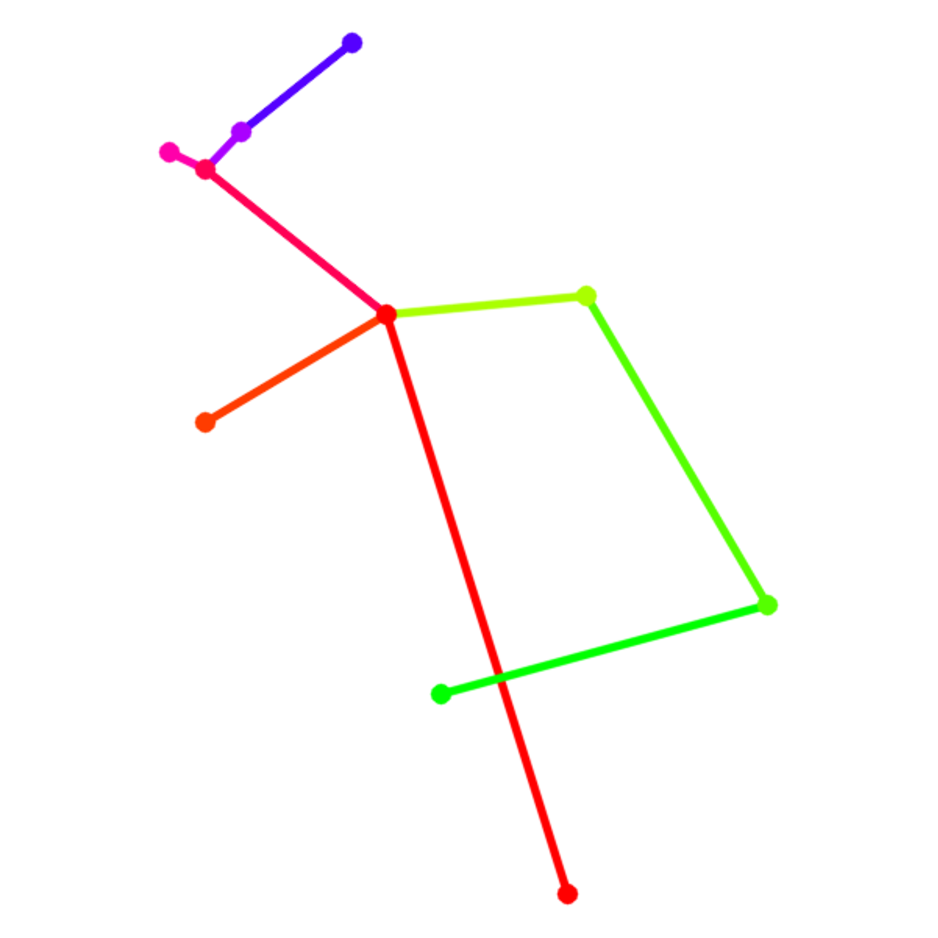}
        \caption{}
        \label{fig:misclassified_hand-raining_2}
    \end{subfigure}
    \caption{Skeleton samples of misclassified windows using LSTM with pose embeddings. True label: (a, b) non-hand-raising; (c, d) hand-raising.}
    \label{fig:misclassified_examples}
\end{figure}

In order to gain a deeper understanding of our model, we investigated misclassified instances. Figure~\ref{fig:misclassified_examples} depicts four misclassified examples, using the LSTM model on pose embeddings. They reveal poses are misclassified as hand-raising when they have a similar skeleton representation, e.g., when students scratch their heads (Fig.~\ref{fig:misclassified_non_hand-raising_3}) or rest their head on their hand (Fig.~\ref{fig:misclassified_non_hand-raising_2}). In turn, subtle hand-raisings that do not involve raising the hand or elbow above the shoulder are more difficult to recognize (Fig.~\ref{fig:misclassified_hand-raining_1}).  Occasionally, students simply indicate hand-raisings by extending their index finger, which can not be represented in the skeleton we used.
Furthermore, the classifier is prone to false predictions if keypoints of the hand-raising arm are not detected throughout the clip (Fig.~\ref{fig:misclassified_hand-raining_2}). 

\begin{table}[h] 
\centering
\caption{Comparison of different classifiers and pose representations.}
\label{tab:classification_results}
\begin{tabular}{p{3.5cm}<{\centering}|p{3.5cm}<{\centering}|p{1.5cm}<{\centering}p{1.5cm}<{\centering}p{1.5cm}<{\centering}}
\toprule
Classifier & Pose Representation & Precision & Recall & $F_{1}$-Score\\
\midrule
RF & \multirow{4}{*}{Pose embeddings} & \textbf{0.950} & 0.540 & 0.688\\
RF (w/ weighting) &  & 0.763 & 0.695 & 0.727\\
LSTM &  & 0.818 & 0.709 & \textbf{0.760}\\
LSTM (w/ weighting) &  & 0.570 & \textbf{0.783} & 0.660\\
\midrule
\midrule
\multirow{3}{*}{LSTM} & Geometric features~\cite{lin2021student} & 0.816 & 0.576 & 0.676\\
 & Geometric features~\cite{zhang2017geometric} & 0.812 & 0.694 & 0.748\\
 & Pose embeddings & \textbf{0.818} & \textbf{0.709} & \textbf{0.760}\\
\hline
\end{tabular}
\end{table}

\subsubsection{Automated Hand-raising Annotation}
After automatically annotating hand-raisings on a student level for the 22 validation videos (see Sect.~\ref{sec:automated_annotation}), we compared the estimated hand-raising counts with the ground truth by calculating the mean absolute error (\gls{mae}).
The automated annotation, employing the best-performing LSTM model achieved a MAE of 3.76. According to the ground truth, a student raised their hand 6.10 times in a lesson on average. 
Besides the moderate classification performance, this result can be attributed to the tracking capability. Some targets cannot be tracked continuously due to occlusion or failing pose estimation. Using a fragmentary tracklet can generate multiple sub-slots for one actual hand-raising instance, resulting in overestimation. By using the subset of tracklets that span at least 90\% of the videos, the MAE decreased to 3.34.
To demonstrate the application of such automated annotations, we did a re-analysis of the relation of manually and automatically annotated hand-raisings to self-reported learning activities on the validation videos. Table \ref{tab:results_pred_correlation} shows that both annotations are significantly related to the three learning activities, showing comparable $r$ values.

\begin{table}[]
\centering
\caption{Pearson correlations of different hand-raising annotations with self-reported learning activities in validation videos (N = 173).}
\label{tab:results_pred_correlation}
\begin{tabular}{m{2.3cm}p{1.3cm}p{1.3cm}p{1.3cm}p{1.3cm}p{1.3cm}p{1.3cm}}
\toprule
\multirow{2}{*}{\shortstack[l]{\\Hand-raising\\annotation}} & \multicolumn{2}{l}{\begin{tabular}[c]{@{}l@{}}Cognitive \\ engagement\end{tabular}} & \multicolumn{2}{l}{\begin{tabular}[c]{@{}l@{}}Situational \\ interest\end{tabular}} & \multicolumn{2}{l}{Involvement} \\ 
\cmidrule{2-7}
& $r$         & $p$        & $r$            & $p$       & $r$    & $p$ \\ 
\midrule
Manually     & 0.222      & 0.003       & 0.326    & 0.000         & 0.213       &  0.005          \\
Automated      & 0.222     & 0.003      & 0.288   & 0.000         & 0.201         & 0.008          \\ \hline
\end{tabular}
\end{table}

\subsection{Discussion}

We found that the frequency of student hand-raisings is significantly related to self-reported learning activities in a diverse real-world classroom dataset covering grades five to twelve and a variety of subjects. These results are in line with previous research \cite{Boheim.2020, boheim2020engagement} and emphasize the important role of hand-raising as an observable cue for students' engagement in classroom discourse. To enable such analyses, we proposed a novel approach for the automated detection of hand-raising events in classroom videos. The employed view-invariant and occlusion-robust pose embeddings outperformed more simplistic geometric features used in previous research~\cite{lin2021student}. When applying the developed classification model for person-specific hand-raising instance annotation, the total amount of hand-raising instances per person was slightly overestimated, mainly stemming from discontinuous pose tracks. The comparison of correlation analysis between manually and automatically annotated hand-raising and learning-related activities showed comparable results for the two methods, despite the imperfect prediction accuracy, suggesting that automated annotation can be a useful proxy for research. It is important to stress that we employed a dataset containing un-staged real-world school lessons and authentic hand-raisings in naturalistic settings. This is why we employ and compare feature representations from previous research in our setting rather than directly comparing the performance results which were based on either scripted videos or extremely small sample sizes \cite{Ahuja.2019,lin2021student}. Furthermore, the data includes a wide span of age groups, ranging from 11- to 18-year-olds. As shown in Table \ref{tab:overview_videos}, the frequency, as well as the manner of raising the hand and indicating the wish to speak in the classroom, differs substantially between those age groups.

These results strengthen the potential of automated hand-raising annotation for research investigating hand-raising as a form of participation in classroom discourse and an indicator of behavioral engagement. Replacing time-consuming manual coding with an automated process will allow cost-effective large-scale research on classroom interaction in the future. The approach is highly robust to camera viewpoints and therefore presumably generalizes well to new classroom setups. Further, the employment of pose estimation allows for privacy-preserving data collection, as poses can be extracted in real-time, thus eliminating the need to store video recordings. However, as students’ body-pose information might be regarded as sensitive data, it is important to note that automated behavior detection should solely be used as a post hoc analytical tool for research purposes instead of real-time classroom monitoring.

One limitation of this work is that the absolute hand-raising frequency is still over- or underestimated by an average of almost four instances per person. Future research should continue to enhance the precision, by further improving the underlying pose estimation and tracking. This can either be achieved by optimizing camera angles to be most suitable for pose estimation as well as applying more advanced pose estimation techniques. Further room for improvement lies in the differentiation of difficult cases as shown in Fig. \ref{fig:misclassified_examples}. Here hand-raisings that are more restrained, i.e., if students do not raise the hand above the head, are not recognized as such. A solution for this could be to include hand keypoint estimation, as the hand posture can possibly provide further information.

\subsection{Conclusion}

Our results indicate that hand-raising  correlates to cognitive engagement, situational interest, and involvement and presents hence an important measure of active student participation in the classroom. We further show that the automated annotation is a viable substitute for manual coding of hand-raisings, opening up new possibilities for researching hand-raising in classroom discourse and student engagement on a large scale. 

\subsection*{Acknowledgements} 
Babette B{\"u}hler is a doctoral candidate and supported by the LEAD Graduate School and Research Network, which is funded by the Ministry of Science, Research and the Arts of the state of Baden-W{\"u}rttemberg within the framework of the sustainability funding for the projects of the Excellence Initiative II. Efe Bozkir and Enkelejda Kasneci acknowledge the funding by the DFG with EXC number 2064/1 and project number 390727645. This work is also supported by Leibniz-WissenschaftsCampus Tübingen “Cognitive Interfaces” by a grant to Ulrich Trautwein, Peter Gerjets, and Enkelejda Kasneci

%% file: references.bib
@inproceedings{buhler2023hr,
  title={Automated hand-raising detection in classroom videos: A view-invariant and occlusion-robust machine learning approach},
  author={B{\"u}hler, Babette and Hou, Ruikun and Bozkir, Efe and Goldberg, Patricia and Gerjets, Peter and Trautwein, Ulrich and Kasneci, Enkelejda},
  booktitle={International Conference on Artificial Intelligence in Education},
  pages={102--113},
  year={2023},
  organization={Springer},

doi = {https://doi.org/10.1007/978-3-031-36272-9_9}                              
}

@article{buhler2024synchr,
author = {B\"{u}hler, Babette and Bozkir, Efe and Deininger, Hannah and Gerjets, Peter and Trautwein, Ulrich and Kasneci, Enkelejda},
title = {On Task and in Sync: Examining the Relationship between Gaze Synchrony and Self-reported Attention During Video Lecture Learning},
year = {2024},
issue_date = {May 2024},
publisher = {Association for Computing Machinery},
address = {New York, NY, USA},
volume = {8},
number = {ETRA},
url = {https://doi.org/10.1145/3655604}   ,  
journal = {Proc. ACM Hum.-Comput. Interact.},
month = {may},
articleno = {230},
numpages = {18},
  doi={https://doi.org/10.1145/3655604}                                 
}

@inproceedings{buhler2024edpsych,
  title={Temporal Dynamics of Meta-Awareness of Mind Wandering During Lecture Viewing: Implications for Learning and Automated Detection Using Machine Learning.},
  author={B{\"u}hler, Babette and Bozkir, Efe and Goldberg, Patricia and Deininger, Hannah and D'Mello, Sidney and Gerjets, Peter and Trautwein, Ulrich and Kasneci, Enkelejda},
  booktitle={Journal of Educational Psychology},
  year={\textit{In Press.}},
}

@article{Seli.2018b,
 author = {Seli, Paul and Kane, Michael J. and Metzinger, Thomas and Smallwood, Jonathan and Schacter, Daniel L. and Maillet, David and Schooler, Jonathan W. and Smilek, Daniel},
 year = {2018},
 title = {The Family-Resemblances Framework for Mind-Wandering Remains Well Clad},
 url = {https://www.cell.com/trends/cognitive-sciences/fulltext/S1364-6613(18)30167-0},
 pages = {959--961},
 volume = {22},
 number = {11},
 issn = {1879-307X},
 journal = {Trends in Cognitive Sciences},
 doi = {10.1016/j.tics.2018.07.007    }
}

@unpublished{buhler2024mm,
  title={Detecting Aware and Unaware Mind Wandering During Lecture Viewing: A Multimodal Machine Learning Approach Using Eye Tracking, Facial Videos and Physiological Data.},
  author={B{\"u}hler, Babette and Bozkir, Efe  and Deininger, Hannah and Goldberg, Patricia and Gerjets, Peter and Trautwein, Ulrich and Kasneci, Enkelejda},
  year={\textit{Submitted to ACM International Conference on Multimodal Interaction 2024. Under Review.}},
}

@article{buhler2024general,
  title={From the Lab to the Wild: Examining Generalizability of Video-based Mind Wandering Detection.},
  author={B{\"u}hler, Babette and Bozkir, Efe and Goldberg, Patricia and S{\"u}mer, {\"O}mer and D’Mello, Sidney and Gerjets, Peter and Trautwein, Ulrich and Kasneci, Enkelejda},
  journal={International Journal of Artificial Intelligence in Education.},
  pages={1--35},
  year={2024},
  publisher={Springer},
doi={https://doi.org/10.1007/s40593-024-00412-2}        
}

@inproceedings{fuhl2023watch,
  title={Watch out for those bananas! gaze based mario kart performance classification},
  author={Fuhl, Wolfgang and Severitt, Bj{\"o}rn and Castner, Nora and B{\"u}hler, Babette and Meyer, Johannes and Weber, Daniel and Lendway, Regine and Hou, Ruikun and Kasneci, Enkelejda},
  booktitle={Proceedings of the 2023 Symposium on Eye Tracking Research and Applications},
  pages={1--2},
  year={2023}
}

@article{zanesco2024,
  title={Mind-wandering increases in frequency over time during task performance: An individual-participant meta-analytic review.},
  author={Zanesco, Anthony P and Denkova, Ekaterina and Jha, Amishi P},
  journal={Psychological Bulletin},
  year={2024},
  publisher={American Psychological Association}
}

@article{kasneci2023chatgpt,
  title={ChatGPT for good? On opportunities and challenges of large language models for education},
  author={Kasneci, Enkelejda and Se{\ss}ler, Kathrin and K{\"u}chemann, Stefan and Bannert, Maria and Dementieva, Daryna and Fischer, Frank and Gasser, Urs and Groh, Georg and G{\"u}nnemann, Stephan and H{\"u}llermeier, Eyke and others},
  journal={Learning and individual differences},
  volume={103},
  pages={102274},
  year={2023},
  publisher={Elsevier}
}

@article{team2023gemini,
  title={Gemini: a family of highly capable multimodal models},
  author={Team, Gemini and Anil, Rohan and Borgeaud, Sebastian and Wu, Yonghui and Alayrac, Jean-Baptiste and Yu, Jiahui and Soricut, Radu and Schalkwyk, Johan and Dai, Andrew M and Hauth, Anja and others},
  journal={arXiv preprint arXiv:2312.11805           }   ,   
  year={2023}
}

@inproceedings{radford2021learning,
  title={Learning transferable visual models from natural language supervision},
  author={Radford, Alec and Kim, Jong Wook and Hallacy, Chris and Ramesh, Aditya and Goh, Gabriel and Agarwal, Sandhini and Sastry, Girish and Askell, Amanda and Mishkin, Pamela and Clark, Jack and others},
  booktitle={International conference on machine learning},
  pages={8748--8763},
  year={2021},
  organization={PMLR}
}

@article{huang2023chatgpt,
  title={ChatGPT for shaping the future of dentistry: the potential of multi-modal large language model},
  author={Huang, Hanyao and Zheng, Ou and Wang, Dongdong and Yin, Jiayi and Wang, Zijin and Ding, Shengxuan and Yin, Heng and Xu, Chuan and Yang, Renjie and Zheng, Qian and others},
  journal={International Journal of Oral Science},
  volume={15},
  number={1},
  pages={29},
  year={2023},
  publisher={Nature Publishing Group UK London}
}

@article{winne1995,
  title={Inherent details in self-regulated learning},
  author={Winne, Philip H},
  journal={Educational psychologist},
  volume={30},
  number={4},
  pages={173--187},
  year={1995},
  publisher={Taylor \& Francis}
}

@book{berger2014leaders,
  title={Leaders of their own learning: Transforming schools through student-engaged assessment},
  author={Berger, Ron and Rugen, Leah and Woodfin, Libby},
  year={2014},
  publisher={John Wiley \& Sons}
}

@inproceedings{wang2023internimage,
  title={Internimage: Exploring large-scale vision foundation models with deformable convolutions},
  author={Wang, Wenhai and Dai, Jifeng and Chen, Zhe and Huang, Zhenhang and Li, Zhiqi and Zhu, Xizhou and Hu, Xiaowei and Lu, Tong and Lu, Lewei and Li, Hongsheng and others},
  booktitle={Proceedings of the IEEE/CVF Conference on Computer Vision and Pattern Recognition},
  pages={14408--14419},
  year={2023}
}

@inproceedings{hou2024automated,
  author="Hou, Ruikun and F{\"u}tterer, Tim and B{\"u}hler, Babette and Bozkir, Efe and Gerjets, Peter and Trautwein, Ulrich and Kasneci, Enkelejda", 
editor="Olney, Andrew M. and Chounta, Irene-Angelica and Liu, Zitao and Santos, Olga C. and Bittencourt, Ig Ibert", 
title="Automated Assessment of Encouragement and Warmth in Classrooms Leveraging Multimodal Emotional Features and ChatGPT",
booktitle="Artificial Intelligence in Education",
year="2024",
publisher="Springer Nature Switzerland",
address="Cham",
pages="60--74",
doi={https://doi.org/10.1007/978-3-031-64302-6_5  }
}

@article{ainley2018teaching,
  title={Teaching and learning international survey (TALIS) 2018 conceptual framework},
  author={Ainley, John and Carstens, Ralph},
  year={2018},
  publisher={oecd}
}

@article{daltoe2024,
  title={Immersive insights: Unveiling the impact of 360-degree videos on preservice teachers’ classroom observation experiences and teaching-quality ratings},
  author={Dalto{\`e}, Tosca and Ruth-Herbein, Evelin and Brucker, Birgit and Jaekel, Ann-Kathrin and Trautwein, Ulrich and Fauth, Benjamin and Gerjets, Peter and G{\"o}llner, Richard},
  journal={Computers \& Education},
  volume={213},
  pages={104976},
  year={2024},
  publisher={Elsevier}
}

@article{wagner2019multimatch,
  title={multimatch-gaze: The MultiMatch algorithm for gaze path comparison in Python},
  author={Wagner, Adina S and Halchenko, Yaroslav O and Hanke, Michael},
  journal={Journal of Open Source Software},
  volume={4},
  number={40},
  pages={1525},
  year={2019}
}

@article{dias2022motor,
  title={Straying off course: The negative impact of mind wandering on fine motor movements},
  author={Dias da Silva, Mariana Rachel and Postma, Marie},
  journal={Journal of Motor Behavior},
  volume={54},
  number={2},
  pages={186--202},
  year={2022},
  publisher={Taylor \& Francis}
}

@article{hadwin2007,
  title={Examining trace data to explore self-regulated learning},
  author={Hadwin, Allyson F and Nesbit, John C and Jamieson-Noel, Dianne and Code, Jillianne and Winne, Philip H},
  journal={Metacognition and Learning},
  volume={2},
  pages={107--124},
  year={2007},
  publisher={Springer}
}

@article{wang2023log,
  title={Which log variables significantly predict academic achievement? A systematic review and meta-analysis},
  author={Wang, Qin and Mousavi, Amin},
  journal={British Journal of Educational Technology},
  volume={54},
  number={1},
  pages={142--191},
  year={2023},
  publisher={Wiley Online Library}
}

@article{kuvar2023partner,
  title={Partner Keystrokes Can Predict Attentional States during Chat-Based Conversations.},
  author={Kuvar, Vishal and Flynn, Lauren and Allen, Laura and Mills, Caitlin},
  journal={International Educational Data Mining Society},
  year={2023},
  publisher={ERIC}
}

@article{praetorius2012observer,
  title={Observer ratings of instructional quality: Do they fulfill what they promise?},
  author={Praetorius, Anna-Katharina and Lenske, Gerlinde and Helmke, Andreas},
  journal={Learning and instruction},
  volume={22},
  number={6},
  pages={387--400},
  year={2012},
  publisher={Elsevier}
}

@article{hoyt2000rater,
  title={Rater bias in psychological research: When is it a problem and what can we do about it?},
  author={Hoyt, William T},
  journal={Psychological methods},
  volume={5},
  number={1},
  pages={64},
  year={2000},
  publisher={American Psychological Association}
}

@article{mandia2023eng,
  title={Recognition of student engagement in classroom from affective states},
  author={Mandia, Sandeep and Singh, Kuldeep and Mitharwal, Rajendra},
  journal={International Journal of Multimedia Information Retrieval},
  volume={12},
  number={2},
  pages={18},
  year={2023},
  publisher={Springer}
}

@article{giambra1989slef,
  title={Task-unrelated thought frequency as a function of age: a laboratory study.},
  author={Giambra, Leonard M},
  journal={Psychology and aging},
  volume={4},
  number={2},
  pages={136},
  year={1989},
  publisher={American Psychological Association}
}

@book{leigh2015neurology,
  title={The neurology of eye movements},
  author={Leigh, R John and Zee, David S},
  year={2015},
  publisher={Oxford University Press, USA}
}

@article{gobert2013log,
  title={From log files to assessment metrics: Measuring students' science inquiry skills using educational data mining},
  author={Gobert, Janice D and Sao Pedro, Michael and Raziuddin, Juelaila and Baker, Ryan S},
  journal={Journal of the Learning Sciences},
  volume={22},
  number={4},
  pages={521--563},
  year={2013},
  publisher={Taylor \& Francis}
}

@article{mcvay2012drifting,
  title={Drifting from slow to “d'oh!”: Working memory capacity and mind wandering predict extreme reaction times and executive control errors.},
  author={McVay, Jennifer C and Kane, Michael J},
  journal={Journal of Experimental Psychology: Learning, Memory, and Cognition},
  volume={38},
  number={3},
  pages={525},
  year={2012},
  publisher={American Psychological Association}
}

@article{kuvar2023keystroke,
  title={Partner Keystrokes Can Predict Attentional States during Chat-Based Conversations.},
  author={Kuvar, Vishal and Flynn, Lauren and Allen, Laura and Mills, Caitlin},
  journal={International Educational Data Mining Society},
  year={2023},
  publisher={ERIC}
}

@article{mills2015log,
  title={Toward a Real-Time (Day) Dreamcatcher: Sensor-Free Detection of Mind Wandering during Online Reading.},
  author={Mills, Caitlin and D'Mello, Sidney},
  journal={International educational data mining society},
  year={2015},
  publisher={ERIC}
}

@article{nguyen2022ethical,
  title={Ethical principles for artificial intelligence in education},
  author={Nguyen, Andy and Ngo, Ha Ngan and Hong, Yvonne and Dang, Belle and Nguyen, Bich-Phuong Thi},
  journal={Education and Information Technologies},
  pages={1--21},
  year={2022},
  publisher={Springer}
}

@article{dmello2023affect,
  title={Affect Detection From Wearables in the “Real” Wild: Fact, Fantasy, or Somewhere In between?},
  author={D’Mello, Sidney K and Booth, Brandon M},
  journal={IEEE Intelligent Systems},
  volume={38},
  number={1},
  pages={76--84},
  year={2023},
  publisher={IEEE}
}

@article{russell1994,
  title={Is there universal recognition of emotion from facial expression? A review of the cross-cultural studies.},
  author={Russell, James A},
  journal={Psychological bulletin},
  volume={115},
  number={1},
  pages={102},
  year={1994},
  publisher={American Psychological Association}
}

@article{ekman1994,
  title={Strong evidence for universals in facial expressions: a reply to Russell's mistaken critique},
  author={Ekman, Paul},
  journal={Psychological Bulletin},
  volume={115},
  pages={268--287},
  year={1994},
  publisher={American Psychological Association (PsycARTICLES)}
}

@INPROCEEDINGS{Benitez2017,
  author={Benitez-Garcia, Gibran and Nakamura, Tomoaki and Kaneko, Masahide},
  booktitle={2017 Fifteenth IAPR International Conference on Machine Vision Applications (MVA)}, 
  title={Analysis of in- and out-group differences between Western and East-Asian facial expression recognition}, 
  year={2017},
  volume={},
  number={},
  pages={402-405},
  doi={10.23919/MVA.2017.7986886  }}

@article{dmello2021,
  title={Mind wandering during reading: An interdisciplinary and integrative review of psychological, computing, and intervention research and theory},
  author={D'Mello, Sidney K and Mills, Caitlin S},
  journal={Language and Linguistics Compass},
  volume={15},
  number={4},
  pages={e12412},
  year={2021},
  publisher={Wiley Online Library}
}

@inproceedings{zhao2017,
  title={Scalable mind-wandering detection for MOOCs: A webcam-based approach},
  author={Zhao, Yue and Lofi, Christoph and Hauff, Claudia},
  booktitle={European Conference on Technology Enhanced Learning},
  pages={330--344},
  year={2017},
  organization={Springer}
}

@article{dmello2017zone,
  title={Zone out No More: Mitigating Mind Wandering during Computerized Reading.},
  author={D'Mello, Sidney K. and Mills, Caitlin and Bixler, Robert and Bosch, Nigel},
  journal={International Educational Data Mining Society},
  year={2017},
  publisher={ERIC}
}

@inproceedings{hutt2021breaking,
  title={Breaking out of the lab: Mitigating mind wandering with gaze-based attention-aware technology in classrooms},
  author={Hutt, Stephen and Krasich, Kristina and R. Brockmole, James and K. D'Mello, Sidney K.},
  booktitle={Proceedings of the 2021 CHI Conference on Human Factors in Computing Systems},
  pages={1--14},
  year={2021}
}

@article{stewart2017generalizability,
  title={Generalizability of Face-Based Mind Wandering Detection across Task Contexts.},
  author={Stewart, Angela and Bosch, Nigel and D'Mello, Sidney K.},
  journal={International Educational Data Mining Society},
  year={2017},
  publisher={ERIC}
}

@article{dong2021eeg,
  title={Detection of mind wandering using EEG: Within and across individuals},
  author={Dong, Henry W and Mills, Caitlin and Knight, Robert T and Kam, Julia WY},
  journal={Plos one},
  volume={16},
  number={5},
  pages={e0251490},
  year={2021},
  publisher={Public Library of Science San Francisco, CA USA}
}

@article{jin2019eeg,
  title={Predicting task-general mind-wandering with EEG},
  author={Jin, Christina Yi and Borst, Jelmer P and Van Vugt, Marieke K},
  journal={Cognitive, Affective, \& Behavioral Neuroscience},
  volume={19},
  number={4},
  pages={1059--1073},
  year={2019},
  publisher={Springer}
}

@article{bosch2022,
  title={Can computers outperform humans in detecting user zone-outs? Implications for intelligent interfaces},
  author={Bosch, Nigel and D'Mello, Sidney K.},
  journal={ACM Transactions on Computer-Human Interaction},
  volume={29},
  number={2},
  pages={1--33},
  year={2022},
  publisher={ACM New York, NY}
}

@article{just1976eye,
  title={Eye fixations and cognitive processes},
  author={Just, Marcel Adam and Carpenter, Patricia A},
  journal={Cognitive psychology},
  volume={8},
  number={4},
  pages={441--480},
  year={1976},
  publisher={Elsevier}
}

@inproceedings{thomas2017predicting,
  title={Predicting student engagement in classrooms using facial behavioral cues},
  author={Thomas, Chinchu and Jayagopi, Dinesh Babu},
  booktitle={Proceedings of the 1st ACM SIGCHI international workshop on multimodal interaction for education},
  pages={33--40},
  year={2017}
}

@article{larson1983experience,
  title={The experience sampling method.},
  author={Larson, Reed and Csikszentmihalyi, Mihaly},
  journal={New directions for methodology of social \& behavioral science},
  year={1983},
  publisher={Jossey-Bass Publishers, Inc.}
}

@inproceedings{stewart2017,
  title={Face forward: Detecting mind wandering from video during narrative film comprehension},
  author={Stewart, Angela and Bosch, Nigel and Chen, Huili and Donnelly, Patrick and D’Mello, Sidney K},
  booktitle={International Conference on Artificial Intelligence in Education},
  pages={359--370},
  year={2017},
  organization={Springer}
}

@article{bevilacqua2019brain,
  title={Brain-to-brain synchrony and learning outcomes vary by student--teacher dynamics: Evidence from a real-world classroom electroencephalography study},
  author={Bevilacqua, Dana and Davidesco, Ido and Wan, Lu and Chaloner, Kim and Rowland, Jess and Ding, Mingzhou and Poeppel, David and Dikker, Suzanne},
  journal={Journal of cognitive neuroscience},
  volume={31},
  number={3},
  pages={401--411},
  year={2019},
  publisher={MIT Press One Rogers Street, Cambridge, MA 02142-1209, USA journals-info~…}
}

@article{dikker2017brain,
  title={Brain-to-brain synchrony tracks real-world dynamic group interactions in the classroom},
  author={Dikker, Suzanne and Wan, Lu and Davidesco, Ido and Kaggen, Lisa and Oostrik, Matthias and McClintock, James and Rowland, Jess and Michalareas, Georgios and Van Bavel, Jay J and Ding, Mingzhou and others},
  journal={Current biology},
  volume={27},
  number={9},
  pages={1375--1380},
  year={2017},
  publisher={Elsevier}
}

@article{poulsen2017eeg,
  title={EEG in the classroom: Synchronised neural recordings during video presentation},
  author={Poulsen, Andreas Trier and Kamronn, Simon and Dmochowski, Jacek and Parra, Lucas C and Hansen, Lars Kai},
  journal={Scientific reports},
  volume={7},
  number={1},
  pages={43916},
  year={2017},
  publisher={Nature Publishing Group UK London}
}

@article{rosengrant2021,
  title={Investigating student sustained attention in a guided inquiry lecture course using an eye tracker},
  author={Rosengrant, David and Hearrington, Doug and O’Brien, Jennifer},
  journal={Educational psychology review},
  volume={33},
  pages={11--26},
  year={2021},
  publisher={Springer}
}

@article{ekman1978facial,
  title={Facial action coding system},
  author={Ekman, Paul and Friesen, Wallace V},
  journal={Environmental Psychology \& Nonverbal Behavior},
  year={1978}
}

@inproceedings{raca2014sleepers,
  title={Sleepers' lag-study on motion and attention},
  author={Raca, Mirko and Tormey, Roland and Dillenbourg, Pierre},
  booktitle={Proceedings of the fourth international conference on learning analytics and knowledge},
  pages={36--43},
  year={2014}
}

@inproceedings{raca2015head,
  title={Translating head motion into attention-towards processing of student’s body-language},
  author={Raca, Mirko and Kidzinski, Lukasz and Dillenbourg, Pierre},
  booktitle={Proceedings of the 8th international conference on educational data mining},
  year={2015}
}

@inproceedings{2021crossed,
  title={Crossed eyes: Domain adaptation for gaze-based mind wandering models},
  author={Bixler, Robert and D'Mello, Sidney K.},
  booktitle={ACM Symposium on Eye Tracking Research and Applications},
  pages={1--12},
  year={2021}
}

@inproceedings{bixler2014,
  title={Toward fully automated person-independent detection of mind wandering},
  author={Bixler, Robert and D’Mello, Sidney K},
  booktitle={International Conference on User Modeling, Adaptation, and Personalization},
  pages={37--48},
  year={2014},
  organization={Springer}
}

@inproceedings{dmello2016,
  title={Attending to attention: Detecting and combating mind wandering during computerized reading},
  author={D'Mello, Sidney K and Kopp, Kristopher and Bixler, Robert Earl and Bosch, Nigel},
  booktitle={Proceedings of the 2016 CHI conference extended abstracts on human factors in computing systems},
  pages={1661--1669},
  year={2016}
}

@article{baldi1999,
  title={Exploiting the past and the future in protein secondary structure prediction},
  author={Baldi, Pierre and Brunak, S{\o}ren and Frasconi, Paolo and Soda, Giovanni and Pollastri, Gianluca},
  journal={Bioinformatics},
  volume={15},
  number={11},
  pages={937--946},
  year={1999},
  publisher={Oxford University Press}
}

@article{schuster1997,
  title={Bidirectional recurrent neural networks},
  author={Schuster, Mike and Paliwal, Kuldip K},
  journal={IEEE transactions on Signal Processing},
  volume={45},
  number={11},
  pages={2673--2681},
  year={1997},
  publisher={Ieee}
}

@article{hochreiter1997,
  title={Long short-term memory},
  author={Hochreiter, Sepp and Schmidhuber, J{\"u}rgen},
  journal={Neural computation},
  volume={9},
  number={8},
  pages={1735--1780},
  year={1997},
  publisher={MIT Press}
}

@article{sumer2021,
  title={Multimodal engagement analysis from facial videos in the classroom},
  author={S{\"u}mer, {\"O}mer and Goldberg, Patricia and D'Mello, Sidney K and Gerjets, Peter and Trautwein, Ulrich and Kasneci, Enkelejda},
  journal={IEEE Transactions on Affective Computing},
  year={2021},
  publisher={IEEE}
}

@inproceedings{retinaface2020,
  title={Retinaface: Single-shot multi-level face localisation in the wild},
  author={Deng, Jiankang and Guo, Jia and Ververas, Evangelos and Kotsia, Irene and Zafeiriou, Stefanos},
  booktitle={Proceedings of the IEEE/CVF conference on computer vision and pattern recognition},
  pages={5203--5212},
  year={2020}
}

@inproceedings{openface2016,
  title={Openface: an open source facial behavior analysis toolkit},
  author={Baltru{\v{s}}aitis, Tadas and Robinson, Peter and Morency, Louis-Philippe},
  booktitle={2016 IEEE Winter Conference on Applications of Computer Vision (WACV)},
  pages={1-10},
  year={2016},
  organization={IEEE}
}

@article{affectnet2017,
  title={Affectnet: A database for facial expression, valence, and arousal computing in the wild},
  author={Mollahosseini, Ali and Hasani, Behzad and Mahoor, Mohammad H},
  journal={IEEE Transactions on Affective Computing},
  volume={10},
  number={1},
  pages={18--31},
  year={2017},
  publisher={IEEE}
}

@article{Smallwood.2007,
 author = {Smallwood, Jonathan and McSpadden, Merrill and Schooler, Jonathan W.},
 year = {2007},
 title = {The lights are on but no one's home: meta-awareness and the decoupling of attention when the mind wanders},
 pages = {527--533},
 volume = {14},
 number = {3},
 issn = {1531-5320},
 journal = {Psychonomic Bulletin {\&} Review},
 doi = {10.3758/BF03194102 }
}

@article{Smallwood.2007b,
 author = {Smallwood, Jonathan and Fishman, Daniel J. and Schooler, Jonathan W.},
 year = {2007},
 title = {Counting the cost of an absent mind: mind wandering as an underrecognized influence on educational performance},
 pages = {230--236},
 volume = {14},
 number = {2},
 issn = {1531-5320},
 journal = {Psychonomic Bulletin {\&} Review},
 doi = {10.3758/BF03194057 }
}

@article{dunn2023minimal,
  title={Minimal reporting guideline for research involving eye tracking (2023 edition)},
  author={Dunn, Matt J and Alexander, Robert G and Amiebenomo, Onyekachukwu M and Arblaster, Gemma and Atan, Denize and Erichsen, Jonathan T and Ettinger, Ulrich and Giardini, Mario E and Gilchrist, Iain D and Hamilton, Ruth and others},
  journal={Behavior research methods},
  pages={1--7},
  year={2023},
  publisher={Springer}
}

@article{dmello2022comp,
  title={Psychological measurement in the information age: Machine-learned computational models},
  author={D’mello, Sidney K and Tay, Louis and Southwell, Rosy},
  journal={Current Directions in Psychological Science},
  volume={31},
  number={1},
  pages={76--87},
  year={2022},
  publisher={Sage Publications Sage CA: Los Angeles, CA}
}

@article{Smallwood.2004,
 author = {Smallwood, Jonathan and Davies, John B. and Heim, Derek and Finnigan, Frances and Sudberry, Megan and O'Connor, Rory and Obonsawin, Marc},
 year = {2004},
 title = {Subjective experience and the attentional lapse: task engagement and disengagement during sustained attention},
 pages = {657--690},
 volume = {13},
 number = {4},
 issn = {1053-8100},
 journal = {Consciousness and Cognition},
 doi = {10.1016/j.concog.2004.06.003}
}

@article{Smallwood.2011,
 author = {Smallwood, Jonathan},
 year = {2011},
 title = {Mind-wandering While Reading: Attentional Decoupling, Mindless Reading and the Cascade Model of Inattention},
 pages = {63--77},
 volume = {5},
 number = {2},
 issn = {1749-818X},
 journal = {Language and Linguistics Compass},
 doi = {10.1111/j.1749-818X.2010.00263.x}
}

@article{smilek2010,
  title={Out of mind, out of sight: Eye blinking as indicator and embodiment of mind wandering},
  author={Smilek, Daniel and Carriere, Jonathan SA and Cheyne, J Allan},
  journal={Psychological science},
  volume={21},
  number={6},
  pages={786--789},
  year={2010},
  publisher={Sage Publications Sage CA: Los Angeles, CA}
}

@article{Seli.2016,
 author = {Seli, Paul and Wammes, Jeffrey D. and Risko, Evan F. and Smilek, Daniel},
 year = {2016},
 title = {On the relation between motivation and retention in educational contexts: The role of intentional and unintentional mind wandering},
 pages = {1280--1287},
 volume = {23},
 number = {4},
 issn = {1531-5320},
 journal = {Psychonomic Bulletin {\&} Review},
 doi = {10.3758/s13423-015-0979-0}
}

@article{Seli.2013,
 author = {Seli, Paul and Carriere, Jonathan S. A. and Levene, Merrick and Smilek, Daniel},
 year = {2013},
 title = {How few and far between? Examining the effects of probe rate on self-reported mind wandering},
 pages = {430},
 volume = {4},
 issn = {1664-1078},
 journal = {Frontiers in Psychology},
 doi = {10.3389/fpsyg.2013.00430 }
}

@article{jack2012facial,
  title={Facial expressions of emotion are not culturally universal},
  author={Jack, Rachael E and Garrod, Oliver GB and Yu, Hui and Caldara, Roberto and Schyns, Philippe G},
  journal={Proceedings of the National Academy of Sciences},
  volume={109},
  number={19},
  pages={7241--7244},
  year={2012},
  publisher={National Acad Sciences}
}

@ARTICLE{Li2020deepfer,
  author={Li, Shan and Deng, Weihong},
  journal={IEEE Transactions on Affective Computing}, 
  title={Deep Facial Expression Recognition: A Survey}, 
  year={2020},
  volume={},
  number={},
  pages={1-1},
  doi={10.1109/TAFFC.2020.2981446}}

@ARTICLE{ali2020artificial,

  author={Ali, Ghulam and Ali, Amjad and Ali, Farman and Draz, Umar and Majeed, Fiaz and Yasin, Sana and Ali, Tariq and Haider, Noman},

  journal={IEEE Access}, 

  title={Artificial Neural Network Based Ensemble Approach for Multicultural Facial Expressions Analysis}, 

  year={2020},

  volume={8},

  number={},

  pages={134950-134963},

  doi={10.1109/ACCESS.2020.3009908  }}

@article{dailey2010evidence,
  title={Evidence and a computational explanation of cultural differences in facial expression recognition.},
  author={Dailey, Matthew N and Joyce, Carrie and Lyons, Michael J and Kamachi, Miyuki and Ishi, Hanae and Gyoba, Jiro and Cottrell, Garrison W},
  journal={Emotion},
  volume={10},
  number={6},
  pages={874},
  year={2010},
  publisher={American Psychological Association}
}

@article{Schooler.2002,
 author = {Schooler, Jonathan W.},
 year = {2002},
 title = {Re-representing consciousness: dissociations between experience and meta-consciousness},
 pages = {339--344},
 volume = {6},
 number = {8},
 issn = {1364-6613},
 journal = {Trends in Cognitive Sciences},
 doi = {10.1016/S1364-6613(02)01949-6 }
}

@article{Smallwood.2007c,
 author = {Smallwood, Jonathan and O'Connor, Rory C. and Sudbery, Megan V. and Obonsawin, Marc},
 year = {2007},
 title = {Mind-wandering and dysphoria},
 pages = {816--842},
 volume = {21},
 number = {4},
 issn = {1464-0600},
 journal = {Cognition {\&} Emotion},
 doi = {10.1080/02699930600911531}
}

@article{Weinstein.2018,
 author = {Weinstein, Yana and de Lima, Henry J. and {van der Zee}, Tim},
 year = {2018},
 title = {Are you mind-wandering, or is your mind on task? The effect of probe framing on mind-wandering reports},
 pages = {754--760},
 volume = {25},
 number = {2},
 issn = {1531-5320},
 journal = {Psychonomic Bulletin {\&} Review},
 doi = {10.3758/s13423-017-1322-8  }
}

@article{rawat2017deep,
  title={Deep convolutional neural networks for image classification: A comprehensive review},
  author={Rawat, Waseem and Wang, Zenghui},
  journal={Neural computation},
  volume={29},
  number={9},
  pages={2352--2449},
  year={2017},
  publisher={MIT Press}
}

@article{Wammes.2016,
 author = {Wammes, Jeffrey D. and Seli, Paul and Cheyne, J. Allan and Boucher, Pierre O. and Smilek, Daniel},
 year = {2016},
 title = {Mind wandering during lectures II: Relation to academic performance},
 pages = {33--48},
 volume = {2},
 number = {1},
 issn = {2332-2101},
 journal = {Scholarship of Teaching and Learning in Psychology},
 doi = {10.1037/stl0000055 }
}

@article{VaraoSousa.2019,
 author = {Varao-Sousa, Trish L. and Kingstone, Alan},
 year = {2019},
 title = {Are mind wandering rates an artifact of the probe-caught method? Using self-caught mind wandering in the classroom to test, and reject, this possibility},
 pages = {235--242},
 volume = {51},
 number = {1},
 issn = {1554-3528},
 journal = {Behavior Research Methods},
 doi = {10.3758/s13428-018-1073-0 }
}

@article{Smallwood.2015,
 author = {Smallwood, Jonathan and Schooler, Jonathan W.},
 year = {2015},
 title = {The science of mind wandering: empirically navigating the stream of consciousness},
 pages = {487--518},
 volume = {66},
 journal = {Annual review of psychology},
 doi = {10.1146/annurev-psych-010814-015331}
}

@article{Hutt.2019,
 author = {Hutt, Stephen and Krasich, Kristina and Mills, Caitlin and Bosch, Nigel and White, Shelby and Brockmole, James R. and D'Mello, Sidney K.},
 year = {2019},
 title = {Automated gaze-based mind wandering detection during computerized learning in classrooms},
 pages = {821--867},
 volume = {29},
 number = {4},
 issn = {1573-1391},
 journal = {User Modeling and User-Adapted Interaction},
 doi = {10.1007/s11257-019-09228-5                }
}

@inproceedings{hutt2017out,
  title={" Out of the Fr-Eye-ing Pan" Towards Gaze-Based Models of Attention during Learning with Technology in the Classroom},
  author={Hutt, Stephen and Mills, Caitlin and Bosch, Nigel and Krasich, Kristina and Brockmole, James and D'mello, Sidney},
  booktitle={Proceedings of the 25th Conference on User Modeling, Adaptation and Personalization},
  pages={94--103},
  year={2017}
}

@article{Hutt.2017,
 author = {Hutt, Stephen and Hardey, Jessica and Bixler, Robert and Stewart, Angela and Risko, Evan and D'Mello, Sidney K.},
 year = {2017},
 title = {Gaze-Based Detection of Mind Wandering during Lecture Viewing},
 journal = {International Educational Data Mining Society}
}

@inproceedings{lime,
  author    = {Marco Tulio Ribeiro and
               Sameer Singh and
               Carlos Guestrin},
  title     = {"Why Should {I} Trust You?": Explaining the Predictions of Any Classifier},
  booktitle = {Proceedings of the 22nd {ACM} {SIGKDD} International Conference on
               Knowledge Discovery and Data Mining, San Francisco, CA, USA, August
               13-17, 2016},
  pages     = {1135--1144},
  year      = {2016},
}

@book{levine1990,
  title={Keeping a Head in School.},
  author={Levine, Melvin D},
  year={1990},
  publisher={Educators Publishing Service, Inc., 75 Moulton St., Cambridge, MA 02138-1104}
}

@article{smallwood2004phys,
  title={Subjective experience and the attentional lapse: Task engagement and disengagement during sustained attention},
  author={Smallwood, Jonathan and Davies, John B and Heim, Derek and Finnigan, Frances and Sudberry, Megan and O'Connor, Rory and Obonsawin, Marc},
  journal={Consciousness and cognition},
  volume={13},
  number={4},
  pages={657--690},
  year={2004},
  publisher={Elsevier}
}

@article{dhawan2020,
  title={Online learning: A panacea in the time of COVID-19 crisis},
  author={Dhawan, Shivangi},
  journal={Journal of educational technology systems},
  volume={49},
  number={1},
  pages={5--22},
  year={2020},
  publisher={Sage Publications Sage CA: Los Angeles, CA}
}

@article{lemay2021,
  title={Transition to online learning during the COVID-19 pandemic},
  author={Lemay, David John and Bazelais, Paul and Doleck, Tenzin},
  journal={Computers in Human Behavior Reports},
  volume={4},
  pages={100130},
  year={2021},
  publisher={Elsevier}
}

@article{nguyen2022,
  title={Emotional Regulation in Synchronous Online Collaborative Learning: A Facial Expression Recognition Study},
  author={Nguyen, Andy and Hong, Yvonne and Dang, Belle and Nguyen, Phuong Thi Bich},
  journal={ICIS 2022 Proceedings.},
  year={2022}
}

@article{ngo2024fer,
  title={Facial Expression Recognition for Examining Emotional Regulation in Synchronous Online Collaborative Learning},
  author={Ngo, Duong and Nguyen, Andy and Dang, Belle and Ngo, Ha},
  journal={International Journal of Artificial Intelligence in Education},
  pages={1--20},
  year={2024},
  publisher={Springer}
}

@article{hutt2023webcam,
  title={Webcam-based eye tracking to detect mind wandering and comprehension errors},
  author={Hutt, Stephen and Wong, Aaron and Papoutsaki, Alexandra and Baker, Ryan S and Gold, Joshua I and Mills, Caitlin},
  journal={Behavior Research Methods},
  pages={1--17},
  year={2023},
  publisher={Springer}
}

@inproceedings{caruso2023,
  title={Do Associations Between Mind Wandering and Learning from Complex Texts Vary by Assessment Depth and Time?},
  author={Caruso, Megan and D'Mello, Sidney},
  booktitle={LAK23: 13th International Learning Analytics and Knowledge Conference},
  pages={230--239},
  year={2023}
}

@misc{jarvela2023,
  title={Advancing SRL Research with Artificial Intelligence},
  author={J{\"a}rvel{\"a}, Sanna and Molenaar, Inge and Nguyen, Andy},
  journal={Computers in Human Behavior},
  pages={107847},
  year={2023},
  publisher={Elsevier}
}

@inproceedings{kuvar2023,
  title={Detecting When the Mind Wanders Off Task in Real-time: An Overview and Systematic Review},
  author={Kuvar, Vishal and Kam, Julia WY and Hutt, Stephen and Mills, Caitlin},
  booktitle={Proceedings of the 25th International Conference on Multimodal Interaction},
  pages={163--173},
  year={2023}
}

@article{bonifacci2022,
  title={The relationship between mind wandering and reading comprehension: A meta-analysis},
  author={Bonifacci, Paola and Viroli, Cinzia and Vassura, Chiara and Colombini, Elisa and Desideri, Lorenzo},
  journal={Psychonomic Bulletin \& Review},
  pages={1--20},
  year={2022},
  publisher={Springer}
}

@article{wong2022,
  title={Task-unrelated thought during educational activities: A meta-analysis of its occurrence and relationship with learning},
  author={Wong, Aaron Y and Smith, Shelby L and McGrath, Catherine A and Flynn, Lauren E and Mills, Caitlin},
  journal={Contemporary Educational Psychology},
  volume={71},
  pages={102098},
  year={2022},
  publisher={Elsevier}
}

@article{Hollis.2016,
 author = {Hollis, R. Benjamin and Was, Christopher A.},
 year = {2016},
 title = {Mind wandering, control failures, and social media distractions in online learning},
 pages = {104--112},
 volume = {42},
 issn = {0959-4752},
 journal = {Learning and Instruction},
 doi = {10.1016/j.    learninstruc.2016.01.007}
}

@article{Feng.2013,
 author = {Feng, Shi and D'Mello, Sidney K and Graesser, Arthur C.},
 year = {2013},
 title = {Mind wandering while reading easy and difficult texts},
 pages = {586--592},
 volume = {20},
 number = {3},
 issn = {1531-5320},
 journal = {Psychonomic Bulletin {\&} Review},
 doi = {10.3758/s13423-012-0367-y   }
}

@article{Faber.2020,
 author = {Faber, Myrthe and Krasich, Kristina and Bixler, Robert E. and Brockmole, James R. and D'Mello, Sidney K.},
 year = {2020},
 title = {The eye-mind wandering link: Identifying gaze indices of mind wandering across tasks},
 pages = {1201--1221},
 volume = {46},
 number = {10},
 issn = {1939-1277},
 journal = {Journal of Experimental Psychology: Human Perception and Performance},
 doi = {10.1037/xhp0000743}
}

@article{Faber.2018,
 author = {Faber, Myrthe and D'Mello, Sidney K.},
 year = {2018},
 title = {How the stimulus influences mind wandering in semantically rich task contexts},
 pages = {35},
 volume = {3},
 number = {1},
 issn = {2365-7464},
 journal = {Cognitive Research: Principles and Implications},
 doi = {10.1186/s41235-018-0129-0   }
}

@article{Faber.2018b,
 author = {Faber, Myrthe and Bixler, Robert and D'Mello, Sidney K.},
 year = {2018},
 title = {An automated behavioral measure of mind wandering during computerized reading},
 pages = {134--150},
 volume = {50},
 number = {1},
 issn = {1554-3528},
 journal = {Behavior Research Methods},
 doi = {10.3758/s13428-017-0857-y    }
}

@article{Conrad.2021,
  title={Measuring mind wandering during online lectures assessed with EEG},
  author={Conrad, Colin and Newman, Aaron},
  journal={Frontiers in Human Neuroscience},
  volume={15},
  pages={697532},
  year={2021},
  publisher={Frontiers Media SA}
}

@article{Christoff.2018,
 author = {Christoff, Kalina and Mills, Caitlin and Andrews-Hanna, Jessica R. and Irving, Zachary C. and Thompson, Evan and Fox, Kieran C. R. and Kam, Julia W. Y.},
 year = {2018},
 title = {Mind-Wandering as a Scientific Concept: Cutting through the Definitional Haze},
 pages = {957--959},
 volume = {22},
 number = {11},
 issn = {1879-307X},
 journal = {Trends in Cognitive Sciences},
 doi = {10.1016/j.tics.2018.07.004}
}

@article{Christoff.2009,
 author = {Christoff, Kalina and Gordon, Alan M. and Smallwood, Jonathan and Smith, Rachelle and Schooler, Jonathan W.},
 year = {2009},
 title = {Experience sampling during fMRI reveals default network and executive system contributions to mind wandering},
 pages = {8719--8724},
 volume = {106},
 number = {21},
 issn = {1091-6490},
 journal = {Proceedings of the National Academy of Sciences},
 doi = {10.1073/pnas.0900234106}
}

@article{Brishtel.2020,
 author = {Brishtel, Iuliia and Khan, Anam Ahmad and Schmidt, Thomas and Dingler, Tilman and Ishimaru, Shoya and Dengel, Andreas},
 year = {2020},
 title = {Mind Wandering in a Multimodal Reading Setting: Behavior Analysis {\&} Automatic Detection Using Eye-Tracking and an EDA Sensor},
 pages = {2546},
 volume = {20},
 number = {9},
 journal = {Sensors (Basel, Switzerland)},
 doi = {10.3390/s20092546     }   
}

@article{Jang.2020,
 author = {Jang, DongMin and Yang, IlHo and Kim, SeoungUn},
 year = {2020},
 title = {Detecting Mind-Wandering from Eye Movement and Oculomotor Data during Learning Video Lecture},
 pages = {51},
 volume = {10},
 number = {3},
 journal = {Education Sciences},
 doi = {10.3390/educsci10030051}
}

@article{Jing.2016,
 author = {Jing, Helen G. and Szpunar, Karl K. and Schacter, Daniel L.},
 year = {2016},
 title = {Interpolated testing influences focused attention and improves integration of information during a video-recorded lecture},
 pages = {305--318},
 volume = {22},
 number = {3},
 issn = {1939-2192},
 journal = {Journal of experimental psychology. Applied},
 doi = {10.1037/xap0000087 }
}

@inproceedings{Chen.2016,
 author = {Chen, Tianqi and Guestrin, Carlos},
 title = {{XGBoost}: A Scalable Tree Boosting System},
 booktitle = {Proceedings of the 22nd ACM SIGKDD International Conference on Knowledge Discovery and Data Mining},
 series = {KDD '16},
 year = {2016},
 isbn = {978-1-4503-4232-2 },
 location = {San Francisco, California, USA},
 pages = {785--794},
 numpages = {10},
 doi = {10.1145/2939672.2939785 },
 acmid = {2939785},
 publisher = {ACM},
 address = {New York, NY, USA},
}

@article{franklin2013,
    author = {Michael S. Franklin and James M. Broadway and Michael D. Mrazek and Jonathan Smallwood and Jonathan W. Schooler},
    title ={Window to the Wandering Mind: Pupillometry of Spontaneous Thought While Reading},
    journal = {Quarterly Journal of Experimental Psychology},
    volume = {66},
    number = {12},
    pages = {2289-2294},
    year = {2013},
    doi = {10.1080/17470218.2013.858170 },
    note ={PMID: 24313285      }
}

@article{randall2014,
  title={Mind-wandering, cognition, and performance: a theory-driven meta-analysis of attention regulation.},
  author={Randall, Jason G and Oswald, Frederick L and Beier, Margaret E},
  journal={Psychological bulletin},
  volume={140},
  number={6},
  pages={1411},
  year={2014},
  publisher={American Psychological Association}
}

@article{schubert2020validity,
  title={The validity of the online thought-probing procedure of mind wandering is not threatened by variations of probe rate and probe framing},
  author={Schubert, Anna-Lena and Frischkorn, Gidon T and Rummel, Jan},
  journal={Psychological research},
  volume={84},
  number={7},
  pages={1846--1856},
  year={2020},
  publisher={Springer}
}

@article{Mills.2020,
 author = {Mills, Caitlin and Gregg, Julie and Bixler, Robert and D'Mello, Sidney K.},
 year = {2020},
 title = {Eye-Mind reader: an intelligent reading interface that promotes long-term comprehension by detecting and responding to mind wandering},
 pages = {1--27},
 issn = {0737-0024},
 journal = {Human--Computer Interaction},
 doi = {10.1080/07370024.2020.1716762  }
}

@article{conrad2021,
  title={Measuring mind wandering during online lectures assessed with EEG},
  author={Conrad, Colin and Newman, Aaron},
  journal={Frontiers in Human Neuroscience},
  pages={455},
  year={2021},
  publisher={Frontiers}
}

@article{dhindsa2019,
  title={Individualized pattern recognition for detecting mind wandering from EEG during live lectures},
  author={Dhindsa, Kiret and Acai, Anita and Wagner, Natalie and Bosynak, Dan and Kelly, Stephen and Bhandari, Mohit and Petrisor, Brad and Sonnadara, Ranil R},
  journal={PloS one},
  volume={14},
  number={9},
  pages={e0222276},
  year={2019},
  publisher={Public Library of Science San Francisco, CA USA}
}

@article{Zhang.2020,
 author = {Zhang, Han and Miller, Kevin F. and Sun, Xin and Cortina, Kai S.},
 year = {2020},
 title = {Wandering eyes: Eye movements during mind wandering in video lectures},
 pages = {449--464},
 volume = {34},
 number = {2},
 issn = {0888-4080},
 journal = {Applied Cognitive Psychology},
 doi = {10.1002/acp.3632}
}

@article{Lindquist.2011,
 author = {Lindquist, Sophie I. and McLean, John P.},
 year = {2011},
 title = {Daydreaming and its correlates in an educational environment},
 pages = {158--167},
 volume = {21},
 number = {2},
 issn = {1041-6080},
 journal = {Learning and Individual Differences},
 doi = {10.1016/j.lindif.2010.12.006}
}

@inproceedings{Zhao.2017,
 author = {Zhao, Yue and Lofi, Christoph and Hauff, Claudia},
 title = {Scalable Mind-Wandering Detection for MOOCs: A Webcam-Based Approach},
 pages = {330--344},
 publisher = {{Springer International Publishing}},
 isbn = {978-3-319-66610-5},
 series = {Lecture Notes in Computer Science},
 editor = {Lavou{\'e}, {\'E}lise and Drachsler, Hendrik and Verbert, Katrien and Broisin, Julien and P{\'e}rez-Sanagust{\'i}n, Mar},
 booktitle = {Data Driven Approaches in Digital Education},
 year = {2017},
 address = {Cham}
}

@inproceedings{Lee.2022,
  title={Predicting Mind-Wandering with Facial Videos in Online Lectures},
  author={Lee, Taeckyung and Kim, Dain and Park, Sooyoung and Kim, Dongwhi and Lee, Sung-Ju},
  booktitle={Proceedings of the IEEE/CVF Conference on Computer Vision and Pattern Recognition (CVPR) Workshops},
  pages={2104--2113},
  year={2022}
}

@InProceedings{He_2016_CVPR_resnet50,
author = {He, Kaiming and Zhang, Xiangyu and Ren, Shaoqing and Sun, Jian},
title = {Deep Residual Learning for Image Recognition},
booktitle = {Proceedings of the IEEE Conference on Computer Vision and Pattern Recognition (CVPR)},
year = {2016},
publisher = {IEEE},
pages = {770--778},
doi = {10.1109/CVPR.2016.90}
}

@Article{Cortes.1995,
author={Cortes, Corinna
and Vapnik, Vladimir},
title={Support-vector networks},
journal={Machine Learning},
year={1995},
month={Sep},
day={01},
volume={20},
number={3},
pages={273-297},
issn={1573-0565},
doi={10.1007/BF00994018 }
}

@article{Chawla.2002,
author = {Nitesh V. Chawla
          and Kevin W. Bowyer
          and Lawrence O. Hall
          and W. Philip Kegelmeyer},
doi = {10.1613/jair.953 },
year = 2002,
month = {jun},
publisher = {{AI} Access Foundation},
volume = {16},
pages = {321--357},
title = {{SMOTE}: Synthetic Minority Over-sampling Technique},
journal = {Journal of Artificial Intelligence Research}
}

@inproceedings{Erkin.2009,
  title={Privacy-preserving face recognition},
  author={Erkin, Zekeriya and Franz, Martin and Guajardo, Jorge and Katzenbeisser, Stefan and Lagendijk, Inald and Toft, Tomas},
  booktitle={Privacy Enhancing Technologies: 9th International Symposium, PETS 2009},
  pages={235--253},
  year={2009},
  organization={Springer}
}

@article{Bozkir.2020,
author = {Bozkir, Efe and \"{U}nal, Ali Burak and Akg\"{u}n, Mete and Kasneci, Enkelejda and Pfeifer, Nico},
title = {Privacy Preserving Gaze Estimation Using Synthetic Images via a Randomized Encoding Based Framework},
year = {2020},
publisher = {ACM},
address = {New York, NY, USA},
doi = {10.1145/3379156.3391364 },
booktitle = {ACM Symposium on Eye Tracking Research and Applications},
pages = {21:1--21:5}
}

@inproceedings{Suemer.2020,
author = {S{\"u}mer, {\"O}mer and Gerjets, Peter and Trautwein, Ulrich and Kasneci, Enkelejda},
title = {Automated Anonymisation of Visual and Audio Data in Classroom Studies},
year = {2020},
doi = {10.48550/arxiv.2001.05080 },
booktitle = {The Workshops of the Thirty-Fourth AAAI Conference on Artificial Intelligence}
}

@article{Li.2020,
  author={Li, Tian and Sahu, Anit Kumar and Talwalkar, Ameet and Smith, Virginia},
  journal={IEEE Signal Processing Magazine}, 
  title={Federated Learning: Challenges, Methods, and Future Directions}, 
  year={2020},
  volume={37},
  number={3},
  pages={50--60},
  doi={10.1109/MSP.2020.2975749 }
}

@article{mcvay2012wmc,
  title={Why does working memory capacity predict variation in reading comprehension? On the influence of mind wandering and executive attention.},
  author={McVay, Jennifer C and Kane, Michael J},
  journal={Journal of experimental psychology: general},
  volume={141},
  number={2},
  pages={302},
  year={2012},
  publisher={American Psychological Association}
}

@article{bixler2016automatic,
  title={Automatic gaze-based user-independent detection of mind wandering during computerized reading},
  author={Bixler, Robert and D’Mello, Sidney},
  journal={User Modeling and User-Adapted Interaction},
  volume={26},
  pages={33--68},
  year={2016},
  publisher={Springer}
}

@article{khosravi2024,
  title={Exploring the Elusive Mind: A Multimodal Wearable Sensor Solution for Measuring Mind Wandering in University Students},
  author={Khosravi, Sara and Li, Haobo and Khan, Ahsan Raza and Zoha, Ahmad and Ghannam, Rami},
  journal={Advanced Sensor Research},
  volume={3},
  number={1},
  pages={2300067},
  year={2024},
  publisher={Wiley Online Library}
}

@inproceedings{jeni13,

  author={Jeni, László A. and Cohn, Jeffrey F. and De La Torre, Fernando},

  booktitle={2013 Humaine Association Conference on Affective Computing and Intelligent Interaction}, 

  title={Facing Imbalanced Data--Recommendations for the Use of Performance Metrics}, 

  year={2013},

  volume={},

  number={},

  pages={245-251},

  doi={10.1109/ACII.2013.47   }}

@inproceedings{bixler2015gazeeda,
author = {Bixler, Robert and Blanchard, Nathaniel and Garrison, Luke and D'Mello, Sidney},
title = {Automatic Detection of Mind Wandering During Reading Using Gaze and Physiology},
year = {2015},
isbn = {9781450339124},
publisher = {Association for Computing Machinery},
address = {New York, NY, USA},
url = {https://doi.org/10.1145/2818346.2820742}               , 
doi = {10.1145/2818346.2820742},
booktitle = {Proceedings of the 2015 ACM on International Conference on Multimodal Interaction},
pages = {299–306},
numpages = {8},
location = {Seattle, Washington, USA},
series = {ICMI '15}
}

@inbook{winnie1998,
    author = {Winne, P. H. \&  Hadwin, A.},
    title = {Studying as Self-Regulated Learning},
    booktitle = {Metacognition in Educational Theory and Practice},
    editor = { Hacker, D. J. \&  Dunlosky, J. },
    pages = {277-304},
    publisher = {Lawrence Erlbaum} ,
    address = {Hillsdale, NJ},
    year = {1998} 
}

@article{schoen1970,
  title={Use of consciousness sampling to study teaching methods},
  author={Schoen, James R},
  journal={The Journal of Educational Research},
  volume={63},
  number={9},
  pages={387--390},
  year={1970},
  publisher={Taylor \& Francis}
}

@incollection{warm2018,
  title={Vigilance and workload in automated systems},
  author={Warm, Joel S and Dember, William N and Hancock, Peter A},
  booktitle={Automation and human performance},
  pages={183--200},
  year={2018},
  publisher={CRC Press}
}

@article{warm2008,
  title={Vigilance requires hard mental work and is stressful},
  author={Warm, Joel S and Parasuraman, Raja and Matthews, Gerald},
  journal={Human factors},
  volume={50},
  number={3},
  pages={433--441},
  year={2008},
  publisher={SAGE Publications Sage CA: Los Angeles, CA}
}

@article{locke1974,
  title={Thought sampling: A study of student attention through self-report},
  author={Locke, Lawrence F and Jensen, Mary K},
  journal={Research Quarterly. American Alliance for Health, Physical Education and Recreation},
  volume={45},
  number={3},
  pages={263--275},
  year={1974},
  publisher={Taylor \& Francis}
}

@article{smallwood2008,
  title={When attention matters: The curious incident of the wandering mind},
  author={Smallwood, Jonathan and McSpadden, Merrill and Schooler, Jonathan W},
  journal={Memory \& cognition},
  volume={36},
  pages={1144--1150},
  year={2008},
  publisher={Springer}
}

@article{jefferies2008,
  title={Emotional valence and arousal interact in attentional control},
  author={Jefferies, Lisa N and Smilek, Daniel and Eich, Eric and Enns, James T},
  journal={Psychological science},
  volume={19},
  number={3},
  pages={290--295},
  year={2008},
  publisher={SAGE Publications Sage CA: Los Angeles, CA}
}

@article{mills2021,
  title={Eye-mind reader: An intelligent reading interface that promotes long-term comprehension by detecting and responding to mind wandering},
  author={Mills, Caitlin and Gregg, Julie and Bixler, Robert and D’Mello, Sidney K.},
  journal={Human--Computer Interaction},
  volume={36},
  number={4},
  pages={306--332},
  year={2021},
  publisher={Taylor \& Francis}
}

@book{begaze,
    author = {{Sensomotoric Instruments}},
    title = {BeGaze Manual, Version 3.7},
    year = {2017}
}

@article{scheiter2014,
  title={Distraction during learning with hypermedia: difficult tasks help to keep task goals on track},
  author={Scheiter, Katharina and Gerjets, Peter and Heise, Elke},
  journal={Frontiers in Psychology},
  volume={5},
  pages={268},
  year={2014},
  publisher={Frontiers Media SA}
}

@book{cohen2014neuropsychology,
  title={The Neuropsychology of Attention},
  author={Cohen, R.A.},
  isbn={9780387726397},
  lccn={2013941376},
  series={Critical issues in neuropsychology},
  year={2014},
  publisher={Springer US}
}

@article{posner2007research,
  title={Research on attention networks as a model for the integration of psychological science},
  author={Posner, Michael I and Rothbart, Mary K},
  journal={Annu. Rev. Psychol.},
  volume={58},
  pages={1--23},
  year={2007},
  publisher={Annual Reviews}
}

@article{corbetta2002control,
  title={Control of goal-directed and stimulus-driven attention in the brain},
  author={Corbetta, Maurizio and Shulman, Gordon L},
  journal={Nature reviews neuroscience},
  volume={3},
  number={3},
  pages={201--215},
  year={2002},
  publisher={Nature Publishing Group UK London}
}

@article{dmello2016giving,
  title={Giving eyesight to the blind: Towards attention-aware AIED},
  author={D’Mello, Sidney K},
  journal={International Journal of Artificial Intelligence in Education},
  volume={26},
  pages={645--659},
  year={2016},
  publisher={Springer}
}

@article{mcdowd2007,
  title={An overview of attention: behavior and brain},
  author={McDowd, Joan M},
  journal={Journal of Neurologic Physical Therapy},
  volume={31},
  number={3},
  pages={98--103},
  year={2007},
  publisher={LWW}
}

@article{partala2003,
  title={Pupil size variation as an indication of affective processing},
  author={Partala, Timo and Surakka, Veikko},
  journal={International journal of human-computer studies},
  volume={59},
  number={1-2},
  pages={185--198},
  year={2003},
  publisher={Elsevier}
}

@incollection{mccarthy2018,
  title={Methods of studying text: Memory, comprehension, and learning},
  author={McCarthy, Kathryn S and Kopp, Kristopher J and Allen, Laura K and McNamara, Danielle S},
  booktitle={Handbook of research methods in human memory},
  pages={104--124},
  year={2018},
  publisher={Routledge}
}

@article{smote2002,
  title={SMOTE: synthetic minority over-sampling technique},
  author={Chawla, Nitesh V and Bowyer, Kevin W and Hall, Lawrence O and Kegelmeyer, W Philip},
  journal={Journal of artificial intelligence research},
  volume={16},
  pages={321--357},
  year={2002}
}

@article{hutt2016eyes,
  title={The Eyes Have It: Gaze-Based Detection of Mind Wandering during Learning with an Intelligent Tutoring System.},
  author={Hutt, Stephen and Mills, Caitlin and White, Shelby and Donnelly, Patrick J and D'Mello, Sidney K},
  journal={International Educational Data Mining Society},
  year={2016},
  publisher={ERIC}
}

@inproceedings{hutt2021,
  title={Breaking out of the lab: Mitigating mind wandering with gaze-based attention-aware technology in classrooms},
  author={Hutt, Stephen and Krasich, Kristina and R. Brockmole, James and K. D'Mello, Sidney},
  booktitle={Proceedings of the 2021 CHI Conference on Human Factors in Computing Systems},
  pages={1--14},
  year={2021}
}

@article{baird2014decoupled,
  title={The decoupled mind: mind-wandering disrupts cortical phase-locking to perceptual events},
  author={Baird, Benjamin and Smallwood, Jonathan and Lutz, Antoine and Schooler, Jonathan W},
  journal={Journal of Cognitive Neuroscience},
  volume={26},
  number={11},
  pages={2596--2607},
  year={2014},
  publisher={MIT Press One Rogers Street, Cambridge, MA 02142-1209, USA journals-info~…}
}

@article{roscher2020,
  title={Explainable machine learning for scientific insights and discoveries},
  author={Roscher, Ribana and Bohn, Bastian and Duarte, Marco F and Garcke, Jochen},
  journal={Ieee Access},
  volume={8},
  pages={42200--42216},
  year={2020},
  publisher={IEEE}
}

@article{abbott1986,
  title={Optimal matching methods for historical sequences},
  author={Abbott, Andrew and Forrest, John},
  journal={The Journal of Interdisciplinary History},
  volume={16},
  number={3},
  pages={471--494},
  year={1986},
  publisher={JSTOR}
}

@article{reichle2012using,
  title={Using EZ Reader to simulate eye movements in nonreading tasks: A unified framework for understanding the eye--mind link.},
  author={Reichle, Erik D and Pollatsek, Alexander and Rayner, Keith},
  journal={Psychological review},
  volume={119},
  number={1},
  pages={155},
  year={2012},
  publisher={American Psychological Association}
}

@article{rayner1998eye,
  title={Eye movements in reading and information processing: 20 years of research.},
  author={Rayner, Keith},
  journal={Psychological bulletin},
  volume={124},
  number={3},
  pages={372},
  year={1998},
  publisher={American Psychological Association}
}

@article{mills2016,
  title={Automatic Gaze-Based Detection of Mind Wandering during Narrative Film Comprehension.},
  author={Mills, Caitlin and Bixler, Robert and Wang, Xinyi and D'Mello, Sidney K},
  journal={International Educational Data Mining Society},
  year={2016},
  publisher={ERIC}
}

@book{schiffman2001,
  author = {Schiffman, H.R.},
  title = {Sensation and Perception. An Integrated Approach},
  year = {2001},
  publisher = {John Wiley and Sons},
  adress = {New York}
}

@ARTICLE{castner2020,
  author = {Castner,Nora and Appel, Tobias and Eder, The´re´se and  Richter, Juliane and Scheiter, Katharina and Keutel, Constanze  and Hütting, Fabian and Duchowski, Andrew and Kasneci, Enkelejda},
  title = {Pupil diameter differentiates expertise in dental radiography visual search},
  journal = {PLoS ONE},
  year = {2020},
  volume = {15},
  number = {5},
  month = May,
  doi = {10.1371/journal.pone.0223941 }
}

@book{frank2014,
  title={Presence messen in laborbasierter Forschung mit Mikrowelten: Entwicklung und erste Validierung eines Fragebogens zur Messung von Presence},
  author={Frank, Barbara},
  year={2014},
  publisher={Springer-Verlag}
}

@article{smallwood2006,
  title={The restless mind.},
  author={Smallwood, Jonathan and Schooler, Jonathan W},
  journal={Psychological bulletin},
  volume={132},
  number={6},
  pages={946},
  year={2006},
  publisher={American Psychological Association}
}

@article{kane2012,
  title={What mind wandering reveals about executive-control abilities and failures},
  author={Kane, Michael J and McVay, Jennifer C},
  journal={Current Directions in Psychological Science},
  volume={21},
  number={5},
  pages={348--354},
  year={2012},
  publisher={Sage Publications Sage CA: Los Angeles, CA}
}

@article{miyake2000,
  title={The unity and diversity of executive functions and their contributions to complex “frontal lobe” tasks: A latent variable analysis},
  author={Miyake, Akira and Friedman, Naomi P and Emerson, Michael J and Witzki, Alexander H and Howerter, Amy and Wager, Tor D},
  journal={Cognitive psychology},
  volume={41},
  number={1},
  pages={49--100},
  year={2000},
  publisher={Elsevier}
}

@article{friedman2017,
  title={Unity and diversity of executive functions: Individual differences as a window on cognitive structure},
  author={Friedman, Naomi P and Miyake, Akira},
  journal={Cortex},
  volume={86},
  pages={186--204},
  year={2017},
  publisher={Elsevier}
}

@article{krumpe2018,
  title={Unity and diversity in working memory load: Evidence for the separability of the executive functions updating and inhibition using machine learning},
  author={Krumpe, Tanja and Scharinger, Christian and Rosenstiel, Wolfgang and Gerjets, Peter and Sp{\"u}ler, Martin},
  journal={Biological psychology},
  volume={139},
  pages={163--172},
  year={2018},
  publisher={Elsevier}
}

@article{broadway2010,
  title={Working memory capacity: self-control is (in) the goal},
  author={Broadway, James M and Redick, Thomas S and Engle, Randall W},
  journal={Self control in society, mind, and brain},
  volume={1},
  pages={163--174},
  year={2010},
  publisher={Oxford University Press}
}

@article{titz2014,
  title={Working memory and executive functions: effects of training on academic achievement},
  author={Titz, Cora and Karbach, Julia},
  journal={Psychological research},
  volume={78},
  pages={852--868},
  year={2014},
  publisher={Springer}
}

@article{rutherford2018,
  title={Links between achievement, executive functions, and self-regulated learning},
  author={Rutherford, Teomara and Buschkuehl, Martin and Jaeggi, Susanne M and Farkas, George},
  journal={Applied Cognitive Psychology},
  volume={32},
  number={6},
  pages={763--774},
  year={2018},
  publisher={Wiley Online Library}
}

@article{debruin2017,
  title={Bridging cognitive load and self-regulated learning research: A complementary approach to contemporary issues in educational research},
  author={de Bruin, Anique BH and van Merri{\"e}nboer, Jeroen JG},
  journal={Learning and Instruction},
  volume={51},
  pages={1--9},
  year={2017},
  publisher={Elsevier}
}

@article{jansen2019,
  title={Self-regulated learning partially mediates the effect of self-regulated learning interventions on achievement in higher education: A meta-analysis},
  author={Jansen, Ren{\'e}e S and Van Leeuwen, Anouschka and Janssen, Jeroen and Jak, Suzanne and Kester, Liesbeth},
  journal={Educational Research Review},
  volume={28},
  pages={100292},
  year={2019},
  publisher={Elsevier}
}

@article{dent2016,
  title={The relation between self-regulated learning and academic achievement across childhood and adolescence: A meta-analysis},
  author={Dent, Amy L and Koenka, Alison C},
  journal={Educational Psychology Review},
  volume={28},
  pages={425--474},
  year={2016},
  publisher={Springer}
}

@article{efklides2011,
  title={Interactions of metacognition with motivation and affect in self-regulated learning: The MASRL model},
  author={Efklides, Anastasia},
  journal={Educational psychologist},
  volume={46},
  number={1},
  pages={6--25},
  year={2011},
  publisher={Taylor \& Francis}
}

@article{greene2007,
  title={A theoretical review of Winne and Hadwin’s model of self-regulated learning: New perspectives and directions},
  author={Greene, Jeffrey Alan and Azevedo, Roger},
  journal={Review of educational research},
  volume={77},
  number={3},
  pages={334--372},
  year={2007},
  publisher={Sage Publications}
}

@article{boekaerts1996,
  title={Self-regulated learning at the junction of cognition and motivation},
  author={Boekaerts, Monique},
  journal={European psychologist},
  volume={1},
  number={2},
  pages={100--112},
  year={1996},
  publisher={Hogrefe \& Huber Publishers}
}

@article{pintrich2004,
  title={A conceptual framework for assessing motivation and self-regulated learning in college students},
  author={Pintrich, Paul R},
  journal={Educational psychology review},
  volume={16},
  pages={385--407},
  year={2004},
  publisher={Springer}
}

@article{schunk2017,
  title={Handbook of self-regulation of learning and performance  (2nd ed.)},
  editor={Schunk, D H and Greene, Jeffrey A},
  year={2017},
  publisher={Routledge}
}

@article{seli2017,
  title={Intentionality and meta-awareness of mind wandering: Are they one and the same, or distinct dimensions?},
  author={Seli, Paul and Ralph, Brandon CW and Risko, Evan F and W Schooler, Jonathan and Schacter, Daniel L and Smilek, Daniel},
  journal={Psychonomic bulletin \& review},
  volume={24},
  pages={1808--1818},
  year={2017},
  publisher={Springer}
}

@article{pekrun2017,
  title={Measuring emotions during epistemic activities: The epistemically-related emotion scales},
  author={Pekrun, Reinhard and Vogl, Elisabeth and Muis, Krista R and Sinatra, Gale M},
  journal={Cognition and Emotion},
  volume={31},
  number={6},
  pages={1268--1276},
  year={2017},
  publisher={Taylor \& Francis}
}

@article{knogler2015,
  title={How situational is situational interest? Investigating the longitudinal structure of situational interest},
  author={Knogler, Maximilian and Harackiewicz, Judith M and Gegenfurtner, Andreas and Lewalter, Doris},
  journal={Contemporary Educational Psychology},
  volume={43},
  pages={39--50},
  year={2015},
  publisher={Elsevier}
}

@article{rimm2015,
  title={To what extent do teacher--student interaction quality and student gender contribute to fifth graders’ engagement in mathematics learning?},
  author={Rimm-Kaufman, Sara E and Baroody, Alison E and Larsen, Ross AA and Curby, Timothy W and Abry, Tashia},
  journal={Journal of educational psychology},
  volume={107},
  number={1},
  pages={170},
  year={2015},
  publisher={American Psychological Association}
}

@article{mrazek2013,
  title={Young and restless: validation of the Mind-Wandering Questionnaire (MWQ) reveals disruptive impact of mind-wandering for youth},
  author={Mrazek, Michael D and Phillips, Dawa T and Franklin, Michael S and Broadway, James M and Schooler, Jonathan W},
  journal={Frontiers in psychology},
  volume={4},
  pages={560},
  year={2013},
  publisher={Frontiers Media SA}
}

@article{marsh2006,
  title={Integration of multidimensional self-concept and core personality constructs: Construct validation and relations to well-being and achievement},
  author={Marsh, Herbert W and Trautwein, Ulrich and L{\"u}dtke, Oliver and K{\"o}ller, Olaf and Baumert, J{\"u}rgen},
  journal={Journal of personality},
  volume={74},
  number={2},
  pages={403--456},
  year={2006},
  publisher={Wiley Online Library}
}

@article{gaspard2017,
  title={Assessing task values in five subjects during secondary school: Measurement structure and mean level differences across grade level, gender, and academic subject},
  author={Gaspard, Hanna and H{\"a}fner, Isabelle and Parrisius, Cora and Trautwein, Ulrich and Nagengast, Benjamin},
  journal={Contemporary Educational Psychology},
  volume={48},
  pages={67--84},
  year={2017},
  publisher={Elsevier}
}

@article{sweller2010,
  title={Element interactivity and intrinsic, extraneous, and germane cognitive load},
  author={Sweller, John},
  journal={Educational psychology review},
  volume={22},
  pages={123--138},
  year={2010},
  publisher={Springer}
}

@ARTICLE{Bosch.2021,
  author={Bosch, Nigel and D'Mello, Sidney K.},
  journal={IEEE Transactions on Affective Computing}, 
  title={Automatic Detection of Mind Wandering from Video in the Lab and in the Classroom}, 
  year={2021},
  volume={12},
  number={4},
  pages={974-988},
  doi={10.1109/TAFFC.2019.2908837}}

@article{barron2011,
  title={Absorbed in thought: The effect of mind wandering on the processing of relevant and irrelevant events},
  author={Barron, Evelyn and Riby, Leigh M and Greer, Joanna and Smallwood, Jonathan},
  journal={Psychological science},
  volume={22},
  number={5},
  pages={596--601},
  year={2011},
  publisher={Sage Publications Sage CA: Los Angeles, CA}
}

@article{kam2011,
  title={Slow fluctuations in attentional control of sensory cortex},
  author={Kam, Julia WY and Dao, Elizabeth and Farley, James and Fitzpatrick, Kevin and Smallwood, Jonathan and Schooler, Jonathan W and Handy, Todd C},
  journal={Journal of cognitive neuroscience},
  volume={23},
  number={2},
  pages={460--470},
  year={2011},
  publisher={MIT Press One Rogers Street, Cambridge, MA 02142-1209, USA journals-info~…}
}

@article{smallwood2008c,
  title={Going AWOL in the brain: Mind wandering reduces cortical analysis of external events},
  author={Smallwood, Jonathan and Beach, Emily and Schooler, Jonathan W and Handy, Todd C},
  journal={Journal of cognitive neuroscience},
  volume={20},
  number={3},
  pages={458--469},
  year={2008},
  publisher={MIT Press One Rogers Street, Cambridge, MA 02142-1209, USA journals-info~…}
}

@article{duncan2005,
  title={The making of the motivated strategies for learning questionnaire.},
  author={Duncan, Teresa Garcia and McKeachie, Wilbert J},
  journal={Educational psychologist},
  volume={40},
  number={2},
  year={2005}
}

@article{pintrich1991manual,
  title={A manual for the use of the Motivated Strategies for Learning Questionnaire (MSLQ).},
  author={Pintrich, Paul R and others},
  year={1991},
  publisher={ERIC}
}

@inproceedings{degreef2009,
  title={Eye movement as indicators of mental workload to trigger adaptive automation},
  author={de Greef, Tjerk and Lafeber, Harmen and van Oostendorp, Herre and Lindenberg, Jasper},
  booktitle={Foundations of Augmented Cognition. Neuroergonomics and Operational Neuroscience: 5th International Conference, FAC 2009 Held as Part of HCI International 2009 San Diego, CA, USA, July 19-24, 2009 Proceedings 5},
  pages={219--228},
  year={2009},
  organization={Springer}
}

@article{breiman2001,
  title={Statistical modeling: The two cultures (with comments and a rejoinder by the author)},
  author={Breiman, Leo},
  journal={Statistical science},
  volume={16},
  number={3},
  pages={199--231},
  year={2001},
  publisher={Institute of Mathematical Statistics}
}

@article{yarkoni2017,
  title={Choosing prediction over explanation in psychology: Lessons from machine learning},
  author={Yarkoni, Tal and Westfall, Jacob},
  journal={Perspectives on Psychological Science},
  volume={12},
  number={6},
  pages={1100--1122},
  year={2017},
  publisher={Sage Publications Sage CA: Los Angeles, CA}
}

@article{toates1974vergence,
  title={Vergence eye movements},
  author={Toates, FM},
  journal={Documenta Ophthalmologica},
  volume={37},
  pages={153--214},
  year={1974},
  publisher={Springer}
}

@inproceedings{huang2019,
  title={Moment-to-moment detection of internal thought during video viewing from eye vergence behavior},
  author={Huang, Michael Xuelin and Li, Jiajia and Ngai, Grace and Leong, Hong Va and Bulling, Andreas},
  booktitle={Proceedings of the 27th ACM International Conference on Multimedia},
  pages={2254--2262},
  year={2019}
}

@article{smallwood2008b,
  title={Segmenting the stream of consciousness: The psychological correlates of temporal structures in the time series data of a continuous performance task},
  author={Smallwood, Jonathan and McSpadden, Merrill and Luus, Bryan and Schooler, Jonathan},
  journal={Brain and cognition},
  volume={66},
  number={1},
  pages={50--56},
  year={2008},
  publisher={Elsevier}
}

@article{pintrich1993,
  title={Reliability and predictive validity of the Motivated Strategies for Learning Questionnaire (MSLQ)},
  author={Pintrich, Paul R and Smith, David AF and Garcia, Teresa and McKeachie, Wilbert J},
  journal={Educational and psychological measurement},
  volume={53},
  number={3},
  pages={801--813},
  year={1993},
  publisher={Sage Publications Sage CA: Thousand Oaks, CA}
}

@article{smallwood2011pupillometric,
  title={Pupillometric evidence for the decoupling of attention from perceptual input during offline thought},
  author={Smallwood, Jonathan and Brown, Kevin S and Tipper, Christine and Giesbrecht, Barry and Franklin, Michael S and Mrazek, Michael D and Carlson, Jean M and Schooler, Jonathan W},
  journal={PloS one},
  volume={6},
  number={3},
  pages={e18298},
  year={2011},
  publisher={Public Library of Science San Francisco, USA}
}

@article{Makowski2021neurokit,
     author = {Dominique Makowski and Tam Pham and Zen J. Lau and Jan C. Brammer and Fran{\c{c}}ois Lespinasse and Hung Pham and Christopher Schölzel and S. H. Annabel Chen},
     title = {{NeuroKit}2: A Python toolbox for neurophysiological signal processing},
     journal = {Behavior Research Methods},
     volume = {53},
     number = {4},
     pages = {1689--1696},
     publisher = {Springer Science and Business Media {LLC}},
     doi = {10.3758/s13428-020-01516-y  },
     url = {https://doi.org/10.3758%2Fs13428-020-01516-y},
     year = 2021,
     month = {feb}
 }

@article{baltruvsaitis2018multimodal,
  title={Multimodal machine learning: A survey and taxonomy},
  author={Baltru{\v{s}}aitis, Tadas and Ahuja, Chaitanya and Morency, Louis-Philippe},
  journal={IEEE transactions on pattern analysis and machine intelligence},
  volume={41},
  number={2},
  pages={423--443},
  year={2018},
  publisher={IEEE}
}

@article{arrieta2020explainable,
  title={Explainable Artificial Intelligence (XAI): Concepts, taxonomies, opportunities and challenges toward responsible AI},
  author={Arrieta, Alejandro Barredo and D{\'\i}az-Rodr{\'\i}guez, Natalia and Del Ser, Javier and Bennetot, Adrien and Tabik, Siham and Barbado, Alberto and Garc{\'\i}a, Salvador and Gil-L{\'o}pez, Sergio and Molina, Daniel and Benjamins, Richard and others},
  journal={Information fusion},
  volume={58},
  pages={82--115},
  year={2020},
  publisher={Elsevier}
}

@inproceedings{bozkir2019assessment,
  title={Assessment of driver attention during a safety critical situation in VR to generate VR-based training},
  author={Bozkir, Efe and Geisler, David and Kasneci, Enkelejda},
  booktitle={ACM Symposium on Applied Perception 2019},
  pages={1--5},
  year={2019}
}

@inproceedings{kasneci2015online,
  title={Online recognition of fixations, saccades, and smooth pursuits for automated analysis of traffic hazard perception},
  author={Kasneci, Enkelejda and Kasneci, Gjergji and K{\"u}bler, Thomas C and Rosenstiel, Wolfgang},
  booktitle={Artificial Neural Networks: Methods and Applications in Bio-/Neuroinformatics},
  pages={411--434},
  year={2015},
  organization={Springer}
}

@article{braunagel2017ready,
  title={Ready for take-over? A new driver assistance system for an automated classification of driver take-over readiness},
  author={Braunagel, Christian and Rosenstiel, Wolfgang and Kasneci, Enkelejda},
  journal={IEEE Intelligent Transportation Systems Magazine},
  volume={9},
  number={4},
  pages={10--22},
  year={2017},
  publisher={IEEE}
}

@article{baldwin2017detecting,
  title={Detecting and quantifying mind wandering during simulated driving},
  author={Baldwin, Carryl L and Roberts, Daniel M and Barragan, Daniela and Lee, John D and Lerner, Neil and Higgins, James S},
  journal={Frontiers in human neuroscience},
  volume={11},
  pages={406},
  year={2017},
  publisher={Frontiers Media SA}
}

@inproceedings{byrne2023leveraging,
  title={Leveraging eye tracking in digital classrooms: A step towards multimodal model for learning assistance},
  author={Byrne, Sean Anthony and Castner, Nora and Kastrati, Ard and P{\l}omecka, Martyna Beata and Schaefer, William and Kasneci, Enkelejda and Bylinskii, Zoya},
  booktitle={Proceedings of the 2023 Symposium on Eye Tracking Research and Applications},
  pages={1--6},
  year={2023}
}

@inproceedings{gao2021digital,
  title={Digital transformations of classrooms in virtual reality},
  author={Gao, Hong and Bozkir, Efe and Hasenbein, Lisa and Hahn, Jens-Uwe and G{\"o}llner, Richard and Kasneci, Enkelejda},
  booktitle={Proceedings of the 2021 CHI Conference on Human Factors in Computing Systems},
  pages={1--10},
  year={2021}
}

@article{khosravi2022explainable,
  title={Explainable artificial intelligence in education},
  author={Khosravi, Hassan and Shum, Simon Buckingham and Chen, Guanliang and Conati, Cristina and Tsai, Yi-Shan and Kay, Judy and Knight, Simon and Martinez-Maldonado, Roberto and Sadiq, Shazia and Ga{\v{s}}evi{\'c}, Dragan},
  journal={Computers and Education: Artificial Intelligence},
  volume={3},
  pages={100074},
  year={2022},
  publisher={Elsevier}
}

@inproceedings{gao2022evaluating,
  title={Evaluating the effects of virtual human animation on students in an immersive vr classroom using eye movements},
  author={Gao, Hong and Hasenbein, Lisa and Bozkir, Efe and G{\"o}llner, Richard and Kasneci, Enkelejda},
  booktitle={Proceedings of the 28th ACM Symposium on Virtual Reality Software and Technology},
  pages={1--11},
  year={2022}
}

@article{gegenfurtner2013transfer,
  title={Transfer of expertise: An eye tracking and think aloud study using dynamic medical visualizations},
  author={Gegenfurtner, Andreas and Sepp{\"a}nen, Marko},
  journal={Computers \& Education},
  volume={63},
  pages={393--403},
  year={2013},
  publisher={Elsevier}
}

@inproceedings{appel2019predicting,
  title={Predicting cognitive load in an emergency simulation based on behavioral and physiological measures},
  author={Appel, Tobias and Sevcenko, Natalia and Wortha, Franz and Tsarava, Katerina and Moeller, Korbinian and Ninaus, Manuel and Kasneci, Enkelejda and Gerjets, Peter},
  booktitle={2019 International Conference on Multimodal Interaction},
  pages={154--163},
  year={2019}
}

@article{borisov2021robust,
  title={Robust cognitive load detection from wrist-band sensors},
  author={Borisov, Vadim and Kasneci, Enkelejda and Kasneci, Gjergji},
  journal={Computers in Human Behavior Reports},
  volume={4},
  pages={100116},
  year={2021},
  publisher={Elsevier}
}

@article{wang2024turbosvm,
  title={TurboSVM-FL: Boosting Federated Learning through SVM Aggregation for Lazy Clients},
  author={Wang, Mengdi and Bodonhelyi, Anna and Bozkir, Efe and Kasneci, Enkelejda},
  journal={arXiv preprint arXiv:2401.12012  } ,
  year={2024}
}

@article{krupinski2006eye,
  title={Eye-movement study and human performance using telepathology virtual slides. Implications for medical education and differences with experience},
  author={Krupinski, Elizabeth A and Tillack, Allison A and Richter, Lynne and Henderson, Jeffrey T and Bhattacharyya, Achyut K and Scott, Katherine M and Graham, Anna R and Descour, Michael R and Davis, John R and Weinstein, Ronald S},
  journal={Human pathology},
  volume={37},
  number={12},
  pages={1543--1556},
  year={2006},
  publisher={Elsevier}
}

@article{yang2013tracking,
  title={Tracking learners' visual attention during a multimedia presentation in a real classroom},
  author={Yang, Fang-Ying and Chang, Chun-Yen and Chien, Wan-Ru and Chien, Yu-Ta and Tseng, Yuen-Hsien},
  journal={Computers \& Education},
  volume={62},
  pages={208--220},
  year={2013},
  publisher={Elsevier}
}

@article{coskun2021investigation,
  title={Investigation of classroom management skills by using eye-tracking technology},
  author={Coskun, Atakan and Cagiltay, Kursat},
  journal={Education and Information Technologies},
  volume={26},
  pages={2501--2522},
  year={2021},
  publisher={Springer}
}

@article{hasenbein2023investigating,
  title={Investigating social comparison behaviour in an immersive virtual reality classroom based on eye-movement data},
  author={Hasenbein, Lisa and Stark, Philipp and Trautwein, Ulrich and Gao, Hong and Kasneci, Enkelejda and G{\"o}llner, Richard},
  journal={Scientific Reports},
  volume={13},
  number={1},
  pages={14672},
  year={2023},
  publisher={Nature Publishing Group UK London}
}

@inproceedings{bozkir2021exploiting,
  title={Exploiting object-of-interest information to understand attention in VR classrooms},
  author={Bozkir, Efe and Stark, Philipp and Gao, Hong and Hasenbein, Lisa and Hahn, Jens-Uwe and Kasneci, Enkelejda and G{\"o}llner, Richard},
  booktitle={2021 IEEE Virtual Reality and 3D User Interfaces (VR)},
  pages={597--605},
  year={2021},
  organization={IEEE}
}

@article{yang2023xai,
  title={Survey on explainable AI: From approaches, limitations and Applications aspects},
  author={Yang, Wenli and Wei, Yuchen and Wei, Hanyu and Chen, Yanyu and Huang, Guan and Li, Xiang and Li, Renjie and Yao, Naimeng and Wang, Xinyi and Gu, Xiaotong and others},
  journal={Human-Centric Intelligent Systems},
  volume={3},
  number={3},
  pages={161--188},
  year={2023},
  publisher={Springer}
}

@article{fredricks2004school,
  title={School engagement: Potential of the concept, state of the evidence},
  author={Fredricks, Jennifer A and Blumenfeld, Phyllis C and Paris, Alison H},
  journal={Review of educational research},
  volume={74},
  number={1},
  pages={59--109},
  year={2004},
  publisher={Sage Publications Sage CA: Thousand Oaks, CA}
}

@article{Seli.2018family,
 author = {Seli, Paul and Kane, Michael J. and Smallwood, Jonathan and Schacter, Daniel L. and Maillet, David and Schooler, Jonathan W. and Smilek, Daniel},
 year = {2018},
 title = {Mind-Wandering as a Natural Kind: A Family-Resemblances View},
 url = {https://www.sciencedirect.com/science/article/pii/S1364661318300718},
 pages = {479--490},
 volume = {22},
 number = {6},
 issn = {1879-307X},
 journal = {Trends in Cognitive Sciences},
 doi = {10.1016/j.tics.2018.03.010    }
}

@article{Seli.2018wellclad,
 author = {Seli, Paul and Kane, Michael J. and Metzinger, Thomas and Smallwood, Jonathan and Schacter, Daniel L. and Maillet, David and Schooler, Jonathan W. and Smilek, Daniel},
 year = {2018},
 title = {The Family-Resemblances Framework for Mind-Wandering Remains Well Clad},
 url = {https://www.cell.com/trends/cognitive-sciences/fulltext/S1364-6613(18)30167-0},
 pages = {959--961},
 volume = {22},
 number = {11},
 issn = {1879-307X},
 journal = {Trends in Cognitive Sciences},
 doi = {10.1016/j.tics.2018.07.007    }
}

@article{Schooler.2011,
 author = {Schooler, Jonathan W. and Smallwood, Jonathan and Christoff, Kalina and Handy, Todd C. and Reichle, Erik D. and Sayette, Michael A.},
 year = {2011},
 title = {Meta-awareness, perceptual decoupling and the wandering mind},
 url = {https://www.sciencedirect.com/science/article/pii/S1364661311000878},
 pages = {319--326},
 volume = {15},
 number = {7},
 issn = {1879-307X},
 journal = {Trends in Cognitive Sciences},
 doi = {10.1016/j.tics.2011.05.006 }
}

@article{Risko.2012,
 author = {Risko, Evan F. and Anderson, Nicola and Sarwal, Amara and Engelhardt, Megan and Kingstone, Alan},
 year = {2012},
 title = {Everyday Attention: Variation in Mind Wandering and Memory in a Lecture},
 pages = {234--242},
 volume = {26},
 number = {2},
 issn = {0888-4080},
 journal = {Applied Cognitive Psychology},
 doi = {10.1002/acp.1814}
}

@article{di2018unobtrusive,
  title={Unobtrusive assessment of students' emotional engagement during lectures using electrodermal activity sensors},
  author={Di Lascio, Elena and Gashi, Shkurta and Santini, Silvia},
  journal={Proceedings of the ACM on Interactive, Mobile, Wearable and Ubiquitous Technologies},
  volume={2},
  number={3},
  pages={1--21},
  year={2018},
  publisher={ACM New York, NY, USA}
}

@article{disalvo2022,
  title={Reading the room: Automated, momentary assessment of student engagement in the classroom: Are we there yet?},
  author={DiSalvo, Betsy and Bandaru, Dheeraj and Wang, Qiaosi and Li, Hong and Pl{\"o}tz, Thomas},
  journal={Proceedings of the ACM on Interactive, Mobile, Wearable and Ubiquitous Technologies},
  volume={6},
  number={3},
  pages={1--26},
  year={2022},
  publisher={ACM New York, NY, USA}
}

@article{zaletelj2017kinect,
  title={Predicting students’ attention in the classroom from Kinect facial and body features},
  author={Zaletelj, Janez and Ko{\v{s}}ir, Andrej},
  journal={EURASIP journal on image and video processing},
  volume={2017},
  pages={1--12},
  year={2017},
  publisher={Springer}
}

@article{gao2020eda,
  title={n-gage: Predicting in-class emotional, behavioural and cognitive engagement in the wild},
  author={Gao, Nan and Shao, Wei and Rahaman, Mohammad Saiedur and Salim, Flora D},
  journal={Proceedings of the ACM on Interactive, Mobile, Wearable and Ubiquitous Technologies},
  volume={4},
  number={3},
  pages={1--26},
  year={2020},
  publisher={ACM New York, NY, USA}
}

@article{vuilleumier2005emotionalattention,
  title={How brains beware: neural mechanisms of emotional attention},
  author={Vuilleumier, Patrik},
  journal={Trends in cognitive sciences},
  volume={9},
  number={12},
  pages={585--594},
  year={2005},
  publisher={Elsevier}
}

@article{pecchinenda1996affective,
  title={The affective significance of skin conductance activity during a difficult problem-solving task},
  author={Pecchinenda, Anna},
  journal={Cognition \& Emotion},
  volume={10},
  number={5},
  pages={481--504},
  year={1996},
  publisher={Taylor \& Francis}
}

@article{Smallwood.2006,
 author = {Smallwood, Jonathan and Schooler, Jonathan W.},
 year = {2006},
 title = {The restless mind},
 pages = {946--958},
 volume = {132},
 number = {6},
 issn = {0033-2909},
 journal = {Psychological Bulletin}
}

@article{Wammes.2017,
 author = {Wammes, Jeffrey D. and Smilek, Daniel},
 year = {2017},
 title = {Examining the Influence of Lecture Format on Degree of Mind Wandering},
 url = {https://www.sciencedirect.com/science/article/pii/S2211368116301413},
 pages = {174--184},
 volume = {6},
 number = {2},
 issn = {2211-3681},
 journal = {Journal of Applied Research in Memory and Cognition},
 doi = {10.1016/j.jarmac.2017.01.015}
}

@article{Szpunar.2013,
 author = {Szpunar, Karl K. and Moulton, Samuel T. and Schacter, Daniel L.},
 year = {2013},
 title = {Mind wandering and education: from the classroom to online learning},
 pages = {495},
 volume = {4},
 issn = {1664-1078},
 journal = {Frontiers in Psychology},
 doi = {10.3389/fpsyg.2013.00495 }
}

@article{Szpunar.2013b,
 author = {Szpunar, Karl K. and Khan, Novall Y. and Schacter, Daniel L.},
 year = {2013},
 title = {Interpolated memory tests reduce mind wandering and improve learning of online lectures},
 url = {https://www.pnas.org/content/110/16/6313.short},
 pages = {6313--6317},
 volume = {110},
 number = {16},
 issn = {1091-6490},
 journal = {Proceedings of the National Academy of Sciences},
 doi = {10.1073/pnas.1221764110}
}

@article{baird2011,
  title={Back to the future: Autobiographical planning and the functionality of mind-wandering},
  author={Baird, Benjamin and Smallwood, Jonathan and Schooler, Jonathan W},
  journal={Consciousness and cognition},
  volume={20},
  number={4},
  pages={1604--1611},
  year={2011},
  publisher={Elsevier}
}

@inproceedings{Pham.2015,
 author = {Pham, Phuong and Wang, Jingtao},
 title = {AttentiveLearner: Improving Mobile MOOC Learning via Implicit Heart Rate Tracking},
 pages = {367--376},
 publisher = {Springer},
 isbn = {978-3-319-19773-9},
 series = {Lecture notes in computer science Lecture notes in artificial intelligence},
 editor = {Conati, Cristina and Heffernan, Neil and Mitrovic, Antonija and Verdejo, M. Felisa},
 booktitle = {Artificial intelligence in education},
 year = {2015},
 address = {Cham}
}

@article{Pan.2020,
 author = {Pan, Steven C. and Sana, Faria and Schmitt, Alexandra G. and Bjork, Elizabeth Ligon},
 year = {2020},
 title = {Pretesting Reduces Mind Wandering and Enhances Learning During Online Lectures},
 url = {https://www.sciencedirect.com/science/article/pii/S2211368120300577},
 pages = {542--554},
 volume = {9},
 number = {4},
 issn = {2211-3681},
 journal = {Journal of Applied Research in Memory and Cognition},
 doi = {10.1016/j.jarmac.2020.07.004}
}

@incollection{shap2017,
title = {A Unified Approach to Interpreting Model Predictions},
author = {Lundberg, Scott M and Lee, Su-In},
booktitle = {Advances in Neural Information Processing Systems 30},
editor = {I. Guyon and U. V. Luxburg and S. Bengio and H. Wallach and R. Fergus and S. Vishwanathan and R. Garnett},
pages = {4765--4774},
year = {2017},
publisher = {Curran Associates, Inc.},
url = {http://papers.nips.cc/paper/7062-a-unified-approach-to-interpreting-model-predictions.pdf}
}

@inproceedings{bixler2021crossed,
  title={Crossed eyes: Domain adaptation for gaze-based mind wandering models},
  author={Bixler, Robert E. and D'Mello, Sidney K.},
  booktitle={ACM Symposium on Eye Tracking Research and Applications},
  pages={1--12},
  year={2021}
}

@article{holmqvist2017,
  title={Eye tracking: A comprehensive guide to methods},
  author={Holmqvist, Kenneth and Andersson, R},
  journal={Paradigms and measures},
  year={2017},
  publisher={Lund Eye-Tracking Research Institute Lund, Sweden}
}

@inproceedings{dmello2013,
  title={Automatic gaze-based detection of mind wandering during reading},
  author={D'Mello, Sidney and Cobian, Jonathan and Hunter, Matthew},
  booktitle={Educational Data Mining 2013},
  year={2013}
}

@article{kahneman1966,
  title={Pupil diameter and load on memory},
  author={Kahneman, Daniel and Beatty, Jackson},
  journal={Science},
  volume={154},
  number={3756},
  pages={1583--1585},
  year={1966},
  publisher={American Association for the Advancement of Science}
}

@article{whitehill2014faces,
  title={The faces of engagement: Automatic recognition of student engagementfrom facial expressions},
  author={Whitehill, Jacob and Serpell, Zewelanji and Lin, Yi-Ching and Foster, Aysha and Movellan, Javier R},
  journal={IEEE Transactions on Affective Computing},
  volume={5},
  number={1},
  pages={86--98},
  year={2014},
  publisher={IEEE}
}

@inproceedings{kohavi1995,
  title={A study of cross-validation and bootstrap for accuracy estimation and model selection},
  author={Kohavi, Ron and others},
  booktitle={Ijcai},
  volume={14},
  number={2},
  pages={1137--1145},
  year={1995},
  organization={Montreal, Canada}
}

@article{darvishi2023,
  title={Impact of AI assistance on student agency},
  author={Darvishi, Ali and Khosravi, Hassan and Sadiq, Shazia and Ga{\v{s}}evi{\'c}, Dragan and Siemens, George},
  journal={Computers \& Education},
  pages={104967},
  year={2023},
  publisher={Elsevier}
}

@article{carter2020best,
  title={Best practices in eye tracking research},
  author={Carter, Benjamin T and Luke, Steven G},
  journal={International Journal of Psychophysiology},
  volume={155},
  pages={49--62},
  year={2020},
  publisher={Elsevier}
}

@Manual{r2021,
    title = {R: A Language and Environment for Statistical Computing},
    author = {{R Core Team}},
    organization = {R Foundation for Statistical Computing},
    address = {Vienna, Austria},
    year = {2021},
    url = {https://www.R-project.org/},
           }

@article{gabadinho2011analyzing,
  title={Analyzing and visualizing state sequences in R with TraMineR},
  author={Gabadinho, Alexis and Ritschard, Gilbert and M{\"u}ller, Nicolas S and Studer, Matthias},
  journal={Journal of statistical software},
  volume={40},
  pages={1--37},
  year={2011}
}

@book{van1995python,
  title={Python reference manual},
  author={Van Rossum, Guido and Drake Jr, Fred L},
  year={1995},
  publisher={Centrum voor Wiskunde en Informatica Amsterdam}
}

@article{scikit-learn,
 title={Scikit-learn: Machine Learning in {P}ython},
 author={Pedregosa, F. and Varoquaux, G. and Gramfort, A. and Michel, V.
         and Thirion, B. and Grisel, O. and Blondel, M. and Prettenhofer, P.
         and Weiss, R. and Dubourg, V. and Vanderplas, J. and Passos, A. and
         Cournapeau, D. and Brucher, M. and Perrot, M. and Duchesnay, E.},
 journal={Journal of Machine Learning Research},
 volume={12},
 pages={2825--2830},
 year={2011}
}

@article{kazak2018apa,
  title={Journal article reporting standards.},
  author={Kazak, Anne E},
  year={2018},
  publisher={American Psychological Association}
}

@article{dasilva2020mouse,
  title={Wandering minds, wandering mice: Computer mouse tracking as a method to detect mind wandering},
  author={da Silva, MR Dias and Postma, Marie},
  journal={Computers in Human Behavior},
  volume={112},
  pages={106453},
  year={2020},
  publisher={Elsevier}
}

@article{Faber.2018read,
 author = {Faber, Myrthe and Bixler, Robert and D'Mello, Sidney K.},
 year = {2018},
 title = {An automated behavioral measure of mind wandering during computerized reading},
 pages = {134--150},
 volume = {50},
 number = {1},
 issn = {1554-3528},
 journal = {Behavior Research Methods},
 doi = {10.3758/s13428-017-0857-y}
}

@inproceedings{Blanchard.2014,
 author = {Blanchard, Nathaniel and Bixler, Robert and Joyce, Tera and D'Mello, Sidney},
 title = {Automated Physiological-Based Detection of Mind Wandering during Learning},
 pages = {55--60},
 publisher = {Springer},
 isbn = {978-3-319-07221-0},
 series = {Lecture Notes in Computer Science},
 editor = {Trău{\c{s}}an-Matu, {\c{S}}tefan and Boyer, Kristy Elizabeth and Crosby, Martha and Panourgia, Kitty},
 booktitle = {Intelligent tutoring systems},
 year = {2014},
 address = {Cham}
}

@article{Kane.2017,
 author = {Kane, M. J. and Smeekens, B. A. and von Bastian, Claudia C. and Lurquin, J. H. and Carruth, N. P. and Miyake, A.},
 year = {2017},
 title = {A Combined Experimental and Individual-Differences Investigation into Mind Wandering During a Video Lecture},
 url = {http://eprints.bournemouth.ac.uk/29494/},
 pages = {1649--1674},
 volume = {146},
 number = {11},
 issn = {0096-3445},
 journal = {Journal of Experimental Psychology: General}
}

@article{sumer2021multimodal,
  title={Multimodal engagement analysis from facial videos in the classroom},
  author={S{\"u}mer, {\"O}mer and Goldberg, Patricia and D'Mello, Sidney and Gerjets, Peter and Trautwein, Ulrich and Kasneci, Enkelejda},
  journal={IEEE Trans. on Affective Computing},
  year={2021}
}

@article{marsh2023synergies,
  title={Disentangling the long-term compositional effects of school-average achievement and SES: A substantive-methodological synergy},
  author={Marsh, Herbert W and Pekrun, Reinhard and Dicke, Theresa and Guo, Jiesi and Parker, Philip D and Basarkod, Geetanjali},
  journal={Educational Psychology Review},
  volume={35},
  number={3},
  pages={70},
  year={2023},
  publisher={Springer}
}

@incollection{majaranta2014eye,
  title={Eye tracking and eye-based human--computer interaction},
  author={Majaranta, P{\"a}ivi and Bulling, Andreas},
  booktitle={Advances in physiological computing},
  pages={39--65},
  year={2014},
  publisher={Springer}
}

@article{georgiou2018hr,
  title={Can wearable devices accurately measure heart rate variability? A systematic review},
  author={Georgiou, Konstantinos and Larentzakis, Andreas V and Khamis, Nehal N and Alsuhaibani, Ghadah I and Alaska, Yasser A and Giallafos, Elias J},
  journal={Folia medica},
  volume={60},
  number={1},
  pages={7--20},
  year={2018},
  publisher={MEDICAL UNIVERSITY-PLOVDIV}
}

@article{andreassi1980human,
  title={Human behavior and physiological response},
  author={Andreassi, JL},
  journal={Psychophysiology},
  year={1980}
}

@article{greco2015cvxeda,
  title={cvxEDA: A convex optimization approach to electrodermal activity processing},
  author={Greco, Alberto and Valenza, Gaetano and Lanata, Antonio and Scilingo, Enzo Pasquale and Citi, Luca},
  journal={IEEE transactions on biomedical engineering},
  volume={63},
  number={4},
  pages={797--804},
  year={2015},
  publisher={IEEE}
}

@article{duchowski2002,
  title={A breadth-first survey of eye-tracking applications},
  author={Duchowski, Andrew T},
  journal={Behavior Research Methods, Instruments, \& Computers},
  volume={34},
  number={4},
  pages={455--470},
  year={2002},
  publisher={Springer}
}

@inproceedings{xgb,
 author = {Chen, Tianqi and Guestrin, Carlos},
 title = {{XGBoost}: A Scalable Tree Boosting System},
 booktitle = {Proceedings of the 22nd ACM SIGKDD International Conference on Knowledge Discovery and Data Mining},
 series = {KDD '16},
 year = {2016},
 isbn = {978-1-4503-4232-2 },
 location = {San Francisco, California, USA},
 pages = {785--794},
 numpages = {10},
 doi = {10.1145/2939672.2939785 },
 acmid = {2939785},
 publisher = {ACM},
 address = {New York, NY, USA},
}

@inproceedings{kubler2014subsmatch,
  title={Subsmatch: Scanpath similarity in dynamic scenes based on subsequence frequencies},
  author={K{\"u}bler, Thomas C and Kasneci, Enkelejda and Rosenstiel, Wolfgang},
  booktitle={Proceedings of the Symposium on Eye Tracking Research and Applications},
  pages={319--322},
  year={2014}
}

@inproceedings{geisler2020minhash,
  title={A MinHash approach for fast scanpath classification},
  author={Geisler, David and Castner, Nora and Kasneci, Gjergji and Kasneci, Enkelejda},
  booktitle={ACM Symposium on Eye Tracking Research and Applications},
  pages={1--9},
  year={2020}
}

@article{voulodimos2018deep,
  title={Deep learning for computer vision: A brief review},
  author={Voulodimos, Athanasios and Doulamis, Nikolaos and Doulamis, Anastasios and Protopapadakis, Eftychios},
  journal={Computational intelligence and neuroscience},
  volume={2018},
  year={2018},
  publisher={Hindawi Limited}
}

@book{szeliski2022cv,
  title={Computer vision: algorithms and applications},
  author={Szeliski, Richard},
  year={2022},
  publisher={Springer Nature}
}

@article{helmke1992,
  title={Das M{\"u}nchener Aufmerksamkeitsinventar (MAI): Ein Instrument zur systematischen Verhaltensbeobachtung der Sch{\"u}leraufmerksamkeit im Unterricht},
  author={Helmke, Andreas and Renkl, Alexander},
  journal={Diagnostica},
  volume={38},
  number={2},
  pages={130--141},
  year={1992}
}

@inproceedings{dutta2019via,
  title={The VIA annotation software for images, audio and video},
  author={Dutta, Abhishek and Zisserman, Andrew},
  booktitle={ACM International Conference on Multimedia},
  pages={2276--2279},
  year={2019}
}

@article{cao2019openpose,
  author = {Z. {Cao} and G. {Hidalgo Martinez} and T. {Simon} and S. {Wei} and Y. A. {Sheikh}},
  journal = {IEEE Trans. on Pattern Analysis and Machine Intelligence},
  title = {OpenPose: Realtime Multi-Person 2D Pose Estimation using Part Affinity Fields},
  year = {2019}
}

@article{liu2022prvipe,
  title={View-invariant, occlusion-robust probabilistic embedding for human pose},
  author={Liu, Ting and Sun, Jennifer J and Zhao, Long and Zhao, Jiaping and Yuan, Liangzhe and Wang, Yuxiao and Chen, Liang-Chieh and Schroff, Florian and Adam, Hartwig},
  journal={International Journal of Computer Vision},
  volume={130},
  number={1},
  pages={111--135},
  year={2022},
  publisher={Springer}
}

@article{goldberg2021,
  title={Attentive or not? Toward a machine learning approach to assessing students’ visible engagement in classroom instruction},
  author={Goldberg, Patricia and S{\"u}mer, {\"O}mer and St{\"u}rmer, Kathleen and Wagner, Wolfgang and G{\"o}llner, Richard and Gerjets, Peter and Kasneci, Enkelejda and Trautwein, Ulrich},
  journal={Educational Psychology Review},
  volume={33},
  pages={27--49},
  year={2021},
  publisher={Springer}
}

@article{lin2021student,
  title={Student behavior recognition system for the classroom environment based on skeleton pose estimation and person detection},
  author={Lin, Feng-Cheng and Ngo, Huu-Huy and Dow, Chyi-Ren and Lam, Ka-Hou and Le, Hung Linh},
  journal={Sensors},
  volume={21},
  number={16},
  pages={5314},
  year={2021},
  publisher={MDPI}
}

@inproceedings{zhang2017geometric,
  title={On geometric features for skeleton-based action recognition using multilayer lstm networks},
  author={Zhang, Songyang and Liu, Xiaoming and Xiao, Jun},
  booktitle={IEEE Winter Conference on Applications of Computer Vision},
  pages={148--157},
  year={2017}
}

@article{h36m_pami,
author = {Ionescu, Catalin and Papava, Dragos and Olaru, Vlad and Sminchisescu,  Cristian},
title = {{Human3.6M}: Large Scale Datasets and Predictive Methods for 3D Human Sensing in Natural Environments},
journal = {IEEE Trans. on Pattern Analysis and Machine Intelligence},
publisher = {IEEE Computer Society},
volume = {36},
number = {7},
pages = {1325-1339},
year = {2014}
}

@inproceedings{JieYao.2002,
 author = {{Jie Yao} and Cooperstock, J. R.},
 title = {Arm gesture detection in a classroom environment},
 isbn = {0769518583   },
 booktitle = {Sixth IEEE Workshop on Applications of Computer Vision},
 year = {2002}
}

@inproceedings{Liao.2019,
 author = {Liao, Wang and Xu, Wei and Kong, SiCong and Ahmad, Fowad and Liu, Wei},
 title = {A Two-stage Method For Hand-Raising Gesture Recognition in Classroom},
 publisher = {ACM},
 booktitle = {International Conference on Educational and Information Technology},
 year = {2019}
}

@article{Si.2019,
 author = {Si, Jiaxin and Lin, Jiaojiao and Jiang, Fei and Shen, Ruimin},
 year = {2019},
 title = {Hand-raising gesture detection in real classrooms using improved R-FCN},
 pages = {69--76},
 volume = {359},
 issn = {0925-2312},
 journal = {Neurocomputing}
}

@inproceedings{TaoLiu.2020,
 author = {{Tao Liu} and {Fei Jiang} and {Ruimin Shen}},
 title = {Fast and Accurate Hand-Raising Gesture Detection in Classroom},
 pages = {232--239},
 booktitle = {International Conference on Neural Information Processing},
 publisher = {{Springer, Cham}},
 year = {2020}
}

@inproceedings{nguyen2022new,
  title={A new dataset and systematic evaluation of deep learning models for student activity recognition from classroom videos},
  author={Nguyen, Phuong-Dung and Nguyen, Hong-Quan and Nguyen, Thuy-Binh and Le, Thi-Lan and Tran, Thanh-Hai and Vu, Hai and Huu, Quynh Nguyen},
  booktitle={International Conference on Multimedia Analysis and Pattern Recognition},
  year={2022},
  organization={IEEE}
}

@inproceedings{YuTe.2019,
 author = {Yu-Te, Ku and Han-Yen, Yu and Yi-Chi, Chou},
 title = {A Classroom Atmosphere Management System for Analyzing Human Behaviors in Class Activities},
 publisher = {IEEE},
 booktitle = {International Conference on Artificial Intelligence in Information and Communication},
 year = {2019}
}

@article{HuayiZhou.2018,
 author = {{Huayi Zhou} and {Fei Jiang} and {Ruimin Shen}},
 year = {2018},
 title = {Who Are Raising Their Hands? Hand-Raiser Seeking Based on Object Detection and Pose Estimation},
 pages = {470--485},
 issn = {2640-3498},
 journal = {Asian Conference on Machine Learning}
}

@article{Ahuja.2019,
 author = {Ahuja, Karan and Kim, Dohyun and Xhakaj, Franceska and Varga, Virag and Xie, Anne and Zhang, Stanley and Townsend, Jay Eric and Harrison, Chris and Ogan, Amy and Agarwal, Yuvraj},
 year = {2019},
 title = {EduSense: Practical Classroom Sensing at Scale},
 pages = {1--26},
 volume = {3},
 number = {3},
 journal = {Proceedings of the ACM on Interactive, Mobile, Wearable and Ubiquitous Technologies}
}

@inproceedings{Bo.2011,
 author = {Bo, Nyan Bo and {van Hese}, Peter and {van Cauwelaert}, Dimitri and Veelaert, Peter and Philips, Wilfried},
 title = {Detection of a hand-raising gesture by locating the arm},
 isbn = {9781457721380                  },
 booktitle = {IEEE International Conference on Robotics and Biomimetics},
 year = {2011}
}

@article{Boheim.2020,
 author = {B{\"o}heim, Ricardo and Knogler, Maximilian and Kosel, Christian and Seidel, Tina},
 year = {2020},
 title = {Exploring student hand-raising across two school subjects using mixed methods: An investigation of an everyday classroom behavior from a motivational perspective},
 pages = {101250},
 volume = {65},
 issn = {0959-4752},
 journal = {Learning and Instruction}
}

@article{finn1995disruptive,
  title={Disruptive and inattentive-withdrawn behavior and achievement among fourth graders},
  author={Finn, Jeremy D and Pannozzo, Gina M and Voelkl, Kristin E},
  journal={The Elementary School Journal},
  volume={95},
  number={5},
  pages={421--434},
  year={1995},
  publisher={University of Chicago Press}
}

@article{boheim2020engagement,
  title={Student hand-raising as an indicator of behavioral engagement and its role in classroom learning},
  author={B{\"o}heim, Ricardo and Urdan, Tim and Knogler, Maximilian and Seidel, Tina},
  journal={Contemporary Educational Psychology},
  volume={62},
  pages={101894},
  year={2020},
  publisher={Elsevier}
}

@article{knogler2015situational,
  title={How situational is situational interest? Investigating the longitudinal structure of situational interest},
  author={Knogler, Maximilian and Harackiewicz, Judith M and Gegenfurtner, Andreas and Lewalter, Doris},
  journal={Contemporary Educational Psychology},
  volume={43},
  pages={39--50},
  year={2015},
  publisher={Elsevier}
}

@article{rimm2015extent,
  title={To what extent do teacher--student interaction quality and student gender contribute to fifth graders’ engagement in mathematics learning?},
  author={Rimm-Kaufman, Sara E and Baroody, Alison E and Larsen, Ross AA and Curby, Timothy W and Abry, Tashia},
  journal={Journal of Educational Psychology},
  volume={107},
  number={1},
  pages={170},
  year={2015},
  publisher={American Psychological Association}
}

@book{frank2014presence,
  title={Presence messen in laborbasierter Forschung mit Mikrowelten: Entwicklung und erste Validierung eines Fragebogens zur Messung von Presence},
  author={Frank, Barbara},
  year={2014},
  publisher={Springer-Verlag}
}

@article{sedova2019,
  title={Do those who talk more learn more? The relationship between student classroom talk and student achievement},
  author={Sedova, Klara and Sedlacek, Martin and Svaricek, Roman and Majcik, Martin and Navratilova, Jana and Drexlerova, Anna and Kychler, Jakub and Salamounova, Zuzana},
  journal={Learning and instruction},
  volume={63},
  pages={101217},
  year={2019},
  publisher={Elsevier}
}

@article{Dorr.2010,
 author = {Dorr, Michael and Martinetz, Thomas and Gegenfurtner, Karl R. and Barth, Erhardt},
 year = {2010},
 title = {Variability of eye movements when viewing dynamic natural scenes},
 url = {https://jov.arvojournals.org/article.aspx?articleid=2121333},
 pages = {28},
 volume = {10},
 number = {10},
 issn = {1534-7362},
 journal = {Journal of Vision},
 doi = {10.1167/10.10.28}
}

@article{Kok.2023,
 author = {Kok, Ellen M. and Jarodzka, Halszka and Sibbald, Matt and {van Gog}, Tamara},
 year = {2023},
 title = {Did You Get That? Predicting Learners' Comprehension of a Video Lecture from Visualizations of Their Gaze Data},
 pages = {e13247},
 volume = {47},
 number = {2},
 issn = {1551-6709},
 journal = {Cognitive Science},
 doi = {10.1111/cogs.13247}
}

@article{Liu.2023,
 author = {Liu, Qing and Yang, Xueyao and Chen, Zekai and Zhang, Wenjuan},
 year = {2023},
 title = {Using synchronized eye movements to assess attentional engagement},
 url = {https://link.springer.com/article/10.1007/s00426-023-01791-2}, 
 pages = {2039--2047},
 volume = {87},
 number = {7},
 issn = {1430-2772},
 journal = {Psychological Research},
 doi = {10.1007/s00426-023-01791-2}
}

@article{Madsen.2021,
 author = {Madsen, Jens and J{\'u}lio, Sara U. and Gucik, Pawel J. and Steinberg, Richard and Parra, Lucas C.},
 year = {2021},
 title = {Synchronized eye movements predict test scores in online video education},
 pages = {e2016980118},
 volume = {118},
 number = {5},
 journal = {Proceedings of the National Academy of Sciences of the United States of America},
 doi = {10.1073/pnas.2016980118}
}
